\documentclass[11pt,onecolumn,twoside,notitlepage,fleqn,letterpaper,final]
	      {article}[]

\usepackage{amsmath}
\usepackage{amssymb}
\usepackage{latexsym}
\usepackage{amsfonts}
\usepackage[mathscr]{eucal}
\usepackage[letterpaper,pdfborder={0 0 0}]{hyperref}
\usepackage[all]{hypcap}
\usepackage{array}
\usepackage{calc}
\usepackage{longtable}
\usepackage{multirow}
\usepackage{bm}
\usepackage{slashed}
\usepackage[dvips]{epsfig}
\usepackage{axodraw}
\usepackage{pstricks}
\usepackage{graphicx,psfrag}
\usepackage{xspace}
\usepackage{mciteplus}
\usepackage{setspace}
\usepackage[T1]{fontenc}
\usepackage{textcomp}
\usepackage{helvet}
\usepackage[font=small,labelfont=bf]{caption}
\usepackage{sectsty}
\allsectionsfont{\sffamily}
\subsubsectionfont{\bfseries\slshape\large}
\DeclareMathAlphabet{\mathitbf}{OML}{cmm}{b}{it}
\usepackage{fancyhdr}
\let\spreprint\empty
\newcommand{\preprint}[1]{\def\spreprint{\protect#1}}
\let\sinstitute\empty
\newcommand{\institute}[1]{\def\sinstitute{\protect#1}}
\makeatletter
\newcommand{\mymaketitle}{\begingroup
  \null\thispagestyle{empty}%
    \ifx\spreprint\empty
      \vskip5ex
    \else
      \vspace*{-10ex}\flushright\small\spreprint\vskip2ex
    \fi
    \vskip7ex
    \begin{center}
      \doublespacing{\sffamily\bfseries\LARGE\@title}\vskip5ex
      {\bf\large\@author\vskip 2ex}
      \ifx\sinstitute\empty
      \else
        {\sl\small\sinstitute}
      \fi
    \end{center}
    \vskip7ex
  \endgroup
}
\makeatother
\renewenvironment{abstract}{\begin{center}
  {\large\sffamily\bfseries Abstract}\\[4mm]
  \begin{minipage}[t]{0.87\textwidth}\small
}{\end{minipage}\end{center}\vskip10ex}

\newcommand{\bit}{\begin{itemize}}
\newcommand{\eit}{\end{itemize}}
\newcommand{\bc}{\begin{center}}
\newcommand{\ec}{\end{center}}
\newcommand{\bq}{\begin{quote}}
\newcommand{\eq}{\end{quote}}
\newcommand{\bi}{\begin{itemize}}
\newcommand{\ei}{\end{itemize}}
\newcommand{\be}{\begin{equation}}
\newcommand{\ee}{\end{equation}}
\newcommand{\bea}{\begin{eqnarray}}
\newcommand{\eea}{\end{eqnarray}}
\newcommand{\bt}{\begin{tabular}}
\newcommand{\et}{\end{tabular}}
\newcommand{\bmp}{\begin{minipage}}
\newcommand{\emp}{\end{minipage}}
\newcommand{\btab}{\begin{tabbing}}
\newcommand{\etab}{\end{tabbing}}

\newcommand{\sherpa}{S\protect\scalebox{0.8}{HERPA}\xspace}
\newcommand{\apacic}{A\protect\scalebox{0.8}{PACIC++}\xspace}

\newcommand{\comix}{C\protect\scalebox{0.8}{OMIX}\xspace}
\newcommand{\amegic}{A\protect\scalebox{0.8}{MEGIC++}\xspace}

\newcommand{\veps}{\varepsilon}

\newcommand{\td}{\tilde}
\def\trm{\textrm}
\def\nn{\nonumber}

\def\mrm{\mathrm}
\def\cal{\mathcal}
\def\cdt{\mbox{$\cdot\;\!\!$}}
\def\eg{\trm{e.g.\ }}
\def\ie{\trm{i.e.\ }}

\def\eq#1{eq.~(\ref{#1})}

\def\Eq#1{Eq.~(\ref{#1})}
\def\Eqs#1{Eqs.~(\ref{#1})}

\def\Fig#1{Figure~\ref{#1}}
\def\Figs#1{Figures~\ref{#1}}

\def\Tab#1{Table~\ref{#1}}
\def\Tabs#1{Tables~\ref{#1}}

\def\Sec#1{Section~\ref{#1}}
\def\App#1{Appendix~\ref{#1}}

\def\muf{\mbox{$\mu_{\mrm F}$\xspace}}
\def\mur{\mbox{$\mu_{\mrm R}$\xspace}}

\newcommand{\abs}[1]{\lvert#1\rvert}

\newcommand{\co}{{\rm cos}\,\theta_j}
\newcommand{\ct}{{\rm cos}\,\theta_\ell}
\newcommand{\so}{{\rm sin}\,\theta_j}
\newcommand{\st}{{\rm sin}\,\theta_\ell}
\renewcommand{\c}{{\rm cos}\,\varphi_\ell}
\newcommand{\s}{{\rm sin}\,\varphi_\ell}

\newcommand{\sosq}{{\rm sin}^2\theta_j}

\preprint{FERMILAB-PUB-11-499-T}
\author{\hphantom{MisterX}\\[0mm]\sffamily\bfseries\normalsize%
  Joseph~D.~Lykken\thanks{lykken@fnal.gov}\,,\ \
  Adam~O.~Martin\thanks{aomartin@fnal.gov}\\[4mm]
  {\small\sl Theoretical Physics Department}\\[-2mm]
  {\small\sl Fermi National Accelerator Laboratory, Batavia, IL 60510, USA}
  \\[10mm]\sffamily\bfseries\normalsize%
  Jan Winter\thanks{jwinter@cern.ch}\\[4mm]
  {\small\sl PH-TH, CERN, CH-1211 Geneva 23, Switzerland}\\[12mm]
}
\title{\sffamily\bfseries\Large%
  Semileptonic decays of the Higgs boson at the Tevatron}
\institute{Theoretical Physics Department, Fermi National Accelerator
  Laboratory, Batavia, IL 60510, USA}

\begin{document}
\renewcommand{\baselinestretch}{1.4}
\maketitle
\thispagestyle{empty}
\begin{flushright}
  \vspace*{-137mm}
  {\small FERMILAB-PUB-11-499-T\\CERN-PH-TH-2011-237}\\[140mm]
\end{flushright}

\renewcommand{\baselinestretch}{1.1}
\begin{abstract}
\noindent
We examine the prospects for extending the Tevatron reach for a
Standard Model Higgs boson by including the semileptonic Higgs boson
decays $h\to WW\to\ell\nu_\ell\,jj$\/ for $M_h\gtrsim2\,M_W$, and
$h\to~Wjj\to\ell\nu_\ell\,jj$\/ for $M_h\lesssim2\,M_W$, where $j$\/
is a hadronic jet. We employ a realistic simulation of the signal and
backgrounds using the \protect\sherpa Monte Carlo event generator. We
find kinematic selections that enhance the signal over the dominant
$W$\!+jets background. The resulting sensitivity could be an important
addition to ongoing searches, especially in the mass range
$120\lesssim M_h\lesssim150\ \mrm{GeV}$. The techniques described
can be extended to Higgs boson searches at the Large Hadron Collider.
\end{abstract}

\clearpage
\renewcommand{\baselinestretch}{0.5}
\tableofcontents
\thispagestyle{empty}
\renewcommand{\baselinestretch}{1.04}
\vspace*{3mm}
\noindent\hrulefill
\vspace*{4mm}


\section{Introduction}

The Standard Model (SM) predicts a neutral Higgs boson particle whose couplings
to other particles are proportional to the particle masses, and that couples to
photons and gluons via one-loop-generated effective couplings. While the Higgs
boson mass is not predicted, the relation between the Higgs boson mass and
its width is fixed from the predicted couplings. Virtual Higgs boson
contributions to electroweak precision observables have been computed,
and precision data favor $M_h<169\ \mrm{GeV}$ at 95\% confidence level
\cite{Baak:2011ze}. Searches at the CERN Large Hadron Collider have
produced 95\% confidence level exclusion of a SM Higgs boson for a
broad mass range above 145 GeV \cite{ATLAS-CONF-2011-112,CMS-PAS-HIG-11-022}.

Because of small signal cross sections and large backgrounds, the
search for the Higgs boson in experiments at the Fermilab Tevatron is
very challenging, even with the large final datasets approaching
$10\ \mrm{fb}^{-1}$ per experiment.
Nevertheless both the CDF and D\O\ experiments have achieved steady
improvements in their sensitivities in multiple channels to a SM Higgs
boson, and their individual results already exclude a SM Higgs boson
in the mass range $156.7$--$173.8\ \mrm{GeV}$ and
$162$--$170\ \mrm{GeV}$, respectively, at 95\% confidence level
\cite{:2011cb,EPS:CDFhiggs,EPS:D0higgs}.
This exclusion relies mainly on sensitivity to the dilepton
final-state decay chain analyzed in
Refs.~\cite{Han:1998ma,Han:1998sp,Carena:2000yx}\,:
$h\to W^+W^-\!\to\ell^+\nu_\ell\;\ell^-\bar\nu_\ell$ where
$\ell^\pm=e^\pm$\/ or $\mu^\pm$.

Here we examine the prospects for extending the Tevatron reach by
including a search for the semileptonic Higgs boson decays
$h\to WW\to\ell\nu_\ell\;jj$\/ for $M_h\gtrsim2\,M_W$, and
$h\to Wjj\to\ell\nu_\ell\;jj$\/ for $M_h\lesssim2\,M_W$,
where $j$\/ is a hadronic jet. This process was first considered as a
potential Higgs boson discovery channel for the SSC
\cite{Stirling:1985bi,Gunion:1985vt,Gunion:1985dj,Gunion:1986cc},
emphasizing the case of a very heavy Higgs boson, where the
$h\to ZZ\to4\,\ell$\/ ``golden mode'' becomes limited by its small
branching fraction and the broad Higgs boson width.
Similar to the golden mode, the semileptonic  $h\to WW$\/ modes are
(almost) fully reconstructible: assuming that the leptonic $W$\/ is
close to on-shell, the mass constraint gives an estimate of the
unmeasured longitudinal momentum of the neutrino, up to a two-fold
ambiguity \cite{Gunion:1986cc}. For $M_h\gtrsim140\ \mrm{GeV}$ the
overall decay rate is 6 times larger than any other SM Higgs boson
decay mode with a triggerable lepton. Including these semileptonic
channels thus offers the distinct possibility of significantly
extending the Tevatron reach over a rather broad mass range.

This channel suffers from large backgrounds from SM processes with a
leptonically decaying $W$\/ boson. These include diboson production,
top quark production, and direct inclusive $W$\!+2-jet production.
There is also a purely QCD background that is difficult to estimate
absent a dedicated analysis with data. The dominant background is
inclusive $W$\!+2-jets; from this background alone we have estimated a
signal to background ratio ($S/B$) of $3\times 10^{-4}$, after nominal
preselections. Though worrisome, this is not smaller than the
analogous $S/B\simeq4\times 10^{-5}$ for the $e^+e^-$ and $\mu^+\mu^-$
modes after preselection in the successful Tevatron analyses of
$h\to W^+W^-\!\to\ell^+\nu_\ell\;\ell^-\bar\nu_\ell$
\cite{Benjamin:2011sv,CDFnote10599:2011,D0Note6219:2011,Aaltonen:2010yv,Aaltonen:2010cm,Abazov:2010ct}.

A drastic reduction in both the $W$\!+2-jet and diboson backgrounds to
semileptonic Higgs boson decay can be achieved by forward jet tagging,
\ie by restricting to Higgs boson production from vector boson fusion
(VBF)~\cite{Stirling:1985bi,Iordanidis:1997vs}; it is estimated that
the additional requirement of forward jet tagging then gives a factor
of $\sim$~100 reduction in these backgrounds. However the reduction in
the Higgs signal, versus inclusive Higgs boson production, is also
severe: a factor of $\sim$~10 at the Tevatron \cite{Djouadi:2005gi}.
Looking at the similar trade-off for the dilepton
$h\to W^+W^-\!\to\ell^+\nu_\ell\;\ell^-\bar\nu_\ell$ channel, a
Tevatron study \cite{Mellado:2007fb} concluded that the overall
sensitivity does not improve by restricting to VBF Higgs boson versus
inclusive Higgs boson production. We do not know of any comparable
analysis for the semileptonic channel.

For inclusive Higgs boson production at the Tevatron, the semileptonic
channels were first studied by Han and Zhang \cite{Han:1998ma,Han:1998sp}.
In a parton-level study with some jet smearing they found that, after
basic acceptance cuts together with a veto on extra energetic jets
designed to suppress the $t\bar t$\/ background, the remaining
background is completely dominated by $W$\!+2-jets. Han and Zhang then
made additional kinematic selections that enhance the signal to
background ratio $S/B$. For $M_h=140\ (160)\ \mrm{GeV}$ they thus
obtained a significance estimate of $S/\sqrt{B}=1.0\ (3.3)$ for
$30\ \mrm{fb}^{-1}$ of integrated Tevatron luminosity. The fully
differential Higgs boson decay width for this process was exhibited by
Dobrescu and Lykken \cite{Dobrescu:2009zf}, who analyzed the basic
kinematics and angular distributions that characterize the Higgs
signal.

We improve on these studies by including realistic parton showering
(since parton-level jet smearing is inadequate), an NLO-rate improved
treatment of the Higgs boson decays (including off-shell effects), and
a resummed NNLO estimate of the $gg\to h$\/ production cross section.
The first two improvements are incorporated by the use of \sherpa
\cite{Gleisberg:2003xi,Gleisberg:2008ta}, a general purpose showering
Monte Carlo program, for simulation of both the signal and the
inclusive $W$\!+2-jets background. The NNLO signal cross section is
modeled by a $K$-factor.

Our purpose is to study these semileptonic Higgs boson decay channels
in a systematic way, but not to mimic a fully-optimized experimental
analysis. The D\O\ experiment has already reported on a semileptonic
Higgs boson search using $5.4\ \mrm{fb}^{-1}$ of Tevatron data
\cite{D0note6095:2010,Abazov:2011bc}; this analysis uses multivariate
decision trees to enhance the significance of the result. Here we will
limit ourselves to simple cuts, in order to make the features of the
analysis and the underlying physics more explicit. We study the Higgs
signal in the mass range $110$--$220\ \mrm{GeV}$ to reasonably cover
the below, near and above threshold regions for Higgs boson decay to
two on-shell $W$\/ bosons.

In \Sec{sec:strategy} we outline the strategy and define several
useful observables.
In \Sec{sec:xsecs} we discuss inclusive Higgs boson production from
the dominant gluon--gluon fusion mechanism, and implement a $K$-factor
correction to the \sherpa result. In \Sec{sec:results} we introduce basic
preselections and develop cuts implemented in \sherpa to enhance
$S/\sqrt{B}$; this section also contains our main results. We conclude
in \Sec{sec:conclusions} with caveats about the limitations of our
analysis and suggestions for further improvements. Cross-checks and
additional material are presented in the appendices.


\section{Strategy and key observables}\label{sec:strategy}

We are interested in the Higgs boson decay
\be
h\to W^*W^*\to e\nu_e\;jj\ ,
\ee
and the similar decay with a muon in the final state. In general we
take both $W$\/ bosons off shell. We will write $e\nu_e\,jj$\/ as
$e\nu_e\,jj'$ where $j$\/ is the jet with higher transverse momentum
($p_T$), and noting that physical observables will be symmetric under
$j\leftrightarrow j'$.
We use a baseline selection adopted from the D\O\ analysis to define
reconstructed jets and leptons and impose realistic acceptance cuts.
We will assume that events with more than one reconstructed lepton
are vetoed, but we want to allow the possibility of extra jets in
order to increase signal efficiency. When more than two jets are
present there is a combinatorial problem; we will define a
\textbf{\em Higgs boson candidate selection} algorithm that assigns
which two jets to use in the Higgs boson reconstruction; these jets
may or may not correspond to the two leading jets in the event. It is
important to note that this algorithm is chosen so as to optimize the
signal sensitivity after the full selection, which is not equivalent
to maximizing the number of correctly reconstructed signal events.

When the leptonically decaying $W$\/ boson is (close to) on-shell,
these decays are fully reconstructible up to a two-fold ambiguity in
the neutrino momentum without making any assumption about the Higgs
boson mass. Here we are assuming that the transverse momentum of the
neutrino is well estimated given a measurement of the missing
transverse energy (MET), as has been demonstrated by both Tevatron
experiments in the determination of the $W$\/ boson mass.

The semileptonic channel's advantage of being, in principle,
completely reconstructible offers a great way to separate signal from
backgrounds. However, when the leptonically decaying $W$\/ boson is
far off shell, a straightforward full reconstruction is not possible.
There are then three generic possibilities for how to proceed:
\begin{quote}\begin{itemize}
\item Use only transverse observables.
\item Perform an approximate event-by-event reconstruction using an
estimate of the off-shell $W$\/ boson mass.
\item Perform an approximate event-by-event reconstruction using a
(hypothesized) Higgs boson mass constraint.
\end{itemize}\end{quote}
Since it is not clear {\em a priori}\/ which of these approaches
maximizes the Higgs boson sensitivity, we will pursue all three and
compare the results.

Given an event-by-event approximate combinatorial full reconstruction
of the putative decaying Higgs boson, one can approximately reproduce
the kinematics in the Higgs boson rest frame. The true Higgs boson
rest frame is given by a longitudinal boost from the lab frame
together with a transverse boost defined by the transverse momentum
$p_{T,h}$ of the Higgs boson. An explicit representation for the
four-momenta in the Higgs boson rest frame is given by:
\bea
p_{e} &=& \frac{1}{2}\;m_{e\nu_e}\biggl(\gamma_{e\nu_e}(1+\beta_{e\nu_e}\ct),\;\st\,\c ,\;\st\,\s ,\;\gamma_{e\nu_e}(\beta_{e\nu_e} +\ct )\biggr)
\; ,\nonumber\\[2mm]
p_{\nu_e} &=& \frac{1}{2}\;m_{e\nu_e}\biggl(\gamma_{e\nu_e}(1-\beta_{e\nu_e}\ct),\;-\st\,\c ,\;-\st\,\s ,\;-\gamma_{e\nu_e}(\beta_{e\nu_e} -\ct )\biggr)
\; ,\nonumber\\[2mm]
p_{j} &=& \frac{1}{2}\;m_{jj'}\biggl(\gamma_{jj'}(1+\beta_{jj'}\co ),\;\so,\;0,\;-\gamma_{jj'}(\beta_{jj'} +\co)\biggr)
\; , \nonumber\\[2mm]
p_{j'} &=& \frac{1}{2}\;m_{jj'}\biggl(\gamma_{jj'}(1-\beta_{jj'}\co ),\;-\so,\;0,\;-\gamma_{jj'}(\beta_{jj'} -\co)\biggr)\; ,
\eea
where we have chosen the dijet plane to coincide to the $x$--$z$ plane,
and have chosen the positive $z$-axis to be the direction of the
leptonically decaying $W$ boson. The boost factors of the two $W$
bosons relative to the Higgs boson rest frame are given by
\bea
\label{eq:boostfactors}
\gamma_{jj'} &=& \frac{M_h}{2\,m_{jj'}}\left( 1 + \frac{m_{jj'}^2-m_{e\nu_e}^2}{M_h^2} \right)
\; , \nonumber\\[2mm]
\gamma_{e\nu_e} &=& \frac{M_h}{2\,m_{e\nu_e}}\left( 1 - \frac{m_{jj'}^2-m_{e\nu_e}^2}{M_h^2} \right)\; ,
\eea
and we note the identities
\bea
M_h &=& m_{jj'}\,\gamma_{jj'} + m_{e\nu_e}\,\gamma_{e\nu_e}
\; , \nonumber\\[2mm]
m_{jj'}\,\beta_{jj'}\,\gamma_{jj'} &=& m_{e\nu_e}\,\beta_{e\nu_e}\,\gamma_{e\nu_e}
\; .
\eea
Note that $\theta_j$ is the angle between jet $j$\/ and the direction
of the hadronic $W$\/ boson, as seen in the $W$\/ rest frame, while
$\theta_e$ is the angle between the charged lepton and the direction
of the leptonic $W$\/ boson as seen in the $W$\/ rest frame. The
azimuthal angle $\varphi_e$ is the angle between the dilepton and
dijet planes. Defining
\bea
r_{jj'} &=& \beta_{jj'}^2\,\gamma_{jj'}^2\,\sosq\ ,
\eea
we can calculate the angle $\theta_{jj'}$ between the two jets as seen
in the Higgs boson rest frame:
\bea
{\rm cos}\,\theta_{jj'} &=& \frac{r_{jj'} -1}{r_{jj'}+1}\ .
\eea
Signal events have a minimum opening angle between the jets as seen in
the Higgs boson rest frame:
\bea
-1\;\leq\;{\rm cos}\,\theta_{jj'}\;\leq\;2\,\beta_{jj'}^2 - 1\ .
\eea

\begin{figure}[t!]
\centering\vskip5mm
\centerline{
  \includegraphics[clip,width=0.397\columnwidth,angle=-90]{%
    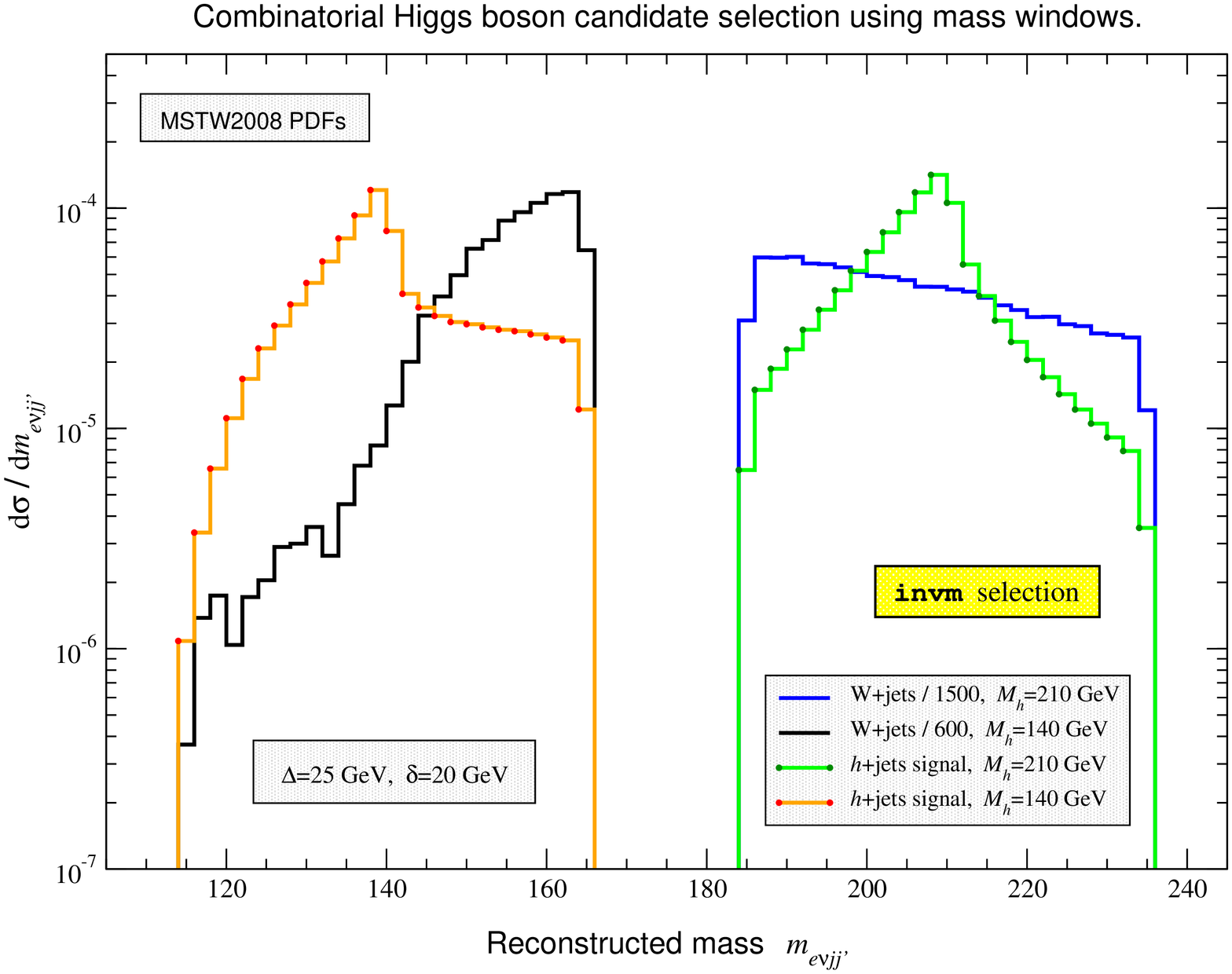}
  \includegraphics[clip,width=0.397\columnwidth,angle=-90]{%
    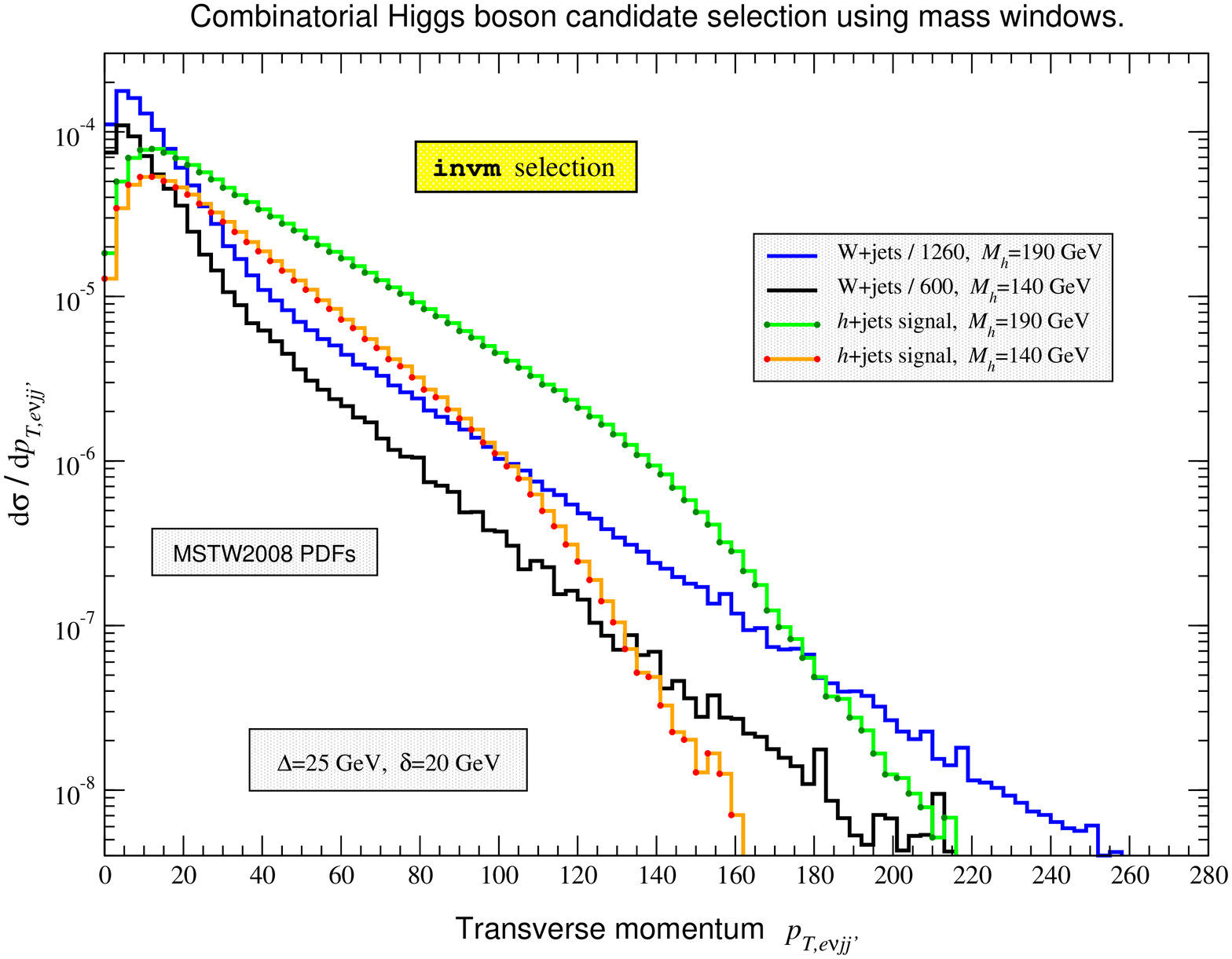}}
\centerline{
  \includegraphics[clip,width=0.397\columnwidth,angle=-90]{%
    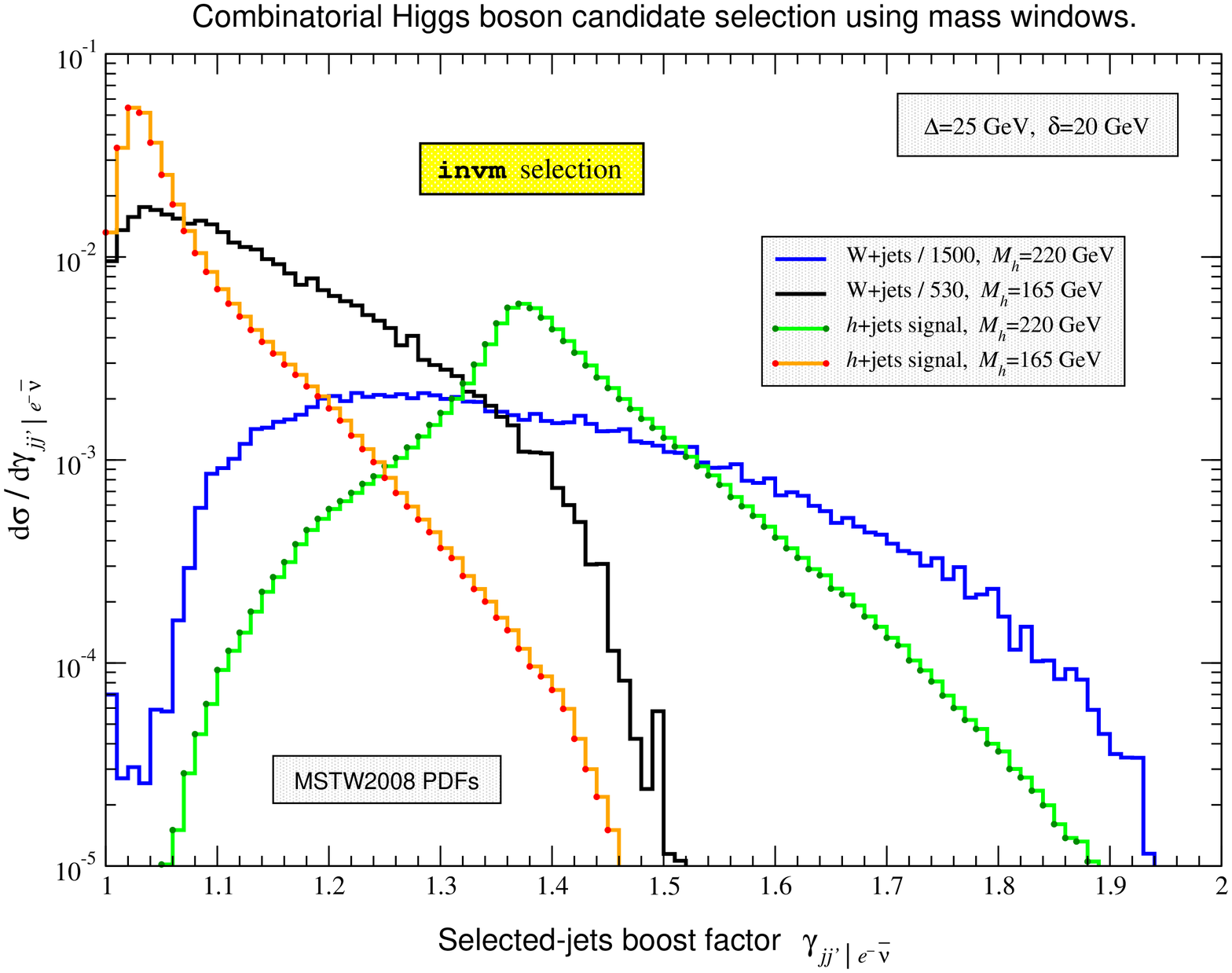}
  \includegraphics[clip,width=0.397\columnwidth,angle=-90]{%
    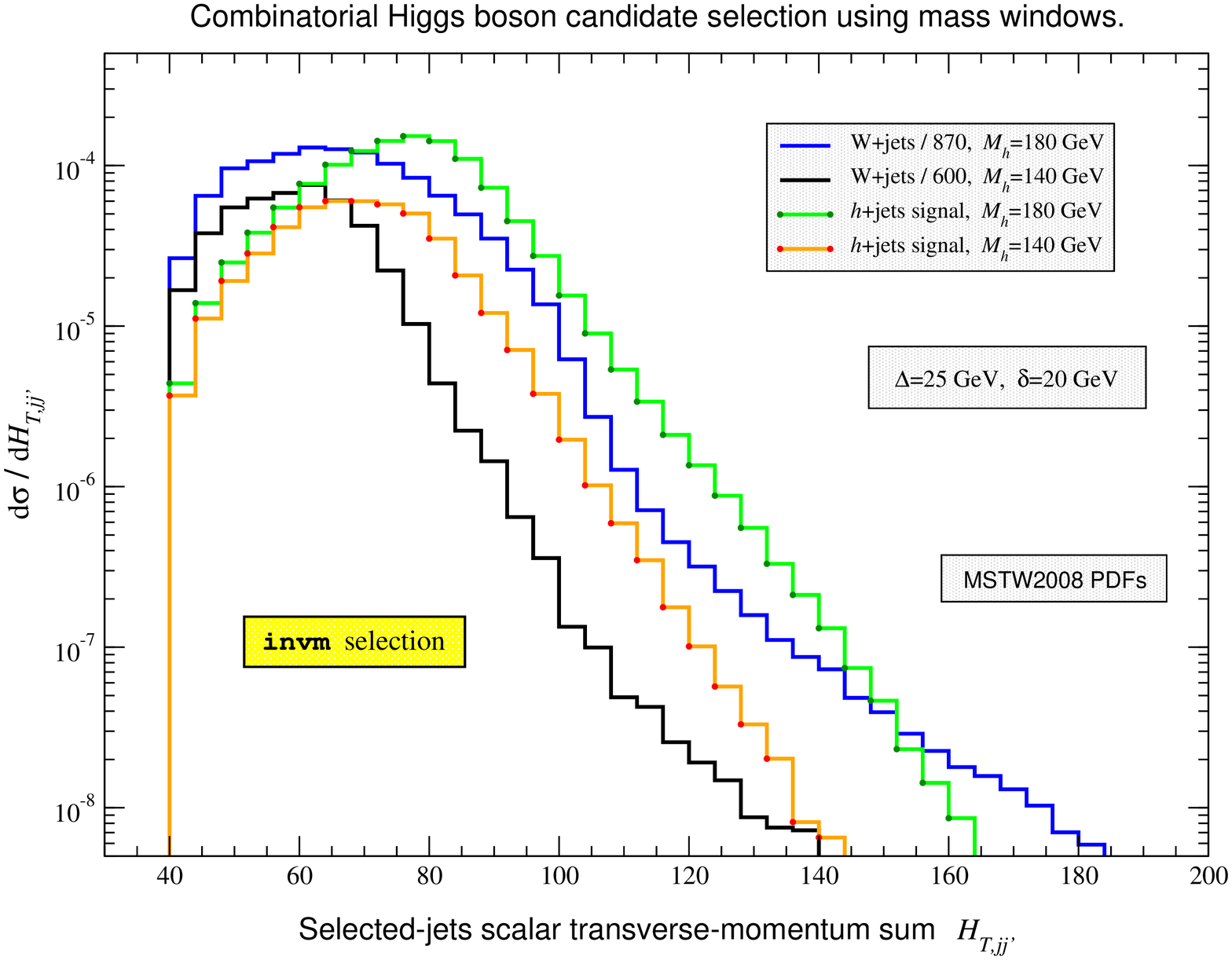}}
  \caption{\label{fig:sec2}
    Examples of observables discriminating between Higgs signal
    and (the dominant $W$\!+jets) backgrounds. Distributions are shown
    from the upper left to the lower right, for the reconstructed
    masses of the $\{e,\nu_e,j,j'\}$ final states, their transverse
    momenta, the boost of the dijet subsystem with respect to the
    parent system and the scalar sum of the $p_T$ of the two jets
    $j$\/ and $j'$. All results are obtained after the combinatorial
    Higgs boson candidate selection based on an ideal mass
    reconstruction of the potential resonance. The selection is
    facilitated by symmetric mass window constraints for both the
    4-object and dijet mass: $|m_{e\nu_ejj'}-M_h|<\Delta$ and
    $|m_{jj'}-M_W|<\delta$, respectively. More detailed explanations
    will follow in \Sec{sec:results}. Note that the background
    differential cross sections are scaled down by the respective
    factors indicated in the legends.}
\end{figure}

In the approximate reconstructions that we will employ in our
analysis, the Higgs boson mass $M_h$ is approximated by a 4-object
invariant mass $m_{e\nu_e jj'}$. The transverse momentum of the Higgs
boson is approximated by the 4-object transverse momentum $p_{T,e\nu_ejj'}$.
The dijet boost $\gamma_{jj'}$ defined in the Higgs boson rest frame is
approximated by $\gamma_{jj'|e\nu_e}$, which is the dijet boost
defined in the 4-object rest frame, in which we can compute this boost
factor via $\gamma_{jj'|e\nu_e}=E_{jj'}/m_{jj'}$. This is equivalent
to using \Eqs{eq:boostfactors} where one inserts for each invariant
mass its reconstructed counterpart.

All three of these observables discriminate between Higgs signal and
backgrounds. As seen in \Fig{fig:sec2} (top left), $m_{e\nu_ejj'}$ for
signal events peaks strongly near the true Higgs boson mass, with a
width determined primarily by parton shower effects. Thus a simple
mass window selection significantly enhances the signal, and since we
are interested in Higgs boson exclusion there is no ``look-elsewhere''
effect associated with imposing mass windows \cite{cms:2011}. Note
that the backgrounds are not necessarily flat in the mass windows:
as seen in \Fig{fig:sec2} the dominant $W$\!+jets background is rather
flat in the high mass window, but is steeply rising in the lower mass
window because of the underlying kinematics. In \Fig{fig:sec2} (top
right) one sees that the 4-object transverse momentum
$p_{T,e\nu_ejj'}$ has a harder spectrum for Higgs signal events than
for the $W$\!+jets backgrounds, independent of the Higgs boson mass.

The reconstructed dijet boost $\gamma_{jj'|e\nu_e}$ has qualitatively
different behaviour depending on the underlying Higgs boson mass. When
the Higgs boson mass is close to $2\,M_W$, the distribution of the
dijet boost for signal events is strongly peaked near one compared
to the distribution for $W$\!+jets, as seen in \Fig{fig:sec2} (lower
left). For larger Higgs boson masses the signal distribution of the
dijet boost is instead rather strongly peaked around the value
$M_h/(2\,M_W)$, as expected from \Eq{eq:boostfactors}.

Other physical observables of interest for signal versus background
discrimination are defined directly in the lab frame. This includes
the 4-object pseudo-rapidity $\eta_{e\nu_ejj'}$, the pseudo-rapidity
difference of the two jets, $\Delta\eta_{j,j'}$, and the scalar sum of
the two selected jet transverse momenta  $H_{T,jj'}$. As seen in
\Fig{fig:sec2} (lower right), the distribution of $H_{T,jj'}$ for
signal events is harder than for the $W$\!+jets background,
independent of the Higgs boson mass. From the distributions shown in
\Fig{fig:detajj} one sees that the dijet pseudo-rapidity difference
has a maximum at zero for signal events (which tend to be confined to
the central region), but peaks at a larger value for the $W$\!+jets
background. Similarly from \Fig{fig:eta.set} one notes that the
4-object pseudo-rapidity distribution is more central for signal than
for the $W$\!+jets background.

We will also employ various transverse masses, both as signal
discriminators and as inputs to the algorithms for the approximate
reconstruction of the Higgs boson candidates. One class of $m_T$
observables is solely constructed out of transverse degrees of
freedom, $\vec p_{T,i}=(p_{x,i},p_{y,i})$; we define these $m_T$
observables as
\be\label{eq:mt}
m_{T,ij..}\;=\;\sqrt{(\abs{\vec p_{T,i}}+\abs{\vec p_{T,j}}+\ldots)^2
  -(\vec p_{T,i}+\vec p_{T,j}+\ldots)^2}
\;\le\;m_{ij..}\ ,
\ee
where, for the purposes of this study, the labels $ij..$ refer to two,
three, or four final-state physics objects (charged lepton, MET, and
the two selected jets). We also investigate how our results change
when we adopt a slightly different definition that includes the full
information from the invariant mass of the visible subset:
\be\label{eq:mtvis}
\left(m^{(k..)}_{T,ik..l..}\right)^2\;=\;m^2_{il..}\,+\;
2\,\bigl(E_{T,il..}\,p_{T,k..}-\vec p_{T,il..}\cdot\vec p_{T,k..}\bigr)
\;\le\;m_{ik..l..}\ ,
\ee
where $E^2_{T,il..}=p^2_{T,il..}+m^2_{il..}$. Here, we have separated
the event into a ``visible'' ($il..$) and ``invisible'' ($k..$) part.
The transverse masses are all approximately bounded from above by a
kinematic edge; this gives us another handle when fully reconstructing
the event. Schematically, we have 
\be\label{eq:mtorder}
m_{T,i[k..]l..}\;\le\;m^{(k..)}_{T,ik..l..}\;\le\;m_{ik..l..}
\ee
where $m_{T,i[k..]l..}$ just indicates that in the presence of
multiple invisible objects the $[k..]$ subsystem enters as a whole
when computing \Eq{eq:mt}. In the 2-particle case, the two
transverse-mass definitions coincide provided the single objects are
massless.

Given the above arsenal of kinematic discriminators and approximate
reconstruction techniques, our basic strategy will be to find the most
promising combinations of selections as a function of the Higgs boson
mass. Since we are only performing a cut and count analysis, and are
lacking a realistic detector description, there is no point in
attempting a complete optimization. Instead we will concentrate on
providing a comprehensive look at the physics that distinguishes
signal from background.


\section{Inclusive cross sections and event generation}\label{sec:xsecs}

We use the multi-purpose Monte Carlo event generator \sherpa
\cite{Gleisberg:2003xi,Gleisberg:2008ta} to pursue our analysis of
semileptonic Higgs boson decays in Higgs boson production via gluon
fusion. This way we can easily include all (hard and soft)
initial-state radiation and final-state radiation (ISR and FSR)
effects and arrive at a fairly realistic description of final states
as used for detector simulations. Furthermore, sophisticated cuts can
be implemented in a straightforward manner owing to the convenient
analysis features that come with the \sherpa package. We also want to
make use of \sherpa's capabilities in providing an enhanced modeling
of multi-jet final states with respect to a treatment by parton
showers only. Apart from a handful of key processes, the \sherpa Monte
Carlo program evaluates cross sections at leading order/tree level
utilizing its integrated automated matrix-element generators
\amegic~\cite{Krauss:2001iv} and/or \comix~\cite{Gleisberg:2008fv}.
However, in a number of studies, \sherpa has been shown to generate
predictions that are in sufficient, often good agreement with the
shapes of kinematic distributions obtained from measurements as well
as higher-order calculations; we will give more details in the
respective subsections that follow up. Hence, we wish to identify
appropriate constant $K$-factors between the most accurate theory
results and the leading-order predictions. We in turn want to apply
these $K$-factors to correct \sherpa's predictions for the inclusion
of exact higher-order rate effects. Therefore, we study signal and
background fixed-order and resummed cross sections at Tevatron Run II
energies for the processes
\be
P\bar P\to\ell\nu_\ell\;pp
\ee
leading to final states consisting of an isolated lepton, missing
transverse energy and at least two jets. The label $\ell$\/ denotes
electrons, $e$, and muons, $\mu$; the parton label $p$\/ contains
light/massless-quark flavours and gluons. Note that final-state
gluons may only occur in background hard processes or through the
inclusion of I/FSR effects. For the respective $K$-factors, it is not
clear a priori at which level of cuts they are defined most
accurately. The most convenient definition should be given in terms of
the total inclusive cross sections, while one may define more
exclusive $K$-factors, if the higher-order tools allow for the
specification of the desired cuts. We do not expect a strong
dependence on the exact $K$-factor definition provided the shapes are
comparable. These issues are examined in more detail below with the
goal of determining reasonable signal and background $K$-factors that
can be used to rescale the respective leading-order cross sections
$\sigma^{(0)}$ of the \sherpa predictions, which we take to pursue our
signal versus background studies.

\subsection{Standard Model Higgs boson production and decay}\label{sec:ggh}

The signal processes for the final states of interest are summarized
by
\be\label{proc:sgnl}
P\bar P\to h\to W^{(\ast)}W^{(\ast)}\to\ell\nu_\ell\;pp
\ee
where the Higgs particles are produced through gluon--gluon fusion and
decay into $W$\/ boson pairs that split further into the desired
semileptonic final states. There are other Higgs boson production
mechanisms that can contribute. In particular, the associated
production $P\bar P\to Vh$\/ with the additional vector boson $V$\/
decaying hadronically and the production via vector-boson fusion (VBF)
ought to be mentioned in this context. For all production channels,
up-to-date theory predictions for the total inclusive cross sections
of these events are needed to arrive at reliable acceptance estimates
for various Higgs boson masses. Refs.~\cite{:2011cb,:2011gs,Dittmaier:2011ti}
give the most recent overview of the theory calculations and results
that are used as input for the ongoing Tevatron (and LHC) SM Higgs
boson searches. For the Higgs boson masses we are interested in, it is
appropriate to separate the Higgs boson production from its subsequent
decays and multiply the production rates by the respective branching
fractions, which we obtain from \textsc{Hdecay}
\cite{Spira:1997dg,Djouadi:1997yw,Butterworth:2010ym,Denner:2011mq}
for $h\to W^\ast W^\ast$ and the Particle Data Group (PDG) listings
\cite{Amsler:2008zzb} for the subsequent decays of the $W$\/ bosons.

For our main production channel, the Standard Model Higgs boson
production via gluon fusion, we want to use the most precise
theoretical predictions that have become available over the last few
years, for a great review, we refer to Ref.~\cite{Dittmaier:2011ti}.
Using an effective theory approach, this production channel is known at
NNLO including electroweak and mixed QCD--electroweak contributions
\cite{Anastasiou:2008tj,Keung:2009bs,Actis:2008ug,Actis:2008ts}.
For a wide range of Higgs boson masses, these NNLO cross sections have
been shown to reproduce the latest results obtained from soft-gluon
resummation up to NNLL accuracy, cf.\
Refs.~\cite{deFlorian:2009hc,Catani:2003zt,Moch:2005ky}. To mimic the
resummation effects, the optimal scale choice at NNLO is found to be
$\muf=\mur=M_h/2$, while for the NNLL calculation one employs common
scales of $\mu=M_h$.
Both higher-order calculations take the most recent parametrization of
PDFs at next-to-next-to-leading order into account where the
corresponding PDF sets have been provided by the MSTW group in 2008
\cite{Martin:2009iq}. The Tevatron Higgs boson searches use these
higher-order $gg\to h$\/ cross section predictions to report their
combined CDF and D\O\ upper limits on Standard Model Higgs boson
production in the $W^+W^-$ decay mode \cite{Benjamin:2011sv,%
  CDFnote10599:2011,D0Note6219:2011,:2011cb,:2010ar,Aaltonen:2010yv,%
  Aaltonen:2010cm,Abazov:2010ct}.
Hence, it makes sense to input the same theory cross sections in our
studies to guarantee a reasonable level of compatibility between our
work and the experimental searches. However, one should keep in mind
that different viewpoints exist concerning the determination of the
best $gg\to h$\/ cross section numbers. For example, in
Ref.~\cite{Baglio:2010um} the authors argue that the 10--15\%
enhancement seen in the inclusive rates is unlikely to survive the
cuts applied in the Tevatron analyses and, therefore, should not be
included in the calculation of the limits \cite{Baglio:2011wn,Baglio:2011hc}.
On the contrary, the renormalization-group improved resummed NNLO
cross sections discussed by Ahrens, Becher, Neubert and Yang in
Refs.~\cite{Ahrens:2008qu,Ahrens:2008nc,Ahrens:2010rs} would yield a
further 5--6\% increase of the NNLL $gg\to h$\/ rates. This is because
in their approach Ahrens et al.\ do not only resum threshold
logarithms from soft-gluon emission but also $\pi^2$-enhanced terms,
which arise in the analytic continuation of the gluon form factor to
timelike momentum transfer.

\begin{table}[t!]
\centering\footnotesize\vskip5mm
\begin{tabular}{ll|rrrrrr|rlrl}\hline\hline
\multicolumn{1}{l}{\rule[-3mm]{0mm}{8mm}$M_h$} &
\multicolumn{1}{l|}{$\Gamma_h$} &
\multicolumn{1}{r}{$\sigma^\mrm{NNLL}_{ggh}$} &
\multicolumn{1}{r}{$B^{\mbox{\tiny Ref.~\cite{:2011cb}}}_{W^\ast W^\ast}$} &
\multicolumn{1}{r}{$\ \sigma^\mrm{NNLO}_{ggh}$} &
\multicolumn{1}{r}{$B_{W^\ast W^\ast}$} &
\multicolumn{1}{r}{$\ \;\sigma^{(0)}_{S,\mrm{all}}$} &
\multicolumn{1}{r|}{$\sigma^{(0)}_{S,\mrm{NNLO}}$} &
\multicolumn{1}{r}{$\ \sigma^{(0)}_S$} &
\multicolumn{1}{r}{$K_S$} &
\multicolumn{1}{r}{$\ \sigma^{(0)}_{S,\mrm{66}}$} &
\multicolumn{1}{r}{$K^\mrm{66}_S$}\\
\rule[-2mm]{0mm}{6mm}$\mrm{[GeV]}$&$\mrm{[GeV]}$&$\mrm{[fb]}$&&
$\mrm{[fb]}$&&$\mrm{[fb]}$&$\mrm{[fb]}$&$\mrm{[fb]}$&&$\mrm{[fb]}$&\\\hline
\rule[-1.5mm]{0mm}{5mm}$110$&$0.002939$& $1385.0$&$0.0482$&$1428$&$0.04633$&
$11.81$ &$9.660$ &$3.350$ &\bf2.88&$2.550$&\bf3.79\\
\rule[-1.5mm]{0mm}{5mm}$120$&$0.003595$& $1072.3$&$0.143$ &$1102$&$0.1380$&
$27.11$ &$22.21$ &$7.717$ &\bf2.88&$5.990$&\bf3.71\\
\rule[-1.5mm]{0mm}{5mm}$130$&$0.004986$& $842.9$ &$0.305$ &$863$ &$0.2976$&
$45.83$ &$37.50$ &$13.06$ &\bf2.87&$10.15$&\bf3.69\\
\rule[-1.5mm]{0mm}{5mm}$140$&$0.008222$& $670.6$ &$0.504$ &$685$ &$0.4959$&
$60.53$ &$49.60$ &$17.29$ &\bf2.87&$13.46$&\bf3.68\\
\rule[-1.5mm]{0mm}{5mm}$150$&$0.01726$ & $539.1$ &$0.699$ &$550$ &$0.6927$&
$67.88$ &$55.63$ &$19.33$ &\bf2.88&$15.08$&\bf3.69\\\hline
\rule[-1.5mm]{0mm}{5mm}$165$&$0.2429$  & $383.7$ &$0.960$ &$389$ &$0.9595$&
$66.72$ &$54.50$ &$19.35$ &\bf2.82&$15.13$&\bf3.60\\
\rule[-1.5mm]{0mm}{5mm}$170$&$0.3759$  & $344.0$ &$0.965$ &$347$ &$0.9642$&
$60.02$ &$48.85$ &$17.77$ &\bf2.75&$13.91$&\bf3.51\\
\rule[-1.5mm]{0mm}{5mm}$180$&$0.6290$  & $279.2$ &$0.932$ &$283$ &$0.9327$&
$47.33$ &$38.54$ &$14.19$ &\bf2.72&$11.15$&\bf3.46\\
\rule[-1.5mm]{0mm}{5mm}$190$&$1.036$   & $228.0$ &$0.786$ &$229$ &$0.7871$&
$32.50$ &$26.32$ &$9.862$ &\bf2.67&$7.775$&\bf3.39\\
\rule[-1.5mm]{0mm}{5mm}$200$&$1.426$   & $189.1$ &$0.741$ &$190$ &$0.7426$&
$25.40$ &$20.60$ &$7.827$ &\bf2.63&$6.191$&\bf3.33\\
\rule[-1.5mm]{0mm}{5mm}$210$&$1.841$   &         &        &$159$ &$0.7250$&
        &$16.83$ &$6.473$ &\bf2.60&$5.131$&\bf3.28\\
\rule[-1.5mm]{0mm}{5mm}$220$&$2.301$   &         &        &$134$ &$0.7160$&
        &$14.01$ &$5.420$ &\bf2.58&$4.317$&\bf3.25\\[1mm]
\hline\hline
\end{tabular}
\caption{\label{tab:sgnlKfacs}
  Signal cross sections $\sigma^{(0)}_S$ at NNLO and LO plus the
  resulting signal $K$-factors for several different SM Higgs boson
  masses. The Higgs boson widths are found from \textsc{Hdecay}
  calculations. Columns 5 and 6 show our input NNLO cross sections
  taken from recent \textsc{Fehip} calculations
  \cite{Anastasiou:2004xq,Anastasiou:2005qj,Petriello.priv:2010,%
    Anastasiou:2008tj} and input branching
  fractions for $h\to W^\ast W^\ast$ obtained from \textsc{Hdecay}
  \cite{Djouadi:1997yw,Butterworth:2010ym}, respectively. They are
  comparable to the values given in the 3rd and 4th columns that are
  used by the Tevatron experimentalists in their ongoing Higgs boson
  searches \cite{:2011cb}. The signal cross sections labelled ``all''
  account for contributions stemming not only from gluon--gluon fusion
  but also from Higgs strahlung and VBF Higgs boson production processes.
  All calculations use MSTW2008 parton distributions except those to
  extract the LO values denoted by ``$\mrm{66}$'', which result from
  using CTEQ6.6 PDFs.}
\end{table}


For various Higgs boson masses, \Tab{tab:sgnlKfacs} summarizes the
signal cross sections $\sigma^{(0)}_S$ that are of relevance for our
studies. Separated from the other entries, the left and right parts of
the table show parameters, which we use as input to our analysis.
We have used the NNLO $gg\to h$\/ inclusive cross sections,
$\sigma^\mrm{NNLO}_{ggh}$, given in the 5th column and multiplied them
with the branching ratios listed in the column to the right of it,
which we have computed with \textsc{Hdecay} version~3.51.\footnote{To
  obtain the values in \Tab{tab:sgnlKfacs}, the NNLO MSTW fit result
  for the strong coupling, $\alpha_\mrm{s}(M_Z)=0.11707$, has been
  employed, the running of quark masses at NNLO has been enabled and
  the top and bottom quark masses have been set to $m_t=173.1$ and
  $m_b=4.8\ \mrm{GeV}$, respectively.}
The Higgs boson widths shown in the 2nd column are also taken from the
\textsc{Hdecay} calculation; they slightly differ from the older
values given in \cite{Djouadi:2005gi}. To arrive at the higher-order
prediction of our signal cross sections depicted in the 8th column,
$\sigma^{(0)}_{S,\mrm{NNLO}}$, we furthermore have accounted for the
$W$\/ decays described by the branching ratios for one light lepton
species, $B(W\to\ell\nu_\ell)=0.108$, and for jets, $B(W\to pp)=0.676$,
and a combinatorial factor of~2 reflecting that either of the $W$\/
bosons may decay leptonically. The NNLO cross sections used
here~\cite{Petriello.priv:2010} are updated values with respect to the
ones published in \cite{Anastasiou:2008tj}. The difference can be traced
back to the addition of the electroweak real-radiation corrections as
encoded in Ref.~\cite{Keung:2009bs} and the change to top masses of
$m_t=173.1\ \mrm{GeV}$. In the 3rd and 4th columns we respectively
also give the $gg\to h$\/ cross sections and $h\to W^\ast W^\ast$\/
branching fractions as used by the Tevatron experimentalists in their
ongoing searches~\cite{:2011cb}. These values are in good agreement
with the respective numbers used in our study. The other signal cross
sections listed in the rightmost part of \Tab{tab:sgnlKfacs} are the
LO rates obtained from \sherpa, where the one labelled
$\sigma^{(0)}_{S,\mrm{66}}$ refers to the use of CTEQ6.6 PDFs
\cite{Nadolsky:2008zw}. We will discuss these LO results and the
corresponding $K$-factors in \Sec{sec:sherpa.sgnl}. In the 7th column
we show an upper estimate for the signal cross sections
$\sigma^{(0)}_{S,\mrm{all}}$ if one were to include the contributions
from the $Wh$, $Zh$\/ and VBF production channels. Over the considered
Higgs boson mass range the extra processes would enhance the signal
rates resulting from gluon--gluon fusion by about 22--23\%. We have
determined these estimates by adding to the $gg\to h$\/ rates the
theory predictions for $\sigma_{Wh}$, $\sigma_{Zh}$ and
$\sigma_\mrm{VBF}$ as presented in the July 2010 CDF and D\O\ Higgs
boson searches combination paper \cite{:2010ar} (for updated values,
cf.\ \cite{:2011cb}) including all necessary branching fractions to
arrive at the $pp\;\ell\nu_\ell\,pp$\/ final states.\footnote{For all
  production channels, we consider the Higgs boson decay as specified
  in (\ref{proc:sgnl}), since our selection and reconstruction
  procedures are tailored to this decay mode, see \Sec{sec:results}.
  In our analysis we deal with $\ell$, MET and multiple jets (from the
  decays as well as I/FSR), thus, the presence of additional jets
  stemming from other hard decays or VBF does not alter our analysis
  procedures considerably, in other words the Higgs boson
  reconstruction and selection procedures are designed in a robust way
  with respect to additional jet activity. The dominant Higgs
  strahlung contributions to the semileptonic final states arise from
  the hadronic decays of the associated vector bosons, where we have
  used the PDG values $B(W\to pp)=0.676$ and $B(Z\to pp)=0.6991$. All
  other combinations are suppressed by about an order of magnitude.
  Also, we do not consider more than one lepton, \ie we implicitly
  assume an exclusive one-lepton cut. We also neglect the $Vh$\/ cases
  where the $Z$\/ boson decays invisibly and the associated $W$\/
  boson splits up leptonically while $h\to pppp$. These modes will
  fail our $h$\/ boson reconstruction.}
The additional 20\% increase resulting from these considerations
should be born in mind when acceptances and significances are
evaluated in a signal versus background study. For the purpose of the
analysis we are pursuing here, we however want to be on the
conservative side and solely concentrate on the gluon--gluon fusion
events.

\subsection{Relevant background processes}\label{sec:rel_bkgd}

Processes that give rise to major background contributions are
$V$\!+jets and multi-jet production, where the latter comes into play
because of jets faking isolated leptons and/or MET. The muon channel
suffers less from jet fakes reducing the multi-jet background by a
factor of 5 with respect to the electron channel. The $Z$\!+jets
background can also contribute in cases where one lepton goes missing
or a jet mimics a lepton while the $Z$\/ decays invisibly.
Minor contributions stem from $VV$\/ and $t\bar t$\/ production. A first
D\O\ search in the semileptonic Higgs boson decay channel shows how
these backgrounds compare to each other after basic selection cuts, see
Ref.~\cite{D0note6095:2010}. The largest fraction of 83\% occurs from
$V$\!+jets, followed by multi-jets, $t\bar t$\/ and $VV$\/ contributing
with 12\%, 3\% and 2\% to the overall background. The $V$\!+jets
contribution is totally dominated by $W$\!+jets production; the
$Z(\to\ell^+\ell^-)$+jets background, where one of the leptons is
missed, is small and makes up less than 1\% of the total background.

To gain a better understanding of the backgrounds, we will have a
closer look at the major contributor $W$\!+jets. The multi-jet
background cannot be simulated straightforwardly, since it requires
detailed knowledge of the experiments and measured fake rates etc.
Regarding the minor background contributors, we will study the
$t\bar t$\/ as well as the $WW$ -- or, more exactly,
electroweak -- background. Even though they enter at a rather low
level after the basic selection compared to $W$\!+jets, it is necessary
to cross-check what number of events remain after more selective cuts
have been applied, as will be discussed in \Sec{sec:results}.

Another background contribution that has been discussed is
gluon-initiated vector boson pair production
\cite{Dixon:2003yb,Binoth:2005ua,Binoth:2006mf}. This
(quark-loop-induced) process occurs at
${\cal O}(\alpha^4_\mrm{ew}\alpha^2_\mrm{s})$, the same order as the
signal. This background formally arises at NNLO, but under realistic
experimental cuts this production channel has been shown to
significantly increase \eg the $WW\to2\,\ell\;2\,\nu_\ell$ background
at the LHC. At the Tevatron the gluon densities are small, so the
impact of $gg\to WW$\/ is expected to be negligible. This expectation
was confirmed in Ref.~\cite{Campbell:2011cu} (a 4\textperthousand\
effect with respect to the NLO cross section for this decay channel).
A more important effect also recently pointed out by Campbell~et~al.\
in Ref.~\cite{Campbell:2011cu} is the interference between $gg\to WW$\/
and $gg\to h\to WW$, which can result in ${\cal O}(0.1)$ corrections
to the Higgs boson signal cross section. However, interference effects
are considerably reduced by requiring the transverse mass of the
leptons plus MET system to be smaller than $M_h$. This type of
transverse cut is frequently used in our analyses, so we can safely
neglect interference effects in our study.

\boldmath
\subsubsection{$W$\/ boson plus jets background}
\unboldmath

For our first study of $W$\!+jets production, we explore the dependence
of inclusive $W$\!+jet cross sections on the number of jets and the
variation of the common scale $\mu$\/ used to specify the
factorization and renormalization scales, $\muf$ and $\mur$,
respectively. This information will help us identify an optimal
definition of the $W$\!+jets $K$-factor, which we take to improve the
rates of the \sherpa predictions. We calculate inclusive
$W$\!+$\,n\le2$-jet cross sections with \textsc{MCFM} version~5.8 to
obtain results that are accurate at NLO in the strong-coupling
constant \cite{Campbell:Ellis,Campbell:2002tg,Campbell:2003hd}. We
also run \textsc{MCFM} at LO to determine explicit NLO-to-LO
theoretical $K$-factors.\footnote{We only consider $W^+$ bosons
  decaying into $e^+\nu_e$ pairs; the charge conjugated process will
  just double the cross section owing to the $P\bar P$\/ initial
  states at the Tevatron. We employ the LO and NLO MSTW2008 PDFs
  \cite{Martin:2009iq} with $\alpha_\mrm{s}(M_Z)=0.13939$ and
  $\alpha_\mrm{s}(M_Z)=0.12018$, respectively, and impose cuts
  according to the parameters given in \Sec{sec:cuts}. Note that we do
  not account for the so-called triangle cut relating the transverse
  mass of the $W$\/ boson and the missing energy. Other parameters,
  such as the electroweak input values of the Standard Model, have
  been taken according to the \textsc{MCFM} default settings.}

\begin{table}[t!]
\centering\footnotesize\vskip5mm
\begin{tabular}{clrrrccrrrc}\hline\hline
\multicolumn{2}{l}{} &
\multicolumn{9}{l}{\rule[-3mm]{0mm}{8mm}\boldmath\textbf{Inclusive
  $W^+$+$\,n$-jet cross sections in pb.}\unboldmath}\\
\multicolumn{1}{c}{\rule[-2mm]{0mm}{6mm}$n$} &
\multicolumn{1}{l}{\hphantom{XXXX}} &
\multicolumn{1}{r}{$\mu=M_{\perp,W}/2$} &
\multicolumn{1}{r}{$=M_{\perp,W}$} &
\multicolumn{1}{r}{$=2\,M_{\perp,W}$} &
\multicolumn{1}{c}{\%} &
\multicolumn{1}{c}{} &
\multicolumn{1}{r}{$\mu=\hat H_T/2$} &
\multicolumn{1}{r}{$=\hat H_T$} &
\multicolumn{1}{r}{$=2\,\hat H_T$} &
\multicolumn{1}{c}{\%}\\\hline
\rule[-1.5mm]{0mm}{5mm}$0$&LO &$457$&$465$&$469$&$^{-1.7}_{+0.9}$&&
                               $453$&$463$&$468$&$^{-2.2}_{+1.1}$\\
\rule[-1.5mm]{0mm}{5mm}   &NLO&$625$&$619$&$616$&$^{+1.0}_{-0.5}$&&
                               $606$&$602$&$602$&$^{+0.7}_{-0.0}$\\
\rule[-1.5mm]{0mm}{5mm}   &$K$&\bf1.37&\bf1.33&\bf1.31&&&
                               \bf1.34&\bf1.30&\bf1.29&\\[2mm]
\rule[-1.5mm]{0mm}{5mm}$1$&LO &$66.2$&$55.6$&$47.3$&$^{+19.1}_{-14.9}$&&
                               $62.5$&$52.7$&$45.1$&$^{+18.6}_{-14.4}$\\
\rule[-1.5mm]{0mm}{5mm}   &NLO&$79.8$&$74.6$&$69.3$&$^{+7.0}_{-7.1}$&&
                               $74.2$&$70.2$&$65.7$&$^{+5.7}_{-6.4}$\\
\rule[-1.5mm]{0mm}{5mm}   &$K$&\bf1.21&\bf1.34&\bf1.47&&&
                               \bf1.19&\bf1.33&\bf1.46&\\
\rule[-1.5mm]{0mm}{5mm}   &$R_\mrm{LO}^{(1,0)}$&
                               $0.145$&$0.120$&$0.101$&&&
                               $0.138$&$0.114$&$0.096$&\\
\rule[-1.5mm]{0mm}{5mm}   &$R_\mrm{NLO}^{(1,0)}$&
                               $0.128$&$0.121$&$0.113$&&&
                               $0.122$&$0.117$&$0.109$&\\[2mm]
\rule[-1.5mm]{0mm}{5mm}$2$&LO &$14.4$&$10.1$&$7.40$&$^{+42.6}_{-26.7}$&&
                               $10.9$&$7.89$&$5.89$&$^{+38.1}_{-25.3}$\\
\rule[-1.5mm]{0mm}{5mm}   &NLO&$12.8$&$11.7$&$10.4$&$^{+9.4}_{-11.1}$&&
                               $12.0$&$10.1$&$8.95$&$^{+18.8}_{-11.4}$\\
\rule[-1.5mm]{0mm}{5mm}   &$K$&\bf0.89&\bf1.16&\bf1.41&&&
                               \bf1.10&\bf1.28&\bf1.52&\\
\rule[-1.5mm]{0mm}{5mm}   &$R_\mrm{LO}^{(2,1)}$&
                               $0.218$&$0.182$&$0.156$&&&
                               $0.174$&$0.150$&$0.131$&\\
\rule[-1.5mm]{0mm}{5mm}   &$R_\mrm{NLO}^{(2,1)}$&
                               $0.160$&$0.157$&$0.150$&&&
                               $0.162$&$0.144$&$0.136$&\\[2mm]
\hline\hline
\end{tabular}
\caption{\label{tab:mcfm}
  Inclusive $W^+$+$\,n$-jet cross sections $\sigma_n$ in pb at LO and
  NLO in QCD for different scale choices and jet multiplicities using
  MSTW2008 PDFs. The variations with respect to the nominal choices,
  $M_{\perp,W}$ with $M^2_{\perp,W}=M^2_W+p^2_{T,W}$ and $\hat H_T$,
  are given in the columns labelled by ``\%''. Numerical integration
  uncertainties are not displayed, since they are at least one order
  of magnitude below the accuracy indicated here. NLO-to-LO
  $K$-factors and $n$-to-($n-1$)-jet cross section ratios are also
  shown for all possible instances.}
\end{table}

We display our \textsc{MCFM} results in \Tab{tab:mcfm} for different
inclusive jet bins $n$\/ and scale choices $\mu$. As expected, for
each $n$-jet multiplicity, the NLO cross sections are more stable
under scale variations with the largest deviations occurring for the
more complex $W^+$+2-jet processes. This is also reflected by the
various NLO-to-LO $K$-factors, which vary from about 0.9 to 1.5 for
$n=2$ while they are rather constant for $n=0$ ranging from about 1.3
to 1.4 only. For illustrative purposes, we also list the LO and NLO
inclusive jet-rate ratios $R^{(n,n-1)}=\sigma_n/\sigma_{n-1}$ starting
with $n=1$. The $W^+$+$\,n\le1$-jet cross sections do not deviate
substantially for the two nominal scales chosen, $\mu\sim M_{\perp,W}$
where $M^2_{\perp,W}=M^2_W+p^2_{T,W}$ and $\mu\sim\hat H_T$, which are
determined dynamically for each event. Note that $\hat H_T$ is the
scalar sum of the transverse momenta of all particles (partons) in the
event, \ie no jet clustering has taken place. The $\mu\sim M_{\perp,W}$
scales lead to slightly larger rates when compared to those obtained
for $\mu\sim\hat H_T$. This can be traced back to the occurrence of
$\mu$-values that are on average larger in the latter case, since
$\langle\hat H_T\rangle\gtrsim\langle M_{\perp,W}\rangle$
for $n\ge1$. For the same reason, the cross section differences become
more manifest for $n=2$. The presence of the second jet gives an extra
$p_T$ contribution to $\hat H_T$ per event whereas $M_{\perp,W}$ is
less affected. This further enhances the deviation of the $\hat H_T$
and $M_{\perp,W}$ averages.

Given the numbers of \Tab{tab:mcfm} we can conclude that our knowledge
of the $W$\!+2-jet background is accurate on the level of
$\lesssim$ 20\%. A $K$-factor of about $1.5$ should be viewed as the
upper limit for correcting LO results; in \Sec{sec:sherpa.bkgd} we
will however compare the \sherpa background rates more closely with
the results of \Tab{tab:mcfm} and determine a $K$-factor accordingly.

\subsection{Monte Carlo simulation of signal and backgrounds using
  \protect\sherpa}\label{sec:sherpa}

For reasons outlined at the beginning of \Sec{sec:xsecs}, we use
\sherpa version~1.1.3 \cite{Gleisberg:2003xi,Gleisberg:2008ta} to
generate the $\ell\nu_\ell$+jets signal and background events that are
needed to understand the potential of a Standard Model Higgs boson
analysis in the lepton + MET + jets channel.\footnote{Version~1.1.3
  was the last of the previous \sherpa generation; for all our
  purposes, it models the necessary physics equally well compared to
  the upgraded versions of the current (1.3.x) generation.
  Cross-comparisons have confirmed this result.}
We will employ the results of the previous two subsections to settle
the inclusive $K$-factors needed to re-scale \sherpa's LO predictions
and include higher-order rate effects.

The signal and background simulations share a number of common
parameters and options that have been set as follows:
we simulate all events at the parton-shower level, \ie we include
initial- and final-state QCD radiation, but do not account for
hadronization effects and corrections owing to the underlying event,
since their impact is considerably smaller with respect to additional
QCD radiation arising from the hard processes. The intrinsic
transverse motion of quarks and gluons inside the colliding hadrons is
however modeled by an intrinsic Gaussian $k_T$-smearing of
$\mu(k_T)=0.2$ and $\sigma(k_T)=0.8\ \mrm{GeV}$. The electroweak
parameters are explicitly given: $M_W=80.419$, $\Gamma_W=2.06$,
$M_Z=91.188$, $\Gamma_Z=2.49\ \mrm{GeV}$; the Higgs boson masses and widths
are mutable, taken according to \Tab{tab:sgnlKfacs}; the couplings are
specified by $\alpha_\mrm{ew}(0)=1/137.036$, $\sin^2_\mrm{W}=0.2222$
and the Higgs field vacuum expectation value and its quartic coupling
are given as $246\ \mrm{GeV}$ and $0.47591$, respectively. The CKM
matrix is simply parametrized by the identity matrix. The bottom and
top quark masses are set to $m_b=4.8$ and $m_t=173.1\ \mrm{GeV}$,
respectively, and all other quark masses are zero. To avoid any bias
owing to the utilization of different PDFs and in order to develop a
consistent picture, signal and background events are generated using
the same parton distributions. Our first choice of PDFs is the LO MSTW
set MSTW2008lo90cl \cite{Martin:2009iq}, because its NNLO version has
been the preferred PDF set used for the recent calculations of the
gluon--gluon fusion Higgs boson production cross sections. The strong
coupling is determined by one-loop running with
$\alpha_\mrm{s}(M_Z)=0.13939$, which is the advertised fit value of
the LO MSTW2008 set.

To gain some understanding of PDF effects, we compare our MSTW2008
results against predictions generated with a different PDF set. To
fully establish the comparison on the same level as for the MSTW2008
PDFs, signal and background rates have to be predicted from theory
using the alternative PDF libraries. We cannot follow this approach
here, instead we start out from the same normalization that has been
used for the \sherpa predictions calculated with MSTW2008 PDFs. After
the application of our cuts we then focus on the differences induced
by the alternative PDF set. As our second choice we employ the CTEQ6.6
PDF libraries \cite{Nadolsky:2008zw} where the strong coupling is set
by $\alpha_\mrm{s}(M_Z)=0.118$ and the running of the coupling is
again computed at one loop. Notice that \sherpa invokes a 6-flavour
running for all strong-coupling evaluations.

\subsubsection{Generation of signal events}\label{sec:sherpa.sgnl}

We simulate signal events with electrons or positrons in the final
state according to
\be
P\bar P\to h\to e\nu_e\;pp\to e\nu_e+\mrm{jets}\ .
\ee
The hard process composed as $gg\to h\to e\nu_e\,pp$ is calculated at
LO. The incoming gluons and the quarks arising from the decay undergo
further parton showering, which automatically is taken care of by the
\sherpa simulation. One ends up with the $e\nu_e+\mrm{jets}$ final
states generated at shower level. The hard-process tree-level matrix
elements and subsequent parton showers needed for the simulation are
provided by the \sherpa modules \amegic and \apacic, respectively. For
our purpose, it is sufficient to treat the muon final states in
exactly the same manner as the electron final states, \ie the muon
decay channel is included by multiplication with the lepton factor
$f_\ell=2$ at the appropriate places.

The Higgs boson production occurs through gluon--gluon fusion via
intermediate heavy-quark loops. In \sherpa this is modeled at LO by
an effective $gg\to h$\/ coupling where the top quarks have been
integrated out. The EHC (Effective Higgs Couplings) implementation of
\sherpa includes all interactions up to 5-point vertices that result
from the effective-theory Lagrangian. These effective vertices can
simply be added to the Standard Model. We do not work in the infinite
top-mass limit, because we also want to consider Higgs bosons heavier
than the top quark, the approximation however is well applicable only
as long as $m_t>M_h$. 
The Higgs boson decays are described by $1\to4$ processes,
\ie we directly consider $h\to e\nu_e\,pp$. We thereby make use of
\sherpa's feature to decompose processes on the amplitude level into
the production and decays of unstable intermediate particles while the
colour and spin correlations are fully preserved between the
production and decay amplitudes \cite{Gleisberg:2008ta}. This way one
can focus on certain resonant contributions instead of calculating the
full set of diagrams contributing to a given final state, which in our
case would lead to the inclusion of contributions from the
backgrounds. The intermediate propagators are allowed to be off-shell,
such that finite-width effects are naturally incorporated into the
simulation. This comes in handy especially for Higgs boson masses
below the $WW$\/ mass threshold as the $1\to4$ decays moreover
guarantee the inclusion of off-shell $W$-boson effects. A consistent
LO treatment would require the use of total Higgs boson widths as
computed at LO. We instead put in the values from the \textsc{Hdecay}
calculations \cite{Djouadi:1997yw,Butterworth:2010ym} as listed in
\Tab{tab:sgnlKfacs}. This modifies the Higgs boson propagators and one
arrives at a more accurate description of the finite-width effects of
the Higgs boson decays. The effect on the total rate,
\be
\sigma^{(0)}_S\;=\;
\frac{\Gamma(h\to e\nu_e\,pp)}{\Gamma_h}\;\sigma^\mrm{LO}_{ggh}\ ,
\ee
is nullified, since we eventually correct for the NNLO rates
$\sigma^{(0)}_{S,\mrm{NNLO}}$ worked out in \Sec{sec:ggh}.

In Ref.~\cite{Butterworth:2010ym} a comparative study has been
presented for Higgs boson production via gluon fusion at the
LHC. Amongst a variety of predictions including those given by
\textsc{HNNLO} \cite{Catani:2007vq,Grazzini:2008tf}, the \sherpa
versions~1.1.3 and 1.2.1 have been validated to produce very
reasonable results for the shapes of distributions like the rapidity
and transverse momentum of the Higgs boson, pseudo-rapidities and transverse
momenta of associated jets and jet--jet $\Delta R$\/ separations. We
hence rely on a well validated approach that works not only for pure
parton showering in addition to the Higgs boson production and decays, but
also beyond in the context of merging higher-order tree-level matrix
elements with parton showers. Nevertheless, we have carried out a
number of cross-checks to convince ourselves of the correctness of the
\sherpa calculations; for the details, we refer the reader to
\App{app:lo_xsecs}.

Finally we turn to the discussion of the $K$-factors. Recalling our
findings of \Sec{sec:ggh}, we want to re-scale \sherpa's leading-order
signal cross sections $\sigma^{(0)}_S$ to the fixed-order NNLO
predictions given by \textsc{Fehip} for Higgs boson production in
$gg\to h$\/ fusion via intermediate heavy-quark loops \cite{Anastasiou:2004xq,%
  Anastasiou:2005qj,Anastasiou:2008tj,Petriello.priv:2010}. To be
consistent, the renormalization and factorization scales of the LO
hard-process evaluations are chosen as for the higher-order
calculations, which employ $\mu=\mur=\muf=M_h/2$. The resulting cross
sections ultimately define our signal $K$-factors:
\be
K_S\;=\;\frac{\sigma^{(0)}_{S,\mrm{NNLO}}}{\sigma^{(0)}_S}\ .
\ee
We have determined two sets of $K$-factors for our two choices of
PDFs where the $K$-factors and LO cross sections labelled by
``$\mrm{66}$'' refer to the case of utilizing the CTEQ6.6 libraries
when calculating the LO cross sections. Our results have already been
summarized in \Sec{sec:ggh}, they are presented in the right part of
\Tab{tab:sgnlKfacs}. The $K$-factors are remarkably stable varying
slowly from 2.8 to 2.6 over the entire Higgs boson mass range when relying
on MSTW2008 PDFs. In the CTEQ6.6 case, where we have employed
$\mu=\mur=\muf=\sqrt{\hat s}/2\approx M_h/2$, they are larger due to
the smaller LO rates but their magnitude still remains
$\lesssim$ 3.6.\footnote{The LO rates calculated with the CTEQ PDF
  libraries are diminished for two reasons mainly, the value of
  $\alpha_\mrm{s}$ at $M_Z$ is considerably lower and the altered
  scale choice entails a further reduction of the cross sections.}

In addition to the default scale choice of $\mu=M_h/2$ that we used
for the MSTW runs, we have explored other options by essentially
varying this default setting for $\mu$\/ by factors of 2. We obtained
results for $\mu=M_h/4$, $\mu=\sqrt{\hat s}/2\approx M_h/2$ and
$\mu=\sqrt{\hat s}\approx M_h$ with the effect that the LO rates were
varied by +20\% to -15\% but -- as expected -- no shape changes were
induced.

\boldmath
\subsubsection{Generation of background events for\ \ $W$\/ boson plus
  jets production}
\unboldmath\label{sec:sherpa.bkgd}

We restrict ourselves to the Monte Carlo simulation of the $e^\pm$
channels. Their final states are generated through
\be
P\bar P\to e\nu_e+0,1,2\,p\to e\nu_e+\mrm{jets}
\ee
using an inclusive $W$\!+2-jets sample obtained from the
Catani--Krauss--Kuhn--Webber (CKKW) merging of the corresponding
tree-level matrix elements with the parton showers (ME+PS)
\cite{Catani:2001cc,Krauss:2002up}. In these $W$\!+2-jets calculations
the electroweak order is tied to $\alpha^2_\mrm{ew}$. Unlike the NLO
calculation we do include matrix elements where the extra partons may
occur as $b$~quarks; effectively, they are however treated as massless
quarks in the evaluation of the matrix elements and generation of the
radiation pattern. The events are corrected for the $b$-quark mass
after the parton showering. This approach generates slightly harder
$p_T$ spectra but as part of being more conservative in estimating
this background it is totally reasonable. Similarly, we simply assume
no effect of a $b$-jet veto in removing $W$\!+jets events.

The parameters of the matrix-element parton-shower merging are the jet
separation scale $Q_\mrm{jet}$ and the $D$-parameter, which is used to
fix the minimal separation of the parton jets. These parameters are
respectively set to $Q_\mrm{jet}=20\ \mrm{GeV}$ and $D=0.4$ in
correspondence to the jet $p_T$ threshold and cone definitions of our
analysis, see \Sec{sec:cuts}. $Q_\mrm{jet}$ denotes the scale at which
-- according to the internal $k_T$-jet measure incorporating the
$D$-parameter -- the multi-jet phase space is divided into the two
domains of $Q>Q_\mrm{jet}$ where the jets are produced through exact
tree-level matrix elements and
$Q_\mrm{jet}>Q>Q_\trm{cut-off}\sim1\ \mrm{GeV}$ where the
parton-shower intra-jet evolution takes place. We generate predictions
from samples that merge matrix elements with up to $n^\mrm{max}_p=2$
partons. Although we could increase this maximum number, at this point
we do not want to include matrix elements with more than two partons
in order to be consistent with our signal event generation where the
jets beyond those arising from the $W$-boson decays are produced by
parton showers only. If one wishes to further improve on the
description of additional hard jets, both background and signal
simulations should be extended on the same footing.

\begin{table}[t!]
\centering\footnotesize\vskip5mm
\begin{tabular}{clrrrcrrrcr}\hline\hline
\multicolumn{1}{c}{\rule[-2mm]{0mm}{6mm}$n$} &
\multicolumn{1}{l}{} &
\multicolumn{1}{r}{$\mu=M_{\perp,W}/2$} &
\multicolumn{1}{r}{$=M_{\perp,W}$} &
\multicolumn{1}{r}{$=2\,M_{\perp,W}$} &
\multicolumn{1}{c}{} &
\multicolumn{1}{r}{$\mu=\hat H_T/2$} &
\multicolumn{1}{r}{$=\hat H_T$} &
\multicolumn{1}{r}{$=2\,\hat H_T$} &
\multicolumn{1}{c}{} &
\multicolumn{1}{r}{$\sigma_\mrm{\scalebox{0.5}{CKKW}}/\mrm{pb}$}\\\hline
\rule[-1.5mm]{0mm}{5mm}LO &$0$&   0.92&0.94&0.95&&   0.91&0.93&0.94&&$496$\\
\rule[-1.5mm]{0mm}{5mm}   &$2$&   1.45&1.02&0.75&&  1.10&0.80&0.59&&$9.90$
\\[1mm]
\rule[-1.5mm]{0mm}{5mm}NLO&$0$&\bf1.26&1.25&1.24&&\bf1.22&1.21&1.21&&\\
\rule[-1.5mm]{0mm}{5mm}   &$2$&\bf1.29&1.18&1.05&&\bf1.21&1.02&0.90&&\\[1mm]
\hline\hline
\end{tabular}
\caption{\label{tab:bkgdKfacs}
  Ratios at LO and QCD NLO taken between rates of \textsc{MCFM} and
  \sherpa CKKW (rightmost column) for inclusive
  $W^+$+$\,n$-jet production at different choices of scales in
  \textsc{MCFM} using MSTW2008 PDFs in all cases. The \textsc{MCFM}
  cross sections are listed in \Tab{tab:mcfm}.}
\end{table}

The $V$\!+jets predictions of \sherpa have been extensively studied
and validated over the last few years. Studies exist for comparisons
against other Monte Carlo tools
\cite{Krauss:2004bs,Krauss:2005nu,Alwall:2007fs,Winter:2007zza,Lenzi:2009fi},
NLO calculations \cite{Krauss:2004bs,Krauss:2005nu,Binoth:2010ra}
and Tevatron Run I and II data \cite{Krauss:2004bs,D0note5066:2006,%
  Abazov:2008ez,Gleisberg:2008ta,Hoeche:2009rj,Abazov:2009av,Abazov:2009pp,%
  Abazov:2010kn}.
They have helped improve \sherpa gradually 
and provided evidence that \sherpa gives a good description of the
shapes of the $V$\!+jet final-state distributions missing a global
scaling factor only, which can be extracted from the data
\cite{Gleisberg:2008ta} or higher-order calculations
\cite{Binoth:2010ra}.
In \App{app:nlo_vs_ckkw} we briefly highlight to what extent the CKKW
ME+PS merging includes important features of NLO computations.

We use the results of \Tab{tab:mcfm} to identify a reasonable
$K$-factor for our simulated $W$\!+jets backgrounds. Relying on MSTW2008
PDFs, the \sherpa numbers for the inclusive $W^+$ and $W^+$+2-jet
cross sections are 496~pb and 9.90~pb, respectively.
The 0-jet \sherpa rate thereby is about 7\% larger than the
corresponding LO rates given by \textsc{MCFM}. The differences occur
because on the one hand \textsc{MCFM} by default invokes a
non-diagonal CKM matrix and a somewhat larger $W$-boson
width\,\footnote{Switching to an unity CKM matrix and using \sherpa's
  input parameters, one finds 486~pb at $\mu=M_W$.}, on the other hand
\sherpa's merged-sample generation relies on a very different
scale-setting procedure compared to the leading fixed-order
calculations. These differences have no effect on the kinematic
distributions -- and are fully absorbed by the $K$-factor, \ie CKM
effects may eventually enter through the correction of \sherpa's rate.
\Tab{tab:bkgdKfacs} summarizes the ratios between the \textsc{MCFM}
predictions of \Tab{tab:mcfm} and \sherpa's CKKW cross sections
mentioned above. This overview neatly points to the two options that
give the most stable ratios; they are found at NLO for
$\mu=M_{\perp,W}/2$ and $\mu=\hat H_T/2$ where the latter scale choice
has been reported to be well suitable for even higher jet
multiplicities \cite{Berger:2009ep,Binoth:2010ra,Berger:2010vm}. Based
on these observations, we can hence conclude that it is fair to apply
a $K$-factor of
\be\label{eq:bkgdK}
K_B\;=\;1.25
\ee
to the $W$\!+jets backgrounds employed in our study. The number found
here compares well to global $K$-factors as reported throughout the
literature.

As outlined at the beginning of \Sec{sec:sherpa}, we want to normalize
the backgrounds obtained with CTEQ6.6 to those computed with MSTW2008
PDFs. In the CTEQ case the \sherpa CKKW cross sections amount to
544~pb and 8.13~pb for the inclusive $W^+$ and $W^+$+2-jet final
states, respectively. Since the latter selection of $W$\!+2-jet events
is more exclusive, we re-scale the CTEQ backgrounds according to
$K^\mrm{66}_B\times8.13\ \mrm{pb}=K_B\times9.90\ \mrm{pb}$ and arrive
at
\be\label{eq:bkgdK66}
K^\mrm{66}_B\;=\;1.52\ .
\ee

\subsubsection{Generation of background events for electroweak and
  top-pair production}

The $WW$\/ background enters at ${\cal O}(\alpha^4_\mrm{ew})$ of the
electroweak coupling constant $\alpha_\mrm{ew}$, \ie it is suppressed
by more than two orders of magnitude with respect to the $W$\!+2-jets
contribution occurring  at ${\cal O}(\alpha^2_\mrm{ew}\alpha^2_\mrm{s})$.
Still, without running the simulation we cannot say for sure whether
the continuum $WW$\/ production remains an 1\% effect after
application of the analysis cuts and -- if necessary -- what handles
exist to distinguish it from the signal. Because of the large
resemblance between the topologies of the Higgs boson decay and the
dominant $WW$\/ production channels, we anticipate some of the cuts to
be equally efficient for both signal and minor background. This makes
it hard to estimate a priori the extent to which the Higgs boson
signal will be diluted by the electroweak production type of
processes. For the same reasons, the $t\bar t$\/ production final
states can be expected to enhance the signal dilution on a similar
level. Certainly, whether we end up with an 1\% or 10\% effect, this
time it is sufficient to apply $K$-factors taken from the literature.

For the simulation of the diboson production background, we take the
complete set of electroweak diagrams occurring at
${\cal O}(\alpha^4_\mrm{ew})$ into account including interference
effects. This way we comprise physics effects beyond the plain $WW$\/
production with subsequent decays of the gauge bosons.%
\footnote{Relying on the full set of electroweak processes is more
  conservative: the rate increases by about 20\%; the effect on the
  shapes is rather small in general, although we observe slightly
  harder tails in $p_T$ distributions.}
As before we only generate the processes regarding the first lepton
family:
\be
P\bar P\to e\nu_e\;pp\to e\nu_e+\mrm{jets}
\ee
where additional jets are produced by the parton shower. Similar
setups have been validated for \sherpa in \cite{Gleisberg:2005qq} and
more recently in \cite{Hoche:2010pf,Hoche:2010kg}. Here, we employ a
dynamic choice, $\mu=\sqrt{\hat s}\sim 2\,M_V$, to calculate the
scales of the LO processes. Parton-level jets are generated as in
\Sec{sec:sherpa.bkgd} using the same jet-finder algorithm and the same
parameters ($Q_\mrm{jet}=20\ \mrm{GeV}$ and $D=0.4$). Processes with
bottom quarks are included; just as in the $W$\!+2-jets case, they are
treated as massless.

The $t\bar t$\/ background events are generated according to
\be
P\bar P\to t\bar t\to b\bar b\;e\nu_e\;pp\to e\nu_e+\mrm{jets}
\ee
again utilizing the parton shower to describe any additional jet
activity beyond that generated by the top quark decays. We only
consider the semileptonic channel. The fully hadronic channel has to
be considered together with the QCD background, and the fully leptonic
channel will suffer from smaller branching fractions, the single
isolated-lepton requirement and any dijet mass window that we impose
around the $W$\/ mass. The LO processes are calculated at the scale
$\mu=m_t$, the mass of the $b$~quarks is fully taken into account and
the partonic phase-space generation is subject to the same jet-finding
constraints as used for the compilation of the electroweak background.
In addition we place mild generation cuts on the $b$~quarks:
$p_{T,b}>10\ \mrm{GeV}$ and $\Delta R_{b,p}>0.3$.

We also examined the impact of $Z$\!+jets production on our analyses,
and found that this contribution makes up less than 1\% of the total
background. Since $Z$\!+jets has kinematics similar to $W$\!+jets, we will
not study it further.


In \sherpa the minor backgrounds are computed at LO. As in all other
cases, we correct the total inclusive cross section for NLO effects by
multiplying with global $K$-factors, which for both electroweak and
$t\bar t$\/ production are larger than 1. Tevatron diboson searches
like \cite{Aaltonen:2009vh,Abazov:2009tr} measure cross sections in
good agreement with the prediction given by Campbell and Ellis
($16.1\pm0.9\ \mrm{pb}$ for $WW$\!+$WZ$). From their work
\cite{Campbell:1999ah} (Table III) we infer an NLO-to-LO $K$-factor
ranging from $1.30$ to $1.35$. For our analysis, we will then use the
conservative estimate\,\footnote{NLO corrections to $VV$\/ production
  can become large, for a recent example, see \cite{Campbell:2011gp}
  where $K$-factors as large as $1.77$ have been reported; taken this
  value, we would certainly overestimate the electroweak contribution,
  since the CDF-type cuts employed in \cite{Campbell:2011gp} are more
  exclusive. As for the shapes, we found them reliably described in a
  cross-check against an electroweak $VV$\!+1-jet merged sample,
  including matrix-element contributions at
  $\cal O(\alpha^4_\mrm{ew}\alpha_\mrm{s})$.}
\be\label{eq:ewK}
K_{B,\mrm{ew}}\;=\;1.35\ .
\ee
For the inclusive $t\bar t$\/ production, we can safely estimate a
conservative $K$-factor of $1.30$ by comparing the cross section
results given for the Tevatron in Ref.~\cite{Cacciari:2008zb}.
Adopting a $b$-tagging efficiency of the order of 50\% would give us a
75\% chance of vetoing $t\bar t$\/ events with at least one $b$-quark
jet, \ie we were able to remove about 3/4 of the $t\bar t$\/
background; again, we will be more conservative here and assume that
about 40\% of the $t\bar t$\/ events will pass; hence, for our
purpose, we finally assign
\be\label{eq:ttK}
K^{b\trm{-}\mrm{veto}}_{B,t\bar t}\;=\;0.52\ .
\ee


\section{Signal versus background studies based on Monte Carlo
  simulations using \protect\sherpa}\label{sec:results}

We report the successive improvements of the $S/\sqrt{B}$
significances when applying a series of cuts that preserve most of the
signal and reduce the inclusive $W$\!+2-jets background significantly.

\subsection{Baseline selection}\label{sec:cuts}

We follow the event-selection procedure as used by the D\O\
collaboration \cite{D0note6095:2010}:\quad hadronic jets $j$\/ are
identified by a seeded midpoint cone algorithm using
the $E$-scheme for recombining the momenta \cite{Blazey:2000qt}. The
cone size is taken as $R=0.5$ and selection cuts of
$p^\mrm{jet}_T>20\ \mrm{GeV}$ and $\abs{\eta^\mrm{jet}}<2.5$ are
imposed. Additionally, we require a lepton--jet isolation of
$\Delta R^{\mrm{lep}\trm{--}\mrm{jet}}>0.4$.
For the leptonic sector, we apply transverse-momentum and
pseudo-rapidity cuts of $p^\mrm{lep}_T>15\ \mrm{GeV}$ and
$\abs{\eta^\mrm{lep}}<1.1$, respectively, supplemented by a
missing-energy cut via $\slashed{p}_T>15\ \mrm{GeV}$. In addition, we
also account for $M_{T,W}+\slashed{E}_T/2>40\ \mrm{GeV}$, which is
known as triangle cut.\footnote{The cut is applied to the leptonic
  $W$\/ boson where $M_{T,W}=
  \sqrt{(\abs{\vec p_{T,\ell}}+\abs{\slashed{\vec p}_T})^2
    -(\vec p_{T,\ell}+\slashed{\vec p}_T)^2}\equiv m_{T,\ell\nu_\ell}$,
  cf.\ \Eq{eq:mt}. For Higgs boson masses above the $WW$\/ threshold,
  the rate reduction and shape changes induced by this cut are
  marginal.}

\subsection{Higgs boson reconstruction based on invariant masses}
\label{sec:hreco}

After the application of the basic cuts, we identify the best-fit
$\{e,\nu_e,j,j'\}$ set from all possible candidates allowed by
combinatorics. The algorithm we use to identify the best-fit object is
referred to as the \textbf{\em Higgs boson candidate selection}. Several
different selection algorithms are possible, however for now, we will
use an invariant mass (or \texttt{invm}) selection: the four particles
(reconstructed in a more or less ideal way) whose combined mass
$m_{e\nu_e jj'}$ is closest to a ``test'' Higgs boson mass $M_h$ are
chosen. Of course, in the context of the analysis, the Higgs boson
mass enters as a hypothesis and, thus, is treated as a parameter.
Regardless of the selection algorithm, we refer to $j$\/ and $j'$ as
the two selected jets, which are not necessarily the hardest jets in
the event.

After selection, we impose a requirement on the absolute difference
between $m_{e\nu_e jj'}$ and the hypothesized Higgs boson mass; events
are kept only if they reconstruct a mass that lies within the window
$M_h-\Delta<m_{e\nu_e jj'}<M_h+\Delta$. This completes our
combinatorial Higgs boson reconstruction, which we label as ``comb.\
$h$-reco'' in our tables. On top of this selection, we may include an
additional dijet mass constraint of $M_W-\delta<m_{jj'}<M_W+\delta$
(marked by $\abs{\td m_{jj'}\!-80}\!<\!\td\delta$ in the tables).
The selection procedure will certainly shape -- to some extent -- the
remaining background to look like the signal, however the primary
effect we are interested in concerns the reduction of the background
rate while we want to preserve as many signal events as possible. 


One may ask whether the reconstruction of the Higgs particle candidate
can be achieved more easily by selecting the set containing the
respective hardest particles, in particular, by choosing the two
hardest jets, $j_1$ and $j_2$, to reconstruct the hadronically
decaying $W$\/ boson. We will refer to this approach as the naive
Higgs boson reconstruction, denoted as ``naive $h$-reco'' later on (as
before we use $\abs{\td m_{j_1j_2}\!-80}\!<\!\td\delta$ in the tables
to indicate that a dijet mass constraint has been imposed in addition).
There are no combinatorial issues in the naive scheme. However, as we
show in our tables, it yields poorer significances than the selection
based on combinatorics.

We calculate the number of $S$\/ signal and $B$\/ background events
for different Higgs boson masses assuming a total integrated
luminosity of ${\cal L}=10\ \mrm{fb}^{-1}$. This seems to be a good
${\cal L}$\/ estimate for what each of the two Tevatron experiments,
CDF and D\O, were able to collect before the eventual Run II shutdown
in September 2011. We compute the numbers according to
\bea\label{eqs:signif}
S&=&K_S\,\veps_S\,\sigma^{(0)}_S\;\times\;2\,f_\ell\,{\cal L}\;=\;
K_S\,\sigma_S\;\times\;2\,f_\ell\,{\cal L}\ ,\nn\\[1mm]
B&=&K_B\,\veps_B\,\sigma^{(0)}_B\;\times\;2\,f_\ell\,{\cal L}\;=\;
K_B\,\sigma_B\;\times\;2\,f_\ell\,{\cal L}
\eea
where $\veps$\/ and $K$\/ respectively denote the total cut
efficiencies and the $K$-factors, which we have worked out in
\Sec{sec:ggh}, cf.\ \Tab{tab:sgnlKfacs}, and \Sec{sec:sherpa}, cf.\
\Eqs{eq:bkgdK},~(\ref{eq:bkgdK66}),~(\ref{eq:ewK})~and~(\ref{eq:ttK}).
The total efficiencies are a product of single-step efficiencies, \ie
$\veps=\prod_i\veps_i$. The factor $f_\ell=2$ accounts for including
the decay channels that involve muons and their associate neutrinos.
Notice that the Higgs boson mass enters in our simulation in
\textbf{\em two}, potentially different ways. In practice, the Higgs
boson mass that we used to generate the signal need not be the same as
the Higgs boson mass we use to formulate the analysis. We refer to the
former as the injected mass $M^\mrm{inj}_h$ in the text, while we have
already introduced the terminology of the latter as the ``test'' or
``hypothesis'' Higgs boson mass $M_h$. However, for simplicity we take
the generation level Higgs boson mass and the analysis level Higgs
boson mass to be equal, $M^\mrm{inj}_h=M_h$. A discussion on how
different generation versus analysis masses would change our results
can be found in the \App{app:hreco}.

\begin{table}[p!]
\centering\footnotesize
\begin{tabular}{|l|c|rrr|rrr|rrr|}\hline
\rule[-2mm]{0mm}{6mm} cuts \& & $2\,\Delta/$
& $\sigma_S/\mrm{fb}$ & $\sigma_B/\mrm{fb}$ & $S/B$
& $\sigma_S/\mrm{fb}$ & $\sigma_B/\mrm{fb}$ & $S/B$
& $\sigma_S/\mrm{fb}$ & $\sigma_B/\mrm{fb}$ & $S/B$\\
\rule[-2mm]{0mm}{6mm} selections & $\mrm{GeV}$
& $\veps_S$ & $\veps_B$ & $S/\sqrt{B}$
& $\veps_S$ & $\veps_B$ & $S/\sqrt{B}$
& $\veps_S$ & $\veps_B$ & $S/\sqrt{B}$\\\hline\hline
\multicolumn{2}{|r|}{
  \rule[-3mm]{0mm}{8mm}$M_h/\mrm{GeV}\quad[\delta/\mrm{GeV}]$}
&\multicolumn{3}{c|}{$165\quad[20]$}
&\multicolumn{3}{c|}{$170\quad[20]$}
&\multicolumn{3}{c|}{$180\quad[20]$}\\\hline\hline
\rule[-1.5mm]{0mm}{5mm} $\sigma^{(0)}$ &
& $19.35$ & $220\cdt10^4$ & 20$\cdt10^{\trm-6}$
& $17.77$ & $220\cdt10^4$ & 18$\cdt10^{\trm-6}$
& $14.19$ & $220\cdt10^4$ & 14$\cdt10^{\trm-6}$\\
\rule[-1.5mm]{0mm}{5mm} &
& 1.0 & 1.0 & \bf 0.21
& 1.0 & 1.0 & \bf 0.19
& 1.0 & 1.0 & \bf 0.15\\\hline
\rule[-1.5mm]{0mm}{5mm} lepton \& &
& $10.66$ & $984\cdt10^3$ & 24$\cdt10^{\trm-6}$
& $9.869$ & $984\cdt10^3$ & 22$\cdt10^{\trm-6}$
& $7.946$ & $984\cdt10^3$ & 18$\cdt10^{\trm-6}$\\
\rule[-1.5mm]{0mm}{5mm} MET cuts &
& 0.551 & 0.45 & \bf 0.17
& 0.555 & 0.45 & \bf 0.15
& 0.560 & 0.45 & \bf 0.12\\\hline
\rule[-1.5mm]{0mm}{5mm} as above \& &
& $8.572$ & $191\cdt10^2$ & 0.0010
& $7.967$ & $191\cdt10^2$ & 92$\cdt10^{\trm-5}$
& $6.471$ & $191\cdt10^2$ & 74$\cdt10^{\trm-5}$\\
\rule[-1.5mm]{0mm}{5mm} $\ge2$ jets &
& 0.443 & 0.0087 & \bf 0.99
& 0.448 & 0.0087 & \bf 0.90
& 0.456 & 0.0087 & \bf 0.72\\\hline
\rule[-1.5mm]{0mm}{5mm} as above \& &
& $5.195$ & $6997$ & 0.0017
& $4.735$ & $6997$ & 0.0015
& $3.691$ & $6997$ & 0.0011\\
\rule[-1.5mm]{0mm}{5mm} $\abs{\td m_{j_1j_2}\!-80}\!<\!\td\delta$ &
& 0.269 & 0.0032 & \bf 0.99
& 0.266 & 0.0032 & \bf 0.88
& 0.260 & 0.0032 & \bf 0.68\\\hline\hline
\rule[-1.5mm]{0mm}{5mm} naive $h$-reco & $50$
& $5.422$ & $6492$ & 0.0019
& $4.983$ & $6492$ & 0.0017
& $3.911$ & $6749$ & 0.0013\\
\rule[-1.5mm]{0mm}{5mm} &
& 0.280 & 0.0030 & \bf 1.07
& 0.280 & 0.0030 & \bf 0.96
& 0.276 & 0.0031 & \bf 0.73\\\hline
\rule[-1.5mm]{0mm}{5mm} naive $h$-reco & $30$
& $3.948$ & $4108$ & 0.0022
& $3.897$ & $4108$ & 0.0021
& $3.039$ & $4199$ & 0.0016\\
\rule[-1.5mm]{0mm}{5mm} &
& 0.204 & 0.0019 & \bf 0.98
& 0.219 & 0.0019 & \bf 0.95
& 0.214 & 0.0019 & \bf 0.72\\\hline
\rule[-1.5mm]{0mm}{5mm} naive $h$-reco & $48$
& $4.657$ & $2965$ & 0.0035
& $4.214$ & $3210$ & 0.0029
& $3.232$ & $3539$ & 0.0020\\
\rule[-1.5mm]{0mm}{5mm} $\abs{\td m_{j_1j_2}\!-80}\!<\!\td\delta$ &
& 0.241 & 0.0013 & \bf 1.36
& 0.237 & 0.0015 & \bf 1.16
& 0.228 & 0.0016 & \bf 0.84\\\hline
\rule[-1.5mm]{0mm}{5mm} naive $h$-reco & $20$
& $3.080$ & $1374$ & 0.0051
& $2.876$ & $1512$ & 0.0042
& $2.219$ & $1676$ & 0.0029\\
\rule[-1.5mm]{0mm}{5mm} $\abs{\td m_{j_1j_2}\!-80}\!<\!\td\delta$ &
& 0.159 & 62$\cdt10^{\trm-5}$ & \bf 1.33
& 0.162 & 69$\cdt10^{\trm-5}$ & \bf 1.15
& 0.156 & 76$\cdt10^{\trm-5}$ & \bf 0.83\\\hline\hline
\rule[-1.5mm]{0mm}{5mm} comb.\ $h$-reco & $50$
& $7.105$ & $6816$ & 0.0024
& $6.557$ & $7117$ & 0.0020
& $5.241$ & $7396$ & 0.0015\\
\rule[-1.5mm]{0mm}{5mm} &
& 0.367 & 0.0031 & \bf 1.37
& 0.369 & 0.0032 & \bf 1.21
& 0.369 & 0.0034 & \bf 0.94\\\hline
\rule[-1.5mm]{0mm}{5mm} comb.\ $h$-reco & $20$
& $4.827$ & $3094$ & 0.0035
& $4.577$ & $3191$ & 0.0032
& $3.657$ & $3255$ & 0.0024\\
\rule[-1.5mm]{0mm}{5mm} &
& 0.249 & 0.0014 & \bf 1.38
& 0.258 & 0.0015 & \bf 1.26
& 0.258 & 0.0015 & \bf 0.99\\\hline
\rule[-1.5mm]{0mm}{5mm} comb.\ $h$-reco & $50$
& $6.346$ & $3336$ & 0.0043
& $5.884$ & $3697$ & 0.0035
& $4.679$ & $4098$ & 0.0025\\
\rule[-1.5mm]{0mm}{5mm} $\abs{\td m_{jj'}\!-80}\!<\!\td\delta$ &
& 0.328 & 0.0015 & \bf 1.75
& 0.331 & 0.0017 & \bf 1.51
& 0.330 & 0.0019 & \bf 1.12\\\hline
\rule[-1.5mm]{0mm}{5mm} comb.\ $h$-reco & $30$
& $5.586$ & $2217$ & 0.0057
& $5.159$ & $2488$ & 0.0046
& $4.083$ & $2756$ & 0.0032\\
\rule[-1.5mm]{0mm}{5mm} $\abs{\td m_{jj'}\!-80}\!<\!\td\delta$ &
& 0.289 & 0.0010 & \bf 1.89
& 0.290 & 0.0011 & \bf 1.61
& 0.288 & 0.0013 & \bf 1.20\\\hline
\rule[-1.5mm]{0mm}{5mm} comb.\ $h$-reco & $20$
& $4.616$ & $1525$ & 0.0068
& $4.280$ & $1731$ & 0.0054
& $3.404$ & $1933$ & 0.0038\\
\rule[-1.5mm]{0mm}{5mm} $\abs{\td m_{jj'}\!-80}\!<\!\td\delta$ &
& 0.239 & 69$\cdt10^{\trm-5}$ & \bf 1.89
& 0.241 & 79$\cdt10^{\trm-5}$ & \bf 1.60
& 0.240 & 88$\cdt10^{\trm-5}$ & \bf 1.19\\\hline
\rule[-1.5mm]{0mm}{5mm} comb.\ $h$-reco & $16$
& $4.075$ & $1235$ & 0.0074
& $3.784$ & $1396$ & 0.0060
& $3.017$ & $1575$ & 0.0042\\
\rule[-1.5mm]{0mm}{5mm} $\abs{\td m_{jj'}\!-80}\!<\!\td\delta$ &
& 0.211 & 56$\cdt10^{\trm-5}$ & \bf 1.85
& 0.213 & 63$\cdt10^{\trm-5}$ & \bf 1.58
& 0.213 & 72$\cdt10^{\trm-5}$ & \bf 1.17\\\hline
\rule[-1.5mm]{0mm}{5mm} comb.\ $h$-reco & $10$
& $3.025$ & $787.9$ & 0.0087
& $2.624$ & $905.0$ & 0.0064
& $2.103$ & $1006$  & 0.0046\\
\rule[-1.5mm]{0mm}{5mm} $\abs{\td m_{jj'}\!-80}\!<\!\td\delta$ &
& 0.156 & 36$\cdt10^{\trm-5}$ & \bf 1.72
& 0.148 & 41$\cdt10^{\trm-5}$ & \bf 1.36
& 0.148 & 46$\cdt10^{\trm-5}$ & \bf 1.02\\\hline\hline
\end{tabular}
\caption{\label{tab:cutimpactA1}
  Impact of the different levels of cuts on the $e\nu_e$+jets final
  states for the $gg\to h\to WW$\/ production and decay signal and the
  $W$\!+jets background as obtained from \sherpa. Cross sections
  $\sigma_S$, $\sigma_B$, acceptances $\veps_S$, $\veps_B$ and $S/B$,
  $S/\sqrt{B}$ ratios are shown for Higgs boson masses of
  $M_h=165\ \mrm{GeV}$, $M_h=170\ \mrm{GeV}$ and $M_h=180\ \mrm{GeV}$.
  Note that $\td m_{ij}=m_{ij}/\mrm{GeV}$ and
  $\td\delta=\delta/\mrm{GeV}$. Significances were calculated using
  \Eqs{eqs:signif} assuming ${\cal L}=10\ \mrm{fb}^{-1}$ of integrated
  luminosity, counting both electrons and muons and combining Tevatron
  experiments.}
\end{table}

We can now go ahead and calculate the $S/B$\/ ratios and $S/\sqrt B$\/
significances. For various Higgs boson mass hypotheses, \Tab{tab:cutimpactA1}
and \Tabs{tab:cutimpactA2}--\ref{tab:cutimpactAx} of \App{app:hreco}
list signal and $W$\!+jet-background cross sections, acceptances,
$S/B$\/ ratios and significances at different levels of cuts for the
selection procedures discussed in this subsection. The \sherpa
simulation runs obtained with the MSTW2008 LO PDFs have been used to
extract the results of all tables except those of
\Tab{tab:cutimpact66} presented in \App{app:hreco} which are based on
a set of runs taken with the CTEQ6.6 PDFs. In \App{app:hreco} we will
then also briefly discuss the differences that can be seen between the
predictions for the two PDF sets.

We now turn to the discussion of the tables. Their setup is as
follows: the rows represent different stages in the cut-flow, Higgs
boson reconstruction strategies, and mass window cuts, while the third
through fifth columns contain the outcomes for different Higgs boson
masses. The second column indicates the mass window cut (in
$\mrm{GeV}$, referred to as $\Delta$ in the text), which has been
applied to all reconstructed Higgs boson candidates. Similarly, the
number in square brackets next to each Higgs boson mass is the dijet
mass window cut (referred to as $\delta$, also in $\mrm{GeV}$). At
every analysis level, six numbers are displayed for each Higgs boson
mass. The top row displays the LO signal cross section (in
$\mrm{fb}$), the LO $W$\!+jets cross section (in $\mrm{fb}$) and $S/B$\/
at $10\ \mrm{fb}^{-1}$ of integrated luminosity, calculated including
$K$-factors and factors of 2 following \Eqs{eqs:signif}. The bottom
three numbers in each table entry are the signal and background
efficiencies and $S/\sqrt B$. Of these entries, $S/\sqrt B$\/ is
displayed in bold. For the first set of tables,
\Tab{tab:cutimpactA1}~and~\ref{tab:cutimpactA2} (see \App{app:hreco}),
we concentrate on Higgs boson masses greater than
$\approx 162\ \mrm{GeV}$ -- above the $WW$\/ threshold. Higgs boson
masses below the $WW$\/ threshold have additional challenges, which we
explore in a later subsection.

The rows are divided into three groups. In the first group, rows 1--4,
the baseline selection cuts, as described in \Sec{sec:cuts}, are
applied.\footnote{Note that the ``lepton \& MET cuts'' level also
  includes a lepton--jet separation of
  $\Delta R^{\mrm{lep}\trm{--}\mrm{jet}}>0.4$ in the presence of
  jets.}
In the second group, rows 5--8, events are selected using the
``naive'' criteria, then retained if their reconstructed sum falls
within various Higgs boson and dijet mass windows. Finally, in the
last set of rows, 9--15, we select events with the ``comb.\ $h$-reco''
algorithm, then apply several different mass windows. The effect of
the mass window cuts, with either the ``naive'' or ``comb.\ $h$-reco"
selection scheme, are fairly intuitive; mass windows always help
because they emphasize the peaks in the signal in comparison to a
featureless $W$\!+jets background. Tighter mass windows are usually, but
not always, better. Clearly, among the three groups the combinatorial
selections give the best significances, followed by the naive ones,
which already improve over the baseline selection cuts.

Comparing rows with identical cuts but different selections (``naive''
versus ``comb. $h$-reco''), such as rows 5 and 9, or 7 and 11, the
combinatorial Higgs boson reconstruction is better across all Higgs
boson masses by roughly 30\%. The difference can be traced to events
where one of the hardest jets comes from I/FSR rather than from one of
the jets of the $W$\/ decay. Had we truncated our treatment of the
background at the matrix-element level (or even at matrix-element
level plus some Gaussian smearing, as in Ref.~\cite{Han:1998ma},
additional jet activity arising from I/FSR would be absent and the
``comb. $h$-reco'' scheme would give the same result as the ``naive
$h$-reco'' scheme. Incorporating these relevant I/FSR jets using a
complete, matrix-element plus parton-shower treatment of the
background, we notice that the ``naive'' scheme is no longer the best
option. The ME+PS merging thereby allows us a fully inclusive
description of $W$\!+2-jet events on almost equal footing with the
related NLO calculation, however with the advantage of accounting for
multiple parton emissions at leading-logarithmic accuracy. These
effects are pivotal to obtain reliable results for the combinatorial
selections.

Showering effects are not just limited to the background. In
particular, the width of the Higgs boson candidates reconstructed from
showered events is much broader than the reconstructed width derived
from parton level. In fact, after showering, the reconstructed Higgs
boson peak is typically so broad that the tightest mass windows used
in the tables ($\Delta=5\ \mbox{and}\ 8\ \mrm{GeV}$) cut out some of
the signal and yield worse significances than broader windows. For
example, the combinatorial selections supplemented by a dijet mass
window yield FWHM of about $10\ \mrm{GeV}$ at the shower level, while
the FWHM at the parton level are reduced down to $2\ \mrm{GeV}$ --
that basically is the width of one bin. If we relied on the
matrix-element level results, we would obtain far too promising
$S/\sqrt B$. Focusing on the $M_h=180\ \mrm{GeV}$ test point and the
``comb.\ $h$-reco'' with a dijet mass window, we would find the
significances increasing from $1.5$ for $\Delta=25\ \mrm{GeV}$, $2.2$
for $\Delta=10\ \mrm{GeV}$ to $3.0$ for $\Delta=5\ \mrm{GeV}$. These
numbers should be contrasted with those in \Tab{tab:cutimpactA1},
namely $1.12$, $1.19$ and $1.02$, respectively.

\begin{figure}[t!]
\psfrag{Significance}[c][c][0.92]{
  \fontfamily{phv}\selectfont{Significance}\quad$S/\sqrt{B}$}
\psfrag{Higgs boson mass}[c][c][0.92]{
  \fontfamily{phv}\selectfont{Higgs boson mass}\quad$M_h$}
\centering\vskip2mm
  \includegraphics[clip,width=0.74\columnwidth,angle=-90]{%
    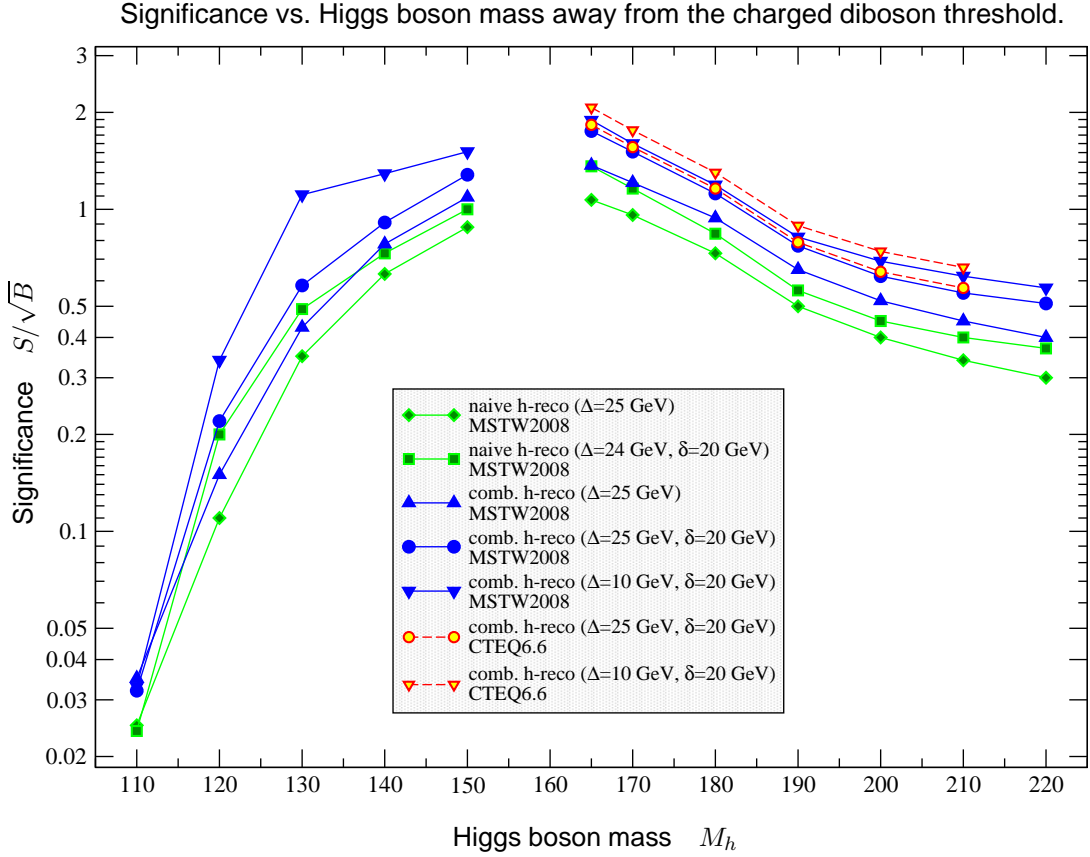}
  \caption{\label{fig:invm_significances}
    $S/\sqrt{B}$ significances for Higgs boson masses varying from
    $M_h=110$ to $220\ \mrm{GeV}$ after different levels of cuts. The
    numbers are taken from \Tabs{tab:cutimpactA1} and
    \ref{tab:cutimpactA2}--\ref{tab:cutimpactAx},
    which reflect in more detail the outcome of the analysis based on
    the \texttt{invm}\ selection procedure for $e\nu_e$+jets final
    states originating from the $gg\to h\to WW$\/ signal and the
    $W$\!+jets background. Results are shown for Higgs boson masses
    below and above the $WW$\/ mass threshold; the threshold region
    has been left out though. All significances were calculated
    according to \Eqs{eqs:signif} under the assumption of an
    integrated luminosity of ${\cal L}=10\ \mrm{fb}^{-1}$ and
    including electron and muon channels, \ie $f_\ell=2$.}
\end{figure}

To conclude this discussion, it is illustrative to show a plot of the
significances versus Higgs boson masses for various selections as
presented in the tables (\Tab{tab:cutimpactA1} and
\Tabs{tab:cutimpactA2}--\ref{tab:cutimpactAx} in \App{app:hreco}), all
of which is summarized in \Fig{fig:invm_significances}. The
significance, at least after this level of analysis, reaches a maximum
of $\sim2.0$. The highest significance occurs, as expected, near the
$WW$\/ threshold. For heavier Higgs bosons, as the $WW$\/ decay mode
becomes subdominant to the $ZZ$\/ mode, the significance drops slowly,
reaching $\sim1.0$ at $M_h\sim 185\ \mrm{GeV}$. By gradually enhancing
the selections the gain in significance remains approximately equal
over the whole region of large $M_h$; this is indicated by the
parallel shifts of the respective significance curves. Hence, the
differences seen in the significances per $M_h$ test point are mainly
driven by the behaviour of the total inclusive cross section for the
signal. Looking back at \Tabs{tab:cutimpactA1}~and~\ref{tab:cutimpactA2},
we in fact realize that the acceptances $\veps_S$ and $\veps_B$ are
rather similar at any selection step (for each row), only mildly
varying across the different Higgs boson test mass points. For Higgs
boson masses below threshold (\Tab{tab:cutimpactBelow} in
\App{app:hreco}), as we will discuss in the next sections, the
drop-off is more severe. Not only does the branching fraction to
$WW^*$ fall rapidly, but the signal becomes more background-like once
the two $W$\/ bosons from the Higgs boson decay cannot both be
produced on-shell. The significances shown in \Fig{fig:invm_significances}
reflect our best estimates, however, we have also performed several
checks on the stability of these significances under slight variations
in the analysis. These checks not only include -- as mentioned earlier
-- varying parton distribution functions, but also varying jet
definitions, etc. and are summarized in \App{app:hreco}.

\subsubsection{Reconstruction below the on-shell diboson mass threshold}

In the above-threshold case there is good hope that the idealized
approach of considering the neutrino as a fully measurable particle
will not lead to results, which are sizably different from those
obtained by a realistic treatment of neutrinos. This is based on the
fact that in most cases the leptonic $W$\/ will be on its mass shell.
The approximation $m_{e\nu_e}\approx M_W$ can in principle be used to
determine the neutrino's longitudinal momentum -- up to a twofold
ambiguity -- by employing the lepton and missing transverse energy
measurements. Below the $WW$\/ mass threshold one of the $W$\/ bosons
will be off-shell, so that the simple ansatz in calculating
$p_{\parallel,\nu_e}$ will be rather inaccurate. Hence, it a priori is
not clear whether an event selection based on invariant-mass windows
will give an overall picture that can be maintained in more realistic
scenarios. Nevertheless here we briefly establish what kind of
significances may be achievable assuming we had knowledge about the
off-shellness (the actual mass) of the leptonic $W$\/ boson. This will
give us a benchmark, which we may use to assess more realistic
reconstruction approaches.

When we apply the same analysis as above the $WW$\/ threshold, we find
significances as summarized in \Tab{tab:cutimpactBelow} of
\App{app:hreco}. They are visualized in \Fig{fig:invm_significances}.
The numbers demonstrate that we quickly lose sensitivity below the
$WW$\/ threshold, in particular for test points
$M_h\lesssim130\ \mrm{GeV}$. This happens for three reasons (which
apply to the signal only): one factor is the decline of the total
inclusive signal cross section $\sigma^{(0)}_{S,\mrm{NNLO}}$ towards
lower $M_h$, which actually is comparable to that seen for large
$M_h$. As shown in \Tab{tab:sgnlKfacs}, this effect is not as drastic
as one would assume from the drop in the $h\to W^*W^*$ branching
ratios; it is partly compensated by the rising gluon--gluon fusion
rate for low $M_h$. In contrast to the above-threshold case, there are
yet two more factors coming into the equation. Firstly, the basic
selection cuts affect the signal more severely;\footnote{For
  $M_h=110\ \mrm{GeV}$, only about 7\% of the events survive, while
  45--49\% of the signal is kept above threshold (cf.\ the respective
  1st rows in \Tab{tab:cutimpactBelow} and 3rd rows in
  \Tabs{tab:cutimpactA1}~and~\ref{tab:cutimpactA2}).}
secondly, the low $M_h$ signals that pass the baseline selection are
often penalized because of substantial off-shell effects. In
particular, the Higgs boson propagator can be pushed far off-shell and
the Higgs boson reconstruction will fall outside the mass window, such
that the event will be discarded. The tendency for lighter Higgs
bosons to go off-shell increases, since the basic cuts make it
extremely unlikely for the leptonic and hadronic $W$\/ masses to drop
below $\sim30\ \mrm{GeV}$.

\begin{figure}[t!]
  \centering\vskip2mm
  \includegraphics[clip,width=0.67\columnwidth,angle=-90]{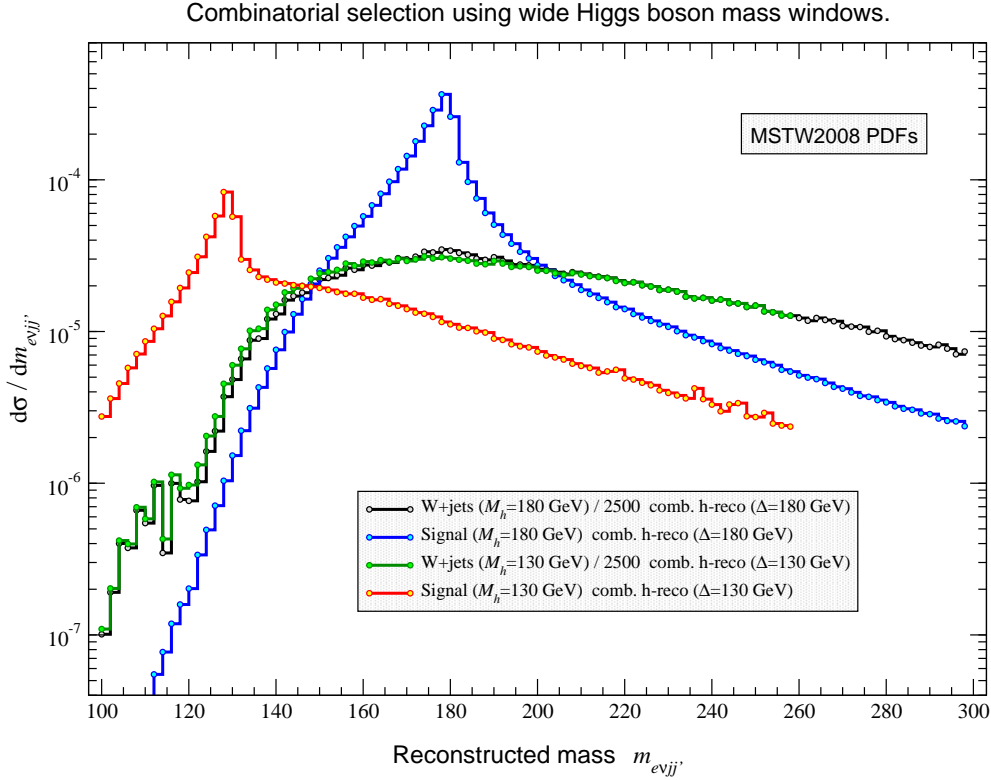}
  \caption{\label{fig:mass_spectrum}
    Mass spectra $m_{e\nu_e jj'}$ after the combinatorial
    reconstruction of Higgs boson candidates for very wide Higgs boson
    mass windows. Results are shown for $M_h=130\ \mrm{GeV}$ and
    $M_h=180\ \mrm{GeV}$ and $e\nu_e$+jets final states originating
    from the $gg\to h\to WW$\/ signal (peaked distributions) and the
    $W$\!+jets background (flat distributions).}
\end{figure}

\Fig{fig:mass_spectrum} shows the $m_{e\nu_e jj'}$ spectra including
shower effects for signals and backgrounds at $M_h=130$ and
$180\ \mrm{GeV}$ after the combinatorial Higgs boson reconstruction
has been applied using wide Higgs boson mass windows
($\Delta\equiv M_h$). The parton showering washes out the peaks,
therefore reduces and broadens them. Both signal distributions develop
a softer tail above $M_h$ as a result of the jet combinatorics. For
$M_h=130\ \mrm{GeV}$, the tail plateaus due to the off-shell effects
mentioned earlier. \Fig{fig:mass_spectrum} also illustrates why the
value of the significance jumps up significantly (as shown in
\Fig{fig:invm_significances}) when we tighten the Higgs boson mass
window from $\Delta=25$ to $10\ \mrm{GeV}$ for $M_h=130\ \mrm{GeV}$.
This effect arises because we place our window cuts in a steeply
rising $W$\!+jets background.

When we studied which choice of mass window gives us the best results
in terms of separating signal from background, it came as somewhat of
a surprise that we did not have to alter the additional dijet mass
constraint of $M_W-\delta<m_{jj'}<M_W+\delta$, $\delta=20\ \mrm{GeV}$.
Our studies indicate that it is helpful to have the hadronically
decaying $W$\/ boson to be close to its on-shell mass $M_W$. The $W$\/
boson decaying leptonically is then forced to go off-shell
($m_{e\nu_e}<M_W$), a kinematic configuration at odds with most
$W$\!+jets events.
Cutting on $m_{jj'}$ therefore helps suppress the dominant background
and, moreover, should also be convenient to demote the production of
multi-jets efficiently.

For the tighter Higgs boson mass windows, our results show that a
simple one-sided lower cut on $m_{jj'}$, \ie $m_{jj'}>M_W-\delta$ is
slightly more efficient than using any type of dijet mass window. The
one-sided cut improves the significances as given in
\Tab{tab:cutimpactBelow} by 1--2\%. The removal of the upper bound on
$m_{jj'}$ has however negligible effects on selections using broad
Higgs boson mass windows. As a consequence of keeping an $m_{jj'}$
constraint the leptonically decaying $W$\/ will almost always be
off-shell, such that the reconstruction of the longitudinal component
of the neutrino's four-momentum cannot succeed without a good guess of
the mass of the $e\nu_e$\/ pair. We will address this issue in
\Sec{sec:realreco}.

\subsubsection{Effect of the subdominant backgrounds}

In this section, we examine to what extent the significances of the
ideal Higgs boson reconstruction will be diluted by contributions from
the electroweak and top-pair production of the $e\nu_e$+jets final
states. To this end we apply the analysis as established so far,
without any modification.

\begin{figure}[t!]
\centering\vskip5mm
\centerline{
  \includegraphics[clip,width=0.397\columnwidth,angle=-90]{%
    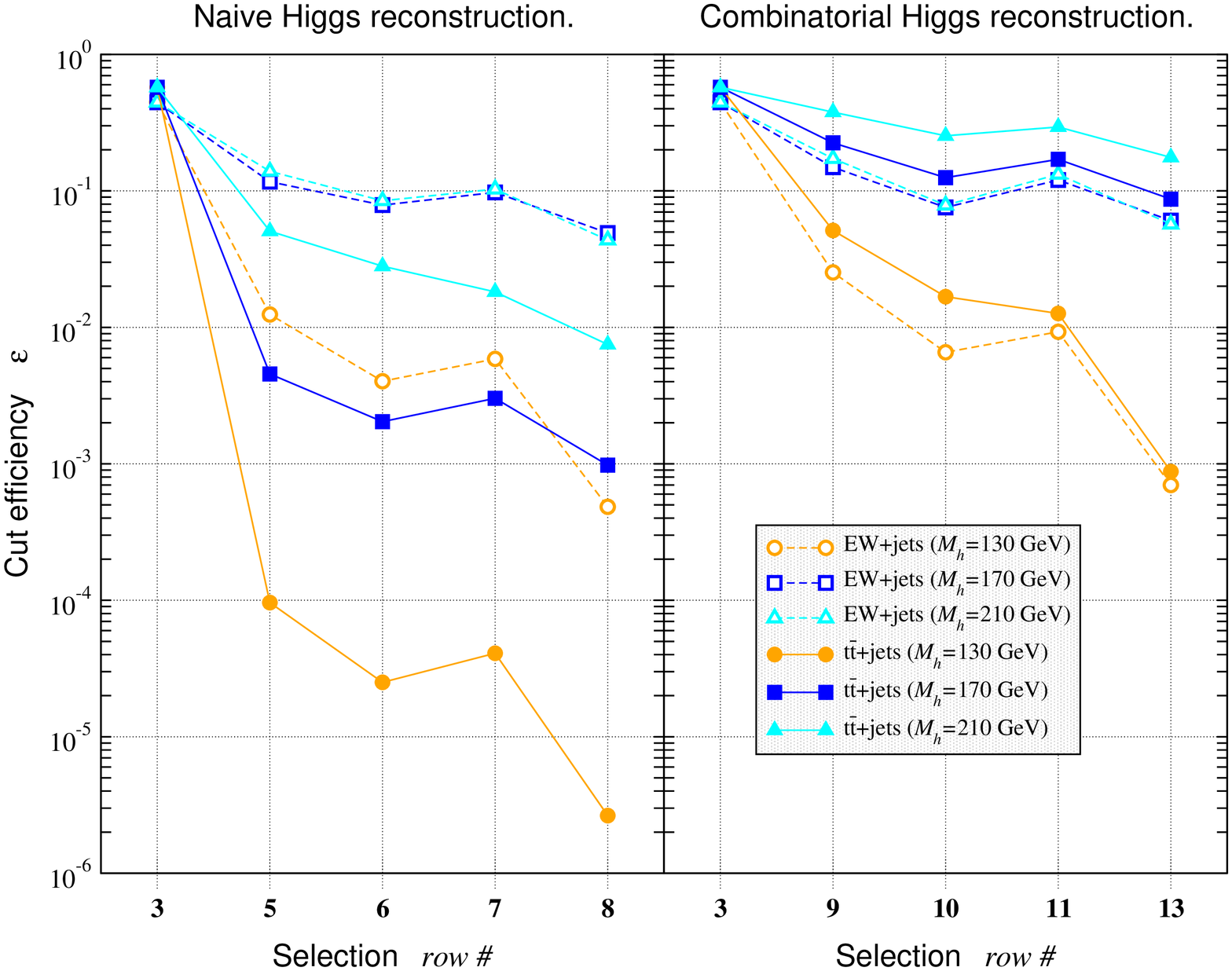}
  \includegraphics[clip,width=0.397\columnwidth,angle=-90]{%
    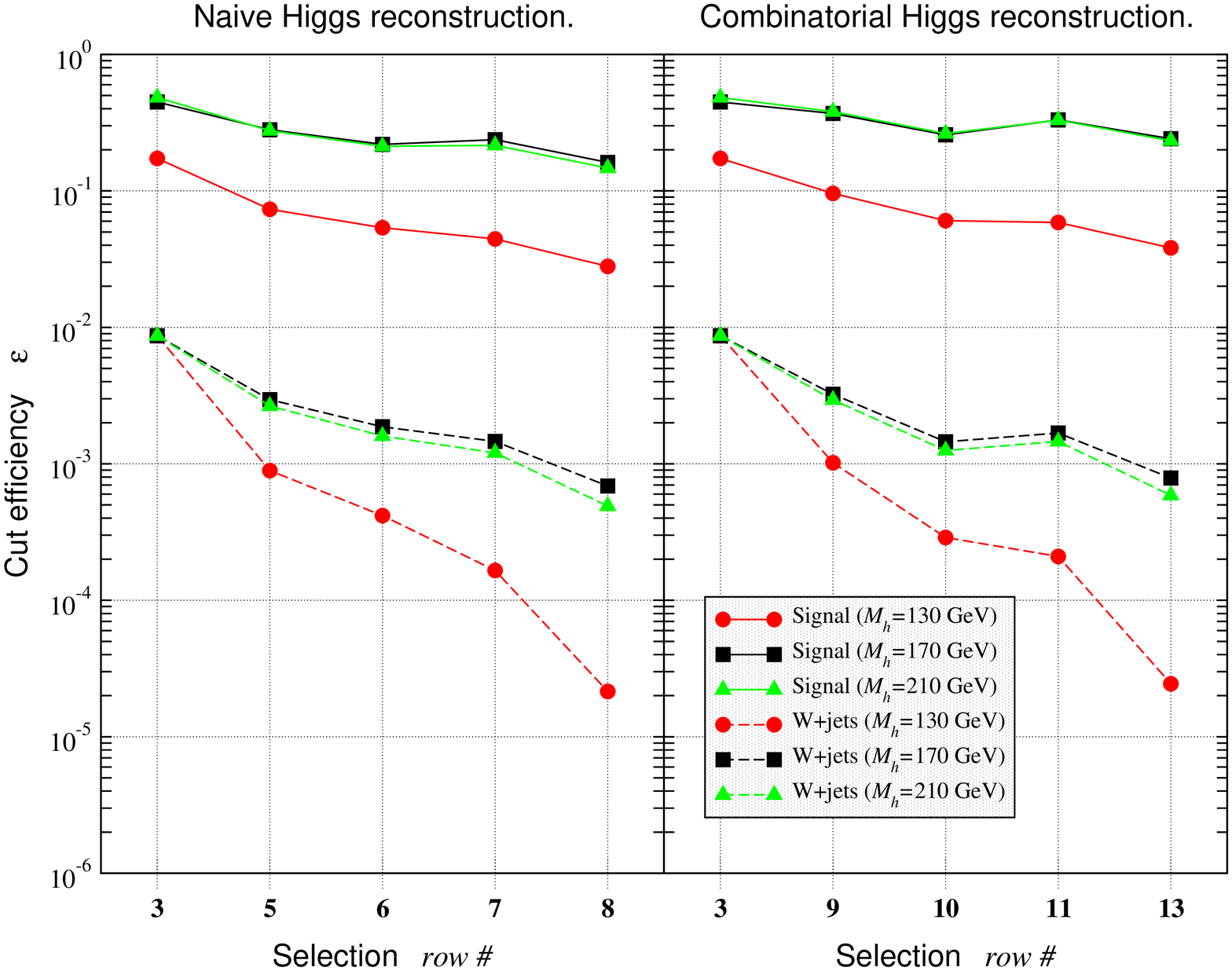}}
\caption{\label{fig:invm_efficiencies}
  Cut efficiencies $\veps_B$ and $\veps_S$ for $e\nu_e$+jets final
  states and different \texttt{invm}\ naive and combinatorial Higgs
  boson reconstructions taking the mass points $M_h=130$, $170$ and
  $210\ \mrm{GeV}$. The selections are labelled by the row numbers as
  assigned in \Tabs{tab:cutimpactA1} and
  \ref{tab:cutimpactA2}--\ref{tab:cutimpactAx}; row 3 marks the
  baseline selection (used as benchmark). The left pane exhibits the
  efficiencies found for the minor backgrounds -- electroweak and
  top--antitop pair production (dashed and solid lines, respectively)
  -- whereas the right pane displays the $\veps_B$ for the $W$\!+jets
  background (dashed lines) as well as the $\veps_S$ of the $gg\to
  h\to WW$\/ signal (solid lines). The two plots to the right compared
  with each other nicely visualize why the combinatorial outperforms
  the naive selection: the signal cut efficiencies get increased,
  while, for $W$\!+jets, the cuts remain about as effective as for the
  naive approach. Also notice the drop of the $M_h=130\ \mrm{GeV}$
  signal curves -- they show the penalty in employing the same basic
  cuts as above the $WW$\/ threshold.}
\end{figure}

\begin{figure}[t!]
\psfrag{Significance}[c][c][0.92]{
  \fontfamily{phv}\selectfont{Significance}\quad$S/\sqrt{B}$}
\psfrag{Higgs boson mass}[c][c][0.92]{
  \fontfamily{phv}\selectfont{Higgs boson mass}\quad$M_h$}
\centering\vskip2mm
  \includegraphics[clip,width=0.74\columnwidth,angle=-90]{%
    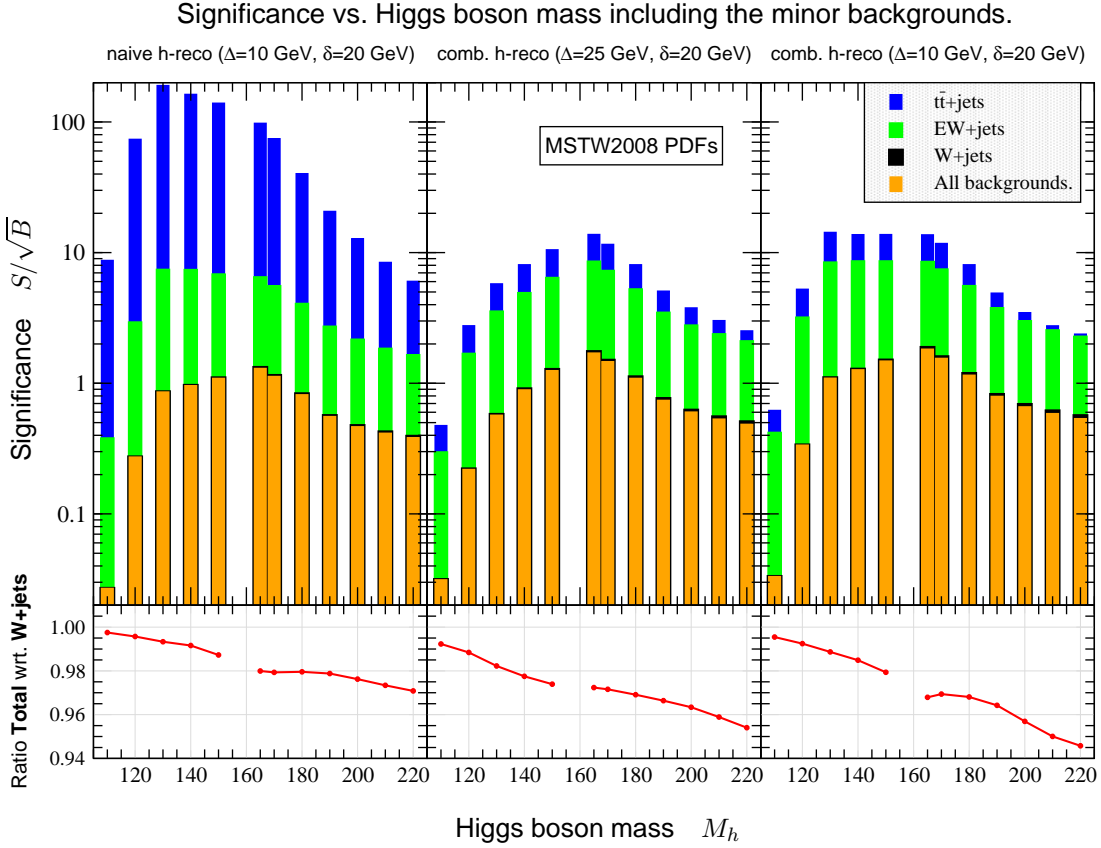}
  \caption{\label{fig:invm_totsignificances}
    Single-background and total significances as a function of $M_h$
    for three different Higgs boson candidate \texttt{invm}\
    selections as denoted on top of each panel. The $e\nu_e$+jets
    final states are generated from the $gg\to h\to WW$\/ signal,
    $W$\!+jets, electroweak and $t\bar t$\/ production backgrounds. All
    $S/\sqrt{B_i}$ were calculated according to \Eqs{eqs:signif}
    assuming an integrated luminosity of ${\cal L}=10\ \mrm{fb}^{-1}$
    and including electron and muon channels, \ie $f_\ell=2$. The
    total-background significances were obtained with \Eq{eq:tot_signif}.
    The lower plots hold the ratios of $S/\sqrt{B_\mrm{tot}}$ over
    $S/\sqrt{B}$ for $W$\!+jets only. Note that the large $t\bar t$\/
    significances obtained by the naive selection (left panel) result
    from the failure of the relatively harder leading-jet pairs to
    satisfy the mass window constraints.}
\end{figure}

The first thing to notice is the total inclusive LO cross sections for
these minor backgrounds are ${\cal O}(1)\ \mrm{pb}$ -- substantially
smaller than the $W$\/ production contribution. After the application
of the basic cuts, the inclusive $e\nu_e$+2-jet cross sections drop to
about $0.5\ \mrm{pb}$, a factor of 40 below the major background.
Including all the various $K$-factors, see \Tab{tab:sgnlKfacs} and
\Eqs{eq:bkgdK},~(\ref{eq:ewK}),~(\ref{eq:ttK}), we find that the total
significance,
\be\label{eq:tot_signif}
\frac{S}{\sqrt{B_\mrm{tot}}}\;=\;
\frac{1}{\sqrt{\,\sum_i\left(\frac{S}{\sqrt{B_i}}\right)^{-2}}\,}\;=\;
\left(\left(\frac{S}{\sqrt{B}}\right)^{-2}+
\left(\frac{S}{\sqrt{B_\mrm{ew}}}\right)^{-2}+
\left(\frac{S}{\sqrt{B^{b\trm{-}\mrm{veto}}_{t\bar t}}}\right)^{-2}
\right)^{-\frac{1}{2}}\ ,
\ee
at the basic selection level is only 2\% smaller compared to the
significance $S/\sqrt{B}$ using only $W$\!+jets.\footnote{The cross
  sections stated are LO-like cross sections as obtained with \sherpa:
  before (after) the basic cuts, we find $1.21$ and $0.886\ \mrm{pb}$
  ($540$ and $508\ \mrm{fb}$) for the electroweak and $t\bar t$\/
  backgrounds, respectively; the resulting single-background
  significances turn out to be almost 6 and 10 times larger than the
  $W$\!+jets $S/\sqrt{B}$.}

Switching from the baseline selection to a combinatorial selection and
finite dijet mass window, the single minor-background significances
improve by up to 50\% above (100\% below) the $WW$\/ threshold. So,
for beyond-baseline $h$\/ reconstructions, the significance
corrections owing to the inclusion of the minor backgrounds will be of
the same order as before. This is documented in both
\Figs{fig:invm_efficiencies}~and~\ref{fig:invm_totsignificances}. In
the former we compare the cut efficiencies between all backgrounds and
the signal (shown together with the $W$\!+jets background in the plots
to the right in \Fig{fig:invm_efficiencies}) for various naive and
combinatorial Higgs boson candidate selections. Firstly, no background
cut efficiency ever exceeds any of the $\veps_S$. In all cases the
background curves decrease more strongly when tightening the
selection. Secondly, the pattern we observe for the minor-background
cut efficiency curves resembles by and large those of the major
background.\footnote{In particular, the minor-background cut
  efficiencies show more pronounced drops, if one enhances the
  baseline to a Higgs boson candidate selection, introduces the dijet
  mass window $\delta$\/ or tightens the $\Delta$ Higgs boson mass
  range. Notice that the naive Higgs boson reconstruction very
  efficiently beats down the $t\bar t$\/ background. This is because
  the two leading jets turn out harder compared to all other cases.
  The presence of a sufficient number of subleading jets however makes
  the selection based on jet combinatorics pick a pair of soft jets,
  and, on the contrary almost too effective for Higgs boson masses
  above the $WW$\/ threshold.}
For these reasons, the single-background significances plotted versus
$M_h$ follow the trend found for the $W$\!+jets contribution, but remain
well above the $W$\!+jets significances.\footnote{Both the electroweak
  and $t\bar t$\/ production significances show the same strong
  enhancement around $M_h=130\ \mrm{GeV}$ as a result of the effect
  discussed around \Fig{fig:mass_spectrum} which is due to the use of
  a tight Higgs boson mass window. As for $W$\!+jets, the minor
  backgrounds fall rapidly for decreasing $M_h$.}
All of which is exemplified in \Fig{fig:invm_totsignificances} using
three Higgs boson candidate selections, which impose a dijet mass
window (corresponding to the rows 8, 11 and 13 in
\Tabs{tab:cutimpactA1},~\ref{tab:cutimpactA2}~and~\ref{tab:cutimpactAx}),
namely the naive method with $\Delta=10\ \mrm{GeV}$ (left panel) and
the combinatorial method with the same and broader window of
$\Delta=25\ \mrm{GeV}$ (middle panel). The total significances
$S/\sqrt{B_\mrm{tot}}$ resulting from combining the three single
backgrounds are also shown. In fact, they only decrease by 1--5\% as
demonstrated by the ratio plots in the lower part of
\Fig{fig:invm_totsignificances}. The high $M_h$ region is found to
receive the larger, ${\cal O}(\mbox{5\%})$ corrections once the
electroweak and $t\bar t$\/ contributions are included in the overall
background. As noted early in \Sec{sec:rel_bkgd}, experimenters have
estimated this fraction of events with 5\% implying a 2.5\% drop in
significance. It is reassuring to be able to confirm this expectation
with our results. Slightly contrary to the expectation, we identify
the electroweak as the leading minor background in all selections.

Based on these results it is easy to conclude; at this stage of our
analysis we do not have to worry about contributions from minor
backgrounds. Although additional handles exist to further reduce these
backgrounds or supplement the (here conservatively chosen) $b$-jet
veto, it is of far more importance to find ways to diminish the
$W$\!+jets background. We postpone this discussion until
\Sec{sec:optimize}.

\subsection{More realistic Higgs boson reconstruction methods}
\label{sec:realreco}

Up to this point we have ignored one big problem, namely the neutrino
problem. In our selection based on the reconstruction of invariant
masses -- which we dubbed \texttt{invm} approach -- we currently treat
neutrinos as if we were able to measure them like leptons. This is, of
course, unrealistic and before we can talk about further significance
improvements, we have to investigate in which way our analysis may
fall short when switching to more experimentally motivated Higgs boson
candidate selections. Under experimental conditions, missing energy is
taken from the $\vec p_T$ imbalance in the event. However, in our
analysis we then make a small simplification and identify the missing
energy with the neutrino's transverse momentum as given by the Monte
Carlo simulation.


There are multiple choices for how to proceed.
Recall that whatever method we pick acts as a selection criterion; we
decide which two jets to keep in the event based on these variables,
therefore we want to design variables, which are best at correctly
picking out the jets from a Higgs boson decay. One way to proceed is
to give up complete reconstruction and to work solely with transverse
quantities; this is clean and unambiguous, but we throw out
information. The second approach is to attempt to guess the
longitudinal neutrino momentum by requiring that some or all of the
final-state objects reconstruct an object we expect, such as a $W$\/
or Higgs boson. Full reconstruction then gives access to a larger set
of observables, therefore keeps more handles and information, but it
is also more ambiguous.

\begin{table}[t!]
\centering\small\vskip5mm
\begin{tabular}{ccrcrcrcr}\hline
  \multicolumn{1}{c}{\rule[-3.5mm]{0mm}{9mm}$M_h/M_W$} &
  \multicolumn{1}{c}{\hphantom{xxx}} &
  \multicolumn{1}{r}{\texttt{pzmw}} &
  \multicolumn{1}{c}{} &
  \multicolumn{1}{r}{\texttt{pzmh}} &
  \multicolumn{1}{c}{} &
  \multicolumn{1}{r}{\texttt{mt}} &
  \multicolumn{1}{c}{\hphantom{xxxx}} &
  \multicolumn{1}{r}{\texttt{mtp}}\\\hline
  \rule[-3mm]{0mm}{8mm}$<2$&&$m^{(\nu_e)}_{T,e\nu_e jj'}$ &&
  $m^{(\nu_e)}_{T,e\nu_e jj'}$ &&
  $m^{(\nu_e)}_{T,e\nu_e jj'}$ && $m^{(\nu_e)}_{T,e\nu_e jj'}$\\[1mm]
  \rule[-3mm]{0mm}{8mm}$>2$&&$m_{T,e\nu_e jj'}$ &&
  $m^{(\nu_e)}_{T,e\nu_e jj'}$ &&
  $m^{(\nu_e)}_{T,e\nu_e jj'}$ && $m_{T,e\nu_e jj'}$\\[1mm]\hline
\end{tabular}
\caption{\label{tab:mtchoice}The preferred choice of definition for
  the 4-particle transverse mass shown for each of the more realistic
  Higgs boson candidate selections. The $m_T$ definitions are given in
  \Eqs{eq:mt}~and~(\ref{eq:mtvis}).}
\end{table}

To remove some of the combinatorial headache, we use $e\nu_e jj'$\/
and $jj'$\/ (transverse) mass windows as before; moreover, we can
impose criteria on subsets of the event. For example, if the
(3-particle) mass of the visible system $m_{ejj'}$ is greater than the
test Higgs boson mass, that particular choice of jets is unphysical
and we can move on to the next choice. A second constraint we often
impose is that the 4-particle transverse mass does not exceed the
upper bound on the Higgs boson mass window:
$m_{T,e\nu_e jj'}\le M_h+\Delta$. As to the definition of $m_T$, we
generally use the definitions stated in \Sec{sec:strategy}, see
\Eqs{eq:mt}~and~(\ref{eq:mtvis}). For our selections, we found that
the distinction of the two $m_T$ definitions in fact only matters when
we calculate the 4-particle transverse masses. Accordingly, each
selection comes in two versions either using $m_{T,e\nu_e jj'}$ or
$m^{(\nu_e)}_{T,e\nu_e jj'}$. In \Tab{tab:mtchoice} we summarize for
each type which version is more appropriate to use and in what
context. Whenever we refer to a specific selection in due course, we
understand it according to the findings listed in \Tab{tab:mtchoice}.

With these criteria in hand, the different selection methods are
specified as follows:
\bi
\item\underline{\texttt{mt}:}\quad
  we want to test a method where the final state of the Higgs boson
  decay will be identified purely with the help of transverse masses
  rather than invariant masses. To this end we calculate
  $m_{T,e\nu_e jj'}$ according to \Eqs{eq:mt}~or~(\ref{eq:mtvis}) and
  prefer the final state giving us the 4-particle transverse mass
  closest to $M_h$. Owing to $m_{T,e\nu_e jj'}\le m_{e\nu_e jj'}$ the
  test mass window is placed on the lower side only,
  $M_h-2\,\Delta<m_{T,e\nu_e jj'}<M_h$, with double the size as
  compared to the other selections. The advantage, but also
  disadvantage of the method is there is no reconstruction. Avoiding
  reconstruction eliminates uncertainties owing to constraining masses
  plus resolving ambiguities, but means we have no access to
  longitudinal and invariant-mass observables involving the neutrino.
\ei
For the next two selections, we aim to approximately determine the
longitudinal momentum of the neutrino, $p_{\parallel,\nu_e}$, using
knowledge about which value for $m_{e\nu_e}$ should likely be
reconstructed by the combined system,$p_e+p_{\nu_e}$.%
\footnote{Provided the MET cut was passed, we assume here that all MET
  in the event has been produced by a single neutrino.}
When we write
\bea\label{eq:recoansatz}
  \lefteqn{m^2_{\ast,i\nu_e k..}\;\approx\;m^2_{i\nu_e k..}\;=\;}\nn\\[1mm]
  && m^2_{ik..}\,+\,2\left(\sqrt{m^2_{ik..}+\,\vec p^2_{ik..}}
  \sqrt{\vec p^2_{T,\nu_e}+p^2_{\parallel,\nu_e}}-
  \vec p_{T,ik..}\cdot\vec p_{T,\nu_e}-
  p_{\parallel,ik..}\,p_{\parallel,\nu_e}\right)\ ,
\eea
using the ``(in)visible'' subsystem notation, we note that such
problems can be solved up to a twofold ambiguity. The difference among
the two selections lies in how particles are grouped in
\Eq{eq:recoansatz} and how the twofold ambiguity is resolved.%
\footnote{If the solutions are complex-valued, we only assign the real
  part to describe $p_{\parallel,\nu_e}$ with no ambiguity left to
  resolve.}
\bi
\item\underline{\texttt{pzmw}:}\quad
  in this selection we use the $W$\/ mass constraint to solve for the
  neutrino momentum: $m^2_{*,e\nu_e}=M^2_W$. The ambiguity is then
  resolved by picking the neutrino $p_z$ solution, which brings the
  reconstructed mass $m_{e\nu_e jj'}$ more closely to the Higgs boson
  test mass $M_h$. For true signal events, it then is more likely to
  find the reconstructed $m_{e\nu_e jj'}$ matching the Higgs boson mass.

  \hskip4.7mm The tricky part is to pick the best choice for the
  $m^2_{*,e\nu_e}$ constraint -- meaning is $M^2_W$ always optimal
  given that the $W$\/ boson may be off-shell? For \texttt{pzmw}, we
  do the following: first, we inspect the transverse mass,
  $m_{T,e\nu_e}$, of the $e\nu_e$ subsystem in each event. If
  $m_{T,e\nu_e}\ge M_W$ we choose $m_{*,e\nu_e}=m_{T,e\nu_e}$,
  otherwise we pick $m_{*,e\nu_e}=M_W$ as long as $M_h>2\,M_W$ or
  $0.9<m_{T,e\nu_e}/M_W<1.0$. That is, above and around the $WW$\/
  threshold, we take $m_{e\nu_e}$ towards $M_W$. If below threshold
  and $m_{T,e\nu_e}/M_W<0.9$, the mass estimate is chosen taking
  various subsystem invariant and transverse masses into account but
  enforcing $m_{*,e\nu_e}$ to lie between $m_{T,e\nu_e}$ and $M_W$.
  For example, if $m_{jj'}>2\,m_{T,jj'}$ we set
  $m_{*,e\nu_e}=m_{T,e\nu_e jj'}-m_{jj'}$ while otherwise
  $m_{*,e\nu_e}=m_{T,e\nu_e jj'}-m_{T,jj'}$ unless
  $m_{ejj'}>m_{T,e\nu_e jj'}$ where we say $m_{*,e\nu_e}=m_{T,e\nu_e}$.
\item\underline{\texttt{pzmh}:}\quad\indent
  we again infer the neutrino's longitudinal momentum from mass
  constraints. Although technically similar to \texttt{pzmw} -- with
  the ``visible'' subsystem entering \Eq{eq:recoansatz} now being
  $\{e,j,j'\}$ -- we here turn the idea around and already require
  $m_{e\nu_e jj'}\approx M_h$ in order to solve for $p_{\parallel,\nu_e}$.
  That is to say we enforce the combined system, $p_{ejj'}+p_{\nu_e}$,
  to mimic a Higgs boson signal mass while leaving us with reasonable
  leptonic $W$\/ masses $m_{e\nu_e}$ at the same time. When
  reconstructing the signal these observables are likely correlated,
  while for the background they are uncorrelated apart from kinematic
  constraints.

  \hskip4.7mm The details of the method are: we specify the target
  mass via $m_{*,e\nu_e jj'}=M_h$ unless we find $m_{T,e\nu_e jj'}/M_h\ge0.94$,
  \ie the 4-particle transverse mass turns out too large already so
  that $m^2_{*,e\nu_e jj'}=m^2_{T,e\nu_e jj'}/0.95$ is the more
  appropriate choice. We approximate the leptonic $W$\/ boson mass by
  $m_{*,e\nu_e}=M_h-m_{jj'}$ freezing it at $m_{*,e\nu_e}=M_W$ if this
  difference exceeds $M_W$. We however require
  $\min(m_{*,e\nu_e})=m_{T,e\nu_e}$. Taking this estimate, we can form
  the absolute difference $\delta m_{e\nu_e}=|m_{e\nu_e}-m_{*,e\nu_e}|$
  using the reconstructed mass $m_{e\nu_e}$ for each possible neutrino
  solution. In the presence of two solutions we define, as a measure
  of the longitudinal activity,
  $b_{ij}=m_{\perp,ij}\exp|y_{ij}|=\max\{E_{ij}\pm p_{\parallel,ij}\}$
  with $m^2_{\perp,ij}=m^2_{ij}+p^2_{T,ij}$ and pick the solution that
  generates the smaller $b_{e\nu_e}$, \ie the $e\nu_e$ subsystem less
  likely going forward. We do so unless the other solution's
  $\delta m_{e\nu'_e}$ drops below $\delta m_{e\nu_e}$ and
  $(b_{jj'}+b_{e\nu'_e})/(m_{jj'}+m_{e\nu'_e})<
  (b_{jj'}+b_{e\nu_e})/(m_{jj'}+m_{e\nu_e})+\delta x$ is satisfied;
  this is when we pick conversely ($\delta x=0.5$ if $M_h<2\,M_W$
  otherwise $\delta x=1.0$). Finally, we ensure that the
  $\{e,\nu_e,j,j\}$ set minimizing $\delta m_{e\nu_e}$ will be
  preferred by the overall selection among all sets reconstructing the
  same 4-particle mass. Note that we do not reject the selected
  ensemble if the $\delta m_{e\nu_e}$ deviation becomes too large; we
  leave this potential to be exploited by supplemental cuts, which we
  discuss in \Sec{sec:optimize}.
\ei

\begin{figure}[t!]
\psfrag{Significance}[c][c][0.92]{
  \fontfamily{phv}\selectfont{Significance}\quad$S/\sqrt{B}$}
\psfrag{Higgs boson mass}[c][c][0.92]{
  \fontfamily{phv}\selectfont{Higgs boson mass}\quad$M_h$}
\centering\vskip2mm
  \includegraphics[clip,width=0.74\columnwidth,angle=-90]{%
    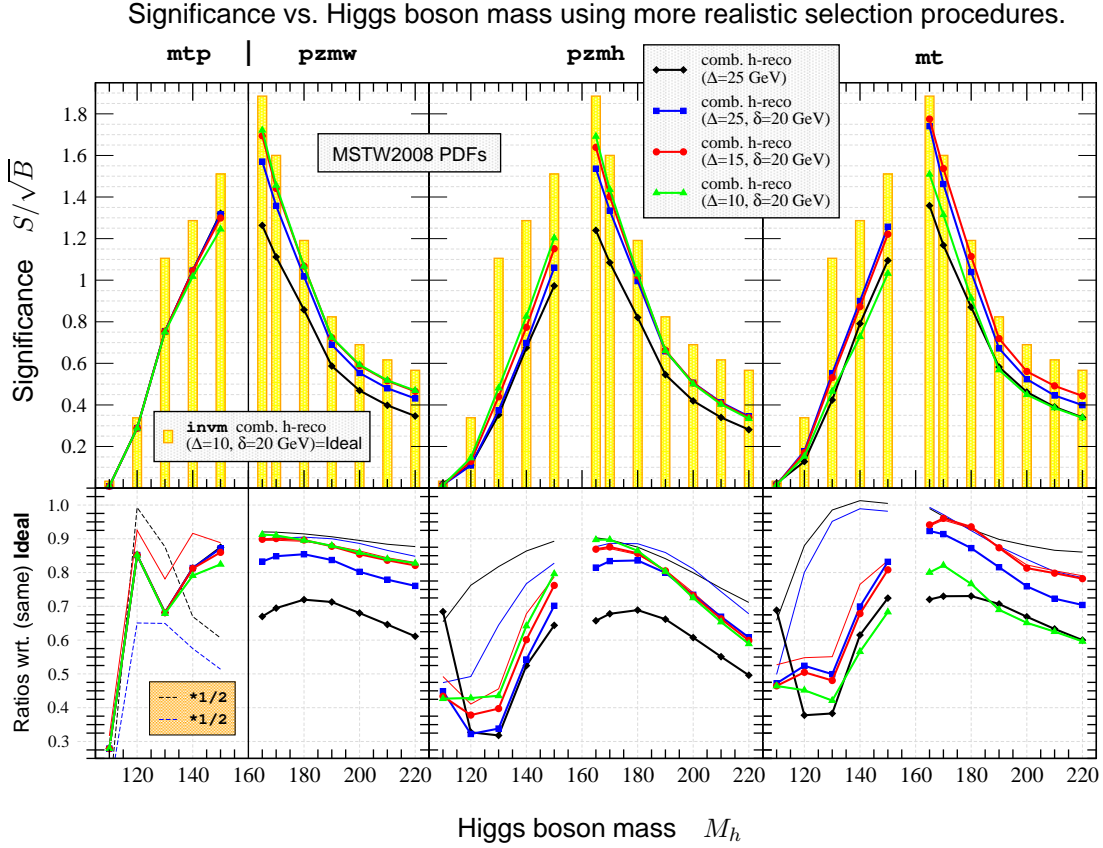}
  \caption{\label{fig:real_significances}
    Single $W$\!+jets background significances as a function of $M_h$
    for 4 different realistic Higgs boson candidate selections using
    jet combinatorics. The selection types are denoted on top of each
    panel. Results are shown for 4 different mass window parameter
    settings each, overlaying the sort of optimal case defined by the
    \texttt{invm} combinatorial $h$\/ reconstruction for $\Delta=10$,
    $\delta=20\ \mrm{GeV}$, which also serves as the main reference.
    The $e\nu_e$+jets final states are generated from the signal,
    $gg\to h\to WW$, and the $W$\!+jets background. All $S/\sqrt{B}$
    were calculated according to \Eqs{eqs:signif} assuming an
    integrated luminosity of ${\cal L}=10\ \mrm{fb}^{-1}$ and
    including electron and muon channels, \ie $f_\ell=2$. The lower
    plots hold the ratios of realistic over ideal $S/\sqrt{B}$. The
    thick lines are drawn with respect to the main reference, while
    the thin lines are taken from comparing to the \texttt{invm}
    combinatorial selection relying on the same window parameters.}
\end{figure}

\begin{figure}[t!]
\centering\vskip2mm
  \includegraphics[clip,width=0.74\columnwidth,angle=-90]{%
    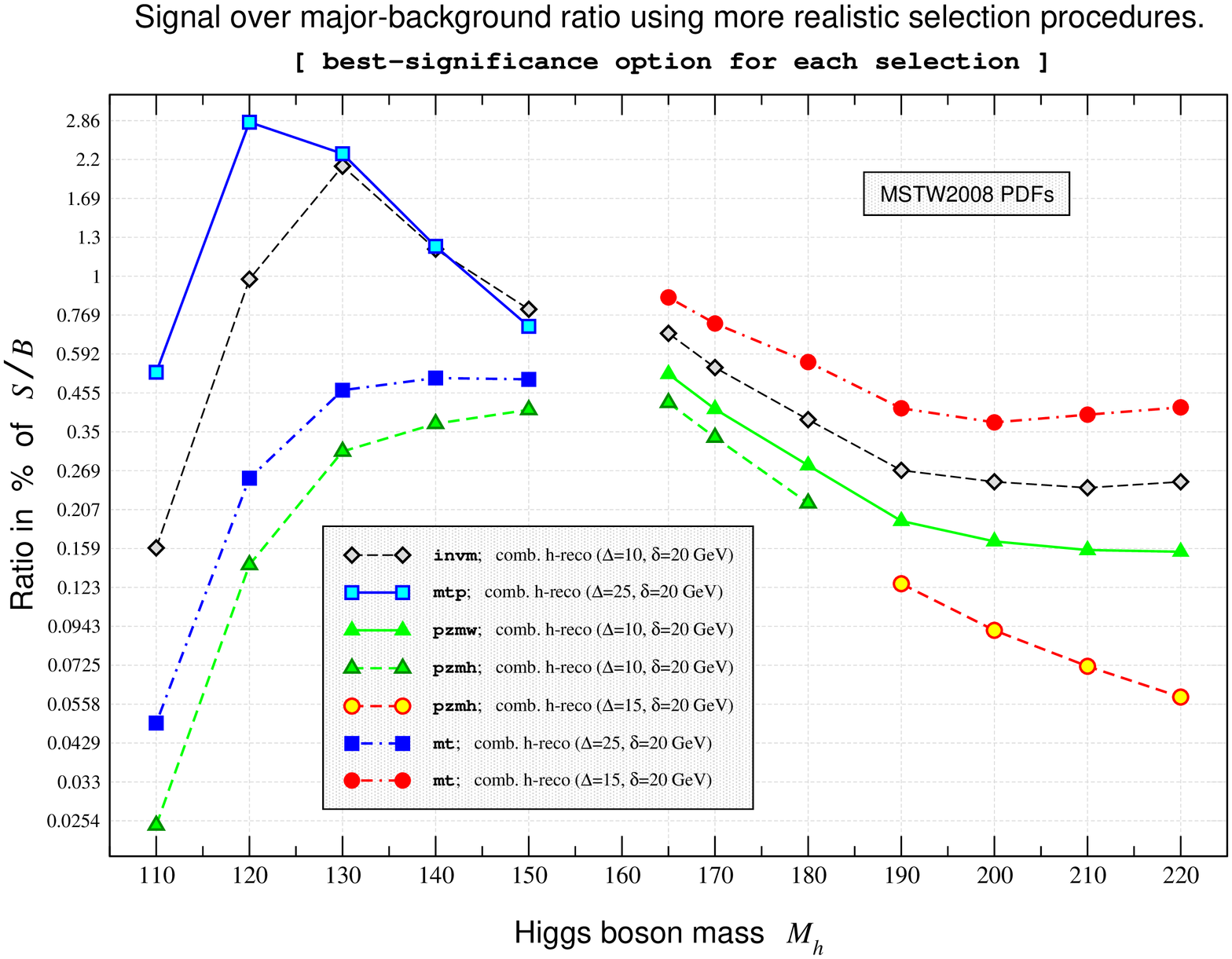}
  \caption{\label{fig:real_soverb}
    $S/B$ ratios as a function of $M_h$ for 4 types of realistic
    combinatorial Higgs boson candidate selections, each with their
    window parameters chosen as to reach maximal significance. The
    $e\nu_e$+jets final states are generated from the $gg\to h\to
    WW$\/ signal and $W$\!+jets background. The $S/B$\/ were calculated
    from the $\sigma_S$ and $\sigma_B$ as obtained after the
    selection, the signal $K$-factors of \Tab{tab:sgnlKfacs} and $K_B$
    as given in \Eq{eq:bkgdK}.}
\end{figure}

We explored in detail how each of these selection criteria compare to
the ideal case. \Fig{fig:real_significances} shows the significances
per $M_h$ test point that we can achieve running the different
combinatorial selections for various, reasonable window parameters. In
the upper panels we display them directly on top of the (quasi
optimal) ideal case, \ie the \texttt{invm} combinatorial $h$\/
candidate selection. The plots on the right and in the center
respectively exhibit the results of the \texttt{mt} and \texttt{pzmh}
methods for the whole $M_h$ test range. The \texttt{pzmw} method
yields similar, yet slightly worse results below the $WW$\/ mass
threshold compared to \texttt{pzmh}. Therefore, we split the leftmost
pane into two subplots: on the right, one finds the \texttt{pzmw}
results for masses above threshold; on the left we then already reveal
the outcome of the \texttt{mtp} selection, whose discussion we
postpone until the next subsection.

The bottom-row plots of \Fig{fig:real_significances} depict the
respective survival fractions: the significance ratios for each
selection and different window parameters always taken with respect to
the quasi optimal case (cf.\ the thick lines with symbols). To
indicate the net effect of the more realistic approaches, we also show
in each case the significance loss relative to the \texttt{invm}
selection using exactly the same mass windows as the realistic one
(cf.\ the thin lines with no symbols). We observe serious losses,
larger than 50\%, if one uses selections with tighter window
parameters and/or runs for $M_h$ points further away from $2\,M_W$.
The reconstruction methods (\texttt{pzwm}, \texttt{pzmh}) do not work
better than 90\% of the quasi optimal case. This only is improved by
the \texttt{mt} approach making use of broad Higgs boson mass windows
where one can reach up to about 100\%. However, lowering $\Delta$ here
quickly results in losing sensitivity. Related to the quasi optimal
\texttt{invm} selection, the various results point us to work with
medium-sized Higgs boson mass windows always imposing the dijet mass
cuts. Tighter $\Delta$ constraints may help improve the outcome of the
reconstruction types, but are of disadvantage in the measurement.

In all selections the below-threshold region is especially
problematic. One might settle for 65--80\% of the ideal significances,
but if we want to get a better handle on the low $h$\/ boson masses,
in particular include the $M_h=130\ \mrm{GeV}$ mass point, we have to
push further -- which we do so in \Sec{sec:realreco_below}.

Focusing on the above-threshold region, we see that \texttt{pzmw}
slightly outperforms \texttt{pzmh} over the whole range; only for the
near-threshold region up to $M_h=180\ \mrm{GeV}$ this is topped by the
\texttt{mt} selection for medium-sized $\Delta$ windows. This is
somewhat surprising, but seems plausible, if one considers that signal
events are central in rapidity and manifest themselves in larger
transverse activity on average. This has to be opposed to the
$W$\!+2-jets background whose events tend to populate phase space more
along the beam direction, which generates $y_{e\nu_e jj'}$
distributions peaking about half an unit away from zero rapidity.
Nevertheless the differences between the 3 methods are not conclusive
per se; to some extent the selections will shape distributions
differently and it is easy to imagine the picture changing if
additional cuts are imposed. But, one has to bear in mind, there is a
second, very important criterion, the ratio $S/B$, which one wants to
maximize. For their optimal window parameters, we show in
\Fig{fig:real_soverb} the $S/B$\/ curves of the more realistic and
ideal selections as functions of $M_h$. Even more surprisingly than
before, the \texttt{mt} outperforms the neutrino reconstruction
methods and, moreover, the ideal $S/B$\/ are also beaten unless
$M_h<2\,M_W$. Based on this observation one may prefer the methods
where the selection utilizes transverse masses, with the only drawback
of having no $p_{\parallel,\nu_e}$ estimate available.

\subsubsection{More realistic reconstruction below the on-shell diboson
  mass threshold}
\label{sec:realreco_below}

We have just seen that the significances achievable in more realistic
scenarios drop off considerably below the $WW$\/ mass threshold,
amplifying the loss already present in the ideal case. Therefore, it
is of great importance to learn how the reconstruction methods
described above can be applied more efficiently in the below-threshold
region.

\begin{figure}[t!]
  \centering\vskip5mm
  \centerline{
    \includegraphics[clip,width=0.397\columnwidth,angle=-90]{%
      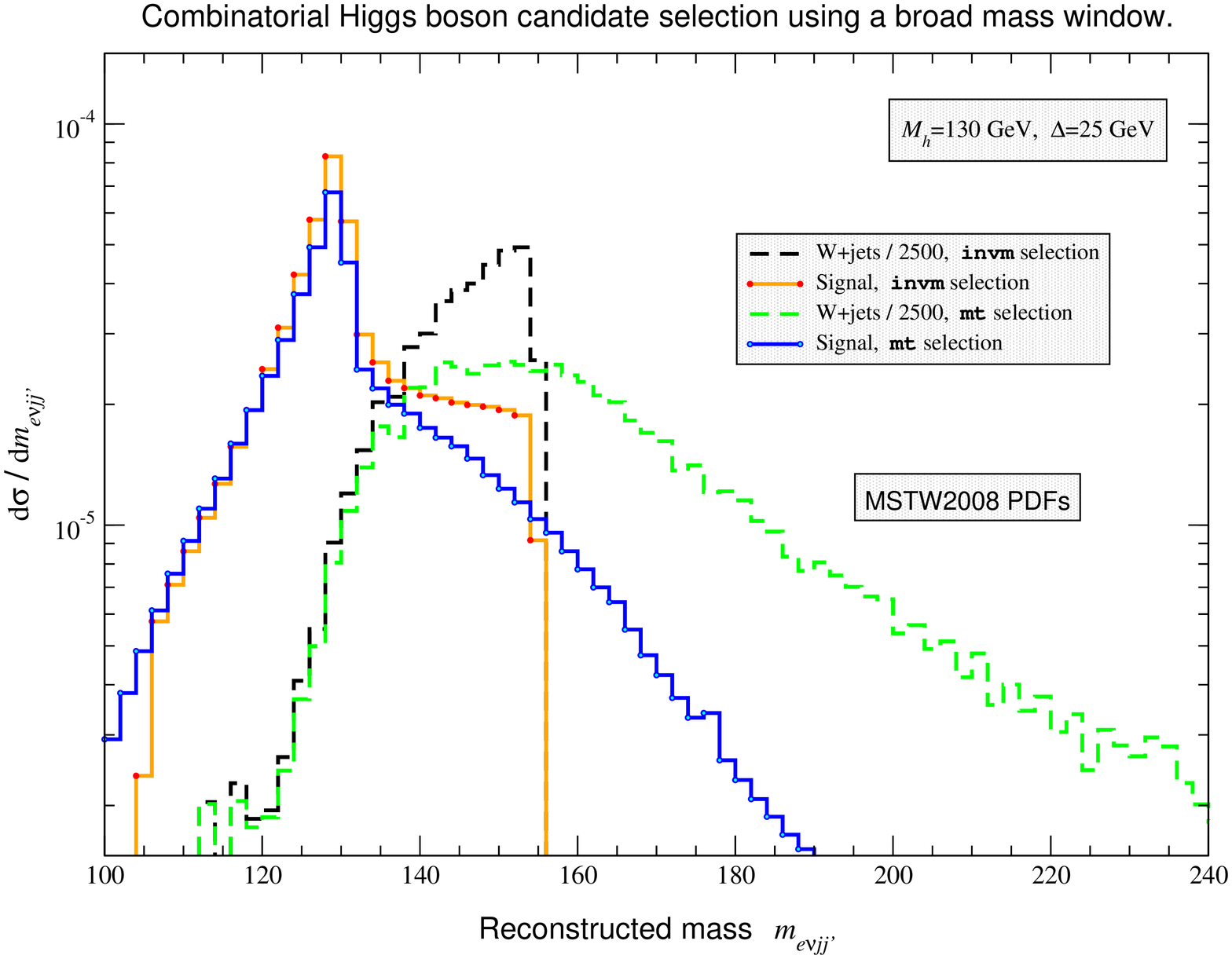}
    \includegraphics[clip,width=0.397\columnwidth,angle=-90]{%
      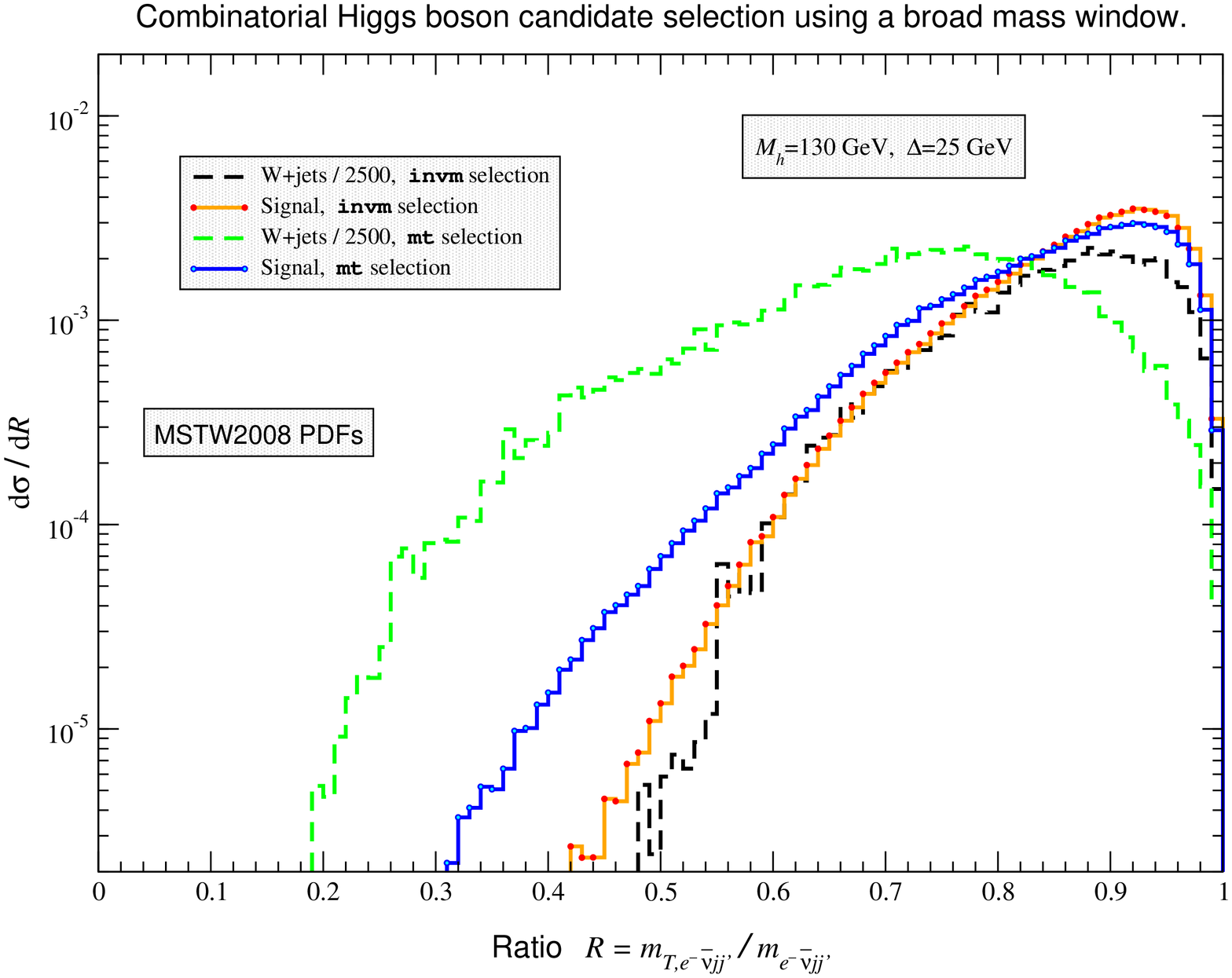}}
  \caption{\label{fig:invm_mt_comparison}
    Differential 4-particle mass spectra $m_{e\nu_e jj'}$ (see left
    panel) and $m_{T,e^-\bar\nu_e jj'}/m_{e^-\bar\nu_e jj'}$ ratio
    distributions after combinatorial selection of Higgs boson
    candidates using a $50\ \mrm{GeV}$ symmetric mass window centered
    at a mass of $M_h=130\ \mrm{GeV}$. The \texttt{mt} selection
    characteristics is compared to the ideal case of choosing
    candidates according to the \texttt{invm} criteria. \Eq{eq:mt} was
    used to compute the \texttt{mt} criteria. All results are shown
    for $e\nu_e$+jets final states originating from the
    $gg\to h\to WW$\/ signal (solid lines) and the $W$\!+jets background
    (dashed lines).}
\end{figure}

We noticed that the \texttt{mt} {\em selection} picks up background
events, which often fall outside (mostly above) the Higgs boson mass
window. As a result, a somewhat different class of background events
survives the \texttt{mt} selection procedure compared to utilizing the
\texttt{invm}, ideal, approach. This is no surprise since we have
already argued that the Higgs boson decays yield an enhanced
transverse production with regard to the $W$\!+2-jet background. Using 
$M_h=130\ \mrm{GeV}$, \Fig{fig:invm_mt_comparison} exemplifies this by
means of the $m_{e\nu_e jj'}$ and $m_{T,e\nu_e jj'}/m_{e\nu_e jj'}$
ratio distributions. We clearly see the large impact on the $W$\!+jets
results being a consequence of enforcing a transverse- rather than
invariant-mass window. Imposing the constraint
$80\le m_{T,e\nu_e jj'}/\mrm{GeV}\le130$ on the background is fairly
equivalent to choosing events with larger longitudinal components.
This drives the associated invariant masses to higher values whereas
the $m_T/m$\/ ratios are shifted to lower ones.

To give a more quantitative example, we consider the case
$M_h=130\ \mrm{GeV}$ for broad (tight) Higgs boson mass windows and
$\delta=20\ \mrm{GeV}$. If we select events using \texttt{mt}, then
discard all those events with an invariant mass $m_{e\nu_e jj'}$
outside $M_h\pm25\ (8)\ \mrm{GeV}$, we find a quite impressive gain
of 82\% (360\%). Similarly, if we use the \texttt{invm} selection and
apply further cuts removing events with $m_{T,e\nu_e jj'}$ values
greater than $M_h$ or less than $M_h-2\,\Delta=M_h-50\ (16)\ \mrm{GeV}$,
we observe that $S/\sqrt{B}$ improves by 12\% (drops by 20\%). As the
\texttt{invm} selection has a better starting $S/\sqrt{B}$ than
\texttt{mt}, the final significances are similar in all cases, however
the essential point is we can improve the significance by combining
the transverse- and invariant-mass selections.

%

To exploit this potential in a more realistic scenario, we use (in a
first phase) the \texttt{mt} selection to pick out the 4-particle
system, then (in a second phase) reconstruct the neutrino momentum
following the \texttt{pzmw} procedure. Compared to the ideal case,
this reconstruction works rather inefficiently in identifying
background events that yield invariant masses
$m_{e\nu_e jj'}>M_h+\Delta$. As it is optimized to the features of the
Higgs boson decay signal, \texttt{pzmw} generates $m_{*,e\nu_e}$
estimates by assuming $m_{T,e\nu_e}/m_{e\nu_e}$ ratios close to 1. For
$W$\!+jets, these choices often turn out to be sufficiently smaller than
the actual masses of the leptonically decaying $W$\/ boson. The
$W$\!+jets background usually contains an on-shell $W$, a lower
$m_{*, e\nu_e}$ is not ideal and tends to bring down the
deconstructed, associated 4-particle masses $m_{e\nu_e jj'}$. As a
result a good fraction of background events possessing true invariant
masses exceeding the upper bound on $m_{e\nu_e jj'}$ (see left plot of
\Fig{fig:invm_mt_comparison}) are shuffled back into the Higgs boson
mass window.


Hence, we make adjustments to the \texttt{pzmw} reconstruction used in
this symbiosis of selections so that it performs better below threshold.
The basic idea is to maximally exploit the differences of signal and
background in the leptonic and hadronic $W$\/ mass distributions. If,
after initial \texttt{mt} selection, we constrain the transverse
masses of the $e\nu_e$ and dijet subsystems by placing cuts that favor
$m_{T,e\nu_e}\approx M_W/2$ and $m_{T,jj'}\approx M_W$, we enforce
off-shell $W\to e\nu_e$ decays while keeping those of $W\to jj'$
on-shell. This is beneficial to the signal and suppresses, at the same
time, the $W$\!+jets background. Here, we only add requirements on the
hadronic subsystem by imposing a minimum value for $m_{T,jj}$. We
leave the $m_{T,e\nu_e}$ potential to be exploited by the
$M_h$-dependent cut optimization (which we discuss in
\Sec{sec:optimize}), since demanding an upper bound on $m_{T,e\nu_e}$
would not only enhance \texttt{mt} but all realistic selections.



With this major adjustment, we can now design a promising
\texttt{mt+pzmw} selection, which we call \texttt{mtp}:
\bi
\item\underline{\texttt{mtp}:}\quad
  the very first part of computing $m_{T,e\nu_e jj'}$ is identical to
  the \texttt{mt} selection. Subsequently, we require $m_{T,jj'}$ to
  be greater than $m^\mrm{min}_{T,jj'}=3.51\,M_h-0.015\,M^2_h-135.4\ \mrm{GeV}$,
  which we have parametrized in terms of $M_h$ for convenience. This
  concludes the phase of testing the transverse criteria.
  Accepted $\{e,\nu_e,j,j'\}$ candidates are subject to the neutrino
  reconstruction, whose implementation deviates from \texttt{pzmw} to
  some extent: we choose $m_{*,e\nu_e}=0.55\,m_{T,e\nu_e}+0.45\,Q$
  freezing it below $m_{T,e\nu_e}$; here we employ $Q=M_h-m_{jj'}$ but
  keep it constant above $M_W+\delta x$\/ ($\delta x=4\ \mrm{GeV}$).
  As in all other reconstruction methods we then solve for the
  longitudinal momentum of the neutrino, cf.\ \Eq{eq:recoansatz}, and
  reject the particular $\{e,\nu_e,j,j'\}$ choice if the solutions are
  degenerate or generate a mean $m_{e\nu_e jj'}$ mass deviating from
  $M_h$ by more than $\Delta'=\max\{\Delta,20\ \mrm{GeV}\}$.%
  \footnote{Here, we do not adopt tight $\Delta$ choices owing to the
    uncertainties intrinsic to the reconstruction of the
    $p_{\parallel,\nu_e}$ component.}
  In all other cases, we pick the solution giving the smaller
  $|m_{e\nu_e jj'}-M_h|$ and accept it if the reconstructed $h$\/ mass
  falls inside the $\Delta'$ window around $M_h$.
\ei

The \texttt{mtp} results we can reach in terms of significance and
sensitivity are known already from
\Figs{fig:real_significances}~and~\ref{fig:real_soverb}. The increase
in $S/\sqrt{B}$ (leftmost plot in \Fig{fig:real_significances}) and
$S/B$\/ (\Fig{fig:real_soverb}) is highly visible for all of the $h$\/
test masses below threshold. For the \texttt{mtp} analyses with
broadest $M_h$ windows, the significance effect is huge compared to
the respective ideal selections using the same $\Delta$ parameter.
Likewise the signal over background ratio turns out similarly or even
better than in the quasi optimal case.

Not only does \texttt{mtp} profit from the better selection
performance, but we end up with a fully reconstructed neutrino vector,
giving us access to a larger set of observables. However, because of
the hard cut on $m_{T,jj'}$, we notice that the \texttt{mtp}-selected
backgrounds are strongly sculpted, washing out a number of shape
differences. The effect of the $m_{T,jj'}$ cut moreover deteriorates
as soon as $M_h>2\,M_W$. We obtain significances similar to the other
methods and since they sculpt the background less, there is no real
advantage to applying \texttt{mtp} above threshold, hence, we restrict
its use to the low-mass $h$\/ region. On that note, it remains to be
studied whether the $S/\sqrt B$, $S/B$\/ improvements came at the
expense of further handles for the $M_h$-dependent optimization. As a
possible consequence the \texttt{mtp} selection actually might be
superseded by an -- at this level -- inferior selection once enhanced
by appropriately designed cuts.

\subsection{Optimized selection -- analyses refinements and (further)
  significance improvements}\label{sec:optimize}

Having established the more realistic overall picture, we here discuss
steps to achieve better signal over background discrimination. After
baseline and combinatorial selections, we are interested in cuts that
help further increase the significances obtained so far, \ie
$\veps_{B,i}<\veps^2_{S,i}$. Our aim is to identify observables for
each $M_h$ test point that are sufficiently uncorrelated such that
simultaneous selections yield a total significance gain in the range:
\be
\max\left\{\frac{\veps_{S,1}}{\sqrt{\veps_{B,1}}}\,,\,
           \frac{\veps_{S,2}}{\sqrt{\veps_{B,2}}}\right\}\;<\;
\frac{\veps_{S}}{\sqrt{\veps_{B}}}\;\le\;
\frac{\veps_{S,1}\,\veps_{S,2}}{\sqrt{\veps_{B,1}\,\veps_{B,2}}}\ .
\ee
Above relation is written out for the example of two extra handles,
but easily extensible to multi-cut scenarios. As we have shown, the
subdominant backgrounds have negligible effects at this level; we
therefore concentrate on reducing the $W$\!+jets background, although
the other backgrounds are still included in computing the total
significance. In order to be conservative about mass resolutions for
hadronic final states at the Tevatron, we will fix the mass window
parameters as $\Delta=\delta=20\ \mrm{GeV}$.

As we have done in earlier sections, we divide the optimized
selection into three broad Higgs boson mass ranges: below threshold,
near threshold, and above threshold. As we vary the Higgs boson mass,
we probe different kinematic configurations for the background. For the 
lowest Higgs boson mass, many background distributions ($H_T$, $p_T$,
etc.) are steeply rising in the region of interest, cut off from below
by baseline kinematics. For intermediate Higgs boson masses, the
backgrounds tend to be flatter or peaked, while the higher Higgs boson
mass region overlaps with backgrounds that are sharply falling. This
basic shape behind (many) background distributions drives which cuts
are optimal for a given Higgs boson mass. We present a number
of distributions in \App{app:hrealreco} to back up the many findings
presented in this subsection.

The optimized analysis employs only observables constructed out
of the momenta of the selected 4-particle system $\{e,\nu_e,j,j'\}$.
More inclusive observables may be useful: for example, the scalar sum
of the two selected-jet $p_T$ versus the two hardest-jet $p_T$,
$H_{T,jj'}\leftrightarrow H_{T,12}$. However such observables may also
be subject to larger uncertainties from, \eg the modeling of hard
initial state radiation. To reduce these uncertainties one could
extend the ME+PS program here, \eg to inclusive $W$\!+3-jets, however
this is beyond the scope of this study.

In the optimized analysis the longitudinal observables,
$\Delta\eta_{j,j'}$ or $\eta_{e\nu_ejj'}$, offer only moderate gains
in significance (typically, we obtain gains on the order of 3--10\%
with larger gains occurring for heavy $M_h$), but they are insensitive
to the exact value of the Higgs boson mass and are therefore more
broadly applicable. The total 4-particle pseudo-rapidity
$\eta_{e\nu_ejj'}$ can only be used in the \texttt{pzmw},
\texttt{pzmh} or \texttt{mtp} selections, since reconstruction of the
neutrino is necessary. By the same logic, because the pseudo-rapidity
difference between the selected jets is independent of the neutrino,
cuts on $\Delta\eta_{j,j'}$ can be used with all selections (though
they are not efficient for \texttt{mtp}). We find that the
requirements $|\Delta\eta_{j,j'}|\lesssim1.5$ and
$|\eta_{e\nu_ejj'}|\lesssim3.0$ work well for the entire range of
Higgs boson masses we are interested in, so we include these cuts into
our optimized (\texttt{pzmh/w} and \texttt{mt}) selection. In
\App{app:hrealreco} we present examples of $|\Delta\eta_{j,j'}|$
distributions (\Fig{fig:detajj}) and $\eta_{e\nu_ejj'}$ spectra
(\Fig{fig:eta.set}) after un-optimized combinatorial selections,
documenting the usefulness of these cuts.

The other useful observables we have found are all transverse, or in
some cases based on invariant masses. Unlike the longitudinal
variables, the optimal transverse variables and cut values depend
strongly on the Higgs boson mass. We also find that transverse and
longitudinal observables are largely uncorrelated in this study, so
any gains in significance from selections in the transverse
observables will add to the gains from $|\Delta\eta_{j,j'}|$ and
$|\eta_{e\nu_ejj'}|$. We refer to
\Figs{fig:2Dpzmw}~and~\ref{fig:2Dmt} of \App{app:hrealreco}
conveniently illustrating this decorrelation for the case
$|\Delta\eta_{j,j'}|$ versus $m_{e\nu_ejj'}$.

\begin{table}[p!]
\centering\footnotesize\vskip0mm
\begin{tabular}{ccrcrccrc}\hline\hline\\[-2mm]
$M_h$         & comb.\ $h$-reco & leading = \textbf{major cut} & gain &
                                  subleading cut               & gain &&
                                  \textbf{minor cut}           & gain \\
$\mrm{[GeV]}$ & selection       & $\mrm{[range\ in\ GeV]}$     & $\mrm{[\%]}$ &
                                  $\mrm{[range\ in\ GeV]}$     & $\mrm{[\%]}$ &&
                                  $\mrm{[range\ in\ GeV]}$     & $\mrm{[\%]}$
\\[1mm]\hline\\[-2mm]
     &\texttt{mtp} & $m_{jj'}\ \ [75,\infty]$ & \bf 17 &
                     $H_{T,jj'}\ \ [76,\infty]$ & 9 &&
                     $p_{T,j}\ \ [38,\infty]$ & \bf 4 \\
$120$&\texttt{mt}  & $m_{T,jj'}\ \ [72,\infty]$ & \bf 70 &
                     $H_{T,jj'}\ \ [72,\infty]$ & 52 &&
                     $H_{T,e\nu_e jj'}\ \ [108,\infty]$ & \bf 6 \\
     &\texttt{pzmw}& $m_{T,e\nu_e}\ \ [0,40]$ & \bf 73 &
                     $m^{(\nu_e)}_{T,e\nu_e jj'}\ \ [0,120]$ & 38 &&
                     $H_{T,jj'}\ \ [64,\infty]$ & \bf 10 \\
     &\texttt{pzmh}& $m^{(\nu_e)}_{T,e\nu_e jj'}\ \ [0,120]$ & \bf 63 &
                     $m_{e\nu_e jj'}\ \ [0,124]$ & 59 &&
                     $m_{\perp,jj'}\ \ [76,\infty]$ & \bf 48
\\[1mm]\hline\\[-2mm]
     &\texttt{mtp} & $p_{T,e\nu_e jj'}\ \ [9,\infty]$ & \bf 6 &
                     $m_{T,e\nu_e}\ \ [0,48]$ & 4 &&
                     $m_{\perp,jj'}\ \ [76,\infty]$ & \bf 4 \\
$130$&\texttt{mt}  & $H_{T,jj'}\ \ [72,\infty]$ & \bf 38 &
                     $m_{T,jj'}\ \ [68,\infty]$ & 31 &&
                     $m_{jj'}\ \ [73,\infty]$ & \bf 4 \\
     &\texttt{pzmw}& $m_{T,e\nu_e}\ \ [0,48]$ & \bf 53 &
                     $m^{(\nu_e)}_{T,e\nu_e jj'}\ \ [0,130]$ & 42 &&
                     $m_{\perp,jj'}\ \ [72,\infty]$ & \bf 11 \\
     &\texttt{pzmh}& $m^{(\nu_e)}_{T,e\nu_e jj'}\ \ [0,130]$ & \bf 48 &
                     $m_{e\nu_e jj'}\ \ [0,134]$ & 46 &&
                     $H_{T,jj'}\ \ [68,\infty]$ & \bf 25
\\[1mm]\hline\\[-2mm]
     &\texttt{mtp} & $p_{T,e\nu_e jj'}\ \ [12,\infty]$ & \bf 8 &
                     $H_{T,jj'}\ \ [68,\infty]$ & 3 &&
                     $m_{T,e\nu_e}\ \ [0,60]$ & \bf 3 \\
$140$&\texttt{mt}  & $H_{T,jj'}\ \ [68,\infty]$ & \bf 24 &
                     $p_{T,e\nu_e jj'}\ \ [16,\infty]$ & 15 &&
                     $p_{T,e\nu_e jj'}\ \ [15,\infty]$ & \bf 6 \\
     &\texttt{pzmw}& $m_{T,e\nu_e}\ \ [0,56]$ & \bf 30 &
                     $m^{(\nu_e)}_{T,e\nu_e jj'}\ \ [0,140]$ & 30 &&
                     $m_{\perp,jj'}\ \ [70,\infty]$ & \bf 6 \\
     &\texttt{pzmh}& $m^{(\nu_e)}_{T,e\nu_e jj'}\ \ [0,140]$ & \bf 29 &
                     $m_{e\nu_e jj'}\ \ [0,144]$ & 28 &&
                     $H_{T,jj'}\ \ [68,\infty]$ & \bf 14
\\[1mm]\hline\\[-2mm]
     &\texttt{mtp} & $p_{T,e\nu_e jj'}\ \ [17,\infty]$ & \bf 14 &
                     $H_{T,e\nu_e jj'}\ \ [116,\infty]$ & 3 &&
                     $H_{T,e\nu_e jj'}\ \ [116,\infty]$ & \bf * \\
$150$&\texttt{mt}  & $p_{T,e\nu_e jj'}\ \ [20,\infty]$ & \bf 18 &
                     $H_{T,jj'}\ \ [60,\infty]$ & 10 &&
                     $H_{T,jj'}\ \ [56,\infty]$ & \bf * \\
     &\texttt{pzmw}& $m^{(\nu_e)}_{T,e\nu_e jj'}\ \ [0,150]$ & \bf 20 &
                     $p_{T,e\nu_e jj'}\ \ [18,\infty]$ & 18 &&
                     $p_{T,e\nu_e jj'}\ \ [18,\infty]$ & \bf 9 \\
     &\texttt{pzmh}& $m^{(\nu_e)}_{T,e\nu_e jj'}\ \ [0,150]$ & \bf 19 &
                     $m_{e\nu_e jj'}\ \ [0,154]$ & 19 &&
                     $p_{T,e\nu_e jj'}\ \ [18,\infty]$ & \bf 9
\\[1mm]\hline\\[-2mm]
     &\texttt{mt}  & $p_{T,e\nu_e jj'}\ \ [18,\infty]$ & \bf 18 &
                     $H_{T,e\nu_e jj'}\ \ [136,\infty]$ & 12 &&
                     $\Delta\phi_{e,\nu_e}\ge1.9$ & \bf 3 \\
$165$&\texttt{pzmw}& $p_{T,e\nu_e jj'}\ \ [18,\infty]$ & \bf 18 &
                     $m_{e\nu_e jj'}\ \ [0,170]$ & 17 &&
                     $\gamma_{jj'|e\nu_e}\le1.06$ & \bf 20 \\
     &\texttt{pzmh}& $p_{T,e\nu_e jj'}\ \ [18,\infty]$ & \bf 18 &
                     $m_{e\nu_e jj'}\ \ [0,170]$ & 13 &&
                     $\gamma_{jj'|e\nu_e}\le1.09$ & \bf 15
\\[1mm]\hline\\[-2mm]
     &\texttt{mt}  & $p_{T,e\nu_e jj'}\ \ [21,\infty]$ & \bf 20 &
                     $H_{T,e\nu_e jj'}\ \ [140,\infty]$ & 16 &&
                     $\Delta\phi_{e,\nu_e}\ge1.7$ & \bf * \\
$170$&\texttt{pzmw}& $p_{T,e\nu_e jj'}\ \ [19,\infty]$ & \bf 20 &
                     $m_{e\nu_e jj'}\ \ [0,176]$ & 13 &&
                     $\gamma_{jj'|e\nu_e}\le1.12$ & \bf 11 \\
     &\texttt{pzmh}& $p_{T,e\nu_e jj'}\ \ [20,\infty]$ & \bf 20 &
                     $m_{e\nu_e jj'}\ \ [0,176]$ & 9 &&
                     $\gamma_{jj'|e\nu_e}\le1.16$ & \bf 9
\\[1mm]\hline\\[-2mm]
     &\texttt{mt}  & $p_{T,e\nu_e jj'}\ \ [22,\infty]$ & \bf 24 &
                     $H_{T,e\nu_e jj'}\ \ [148,\infty]$ & 22 &&
                     $H_{T,e\nu_e jj'}\ \ [140,\infty]$ & \bf * \\
$180$&\texttt{pzmw}& $p_{T,e\nu_e jj'}\ \ [21,\infty]$ & \bf 23 &
                     $H_{T,jj'}\ \ [64,\infty]$ & 11 &&
                     $1.06\le\gamma_{jj'|e\nu_e}\le1.22$ & \bf 5\\
     &\texttt{pzmh}& $p_{T,e\nu_e jj'}\ \ [22,\infty]$ & \bf 24 &
                     $H_{T,e\nu_e jj'}\ \ [140,\infty]$ & 10 &&
                     $m^{(\nu_e)}_{T,e\nu_e jj'}\ \ [136,182]$ & \bf 5
\\[1mm]\hline\\[-2mm]
     &\texttt{mt}  & $p_{T,e\nu_e jj'}\ \ [24,\infty]$ & \bf 28 &
                     $H_{T,e\nu_e jj'}\ \ [156,\infty]$ & 27 &&
                     $H_{T,e\nu_e jj'}\ \ [148,\infty]$ & \bf * \\
$190$&\texttt{pzmw}& $p_{T,e\nu_e jj'}\ \ [23,\infty]$ & \bf 24 &
                     $H_{T,e\nu_e jj'}\ \ [148,\infty]$ & 15 &&
                     $1.12\le\gamma_{jj'|e\nu_e}\le1.30$ & \bf 5 \\
     &\texttt{pzmh}& $p_{T,e\nu_e jj'}\ \ [24,\infty]$ & \bf 29 &
                     $H_{T,e\nu_e jj'}\ \ [144,\infty]$ & 17 &&
                     $m^{(\nu_e)}_{T,e\nu_e jj'}\ \ [142,194]$ & \bf 3
\\[1mm]\hline\\[-2mm]
     &\texttt{mt}  & $H_{T,e\nu_e jj'}\ \ [164,\infty]$ & \bf 31 &
                     $p_{T,e\nu_e jj'}\ \ [24,\infty]$ & 28 &&
                     $p_{T,e\nu_e jj'}\ \ [15,\infty]$ & \bf 9 \\
$200$&\texttt{pzmw}& $p_{T,e\nu_e jj'}\ \ [24,\infty]$ & \bf 25 &
                     $H_{T,e\nu_e jj'}\ \ [156,\infty]$ & 20 &&
                     $1.18\le\gamma_{jj'|e\nu_e}\le1.40$ & \bf 6 \\
     &\texttt{pzmh}& $p_{T,e\nu_e jj'}\ \ [27,\infty]$ & \bf 32 &
                     $H_{T,e\nu_e jj'}\ \ [156,\infty]$ & 25 &&
                     $H_{T,e\nu_e jj'}\ \ [144,\infty]$ & \bf 4
\\[1mm]\hline\\[-2mm]
     &\texttt{mt}  & $H_{T,e\nu_e jj'}\ \ [172,\infty]$ & \bf 36 &
                     $p_{T,e\nu_e jj'}\ \ [25,\infty]$ & 27 &&
                     $p_{T,e\nu_e jj'}\ \ [15,\infty]$ & \bf 8 \\
$210$&\texttt{pzmw}& $H_{T,e\nu_e jj'}\ \ [160,\infty]$ & \bf 24 &
                     $p_{T,e\nu_e jj'}\ \ [24,\infty]$ & 23 &&
                     $1.25\le\gamma_{jj'|e\nu_e}\le1.45$ & \bf 14 \\
     &\texttt{pzmh}& $H_{T,e\nu_e jj'}\ \ [162,\infty]$ & \bf 36 &
                     $p_{T,e\nu_e jj'}\ \ [30,\infty]$ & 36 &&
                     $1.25\le\gamma_{jj'|e\nu_e}\le1.54$ & \bf 7
\\[1mm]\hline\\[-2mm]
     &\texttt{mt}  & $H_{T,e\nu_e jj'}\ \ [180,\infty]$ & \bf 39 &
                     $m_{T,e\nu_e jj'}\ \ [174,\infty]$ & 26 &&
                     $p_{T,e\nu_e jj'}\ \ [12,\infty]$ & \bf 8 \\
$220$&\texttt{pzmw}& $H_{T,e\nu_e jj'}\ \ [168,\infty]$ & \bf 29 &
                     $p_{T,e\nu_e jj'}\ \ [24,\infty]$ & 22 &&
                     $1.31\le\gamma_{jj'|e\nu_e}\le1.53$ & \bf 15 \\
     &\texttt{pzmh}& $H_{T,e\nu_e jj'}\ \ [172,\infty]$ & \bf 49 &
                     $p_{T,e\nu_e}\ \ [56,\infty]$ & 43 &&
                     $1.30\le\gamma_{jj'|e\nu_e}\le1.58$ & \bf 8
\\[1mm]\hline\hline
\end{tabular}
\caption{\label{tab:optsig}
  Leading (optimal/major) and subleading cuts for each Higgs boson
  mass and selection criteria (all selections are combinatorial with
  window parameters $\Delta=\delta=20\ \mrm{GeV}$). Gain in $S/\sqrt B$,
  in percent, is shown after each cut. Having used the major
  discriminators including pseudo-rapidity cuts (see text), next in
  the cut hierarchy are the minor cuts shown in the two rightmost
  columns. The significance gains associated with them are understood
  in addition to the major cut improvements. Cuts marked with an
  asterisk have less than 2\% improvement.}
\end{table}

\paragraph{Below-threshold region:}
For \texttt{pzmw} and \texttt{pzmh}, the reconstruction selections,
the transverse mass of the 4-particle system,
$m^{(\nu_e)}_{T,e\nu_e jj'}$, and (for \texttt{pzmw} at low $M_h$) the
transverse mass of the leptonic system, $m_{T,e\nu_e}$, are the best
observables. This result is really just a reiteration of the fact that
simple reconstruction selections work poorly for below-threshold Higgs
bosons. In our effort to make \texttt{pzmh/w} more flexible and apply
them to below-threshold scenarios, we have allowed the possibility for
background configurations that are inconsistent with a single parent
resonance -- such as 4-particle systems with
$m^{(\nu_e)}_{T,e\nu_e jj'}>M_h$. Removing this inconsistent region
results in gains of ${\cal O}(\mbox{40\%})$.

For \texttt{mt}, the transverse mass of the dijet system or the scalar
$p_T$ sum formed with the selected jets, $H_{T,jj'}$, are the most
optimal, with improvements of ${\cal O}(\mbox{50\%})$. The former cut
takes advantage of the fact that the signal jets originate in a (real
or virtual) $W$\/ boson, while the background jets come primarily from
ISR -- information that is not exploited in the initial \texttt{mt}
selection. Somewhat smaller gains come from cutting on the total
selected system's transverse momentum, $p_{T,e\nu_e jj'}$.

For \texttt{mtp}, there is little optimization to be done. Much of the
physics that is behind the optimal cuts in the \texttt{pzmh/w} or
\texttt{mt} cases has already been incorporated into the selection
process. Some improvement is possible by cutting out the region with
very low 4-particle transverse momentum, $p_{T,e\nu_e jj'}$.

We illustrate our findings by showing and commenting on a small
collection of spectra resulting from the baseline combinatorial
selections; for more details, we refer the reader to the discussion
around \Fig{fig:cuts.below} of \App{app:hrealreco}.

\paragraph{Near-threshold region:}
For Higgs boson masses close to the $WW$\/ threshold, the
4-particle $p_T$ is the most powerful additional handle. In Higgs
boson production, as in other colour-singlet resonance production, the
$p_{T,e\nu_e jj'}$ distribution is cut off at low values by soft-gluon
resummation, and falls off at high values because of parton
kinematics. The result is a peaked distribution. The hard scale of the
process, dictated by the Higgs boson mass, sets the initial ISR scale,
thereby influencing where $p_{T,e\nu_e jj'}$ peaks. Since the Higgs
boson is heavier than the $W$\/ boson, $p_{T,e\nu_e jj'}$ always peaks
at higher values for the signal compared to $W$\/ production. Even
though the dominant background for our study is $W$\!+2-jets, rather than
$W$\!+0-jets, the argument still holds. The peak in $p_{T,e\nu_e jj'}$
for $W$\!+2-jets is still governed by $M_W$ and continues to peak at
lower values than the Higgs boson signal. Selection criteria change
the tails of the background $p_{T,e\nu_e jj'}$ distribution, but do
not affect the location of the peak. Cutting out the
low-$p_{T,e\nu_e jj'}$ region, improvements on the order of 20\% are
possible. Distributions such as $H_{T,jj'}$, the scalar $p_T$ sum of
the two selected jets, or $H_{T,e\nu_e jj'}$, the $H_T$ of the
4-particle system, also have potential discriminating power. Signal
versus background distributions in the relevant variables after
baseline combinatorial selections are shown in \Fig{fig:cuts.near},
see \App{app:hrealreco}.

\paragraph{Above-threshold region:}
For higher Higgs boson masses, the total amount of (transverse) energy
in the $W$\!+$jj'$ system becomes the most powerful discriminator
between the signal and the background. Specifically, once
$M_h\gtrsim200\ \mrm{GeV}$, the $H_T$ of the selected 4-particle
system peaks at significantly higher values than the background,
regardless of the selection technique. By cutting away the low-$H_T$
region, we find gains of order 25\% are possible. The 4-object $p_T$
remains a very useful observable, as does the 4-particle transverse
mass. Examples of signal versus background distributions are shown in
\Fig{fig:cuts.above}, see \App{app:hrealreco}.

\paragraph{Results:}
The optimal or ``leading'' or ``major'' cuts for the different Higgs
boson mass categories and selection methods are summarized in
\Tab{tab:optsig}. To give some idea how useful the single best
discriminator is compared to other observables, we also show the
percent increase in significance for the second best or ``subleading''
single discriminator. Separately we have also determined which
combinations of the leading discriminator (supplemented by the
respective pseudo-rapidity cuts discussed above) with a second
observable give the largest (additional) increase in significance. The
second ``minor'' cuts of these optimal two-variable selections are
summarized in the last two columns of \Tab{tab:optsig}. Note that in
most cases these minor cuts do not involve the same observables as the
subleading cut; this is because the subleading discriminator is
typically strongly correlated with the leading discriminator, and thus
does not add much to the combined significance. Some ideas beyond the
application of minor cuts exist; we comment in \App{app:addimprovs} on
a few possible routes one can take to enhance the optimized analyses
presented here.

When looking for minor cuts, we found in addition to variables we have
already discussed, such as $\gamma_{jj'|e\nu_e}$ (in
\Sec{sec:strategy}), $H_{T,(e\nu_e)jj'}$ and the 4-object $p_T$, a few
other observables, namely $p_{T,j}$, $\Delta\phi_{e,\nu_e}$ and
$m_{\perp,jj'}$ to be beneficial. The first two, $p_{T,j}$ and
$\Delta\phi_{e,\nu_e}$ are common, so we do not repeat their
definitions here. The last minor cut observable, $m_{\perp,jj'}$ is
defined through $m^2_{\perp,jj'}=m^2_{jj'}+p^2_{T,jj'}$ exhibiting yet
another way of defining a transverse mass. The additional gains from
the minor cuts are typically small, except close to the $WW$\/
threshold and for the largest Higgs boson masses considered here. In
particular for $M_h>2\,M_W$, the boost of the $jj'$ system in the
reconstructed 4-object rest frame stands out as a helpful
discriminator of secondary order; for more details, we refer again to
\App{app:hrealreco} and the discussion around \Fig{fig:cuts.minor}.
Also, the dijet-system based handles, $m_{\perp,jj'}$ and $H_{T,jj'}$
yield fairly substantial extra gains, but only at low $M_h$ if we rely
on the \texttt{pzmh} method. Once jets are picked stemming from the
backgrounds, the strict $M_h$ mass reconstruction of the selected
4-object, as encoded in \texttt{pzmh} and amplified by the major cut
given through $m^{(\nu_e)}_{T,e\nu_e jj'}$, gives rise to the
selection of less energetic $j$, $j'$ jets with preferred pair masses
of $m_{jj'}\sim M_W-\delta$. The observables $m_{\perp,jj'}$ and
$H_{T,jj'}$ exploit this fact conveniently, hence facilitate such
secondary improvements, as shown in \Fig{fig:cuts.minor} for the
example of $M_h=130\ \mrm{GeV}$.

\begin{figure}[p!]
\psfrag{Significance}[c][c][0.92]{
  \fontfamily{phv}\selectfont{Significance}\quad$S/\sqrt{B}$}
\psfrag{Higgs boson mass}[c][c][0.92]{
  \fontfamily{phv}\selectfont{Higgs boson mass}\quad$M_h$}
\centering\vskip4mm
  \includegraphics[clip,width=0.74\columnwidth,angle=-90]{%
    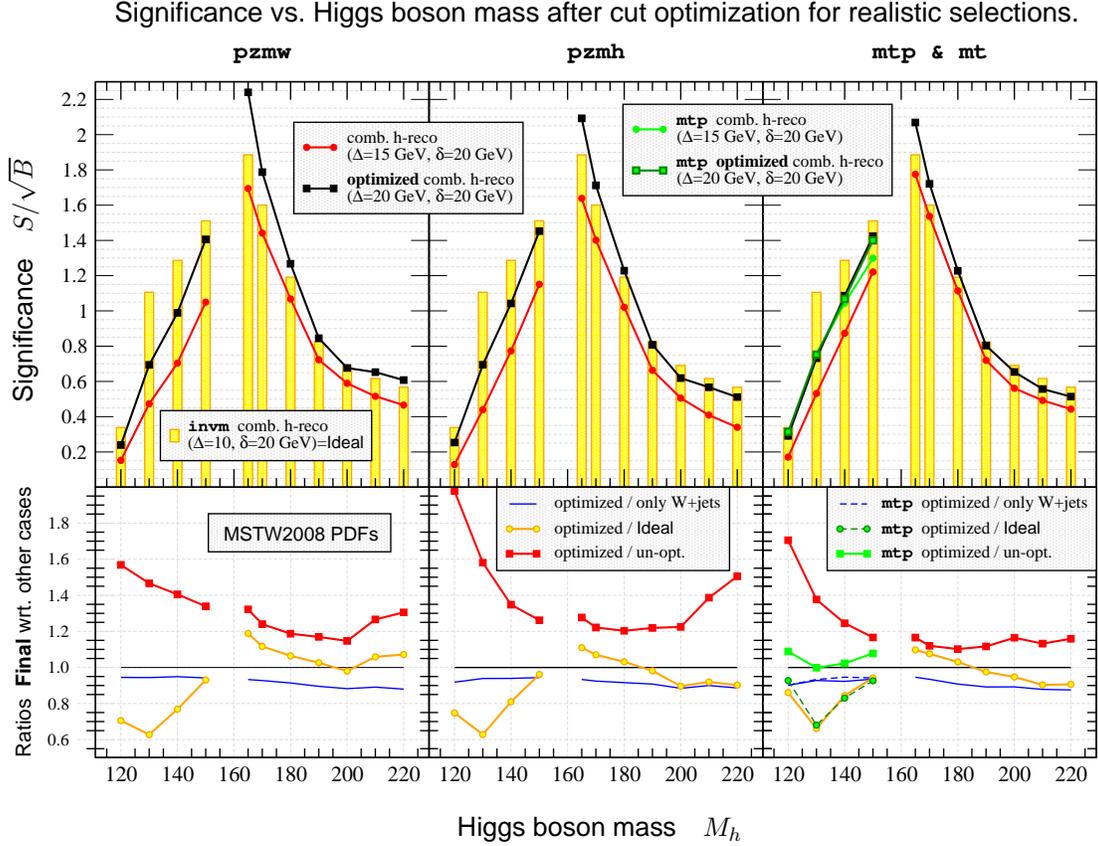}
  \caption{\label{fig:opt_significances}
    Total significances as a function of $M_h$ after including the
    subdominant backgrounds and all major plus minor cuts (as
    specified in the text and \Tab{tab:optsig}). The results (lines
    with squares) are shown for the four types of more realistic Higgs
    boson candidate selections, which have been advertised in this
    work. They are denoted on top of each panel. In all cases the
    combinatorial approach using window parameters
    $\Delta=\delta=20\ \mrm{GeV}$ has been applied for selecting the
    candidate set of particles. As before in \Fig{fig:real_significances},
    the ideal case (\ie the \texttt{invm} combinatorial reconstruction
    of the Higgs boson candidates using $2\,\Delta=\delta=20\
    \mrm{GeV}$) is taken as the main reference to compare the
    different results. For each combinatorial selection, the outcome
    (lines with circles) with no cuts applied (but using a slightly
    smaller mass window, $\Delta=15\ \mrm{GeV}$) is also displayed to
    emphasize the effect of the cut optimization. Note that the effect
    of the minor backgrounds has been neglected in computing each of
    these reference curves.
    The $e\nu_e$+jets final states are generated from the signal,
    $gg\to h\to WW$, the $W$\!+jets, electroweak and $t\bar t$\/
    backgrounds. All $S/\sqrt{B_i}$ were calculated using
    \Eqs{eqs:signif} and combined according to \Eq{eq:tot_signif}
    assuming an integrated luminosity of ${\cal L}=10\ \mrm{fb}^{-1}$
    and including electron and muon channels, \ie $f_\ell=2$. The
    lower plots depict for each selection the ratios of optimized over
    un-optimized $S/\sqrt{B}$ (lines with squares) and optimized over
    ideal-case reference $S/\sqrt{B}$ (lines with circles). The thin
    blue lines visualize in how far the final $S/\sqrt{B_\mrm{(tot)}}$
    results suffer from the presence of the minor backgrounds.}
\end{figure}

\begin{figure}[t!]
\centering\vskip2mm
  \includegraphics[clip,width=0.74\columnwidth,angle=-90]{%
    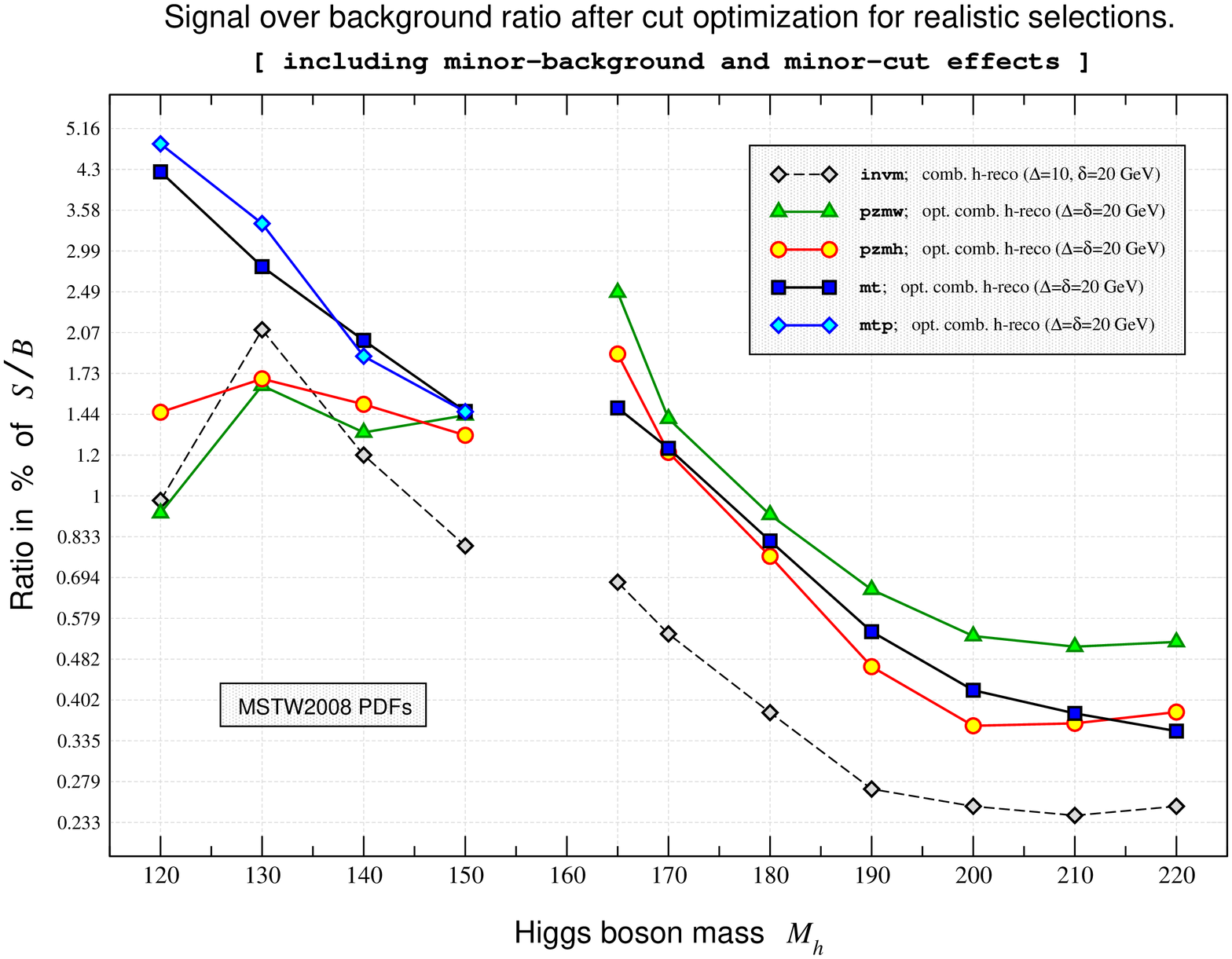}
  \caption{\label{fig:opt_soverb}
    The $S/B$ ratios associated with the total significances presented
    in \Fig{fig:opt_significances}. The ratios are shown as a function
    of $M_h$ for the final selections: the combinatorial
    \texttt{pzmw} (triangles), \texttt{pzmh} (circles), \texttt{mt}
    (squares) and \texttt{mtp} (diamonds) selections using the window
    parameters $\Delta=\delta=20\ \mrm{GeV}$ each supplemented by
    their specific optimization cuts, see \Tab{tab:optsig}. The
    reference curve (dashed line with diamonds) is the same as used in
    \Fig{fig:real_soverb} representing the \texttt{invm} combinatorial
    reconstruction with $2\,\Delta=\delta=20\ \mrm{GeV}$, no extra
    cuts applied and neglecting the impact of the minor backgrounds.
    The $e\nu_e$+jets final states were generated from $gg\to h\to WW$\/
    signal events, the $W$\!+jets, electroweak and $t\bar t$\/
    backgrounds. The $S/B_i$\/ were calculated using the $\sigma_S$ and
    $\sigma_B$ as obtained after the final selections, the signal
    $K$-factors of \Tab{tab:sgnlKfacs} and the $K_B$ as given in
    \Eqs{eq:bkgdK},~(\ref{eq:ewK})~and~(\ref{eq:ttK}). Single ratios
    were combined according to $S/B_\mrm{(tot)}=1/\sum_i(S/B_i)^{-1}$.}
\end{figure}

Finally, the baseline combinatorial and optimized combinatorial
significances are displayed as a function of the Higgs boson mass in
\Fig{fig:opt_significances}. The ratio plots associated with each
selection in the lower part of the figure visualize the significance
increase achieved by the optimization. Independent of the selection,
they also indicate a $\cal O(\mbox{10\%})$ drop of significances
caused by the subdominant backgrounds. Focusing on the $S/\sqrt{B}$
ratios taken with respect to the ideal case (orange lines with
circles), these ratio plots emphasize that the optimized
\texttt{mt(p)} (transverse) and \texttt{pzmw} selections work best
below and above the $WW$\/ mass threshold, respectively. The related
$S/B$\/ ratios presented in \Fig{fig:opt_soverb} confirm these
findings. They turn out to be rather small, as a consequence of
maximizing the significance and trying to preserve most of the signal;
both of which does not allow for imposing too restrictive cuts.
Advantageously, the actual number of signal events, $S$, present in
this $h\to WW$\/ channel is not small. Except for the Higgs signal at
$M_h=120\ \mrm{GeV}$ (with $\cal O(\mbox{4})$ expected events), the
optimized analyses usually leave us with hundreds of signal events
(50--300), if we assume an integrated luminosity of
${\cal L}=10\ \mrm{fb}^{-1}$. Even the $\sim$~25 signal events for
$M_h=130\ \mrm{GeV}$ are sufficient, particularly as $S/B$\/ increases
to $\sim0.04$.

Clearly, as seen from \Fig{fig:opt_significances}, the optimized
significances for the four different Higgs boson reconstruction
methods are very similar. The best significance is for Higgs boson
masses close to the $WW$\/ threshold, suggesting that it would be
possible to achieve 95\% confidence level exclusion in a stand-alone
analysis. For a Higgs boson mass in the range
$130\lesssim M_h\lesssim150\ \mrm{GeV}$, the optimized significance is
between $0.7$ and $1.4$. Given the additional improvements expected
from a full multivariate analysis, this indicates that the
semileptonic Higgs boson decay channel can make a significant
contribution to Higgs boson exclusion in this mass range.


\section{Conclusions, caveats, and prospects}\label{sec:conclusions}

We have presented a systematic study of the prospects for extending
the Tevatron exclusion reach for a Standard Model Higgs boson by
including the final states arising from semileptonic Higgs boson
decays. We have used a realistic simulation of the Higgs signal and
the relevant Standard Model background processes to exhibit the
kinematic differences between the signal and background. We have used
three qualitatively different approaches to extracting the event
kinematics, one based on transverse observables and the others based
on approximate even-by-event full reconstruction. We have shown that
all three approaches give similar results when one optimizes
selections based on several discriminating observables. The details of
the optimization depend on the Higgs boson mass, and in particular on
whether it is below, near, or above the threshold for decay to two
on-shell $W$\/ bosons.

The optimized significances that we have achieved are not sufficient
for stand-alone Higgs boson exclusion except in the most favorable
case where the Higgs boson mass is close to threshold. However the
sensitivities shown here are certainly promising as ingredients to a
combined multi-channel analysis.

One important caveat is that the signal to background ratios for this
type of analysis are fairly small, on the order of a percent, as
illustrated in \Fig{fig:opt_soverb}. This means that Higgs boson
exclusion is sensitive to relatively small systematic errors in the
modeling of the backgrounds, notably the dominant background from
$W$\!+jets. However we have shown here that an experimental analysis has
multiple over-constrained handles on the kinematic features of the
data, providing extra cross-checks. In addition, the Higgs boson
candidate selection employed here to identify the two jets from the
Higgs boson decay is by design rather stable against effects from
extra hard radiation; this reduces the uncertainty in the background
modeling.

The techniques described here are applicable to Higgs boson searches
at the CERN Large Hadron Collider. At 7 TeV center-of-mass collision
energy, the Higgs boson production cross section increases by a factor
of $\sim$~30, while the $W$\!+jets background increases by a factor of
$\sim$~20. With no hard upper limit of the amount of data available,
one can use more restrictive selections and improve both the signal to
background ratio and the overall significance. A recent analysis by
the ATLAS collaboration \cite{Neubauer:2011pp} employed a 4-object
invariant mass reconstruction and a dijet mass window similar to the
baseline naive analysis used here, but applied to the heavy Higgs
boson mass region $M_h>240\ \mrm{GeV}$. A CMS study \cite{Ball:2007zza}
looked at the enhanced signal to background ratio provided by focusing
on VBF production (and thus requiring two extra forward jets). Because
of the large branching fraction, the $h\to WW$\/ semileptonic mode
might prove to be observable at the LHC over a larger mass range than
the $h\to WW$\/ dilepton mode; thus it could play a critical role in
establishing the nature of electroweak symmetry breaking, in the event
that the Higgs boson mass is $\sim$~120~GeV.


\section*{Acknowledgments}

We would like to thank Bogdan Dobrescu for many lively and stimulating
discussions. We would also like to thank Thomas Becher, Li Lin Yang,
Tanju Gleisberg, Frank Petriello, John Campbell and Ciaran Williams
for their help in accomplishing necessary cross-checks. We are
grateful to Frank Krauss and Frank Siegert, as well as all other
members of the \sherpa collaboration for their continuous support.
We gratefully acknowledge useful discussions with Bob Hirosky and
Lidija Zivkovic, as well as Lance Dixon, Gavin Salam, Walter Giele,
Rakhi Mahbubani, Patrick Fox and Agostino Patella.

Fermilab is operated by Fermi Research Alliance, LLC, under contract
DE-AC02-07CH11359 with the United States Department of Energy.


\appendix

\section{Appendix: Monte Carlo event generation}
\subsection{Leading-order cross sections}\label{app:lo_xsecs}

Here we briefly report on the tests we have done to convince ourselves
of the correctness of the \sherpa leading-order cross section
calculations. We found satisfactory or better agreement in all our
cross-checks, which we briefly summarize here:
\bi
\item For the case of $h\to e^-\bar\nu_e\,pp$\/ decays, we have
  compared \sherpa's branching ratios with those obtained by
  multiplying the \textsc{Hdecay} results for $B_{W^\ast W^\ast}$
  given in \Tab{tab:sgnlKfacs} times the PDG literature numbers for
  $B(W\to e\nu_e)\times B(W\to pp)=0.1075\times0.676$.
  Using \amegic's mode of calculating partial widths of $1\to N$\/
  processes, we determined
  $B_{e^-\bar\nu_e pp}=\Gamma(h\to e^-\bar\nu_e\,pp)/\Gamma_h$ for the
  various Higgs boson masses and respective widths of \Tab{tab:sgnlKfacs}.
  The differences seen are at most on the few-percent level.
\item With the explicit knowledge of the \sherpa branching fractions
  we were able to extract Higgs boson production rates at LO from the
  \sherpa signal cross section calculations according to
  \be
  \sigma^\mrm{LO}_{ggh}\;=\;\frac{\Gamma_h}{\Gamma(h\to e^-\bar\nu_e\,pp)}
                          \;\;\sigma^{(0)}_{e^-\bar\nu_e pp}
                       \;=\;\frac{\sigma^{(0)}_{e^-\bar\nu_e pp}}
				 {B_{e^-\bar\nu_e pp}}\ .
  \ee
  The numbers that we obtained from this procedure compare well to
  numbers of other LO calculations, for example the LO rates as
  evaluated in \textsc{MCFM} or provided by Becher and Yang for
  verification purposes \cite{Yang.priv:2010}.
\item \sherpa LO rates were computed for both finite top masses and in
  the infinite top-mass limit. The ratio of the former over the latter
  cross section given as
  \be
  \frac{\sigma^{(0)}_S}{\sigma^{(0)}_{S,m_t\to\infty}}\;=\;
  \left|\,I\left(\frac{m^2_t}{M^2_h}\right)\right|^2
  \ee
  singles out the dependence on the top mass versus Higgs boson mass
  ratio, which is encoded by the function
  \bea
  I(x)\;=\;6x+3x\,(4x-1)&\Biggl\{&\frac{\Theta(1-4x)}{2}\left[
    \ln\left(\frac{1+\sqrt{1-4x}}{1-\sqrt{1-4x}}\right)-i\pi\right]^2-
  \nn\\[2mm]&&
  2\,\Theta(4x-1)\arcsin^2\left(\frac{1}{2\sqrt x}\right)\;\,\Biggl\}\ .
  \eea
  Note that $|I(x)|^2$ attains unity as $x\to\infty$, while it
  vanishes for $x\to0$. Comparing the numerical cross section
  ratios with the analytical values for $|I(x)|^2$, we found excellent
  agreement over the entire Higgs boson mass range considered in this
  study.
\ei

\subsection{NLO calculations versus CKKW ME+PS merging}
\label{app:nlo_vs_ckkw}

When compared to NLO calculations, it is evident that \sherpa's CKKW
merging approach does not account for the virtual corrections to
$V$+jets in their entirety.\footnote{For a brief summary of the basics
  of the CKKW merging, see Ref.~\cite{Alwall:2007fs}. For the current
  generation of \sherpa Monte Carlo programs, the ME+PS facilities
  have been extended to allow for truncated showering, which is a
  major refinement over the CKKW approach, see
  Refs.~\cite{Hoeche:2009rj,Hoeche:2009xc,Carli:2010cg}.}
The only contributions enter through Sudakov form-factor terms at
leading-logarithmic accuracy used in the parton shower and to reweight
the tree-level matrix elements. The real-emission corrections however
are included on a fairly comparable level with respect to full NLO
calculations.\footnote{More recent versions of \sherpa, from
  version~1.2.3 on, have been enhanced by the means to generate, for a
  number of important processes, events at the hadron level with a
  rate correct at next-to-leading order in $\alpha_\mrm{s}$
  \cite{Hoche:2010pf,Hoche:2010kg,Hoeche:2011fd}. To make this work,
  \sherpa relies on interfacing external one-loop amplitude generators
  like \textsc{MCFM}~\cite{Campbell:Ellis,Campbell:2010ff},
  \textsc{BlackHat}~\cite{Berger:2008sj,Berger:2010gf} or, more
  generally, via the Binoth Les Houches Accord \cite{Binoth:2010xt}.
  Because of the complexity of the procedure, such improvements are
  not yet available for arbitrary processes, in particular the
  multi-jet final states we are interested in.}
Unless one decides for a fixed-scale choice at NLO, both approaches
determine the strong-coupling scales dynamically, \ie on an
event-by-event basis, taking the kinematic configuration of the event
into account. For all these reasons, it then occurs that the CKKW
shapes of distributions emerge in many cases quite similarly to those
evaluated at NLO, making an application of global $K$-factors
feasible.\footnote{For example, in \cite{Gleisberg:2008ta} a global
  $K$-factor of magnitude $1.33$ with respect to the total inclusive
  cross section as measured by CDF \cite{Aaltonen:2007ip} was applied
  to achieve a good agreement between the data and the \sherpa
  predictions for inclusive jet multiplicity and transverse momentum
  distributions.}
The treatment to fix the strong couplings is different when multiple
scales are present. While at NLO the scales are set uniformly such
that all $\alpha_\mrm{s}$ factors obtain the same value, in the CKKW
method they are set locally by the procedure itself, cf.\
\cite{Krauss:2004bs,Schalicke:2005nv,Alwall:2007fs} for example. This
is known as $\alpha_\mrm{s}$ reweighting, constituting the second
component of the matrix-element reweighting of the CKKW method. The
assignment of the scales proceeds hierarchically based on the
splitting history, which is identified by the $k_T$-jet cluster
algorithm when applied to the initial matrix-element configuration
considering physical parton combinations only. The nodal $k_T$ values
found by the clustering can be interpreted as the relative transverse
momenta of the identified splittings. They are then used as the scales
for the strong-coupling constants replacing the predefined choice of
the initial matrix-element generation. It would be interesting to see
if a hierarchical scale setting can further stabilize NLO results, but
no such $\alpha_\mrm{s}$ reweighting has been completely worked out
yet for NLO calculations.

\section{Appendix: Analysis side studies and additional material}

\begin{table}[p!]
\centering\footnotesize
\begin{tabular}{|l|c|rrr|rrr|rrr|}\hline
\rule[-2mm]{0mm}{6mm} cuts \& & $2\,\Delta/$
& $\sigma_S/\mrm{fb}$ & $\sigma_B/\mrm{fb}$ & $S/B$
& $\sigma_S/\mrm{fb}$ & $\sigma_B/\mrm{fb}$ & $S/B$
& $\sigma_S/\mrm{fb}$ & $\sigma_B/\mrm{fb}$ & $S/B$\\
\rule[-2mm]{0mm}{6mm} selections & $\mrm{GeV}$
& $\veps_S$ & $\veps_B$ & $S/\sqrt{B}$
& $\veps_S$ & $\veps_B$ & $S/\sqrt{B}$
& $\veps_S$ & $\veps_B$ & $S/\sqrt{B}$\\\hline\hline
\multicolumn{2}{|r|}{
  \rule[-3mm]{0mm}{8mm}$M_h/\mrm{GeV}\quad[\delta/\mrm{GeV}]$}
&\multicolumn{3}{c|}{$190\quad[20]$}
&\multicolumn{3}{c|}{$200\quad[20]$}
&\multicolumn{3}{c|}{$210\quad[20]$}\\\hline\hline
\rule[-1.5mm]{0mm}{5mm} $\sigma^{(0)}$ &
& $9.862$ & $220\cdt10^4$ & 96$\cdt10^{\trm-7}$
& $7.827$ & $220\cdt10^4$ & 75$\cdt10^{\trm-7}$
& $6.473$ & $220\cdt10^4$ & 61$\cdt10^{\trm-7}$\\
\rule[-1.5mm]{0mm}{5mm} &
& 1.0 & 1.0 & \bf 0.10
& 1.0 & 1.0 & \bf 0.08
& 1.0 & 1.0 & \bf 0.06\\\hline
\rule[-1.5mm]{0mm}{5mm} lepton \& &
& $5.561$ & $984\cdt10^3$ & 12$\cdt10^{\trm-6}$
& $4.433$ & $984\cdt10^3$ & 95$\cdt10^{\trm-7}$
& $3.689$ & $984\cdt10^3$ & 78$\cdt10^{\trm-7}$\\
\rule[-1.5mm]{0mm}{5mm} MET cuts &
& 0.564 & 0.45 & \bf 0.08
& 0.566 & 0.45 & \bf 0.07
& 0.570 & 0.45 & \bf 0.05\\\hline
\rule[-1.5mm]{0mm}{5mm} as above \& &
& $4.586$ & $191\cdt10^2$ & 51$\cdt10^{\trm-5}$
& $3.709$ & $191\cdt10^2$ & 41$\cdt10^{\trm-5}$
& $3.128$ & $191\cdt10^2$ & 34$\cdt10^{\trm-5}$\\
\rule[-1.5mm]{0mm}{5mm} $\ge2$ jets &
& 0.465 & 0.0087 & \bf 0.50
& 0.474 & 0.0087 & \bf 0.40
& 0.483 & 0.0087 & \bf 0.33\\\hline
\rule[-1.5mm]{0mm}{5mm} as above \& &
& $2.533$ & $6997$ & 77$\cdt10^{\trm-5}$
& $2.007$ & $6997$ & 60$\cdt10^{\trm-5}$
& $1.671$ & $6997$ & 50$\cdt10^{\trm-5}$\\
\rule[-1.5mm]{0mm}{5mm} $\abs{\td m_{j_1j_2}\!-80}\!<\!\td\delta$ &
& 0.257 & 0.0032 & \bf 0.46
& 0.256 & 0.0032 & \bf 0.36
& 0.258 & 0.0032 & \bf 0.29\\\hline\hline
\rule[-1.5mm]{0mm}{5mm} naive $h$-reco & $50$
& $2.699$ & $6644$ & 87$\cdt10^{\trm-5}$
& $2.146$ & $6303$ & 72$\cdt10^{\trm-5}$
& $1.789$ & $5837$ & 64$\cdt10^{\trm-5}$\\
\rule[-1.5mm]{0mm}{5mm} &
& 0.274 & 0.0030 & \bf 0.50
& 0.274 & 0.0029 & \bf 0.40
& 0.276 & 0.0027 & \bf 0.34\\\hline
\rule[-1.5mm]{0mm}{5mm} naive $h$-reco & $30$
& $2.082$ & $4062$ & 0.0011
& $1.649$ & $3805$ & 91$\cdt10^{\trm-5}$
& $1.374$ & $3521$ & 81$\cdt10^{\trm-5}$\\
\rule[-1.5mm]{0mm}{5mm} &
& 0.211 & 0.0018 & \bf 0.49
& 0.211 & 0.0017 & \bf 0.40
& 0.212 & 0.0016 & \bf 0.34\\\hline
\rule[-1.5mm]{0mm}{5mm} naive $h$-reco & $48$
& $2.177$ & $3483$ & 0.0013
& $1.699$ & $3151$ & 0.0011
& $1.397$ & $2649$ & 0.0011\\
\rule[-1.5mm]{0mm}{5mm} $\abs{\td m_{j_1j_2}\!-80}\!<\!\td\delta$ &
& 0.221 & 0.0016 & \bf 0.56
& 0.217 & 0.0014 & \bf 0.45
& 0.216 & 0.0012 & \bf 0.40\\\hline
\rule[-1.5mm]{0mm}{5mm} naive $h$-reco & $20$
& $1.488$ & $1565$ & 0.0020
& $1.159$ & $1312$ & 0.0019
& $0.9518$& $1080$ & 0.0018\\
\rule[-1.5mm]{0mm}{5mm} $\abs{\td m_{j_1j_2}\!-80}\!<\!\td\delta$ &
& 0.151 & 71$\cdt10^{\trm-5}$ & \bf 0.57
& 0.148 & 60$\cdt10^{\trm-5}$ & \bf 0.48
& 0.147 & 49$\cdt10^{\trm-5}$ & \bf 0.43\\\hline\hline
\rule[-1.5mm]{0mm}{5mm} comb.\ $h$-reco & $50$
& $3.666$ & $7296$ & 0.0011
& $2.937$ & $6946$ & 89$\cdt10^{\trm-5}$
& $2.464$ & $6465$ & 79$\cdt10^{\trm-5}$\\
\rule[-1.5mm]{0mm}{5mm} &
& 0.372 & 0.0033 & \bf 0.65
& 0.375 & 0.0032 & \bf 0.52
& 0.381 & 0.0029 & \bf 0.45\\\hline
\rule[-1.5mm]{0mm}{5mm} comb.\ $h$-reco & $20$
& $2.545$ & $3145$ & 0.0017
& $2.031$ & $2964$ & 0.0014
& $1.699$ & $2756$ & 0.0013\\
\rule[-1.5mm]{0mm}{5mm} &
& 0.258 & 0.0014 & \bf 0.69
& 0.260 & 0.0013 & \bf 0.56
& 0.262 & 0.0013 & \bf 0.48\\\hline
\rule[-1.5mm]{0mm}{5mm} comb.\ $h$-reco & $50$
& $3.243$ & $4088$ & 0.0017
& $2.573$ & $3755$ & 0.0014
& $2.138$ & $3211$ & 0.0014\\
\rule[-1.5mm]{0mm}{5mm} $\abs{\td m_{jj'}\!-80}\!<\!\td\delta$ &
& 0.329 & 0.0019 & \bf 0.77
& 0.329 & 0.0017 & \bf 0.62
& 0.330 & 0.0015 & \bf 0.55\\\hline
\rule[-1.5mm]{0mm}{5mm} comb.\ $h$-reco & $30$
& $2.806$ & $2662$ & 0.0023
& $2.205$ & $2314$ & 0.0020
& $1.823$ & $1925$ & 0.0020\\
\rule[-1.5mm]{0mm}{5mm} $\abs{\td m_{jj'}\!-80}\!<\!\td\delta$ &
& 0.284 & 0.0012              & \bf 0.82
& 0.282 & 0.0011              & \bf 0.68
& 0.282 & 88$\cdt10^{\trm-5}$ & \bf 0.61\\\hline
\rule[-1.5mm]{0mm}{5mm} comb.\ $h$-reco & $20$
& $2.333$ & $1830$ & 0.0027
& $1.829$ & $1555$ & 0.0025
& $1.507$ & $1293$ & 0.0024\\
\rule[-1.5mm]{0mm}{5mm} $\abs{\td m_{jj'}\!-80}\!<\!\td\delta$ &
& 0.237 & 83$\cdt10^{\trm-5}$ & \bf 0.82
& 0.234 & 71$\cdt10^{\trm-5}$ & \bf 0.69
& 0.233 & 59$\cdt10^{\trm-5}$ & \bf 0.62\\\hline
\rule[-1.5mm]{0mm}{5mm} comb.\ $h$-reco & $16$
& $2.066$ & $1484$ & 0.0030
& $1.617$ & $1255$ & 0.0027
& $1.331$ & $1029$ & 0.0027\\
\rule[-1.5mm]{0mm}{5mm} $\abs{\td m_{jj'}\!-80}\!<\!\td\delta$ &
& 0.209 & 67$\cdt10^{\trm-5}$ & \bf 0.81
& 0.207 & 57$\cdt10^{\trm-5}$ & \bf 0.68
& 0.206 & 47$\cdt10^{\trm-5}$ & \bf 0.61\\\hline
\rule[-1.5mm]{0mm}{5mm} comb.\ $h$-reco & $10$
& $1.436$ & $921.6$ & 0.0033
& $1.121$ & $773.9$ & 0.0030
& $0.9215$& $632.3$ & 0.0030\\
\rule[-1.5mm]{0mm}{5mm} $\abs{\td m_{jj'}\!-80}\!<\!\td\delta$ &
& 0.146 & 42$\cdt10^{\trm-5}$ & \bf 0.71
& 0.143 & 35$\cdt10^{\trm-5}$ & \bf 0.60
& 0.142 & 29$\cdt10^{\trm-5}$ & \bf 0.54\\\hline\hline
\end{tabular}
\caption{\label{tab:cutimpactA2}
  Impact of the different levels of cuts on the $e\nu_e$+jets final
  states for the $gg\to h\to WW$\/ production and decay signal and the
  $W$\!+jets background as obtained from \sherpa. Cross sections
  $\sigma_S$, $\sigma_B$, acceptances $\veps_S$, $\veps_B$ and $S/B$,
  $S/\sqrt{B}$ ratios are shown for Higgs boson masses of
  $M_h=190$, $200$ and $210\ \mrm{GeV}$. Note that
  $\td m_{ij}=m_{ij}/\mrm{GeV}$ and $\td\delta=\delta/\mrm{GeV}$.
  Significances were calculated using \Eqs{eqs:signif} assuming
  ${\cal L}=10\ \mrm{fb}^{-1}$ of integrated luminosity, counting both
  electrons and muons and combining Tevatron experiments.}
\end{table}

\begin{table}[p!]
\centering\footnotesize\vskip5mm
\begin{tabular}{|l|c|rrr|rrr|rrr|}\hline
\rule[-2mm]{0mm}{6mm} cuts \& & $2\,\Delta/$
& $\sigma_S/\mrm{fb}$ & $\sigma_B/\mrm{fb}$ & $S/B$
& $\sigma_S/\mrm{fb}$ & $\sigma_B/\mrm{fb}$ & $S/B$
& $\sigma_S/\mrm{fb}$ & $\sigma_B/\mrm{fb}$ & $S/B$\\
\rule[-2mm]{0mm}{6mm} selections & $\mrm{GeV}$
& $\veps_S$ & $\veps_B$ & $S/\sqrt{B}$
& $\veps_S$ & $\veps_B$ & $S/\sqrt{B}$
& $\veps_S$ & $\veps_B$ & $S/\sqrt{B}$\\\hline\hline
\multicolumn{2}{|r|}{
  \rule[-3mm]{0mm}{8mm}$M_h/\mrm{GeV}\quad[\delta/\mrm{GeV}]$}
&\multicolumn{3}{c|}{$110\quad[20]$}
&\multicolumn{3}{c|}{$120\quad[20]$}
&\multicolumn{3}{c|}{$130\quad[20]$}\\\hline\hline
\rule[-1.5mm]{0mm}{5mm} lepton \& MET &
& $0.2254$ & $19080$ & 27$\cdt10^{\trm-6}$
& $0.8017$ & $19080$ & 97$\cdt10^{\trm-6}$
& $2.260$  & $19080$ & 27$\cdt10^{\trm-5}$\\
\rule[-1.5mm]{0mm}{5mm} cuts \& $\ge2$ jets &
& 0.0673 & 0.0087 & \bf 0.027
& 0.104  & 0.0087 & \bf 0.095
& 0.173  & 0.0087 & \bf 0.27\\\hline
\rule[-1.5mm]{0mm}{5mm} naive $h$-reco & $50$
& $0.02723$ & $321.2$ & 20$\cdt10^{\trm-5}$
& $0.2132$  & $969.2$ & 51$\cdt10^{\trm-5}$
& $0.9580$  & $1963$  & 0.0011\\
\rule[-1.5mm]{0mm}{5mm} &
& 0.00813 & 15$\cdt10^{\trm-5}$ & \bf 0.025
& 0.0276  & 44$\cdt10^{\trm-5}$ & \bf 0.11
& 0.0733  & 89$\cdt10^{\trm-5}$ & \bf 0.35\\\hline
\rule[-1.5mm]{0mm}{5mm} naive $h$-reco & $48$
& $0.00873$ & $34.23$ & 59$\cdt10^{\trm-5}$
& $0.1160$  & $90.62$ & 0.0029
& $0.5804$  & $363.7$ & 0.0037\\
\rule[-1.5mm]{0mm}{5mm} $\abs{\td m_{j_1j_2}\!-80}\!<\!\td\delta$ &
& 0.00261 & 16$\cdt10^{\trm-6}$ & \bf 0.024
& 0.0150  & 41$\cdt10^{\trm-6}$ & \bf 0.20
& 0.0444  & 17$\cdt10^{\trm-5}$ & \bf 0.49\\\hline
\rule[-1.5mm]{0mm}{5mm} comb.\ $h$-reco & $50$
& $0.04236$ & $392.8$ & 25$\cdt10^{\trm-5}$
& $0.3000$  & $1131$  & 61$\cdt10^{\trm-5}$
& $1.253$   & $2239$  & 0.0013\\
\rule[-1.5mm]{0mm}{5mm} &
& 0.0126 & 18$\cdt10^{\trm-5}$ & \bf 0.035
& 0.0389 & 51$\cdt10^{\trm-5}$ & \bf 0.15
& 0.0959 & 0.0010              & \bf 0.43\\\hline
\rule[-1.5mm]{0mm}{5mm} comb.\ $h$-reco & $50$
& $0.01252$ & $41.87$ & 69$\cdt10^{\trm-5}$
& $0.1515$  & $124.0$ & 0.0028
& $0.7673$  & $461.3$ & 0.0038\\
\rule[-1.5mm]{0mm}{5mm} $\abs{\td m_{jj'}\!-80}\!<\!\td\delta$ &
& 0.00374 & 19$\cdt10^{\trm-6}$ & \bf 0.032
& 0.0196  & 56$\cdt10^{\trm-6}$ & \bf 0.22
& 0.0587  & 21$\cdt10^{\trm-5}$ & \bf 0.58\\\hline
\rule[-1.5mm]{0mm}{5mm} comb.\ $h$-reco & $20$
& $0.00607$ & $8.805$ & 0.0016
& $0.1017$  & $23.97$ & 0.0098
& $0.4995$  & $53.81$ & 0.021\\
\rule[-1.5mm]{0mm}{5mm} $\abs{\td m_{jj'}\!-80}\!<\!\td\delta$ &
& 0.00181 & 40$\cdt10^{\trm-7}$ & \bf 0.033
& 0.0132  & 11$\cdt10^{\trm-6}$ & \bf 0.34
& 0.0382  & 24$\cdt10^{\trm-6}$ & \bf 1.11\\\hline\hline
\multicolumn{2}{|r|}{
  \rule[-3mm]{0mm}{8mm}$M_h/\mrm{GeV}\quad[\delta/\mrm{GeV}]$}
&\multicolumn{3}{c|}{}
&\multicolumn{3}{c|}{$140\quad[20]$}
&\multicolumn{3}{c|}{$150\quad[20]$}\\\hline\hline
\rule[-1.5mm]{0mm}{5mm} lepton \& MET &
&         &        &
& $4.316$ & $19080$ & 52$\cdt10^{\trm-5}$
& $6.343$ & $19080$ & 77$\cdt10^{\trm-5}$\\
\rule[-1.5mm]{0mm}{5mm} cuts \& $\ge2$ jets &
&         &      &
& 0.250 & 0.0087 & \bf 0.51
& 0.328 & 0.0087 & \bf 0.75\\\hline
\rule[-1.5mm]{0mm}{5mm} naive $h$-reco & $50$
&         &        &
& $2.213$ & $3230$ & 0.0016
& $3.642$ & $4593$ & 0.0018\\
\rule[-1.5mm]{0mm}{5mm} &
&         &      &
& 0.128 & 0.0015 & \bf 0.63
& 0.188 & 0.0021 & \bf 0.88\\\hline
\rule[-1.5mm]{0mm}{5mm} naive $h$-reco & $48$
&         &        &
& $1.374$ & $924.4$ & 0.0034
& $2.509$ & $1662$  & 0.0035\\
\rule[-1.5mm]{0mm}{5mm} $\abs{\td m_{j_1j_2}\!-80}\!<\!\td\delta$ &
&        &        &
& 0.0795 & 42$\cdt10^{\trm-5}$ & \bf 0.73
& 0.130  & 76$\cdt10^{\trm-5}$ & \bf 1.00\\\hline
\rule[-1.5mm]{0mm}{5mm} comb.\ $h$-reco & $50$
&         &        &
& $2.902$ & $3637$ & 0.0018
& $4.778$ & $5103$ & 0.0022\\
\rule[-1.5mm]{0mm}{5mm} &
&       &        &
& 0.168 & 0.0017 & \bf 0.78
& 0.247 & 0.0023 & \bf 1.09\\\hline
\rule[-1.5mm]{0mm}{5mm} comb.\ $h$-reco & $50$
&         &        &
& $1.877$ & $1121$ & 0.0038
& $3.487$ & $1968$ & 0.0041\\
\rule[-1.5mm]{0mm}{5mm} $\abs{\td m_{jj'}\!-80}\!<\!\td\delta$ &
&       &        &
& 0.109 & 51$\cdt10^{\trm-5}$ & \bf 0.91
& 0.180 & 89$\cdt10^{\trm-5}$ & \bf 1.28\\\hline
\rule[-1.5mm]{0mm}{5mm} comb.\ $h$-reco & $20$
&         &         &
& $1.213$ & $234.1$ & 0.012
& $2.462$ & $704.8$ & 0.0080\\
\rule[-1.5mm]{0mm}{5mm} $\abs{\td m_{jj'}\!-80}\!<\!\td\delta$ &
&        &                     &
& 0.0701 & 11$\cdt10^{\trm-5}$ & \bf 1.29
& 0.127  & 32$\cdt10^{\trm-5}$ & \bf 1.51\\\hline
\end{tabular}
\caption{\label{tab:cutimpactBelow}
  Impact of the different levels of cuts on the $e\nu_e$+jets final
  states for the $gg\to h\to WW$\/ production and decay signal and the
  $W$\!+jets background as obtained from \sherpa. Cross sections
  $\sigma_S$, $\sigma_B$, acceptances $\veps_S$, $\veps_B$ and $S/B$,
  $S/\sqrt{B}$ ratios are shown for Higgs boson masses below the
  on-shell diboson mass threshold from $M_h=110\ \mrm{GeV}$ to
  $M_h=150\ \mrm{GeV}$. Note that $\td m_{ij}=m_{ij}/\mrm{GeV}$ and
  $\td\delta=\delta/\mrm{GeV}$. The significances were calculated
  according to \Eqs{eqs:signif} assuming ${\cal L}=10\ \mrm{fb}^{-1}$
  of integrated luminosity, counting both electrons and muons and
  combining Tevatron experiments. The layout of the table is the same
  as in \Tabs{tab:cutimpactA1}~and~\ref{tab:cutimpactA2}, however a
  smaller number of Higgs boson candidate selections is shown.}
\end{table}

\subsection{Ideal Higgs boson reconstruction analyses}
\label{app:hreco}

We first complete the presentation of our \Sec{sec:hreco} main results
by showing \Tabs{tab:cutimpactA2}~and~\ref{tab:cutimpactBelow} where
we display the numbers associated with the high and low Higgs boson
mass region, respectively. As in \Tab{tab:cutimpactA1} we list the
signal and $W$\!+jet background cross sections, selection efficiencies,
$S/B$\/ ratios and significances at different analysis levels. All
cuts, the selection procedures, the layout of the tables and the
interpretation of the results given in the tables have been discussed
in detail in \Sec{sec:hreco}. Note that the rightmost column of
\Tab{tab:cutimpactAx} carries the outcomes for the test mass point
$M_h=220\ \mrm{GeV}$.

One remark shall be added regarding the magnitude of off-shell effects.
The loss one faces due to off-shell Higgs bosons can be read off
\Tab{tab:cutimpactBelow} by comparing the acceptances after baseline
(1st rows) and combinatorial Higgs boson selection (4th rows). While
above the $WW$\/ mass threshold the loss on the signal (background) is
mild (significant) ranging from $1.2$--$1.3$ ($2.6$--$3.2$), it steadily
increases for decreasing $M_h$, approaching $1.8$ and $5.3$ at
$M_h=130\ \mrm{GeV}$ and $M_h=110\ \mrm{GeV}$, respectively. The
background loss factors turn huge (up to 48) because of the steeply
falling $m_{e\nu_e jj'}$ spectrum (cf.\ \Fig{fig:mass_spectrum}), but
this cannot overcome the smallness of $S/\sqrt{B}$ due to the signal
reduction.

We now present the results of our side studies, which we decided to
put in an appendix in order to not distract the flow of the main body.

\begin{figure}[t!]
\psfrag{Significance}[c][c][0.92]{
  \fontfamily{phv}\selectfont{Significance}\quad$S/\sqrt{B}$}
\psfrag{Higgs boson mass}[c][c][0.92]{
  \fontfamily{phv}\selectfont{Higgs boson mass}\quad$M^\mrm{inj}_h$}
\centering\vskip2mm
  \includegraphics[clip,width=0.74\columnwidth,angle=-90]{%
    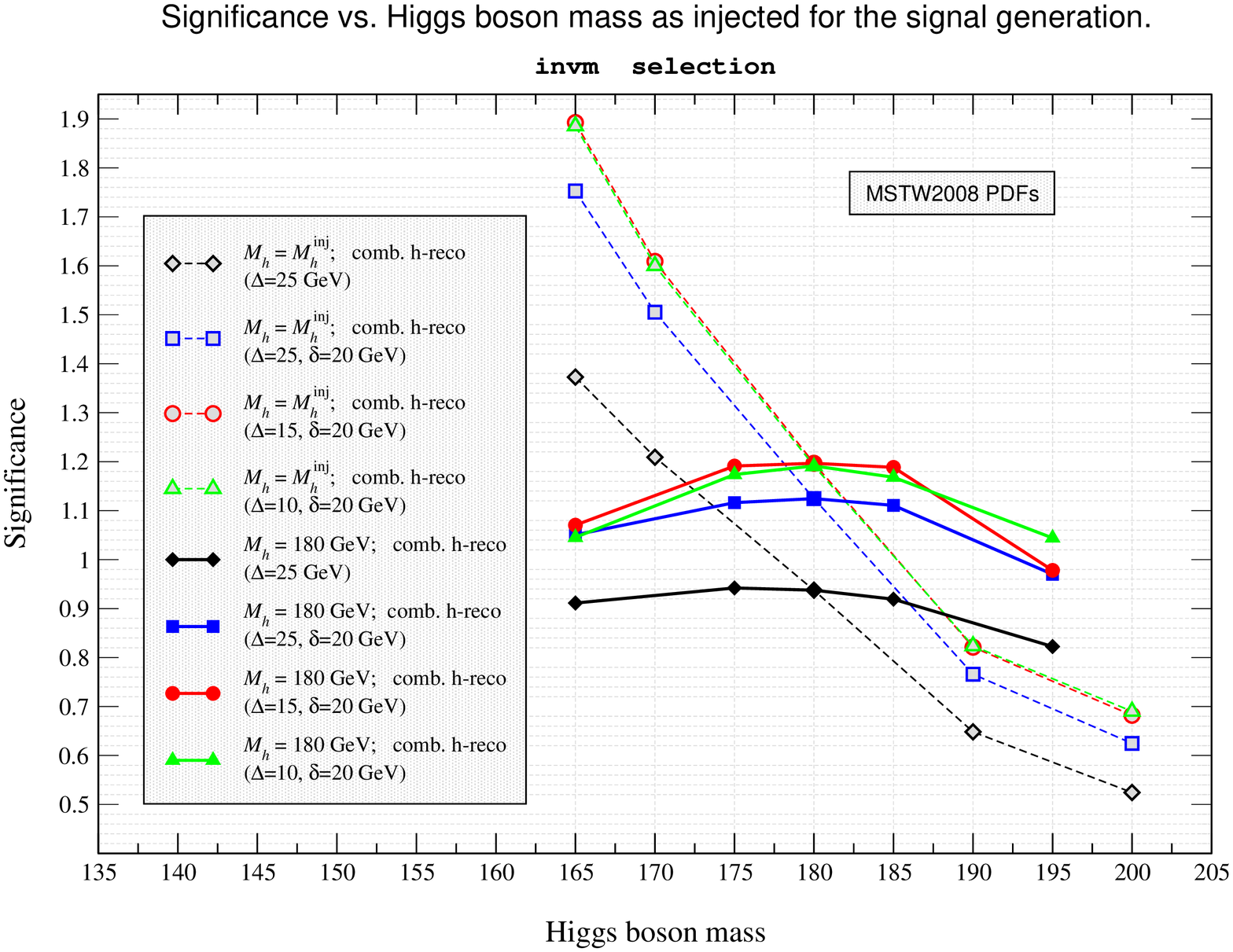}
  \caption{\label{fig:inj_significances}
    $S/\sqrt{B}$ significances as a function of injected Higgs boson
    masses varying from $M^\mrm{inj}_h=165\ \mrm{GeV}$ to $200\ \mrm{GeV}$
    for different mass window parameters $\Delta$ and $\delta$.
    Results of two combinatorial analyses based on the \texttt{invm}
    reconstruction are shown: one using the default setting
    $M_h=M^\mrm{inj}_h$ (dashed lines), the other where the
    hypothesized Higgs boson mass is fixed at $M_h=180\ \mrm{GeV}$
    (solid lines). For the different injected Higgs boson masses, the
    $e\nu_e$+jets final states are generated from the $gg\to h\to WW$\/
    signal and the $W$\!+jets production background. All significances
    were calculated according to \Eqs{eqs:signif} taking only the
    dominant background into account and under the assumption
    of an integrated luminosity of ${\cal L}=10\ \mrm{fb}^{-1}$,
    including electron and muon channels, \ie $f_\ell=2$.}
\end{figure}

\begin{table}[t!]
\centering\footnotesize\vskip5mm
\begin{tabular}{|l|c|rrr|rrr|rrr|}\hline
\rule[-2mm]{0mm}{6mm} cuts \& & $2\,\Delta/$
& $\sigma_S/\mrm{fb}$ & $\sigma_B/\mrm{fb}$ & $S/B$
& $\sigma_S/\mrm{fb}$ & $\sigma_B/\mrm{fb}$ & $S/B$
& $\sigma_S/\mrm{fb}$ & $\sigma_B/\mrm{fb}$ & $S/B$\\
\rule[-2mm]{0mm}{6mm} selections & $\mrm{GeV}$
& $\veps_S$ & $\veps_B$ & $S/\sqrt{B}$
& $\veps_S$ & $\veps_B$ & $S/\sqrt{B}$
& $\veps_S$ & $\veps_B$ & $S/\sqrt{B}$\\\hline\hline
\multicolumn{2}{|r|}{
  \rule[-3mm]{0mm}{8mm}$M_h/\mrm{GeV}\quad[\delta/\mrm{GeV}]$}
&\multicolumn{3}{c|}{$165\quad[20]$}
&\multicolumn{3}{c|}{$170\quad[20]$}
&\multicolumn{3}{c|}{$180\quad[20]$}\\\hline\hline
\rule[-1.5mm]{0mm}{5mm} naive $h$-reco & $50$
& $4.495$ & $5317$ & 0.0020
& $4.121$ & $5317$ & 0.0018
& $3.232$ & $5527$ & 0.0013\\
\rule[-1.5mm]{0mm}{5mm} &
& 0.297 & 0.0022 & \bf 1.14
& 0.296 & 0.0022 & \bf 1.02
& 0.290 & 0.0023 & \bf 0.77\\\hline
\rule[-1.5mm]{0mm}{5mm} naive $h$-reco & $48$
& $3.965$ & $2371$ & 0.0040
& $3.585$ & $2571$ & 0.0032
& $2.759$ & $2849$ & 0.0022\\
\rule[-1.5mm]{0mm}{5mm} $\abs{\td m_{j_1j_2}\!-80}\!<\!\td\delta$ &
& 0.262 & 0.0010 & \bf 1.50
& 0.258 & 0.0011 & \bf 1.27
& 0.248 & 0.0012 & \bf 0.92\\\hline
\rule[-1.5mm]{0mm}{5mm} comb.\ $h$-reco & $50$
& $5.603$ & $5511$ & 0.0024
& $5.149$ & $5758$ & 0.0021
& $4.105$ & $5982$ & 0.0016\\
\rule[-1.5mm]{0mm}{5mm} &
& 0.370 & 0.0023 & \bf 1.39
& 0.370 & 0.0024 & \bf 1.22
& 0.368 & 0.0025 & \bf 0.94\\\hline
\rule[-1.5mm]{0mm}{5mm} comb.\ $h$-reco & $50$
& $5.081$ & $2632$ & 0.0046
& $4.690$ & $2923$ & 0.0037
& $3.722$ & $3251$ & 0.0026\\
\rule[-1.5mm]{0mm}{5mm} $\abs{\td m_{jj'}\!-80}\!<\!\td\delta$ &
& 0.336 & 0.0011 & \bf 1.83
& 0.337 & 0.0012 & \bf 1.56
& 0.334 & 0.0014 & \bf 1.16\\\hline
\rule[-1.5mm]{0mm}{5mm} comb.\ $h$-reco & $20$
& $3.890$ & $1206$ & 0.0076
& $3.592$ & $1357$ & 0.0061
& $2.846$ & $1518$ & 0.0043\\
\rule[-1.5mm]{0mm}{5mm} $\abs{\td m_{jj'}\!-80}\!<\!\td\delta$ &
& 0.257 & 51$\cdt10^{\trm-5}$ & \bf 2.07
& 0.258 & 57$\cdt10^{\trm-5}$ & \bf 1.76
& 0.255 & 64$\cdt10^{\trm-5}$ & \bf 1.30\\\hline\hline
\multicolumn{2}{|r|}{
  \rule[-3mm]{0mm}{8mm}$M_h/\mrm{GeV}\quad[\delta/\mrm{GeV}]$}
&\multicolumn{3}{c|}{$190\quad[20]$}
&\multicolumn{3}{c|}{$200\quad[20]$}
&\multicolumn{3}{c|}{$210\quad[20]$}\\\hline\hline
\rule[-1.5mm]{0mm}{5mm} naive $h$-reco & $50$
& $2.231$ & $5435$ & 92$\cdt10^{\trm-5}$
& $1.779$ & $5158$ & 76$\cdt10^{\trm-5}$
& $1.480$ & $4787$ & 67$\cdt10^{\trm-5}$\\
\rule[-1.5mm]{0mm}{5mm} &
& 0.287 & 0.0023 & \bf 0.53
& 0.287 & 0.0022 & \bf 0.42
& 0.288 & 0.0020 & \bf 0.36\\\hline
\rule[-1.5mm]{0mm}{5mm} naive $h$-reco & $48$
& $1.867$ & $2821$ & 0.0015
& $1.465$ & $2560$ & 0.0013
& $1.203$ & $2161$ & 0.0012\\
\rule[-1.5mm]{0mm}{5mm} $\abs{\td m_{j_1j_2}\!-80}\!<\!\td\delta$ &
& 0.240 & 0.0012              & \bf 0.61
& 0.237 & 0.0011              & \bf 0.49
& 0.234 & 91$\cdt10^{\trm-5}$ & \bf 0.44\\\hline
\rule[-1.5mm]{0mm}{5mm} comb.\ $h$-reco & $50$
& $2.865$ & $5892$ & 0.0011
& $2.302$ & $5607$ & 90$\cdt10^{\trm-5}$
& $1.925$ & $5221$ & 80$\cdt10^{\trm-5}$\\
\rule[-1.5mm]{0mm}{5mm} &
& 0.368 & 0.0025 & \bf 0.65
& 0.372 & 0.0024 & \bf 0.53
& 0.375 & 0.0022 & \bf 0.45\\\hline
\rule[-1.5mm]{0mm}{5mm} comb.\ $h$-reco & $50$
& $2.575$ & $3243$ & 0.0018
& $2.052$ & $2988$ & 0.0015
& $1.702$ & $2557$ & 0.0014\\
\rule[-1.5mm]{0mm}{5mm} $\abs{\td m_{jj'}\!-80}\!<\!\td\delta$ &
& 0.331 & 0.0014 & \bf 0.79
& 0.331 & 0.0013 & \bf 0.64
& 0.332 & 0.0011 & \bf 0.57\\\hline
\rule[-1.5mm]{0mm}{5mm} comb.\ $h$-reco & $20$
& $1.949$ & $1445$ & 0.0030
& $1.536$ & $1242$ & 0.0027
& $1.265$ & $1029$ & 0.0027\\
\rule[-1.5mm]{0mm}{5mm} $\abs{\td m_{jj'}\!-80}\!<\!\td\delta$ &
& 0.251 & 61$\cdt10^{\trm-5}$ & \bf 0.89
& 0.248 & 53$\cdt10^{\trm-5}$ & \bf 0.74
& 0.247 & 44$\cdt10^{\trm-5}$ & \bf 0.66\\\hline
\end{tabular}
\caption{\label{tab:cutimpact66}
Impact of the different levels of cuts on the $e\nu_e$+jets final
states for the $gg\to h\to WW$\/ production and decay signal and the
$W$\!+jets background as obtained from \sherpa when using the CTEQ6.6
PDF libraries. Cross sections $\sigma_S$, $\sigma_B$, acceptances
$\veps_S$, $\veps_B$ and $S/B$, $S/\sqrt{B}$ ratios are shown for
Higgs boson masses from $M_h=165\ \mrm{GeV}$ to $M_h=210\ \mrm{GeV}$.
Note that $\td m_{ij}=m_{ij}/\mrm{GeV}$ and
$\td\delta=\delta/\mrm{GeV}$. All significances were calculated
according to \Eqs{eqs:signif} assuming ${\cal L}=10\ \mrm{fb}^{-1}$ of
integrated luminosity, counting both electron and muon channels and
combining Tevatron experiments.}
\end{table}

The Higgs boson masses used in the analyses are, of course,
hypothetical. However, we only considered the obvious scenario where
the test mass $M_h$ has been set equal to the Higgs boson mass
$M^\mrm{inj}_h$ injected while generating the signal predictions. It
is clear that one scans over a range of masses when pursuing an
analysis, and here we do it in steps of 10~GeV; still the actual Higgs
boson mass could deviate as much as 5~GeV from the assumed mass.
Hence, we want to briefly study how strongly the \texttt{invm} Higgs
boson candidate selections and their related significances depend on
the match between the test and injected Higgs boson mass. To this end
we generated other than default signal predictions for Higgs boson
masses of $M^\mrm{inj}_h=165, 175, 185, 195\ \mrm{GeV}$ and input them
into the analyses using $M_h=180\ \mrm{GeV}$.
\Fig{fig:inj_significances} shows the outcome of this side study where,
for both types of analyses, we collected results for several mass
window settings. We learn two things from plotting the significance as
a function of the Higgs boson generation mass. First, the
significances that we attain if we keep the selection parameters
($M_h$, $\Delta$, $\delta$) constant are fairly robust over a broader
range of generation masses. Yet the maximum $S/\sqrt B$\/ occur for
$M_h=M^\mrm{inj}_h$. Secondly we learn, asymmetric window placements
such that $M_h<M^\mrm{inj}_h$ are beneficial to achieve significance
gains. By focusing on a single generation mass, \eg
$M^\mrm{inj}_h=190\ \mrm{GeV}$, we see that the response in
significance can easily get as large as 40\%. The strong sensitivity
can be understood by comparing the signal and background shapes as
visualized in \Fig{fig:mass_spectrum} or the upper left plot of
\Fig{fig:sec2}.

\begin{table}[p!]
\centering\footnotesize\vskip0mm
\begin{tabular}{|l|c|rrr|rrr|rrr|}\hline
\rule[-2mm]{0mm}{6mm} cuts \& & $2\,\Delta/$
& $\sigma_S/\mrm{fb}$ & $\sigma_B/\mrm{fb}$ & $S/B$
& $\sigma_S/\mrm{fb}$ & $\sigma_B/\mrm{fb}$ & $S/B$
& $\sigma_S/\mrm{fb}$ & $\sigma_B/\mrm{fb}$ & $S/B$\\
\rule[-2mm]{0mm}{6mm} selections & $\mrm{GeV}$
& $\veps_S$ & $\veps_B$ & $S/\sqrt{B}$
& $\veps_S$ & $\veps_B$ & $S/\sqrt{B}$
& $\veps_S$ & $\veps_B$ & $S/\sqrt{B}$\\\hline\hline
\multicolumn{2}{|r|}{
  \rule[-3mm]{0mm}{8mm}$M_h/\mrm{GeV}\quad[\delta/\mrm{GeV}]$}
&\multicolumn{3}{c|}{$180\quad[15]$}
&\multicolumn{3}{c|}{$180\quad[20]\,[p^\mrm{jet}_T/\mrm{GeV}\!>\!30]$}
&\multicolumn{3}{c|}{$220\quad[20]$}\\\hline\hline
\rule[-1.5mm]{0mm}{5mm} $\sigma^{(0)}$ &
& $14.19$ & $220\cdt10^4$ & 14$\cdt10^{\trm-6}$
& $14.19$ & $220\cdt10^4$ & 14$\cdt10^{\trm-6}$
& $5.420$ & $220\cdt10^4$ & 51$\cdt10^{\trm-7}$\\
\rule[-1.5mm]{0mm}{5mm} &
& 1.0 & 1.0 & \bf 0.15
& 1.0 & 1.0 & \bf 0.15
& 1.0 & 1.0 & \bf 0.05\\\hline
\rule[-1.5mm]{0mm}{5mm} lepton \& &
& $7.946$ & $984\cdt10^3$ & 18$\cdt10^{\trm-6}$
& $8.127$ & $987\cdt10^3$ & 18$\cdt10^{\trm-6}$
& $3.101$ & $984\cdt10^3$ & 65$\cdt10^{\trm-7}$\\
\rule[-1.5mm]{0mm}{5mm} MET cuts &
& 0.560 & 0.45 & \bf 0.12
& 0.573 & 0.45 & \bf 0.13
& 0.572 & 0.45 & \bf 0.05\\\hline
\rule[-1.5mm]{0mm}{5mm} as above \& &
& $6.471$ & $191\cdt10^2$ & 74$\cdt10^{\trm-5}$
& $3.907$ & $6715$        & 0.0013
& $2.661$ & $191\cdt10^2$ & 29$\cdt10^{\trm-5}$\\
\rule[-1.5mm]{0mm}{5mm} $\ge2$ jets &
& 0.456 & 0.0087 & \bf 0.72
& 0.275 & 0.0031 & \bf 0.73
& 0.491 & 0.0087 & \bf 0.28\\\hline
\rule[-1.5mm]{0mm}{5mm} as above \& &
& $3.169$ & $5272$ & 0.0013
& $1.932$ & $2194$ & 0.0019
& $1.413$ & $6997$ & 42$\cdt10^{\trm-5}$\\
\rule[-1.5mm]{0mm}{5mm} $\abs{\td m_{j_1j_2}\!-80}\!<\!\td\delta$ &
& 0.223 & 0.0024 & \bf 0.67
& 0.136 & 0.0010 & \bf 0.63
& 0.261 & 0.0032 & \bf 0.25\\\hline\hline
\rule[-1.5mm]{0mm}{5mm} naive $h$-reco & $50$
& $3.911$ & $6749$ & 0.0013
& $1.948$ & $1429$ & 0.0030
& $1.510$ & $5342$ & 58$\cdt10^{\trm-5}$\\
\rule[-1.5mm]{0mm}{5mm} &
& 0.276 & 0.0031              & \bf 0.73
& 0.137 & 65$\cdt10^{\trm-5}$ & \bf 0.79
& 0.279 & 0.0024              & \bf 0.30\\\hline
\rule[-1.5mm]{0mm}{5mm} naive $h$-reco & $30$
& $3.039$ & $4199$  & 0.0016
& $1.551$ & $907.0$ & 0.0037
& $1.161$ & $3196$  & 75$\cdt10^{\trm-5}$\\
\rule[-1.5mm]{0mm}{5mm} &
& 0.214 & 0.0019              & \bf 0.72
& 0.109 & 41$\cdt10^{\trm-5}$ & \bf 0.79
& 0.214 & 0.0015              & \bf 0.30\\\hline
\rule[-1.5mm]{0mm}{5mm} naive $h$-reco & $48$
& $2.857$ & $2778$  & 0.0022
& $1.631$ & $925.5$ & 0.0038
& $1.169$ & $2167$  & 0.0011\\
\rule[-1.5mm]{0mm}{5mm} $\abs{\td m_{j_1j_2}\!-80}\!<\!\td\delta$ &
& 0.201 & 0.0013              & \bf 0.83
& 0.115 & 42$\cdt10^{\trm-5}$ & \bf 0.82
& 0.216 & 99$\cdt10^{\trm-5}$ & \bf 0.37\\\hline
\rule[-1.5mm]{0mm}{5mm} naive $h$-reco & $20$
& $2.118$ & $1351$  & 0.0034
& $1.170$ & $454.3$ & 0.0056
& $0.7991$& $874.9$ & 0.0019\\
\rule[-1.5mm]{0mm}{5mm} $\abs{\td m_{j_1j_2}\!-80}\!<\!\td\delta$ &
& 0.149  & 61$\cdt10^{\trm-5}$ & \bf 0.89
& 0.0825 & 21$\cdt10^{\trm-5}$ & \bf 0.84
& 0.147  & 40$\cdt10^{\trm-5}$ & \bf 0.39\\\hline\hline
\rule[-1.5mm]{0mm}{5mm} comb.\ $h$-reco & $50$
& $5.241$ & $7396$ & 0.0015 
& $2.484$ & $1534$ & 0.0035
& $2.088$ & $5942$ & 73$\cdt10^{\trm-5}$\\
\rule[-1.5mm]{0mm}{5mm} &
& 0.369 & 0.0034              & \bf 0.94 
& 0.175 & 70$\cdt10^{\trm-5}$ & \bf 0.98
& 0.385 & 0.0027              & \bf 0.40\\\hline
\rule[-1.5mm]{0mm}{5mm} comb.\ $h$-reco & $20$
& $3.657$ & $3255$  & 0.0024 
& $1.718$ & $671.7$ & 0.0056
& $1.440$ & $2508$ & 0.0012\\
\rule[-1.5mm]{0mm}{5mm} &
& 0.258 & 0.0015              & \bf 0.99 
& 0.121 & 31$\cdt10^{\trm-5}$ & \bf 1.02
& 0.266 & 0.0011              & \bf 0.42\\\hline
\rule[-1.5mm]{0mm}{5mm} comb.\ $h$-reco & $50$
& $4.248$ & $3241$ & 0.0029
& $2.185$ & $1024$ & 0.0046
& $1.798$ & $2662$ & 0.0014\\
\rule[-1.5mm]{0mm}{5mm} $\abs{\td m_{jj'}\!-80}\!<\!\td\delta$ &
& 0.299 & 0.0015              & \bf 1.15
& 0.154 & 47$\cdt10^{\trm-5}$ & \bf 1.05
& 0.332 & 0.0012              & \bf 0.51\\\hline
\rule[-1.5mm]{0mm}{5mm} comb.\ $h$-reco & $30$
& $3.845$ & $2220$  & 0.0038
& $1.916$ & $699.8$ & 0.0060
& $1.528$ & $1581$  & 0.0020\\
\rule[-1.5mm]{0mm}{5mm} $\abs{\td m_{jj'}\!-80}\!<\!\td\delta$ &
& 0.271 & 0.0010              & \bf 1.26
& 0.135 & 32$\cdt10^{\trm-5}$ & \bf 1.11
& 0.282 & 72$\cdt10^{\trm-5}$ & \bf 0.56\\\hline
\rule[-1.5mm]{0mm}{5mm} comb.\ $h$-reco & $20$
& $3.260$ & $1564$  & 0.0045
& $1.617$ & $492.9$ & 0.0071
& $1.264$ & $1058$  & 0.0025\\
\rule[-1.5mm]{0mm}{5mm} $\abs{\td m_{jj'}\!-80}\!<\!\td\delta$ &
& 0.230 & 71$\cdt10^{\trm-5}$ & \bf 1.27
& 0.114 & 22$\cdt10^{\trm-5}$ & \bf 1.12
& 0.233 & 48$\cdt10^{\trm-5}$ & \bf 0.57\\\hline
\rule[-1.5mm]{0mm}{5mm} comb.\ $h$-reco & $16$
& $2.895$ & $1273$  & 0.0049
& $1.444$ & $399.0$ & 0.0079
& $1.115$ & $846.6$ & 0.0027\\
\rule[-1.5mm]{0mm}{5mm} $\abs{\td m_{jj'}\!-80}\!<\!\td\delta$ &
& 0.204 & 58$\cdt10^{\trm-5}$ & \bf 1.25
& 0.102 & 18$\cdt10^{\trm-5}$ & \bf 1.11
& 0.206 & 38$\cdt10^{\trm-5}$ & \bf 0.56\\\hline
\rule[-1.5mm]{0mm}{5mm} comb.\ $h$-reco & $10$
& $2.010$ & $805.0$ & 0.0054
& $1.038$ & $258.5$ & 0.0087
& $0.7722$& $525.2$ & 0.0030\\
\rule[-1.5mm]{0mm}{5mm} $\abs{\td m_{jj'}\!-80}\!<\!\td\delta$ &
& 0.142  & 37$\cdt10^{\trm-5}$ & \bf 1.09
& 0.0731 & 12$\cdt10^{\trm-5}$ & \bf 0.99
& 0.142  & 24$\cdt10^{\trm-5}$ & \bf 0.49\\\hline\hline
\end{tabular}
\caption{\label{tab:cutimpactAx}
Impact of the different levels of cuts on the $e\nu_e$+jets final
states for the $gg\to h\to WW$\/ production and decay signal and the
$W$\!+jets background as obtained from \sherpa. Cross sections
$\sigma_S$, $\sigma_B$, acceptances $\veps_S$, $\veps_B$ and $S/B$,
$S/\sqrt{B}$ ratios are given for Higgs boson masses of
$M_h=180\ \mrm{GeV}$ and $M_h=220\ \mrm{GeV}$. For the former mass
point, the central column shows the values for a jet $p_T$ threshold
increased by $10\ \mrm{GeV}$, while the left column has the values
for a smaller dijet mass window of $\delta=15\ \mrm{GeV}$. A further
decrease of the dijet mass window to $\delta=10\ \mrm{GeV}$ yields
$S/B=0.0034$ as well as $S/\sqrt{B}=1.14$ and $S/B=0.0058$ as well as
$S/\sqrt{B}=1.36$ for the combinatorial Higgs boson reconstruction and
mass windows of $\td\Delta=50$ and $\td\Delta=20$, respectively.
Note that $\td m_{ij}=m_{ij}/\mrm{GeV}$; all other mass variables
denoted by a tilde are understood in the same way. All significances
were calculated according to \Eqs{eqs:signif} assuming
${\cal L}=10\ \mrm{fb}^{-1}$ of integrated luminosity, counting both
electron and muon channels and combining Tevatron experiments.}
\end{table}

In the remainder of this appendix, we outline the impact of PDF
variations and parameter variations other than $M_h$, $\Delta$ and
$\delta$ on our analyses.

The \sherpa calculations resulting from using the CTEQ6.6 PDF
libraries give a similar pattern with significances that are about
1--10\% larger. This can be read off \Tab{tab:cutimpact66} and seen in
\Fig{fig:invm_significances}. By normalizing the Monte Carlo
predictions for both PDF choices, MSTW2008 and CTEQ6.6, to the same
respective theory cross sections, it is altogether reassuring to see
that the cut and event selection procedures only induce deviations
of the order of 10\% or below. For the combinatorial Higgs boson
candidate selection, one finds almost the same $S/\sqrt{B}$ ratios,
provided a wide Higgs boson mass window is used. The significances of
the CTEQ6.6 calculations outperform those obtained with MSTW2008, once
the mass windows for the Higgs bosons ($\Delta$) and dijets ($\delta$)
are tightened. We find the differences being more pronounced for Higgs
boson masses just above the $WW$\/ threshold.

Speaking of shape differences triggered by the use of different PDFs,
we note that the MSTW2008 PDF set accounts for a larger transverse
activity, \ie rapidity distributions turn out steeper while the $p_T$
spectra develop 10--25\% harder tails compared to the CTEQ6.6
predictions. Also, for MSTW2008, mass peaks are more washed out, again
resulting in differences of the order of 25\%.

Lastly, apart from the rightmost column, \Tab{tab:cutimpactAx} gives
more details for the choice $M_h=M^\mrm{inj}_h=180\ \mrm{GeV}$
including the cases where either the dijet mass window has been
tightened from $\delta=20\ \mrm{GeV}$ to $\delta=15\ \mrm{GeV}$ or the
threshold of the jet transverse momenta has been enhanced to
$p^\mrm{jet}_T>30\ \mrm{GeV}$.

\subsection{More realistic Higgs boson reconstruction analyses}
\label{app:hrealreco}

\begin{figure}[p!]
  \centering\vskip-2mm\hskip15pt
  \includegraphics[clip,width=0.46\columnwidth,angle=0]{%
    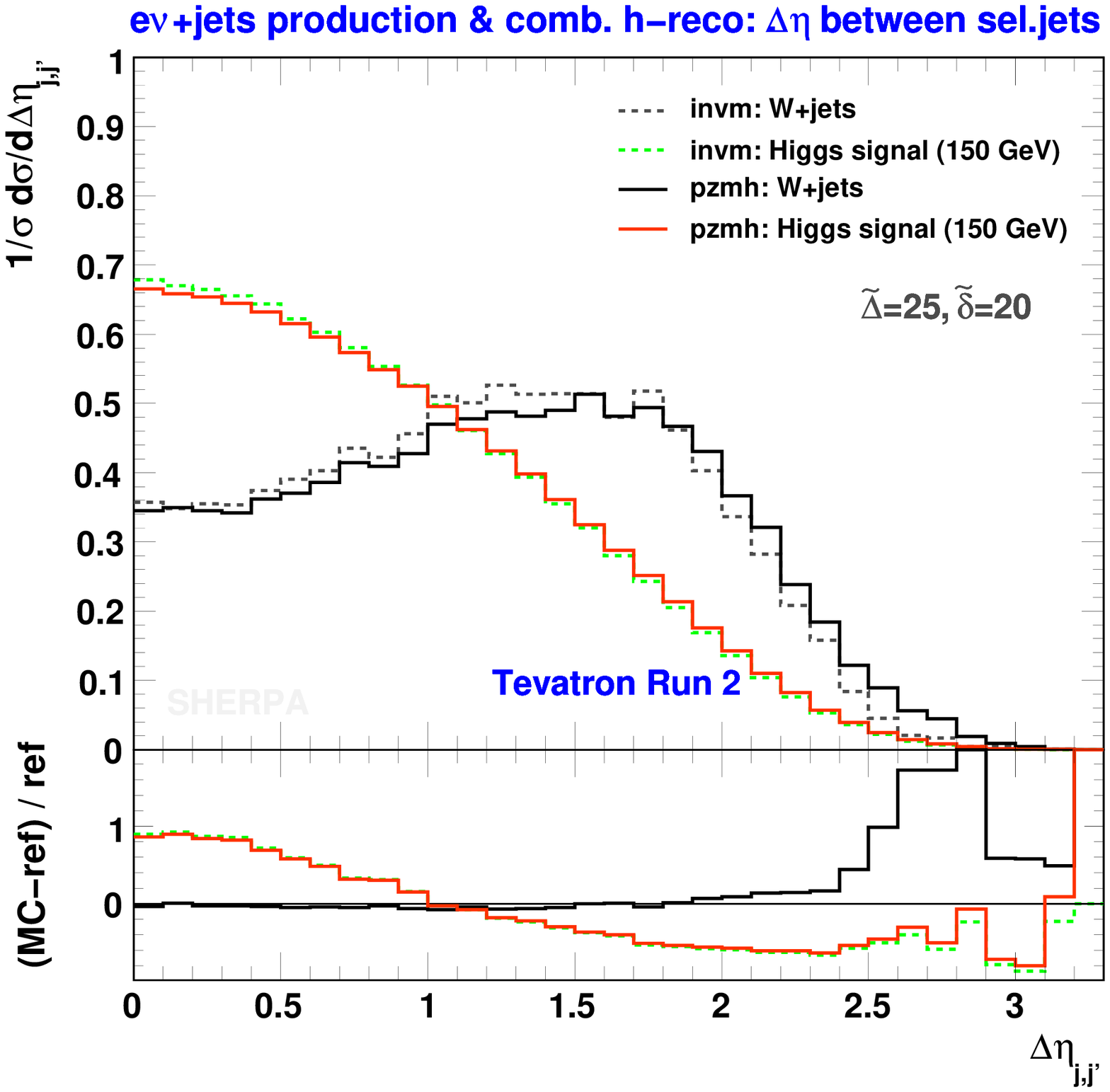}
  \includegraphics[clip,width=0.46\columnwidth,angle=0]{%
    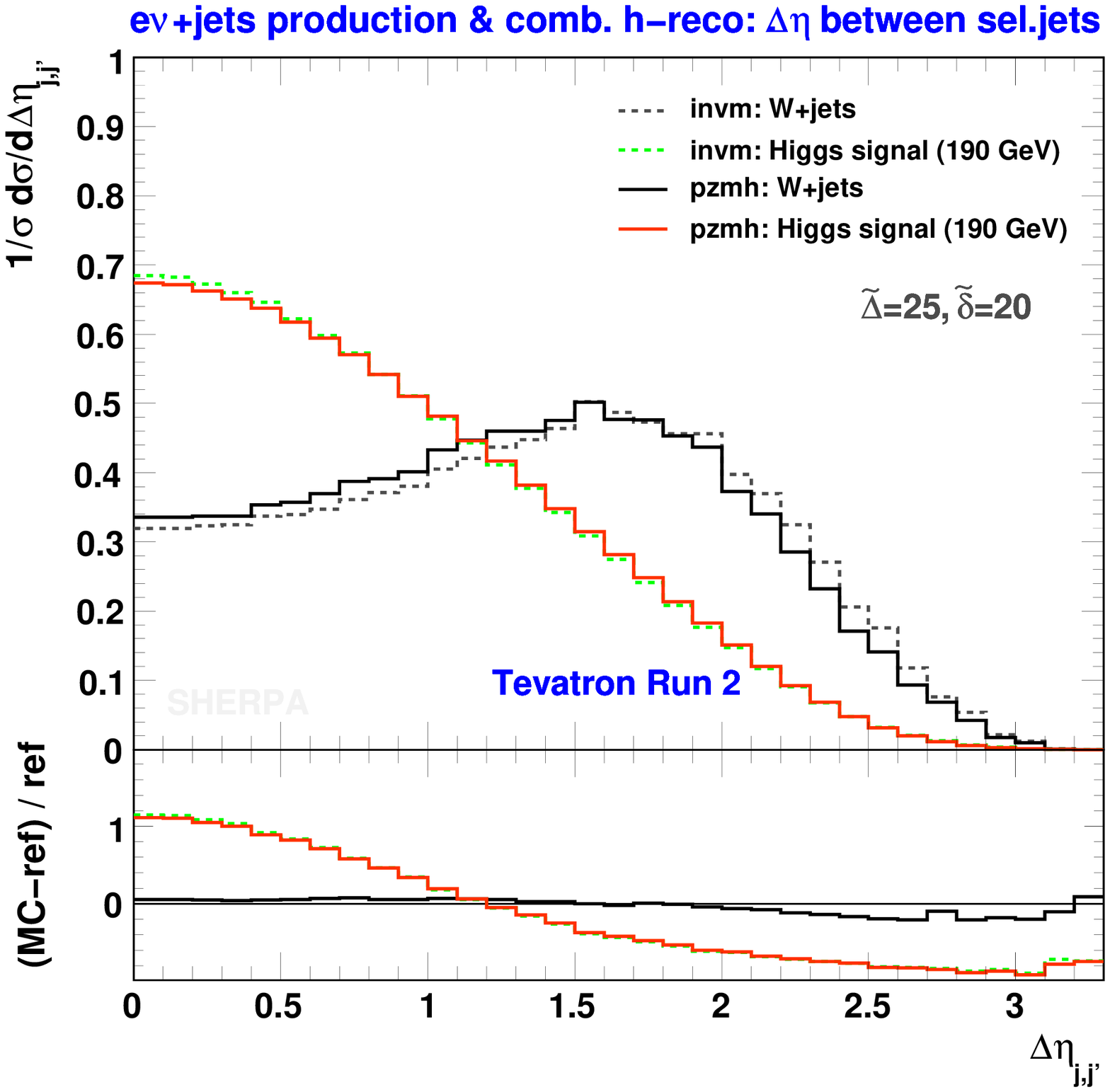}\\[-4mm]\hskip15pt
  \includegraphics[clip,width=0.46\columnwidth,angle=0]{%
    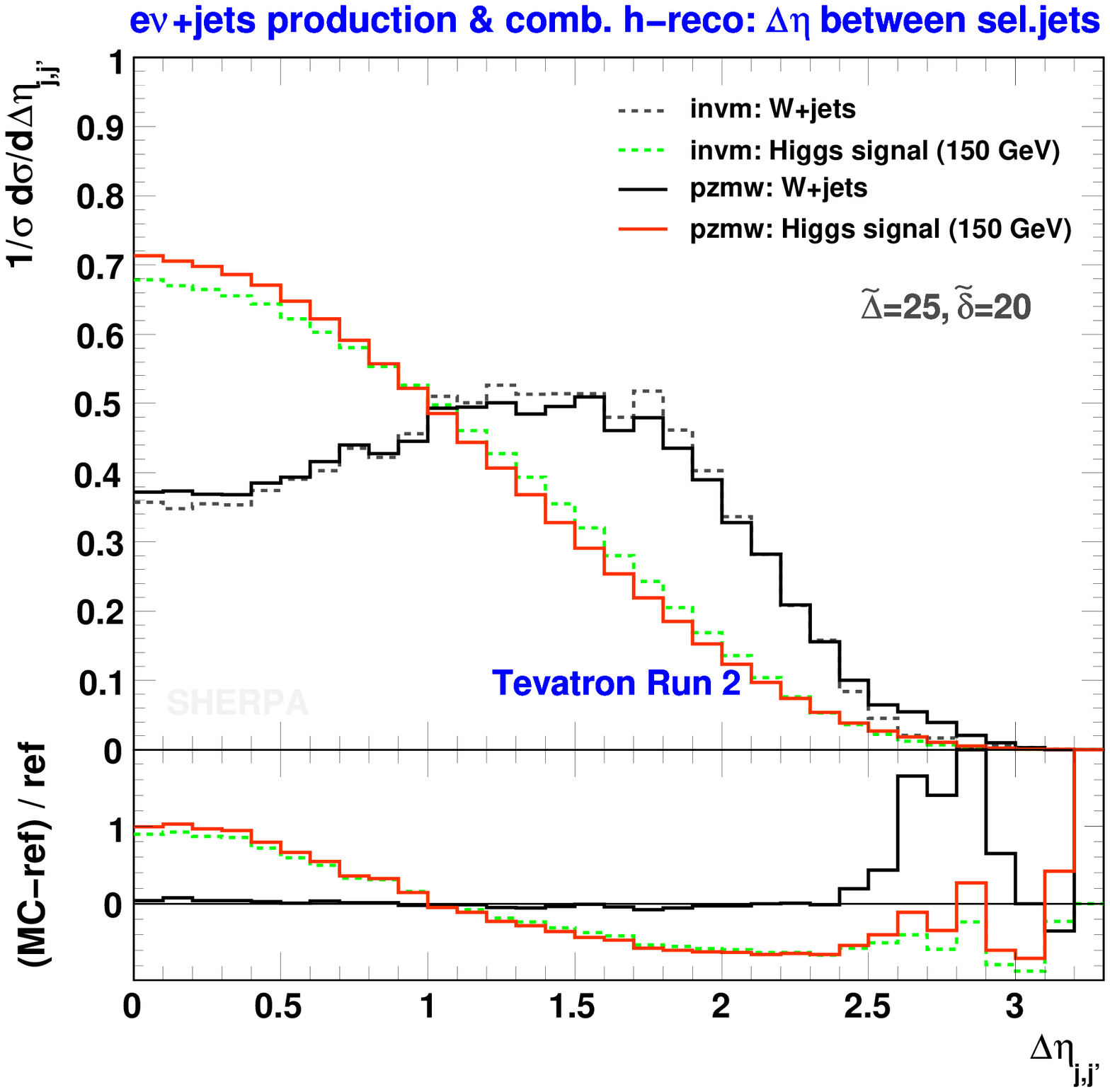}
  \includegraphics[clip,width=0.46\columnwidth,angle=0]{%
    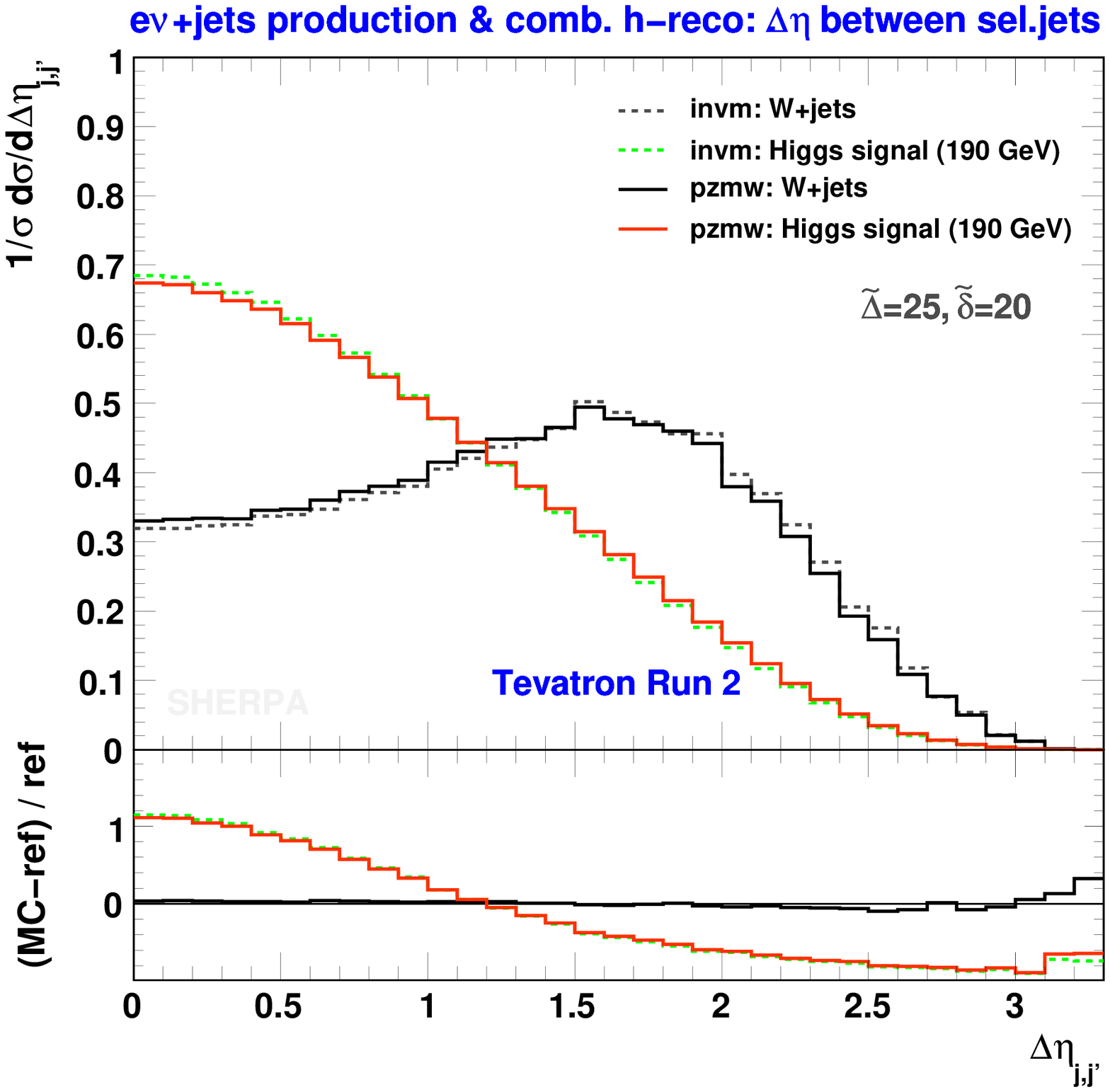}\\[-4mm]\hskip15pt
  \includegraphics[clip,width=0.46\columnwidth,angle=0]{%
    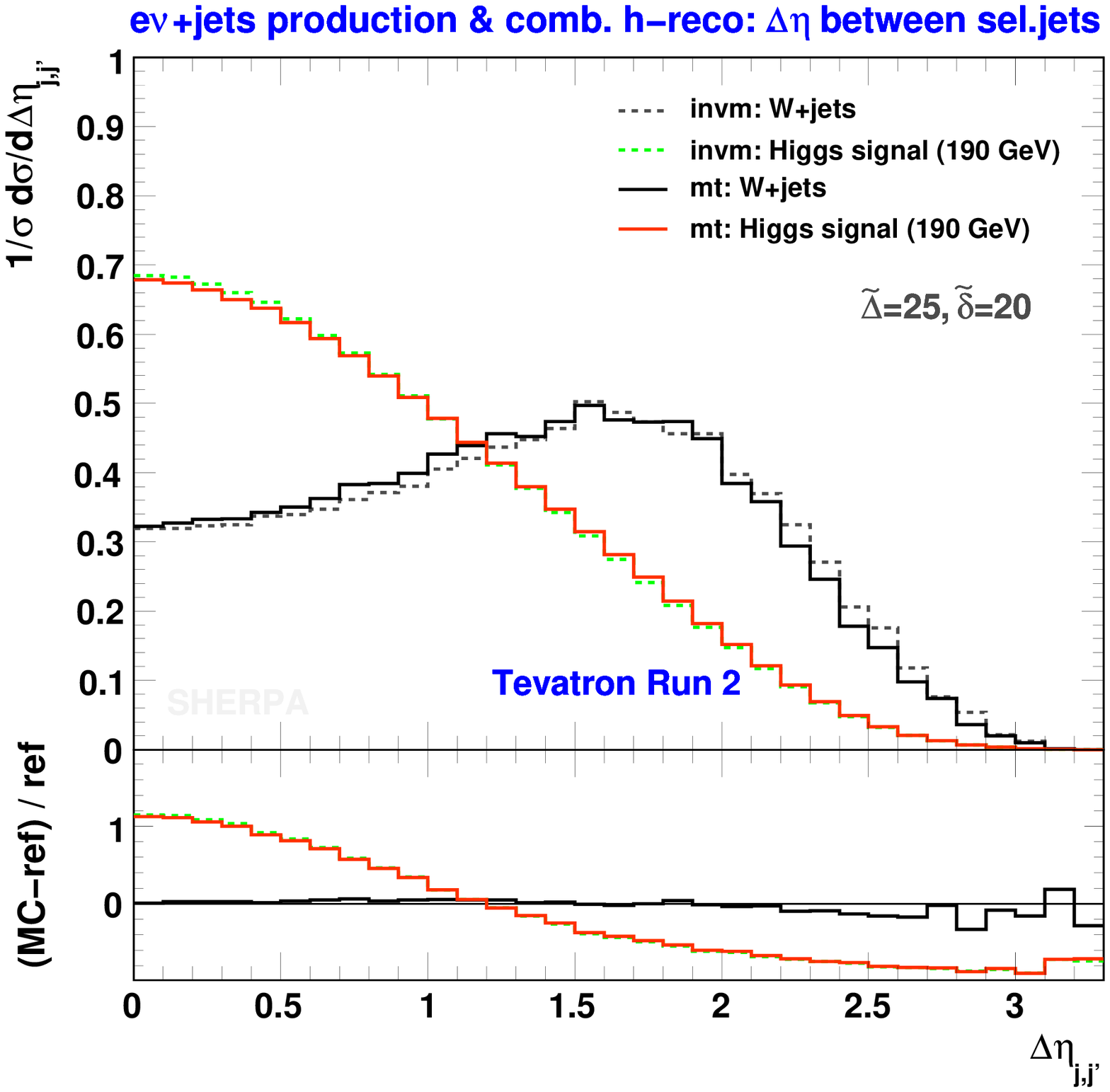}
  \caption{\label{fig:detajj}
    Pseudo-rapidity difference between the two selected jets.
    Predictions for $gg\to h\to e\nu_e$+jets production obtained after
    \texttt{invm} (dashed) and more realistic (solid) combinatorial
    selections (top: \texttt{pzmh}, center: \texttt{pzmw}, bottom:
    \texttt{mt}) are compared with each other and to the corresponding
    predictions for the $W(\to e\nu_e)$+jets background. Left panes
    show results for $M_h=150\ \mrm{GeV}$; in all other cases
    $M_h=190\ \mrm{GeV}$ was used.}
\end{figure}

\begin{figure}[t!]
  \centering\vskip1mm\hskip11pt
  \includegraphics[clip,width=0.485\columnwidth,angle=0]{%
    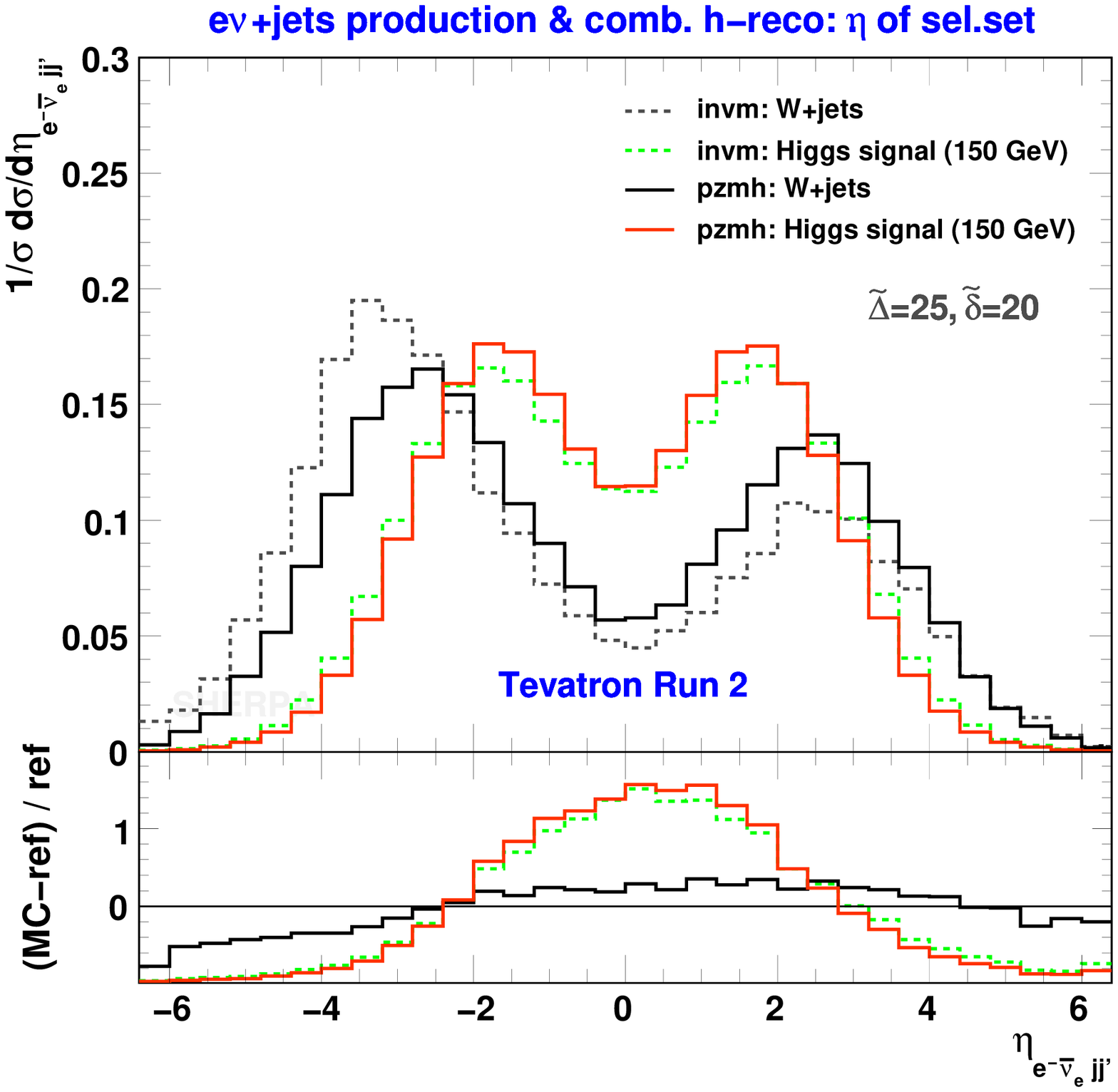}
  \includegraphics[clip,width=0.485\columnwidth,angle=0]{%
    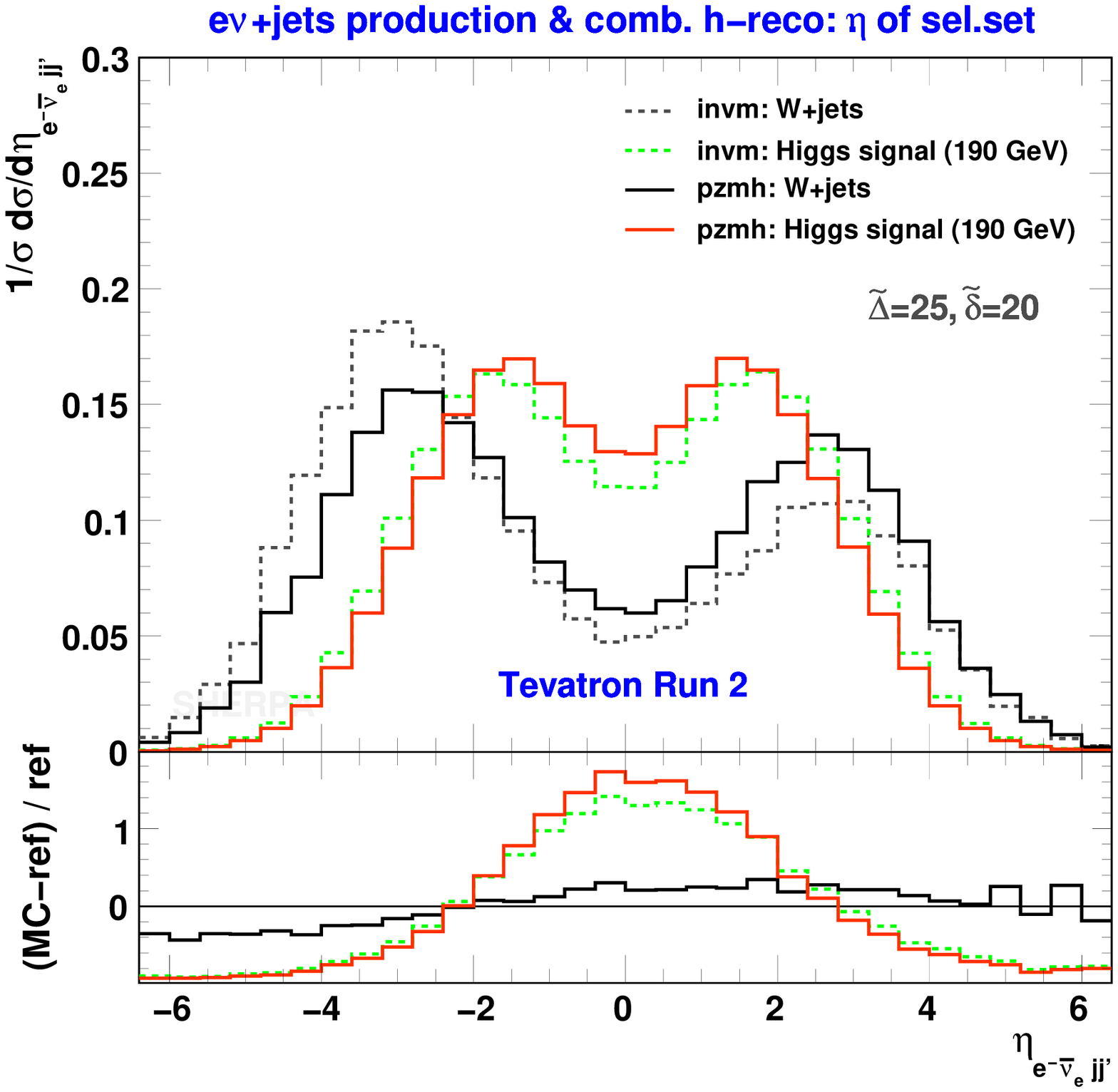}
  \\\hskip11pt
  \includegraphics[clip,width=0.485\columnwidth,angle=0]{%
    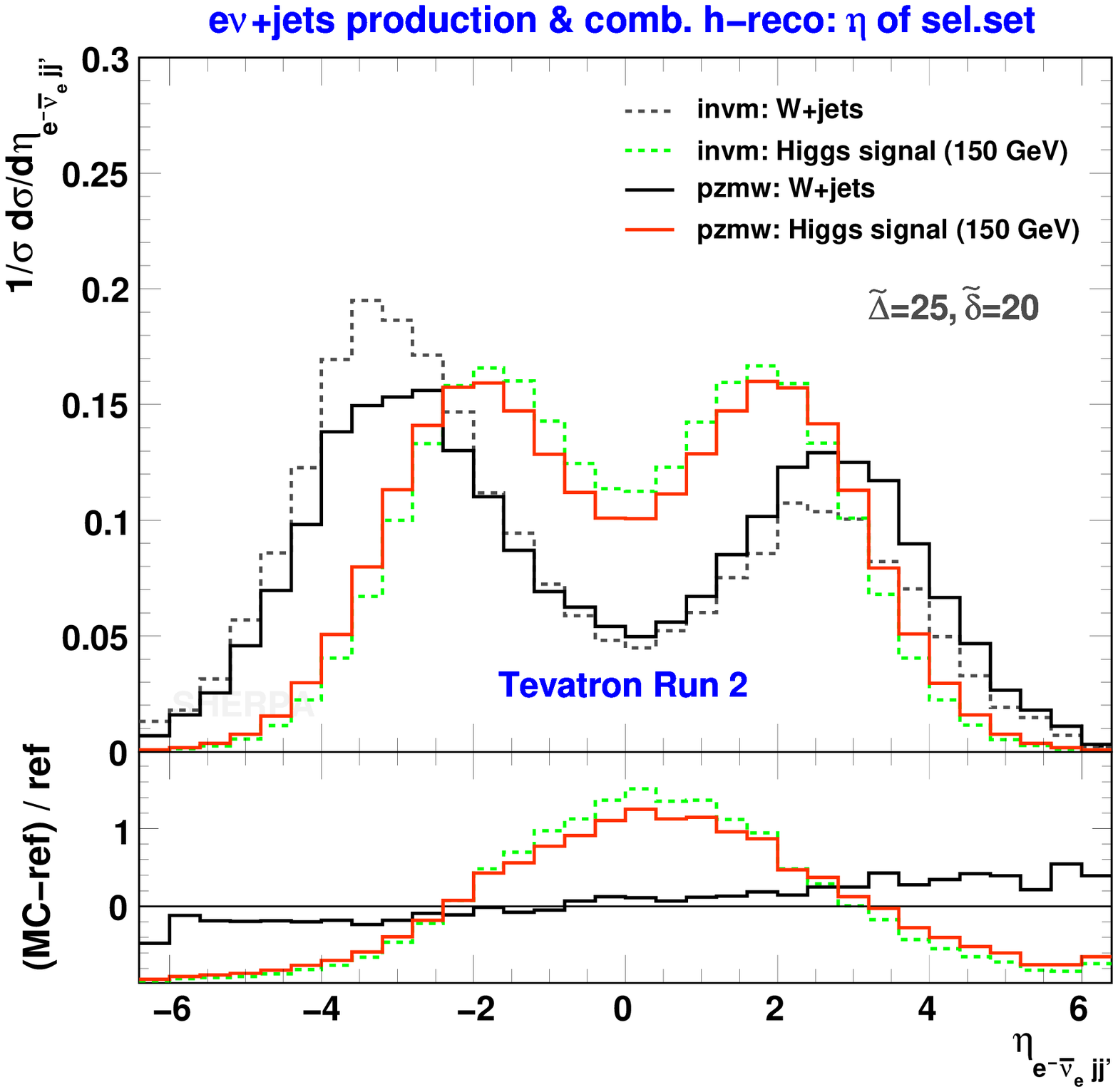}
  \includegraphics[clip,width=0.485\columnwidth,angle=0]{%
    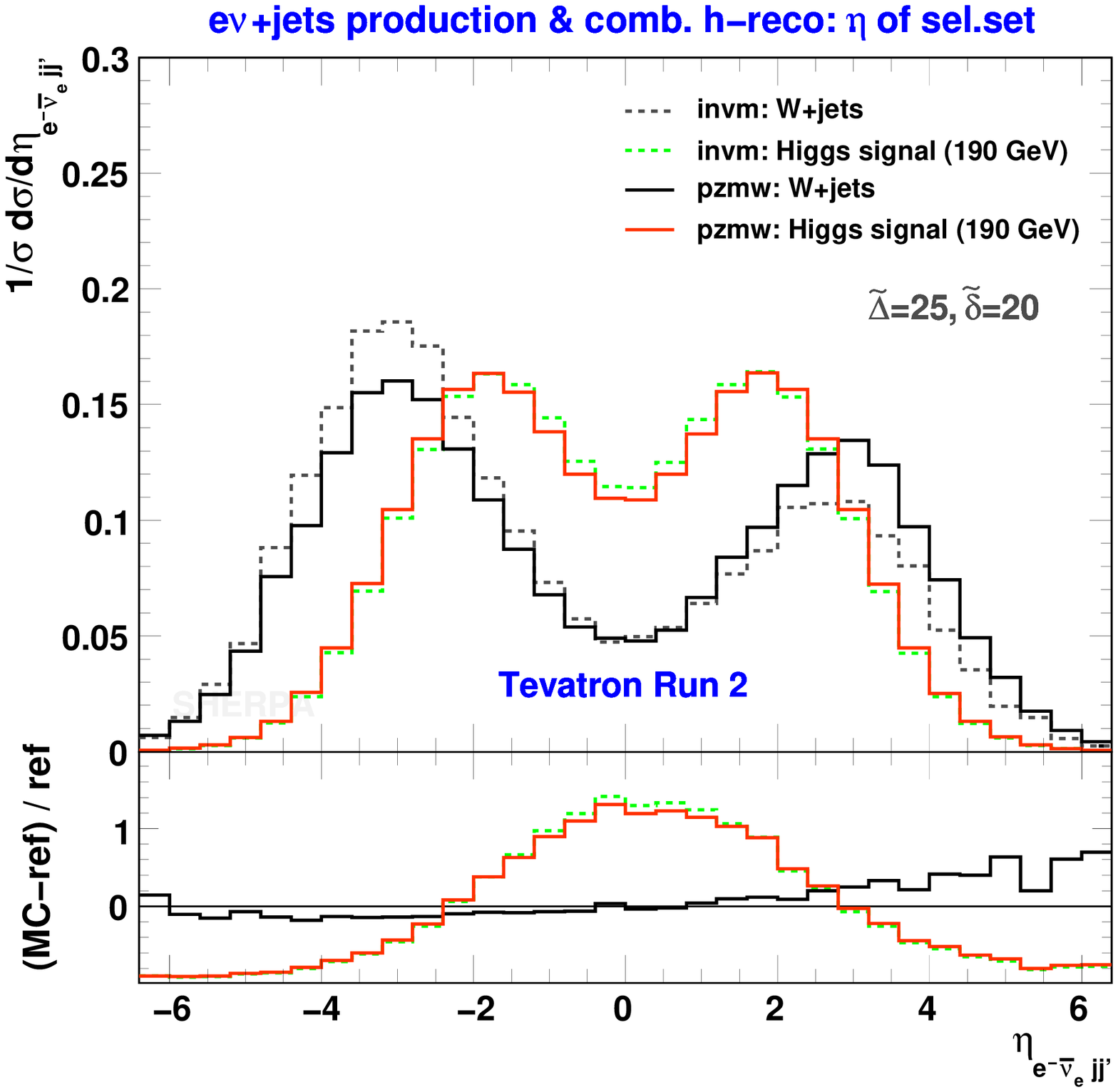}
  \caption{\label{fig:eta.set}
    Pseudo-rapidity of the selected $\{e,\nu_e,j,j'\}$ set, the Higgs
    boson candidate. The predictions for $gg\to h\to e\nu_e$+jets
    production (coloured lines) obtained after the combinatorial
    \texttt{invm} (dashed) and more realistic $h$\/ reconstruction
    (solid) selections (upper: \texttt{pzmh}, lower: \texttt{pzmw})
    are compared with each other and to the corresponding predictions
    of the $W(\to e\nu_e)$+jets background (black lines). Left panes
    show results for $M_h=150\ \mrm{GeV}$, while in the right panes
    the outcomes for $M_h=190\ \mrm{GeV}$ are shown.}
\end{figure}

\begin{figure}[t!]
\centering\vskip2mm
\includegraphics[clip,width=0.384\columnwidth,angle=-90]{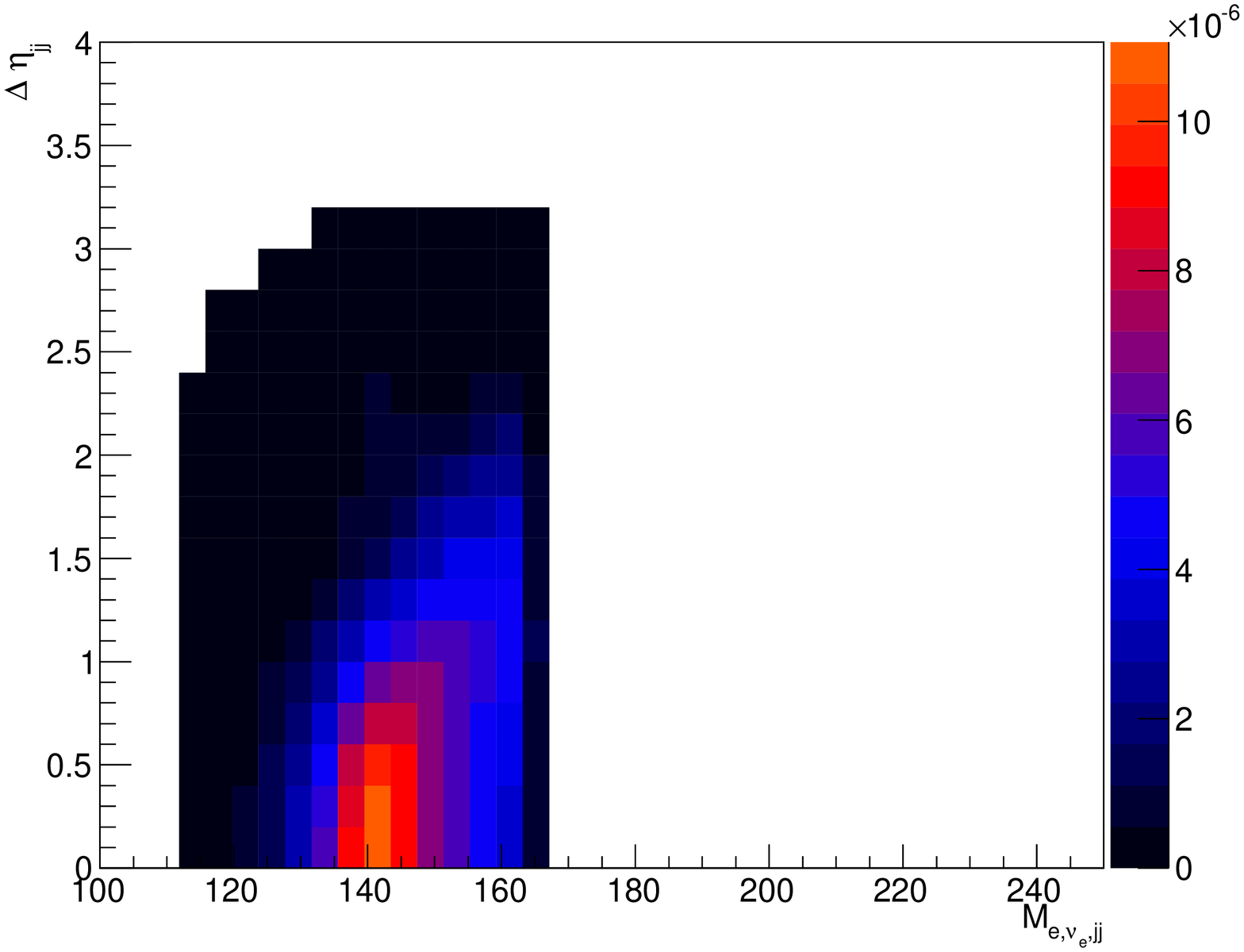}
\includegraphics[clip,width=0.384\columnwidth,angle=-90]{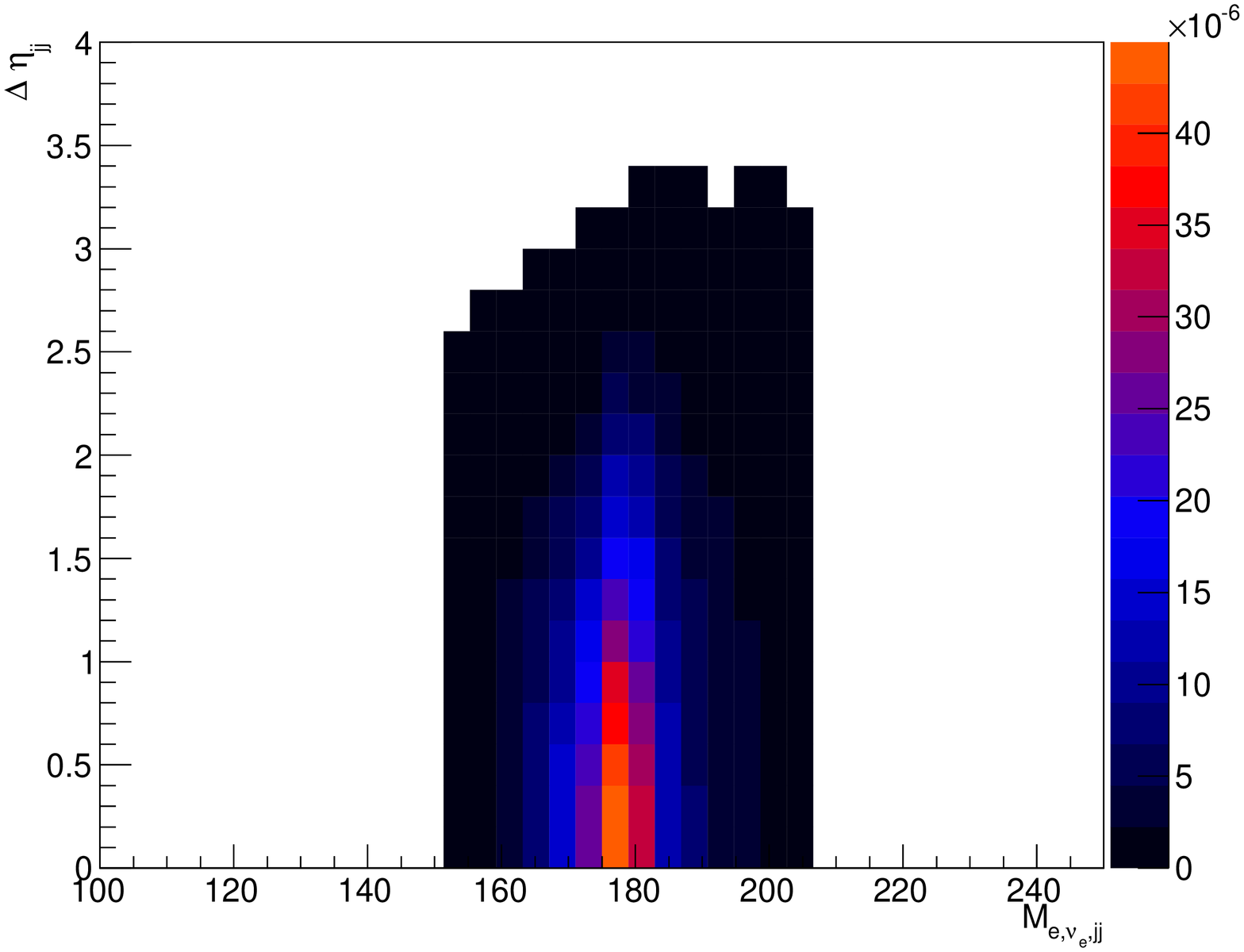}
\includegraphics[clip,width=0.384\columnwidth,angle=-90]{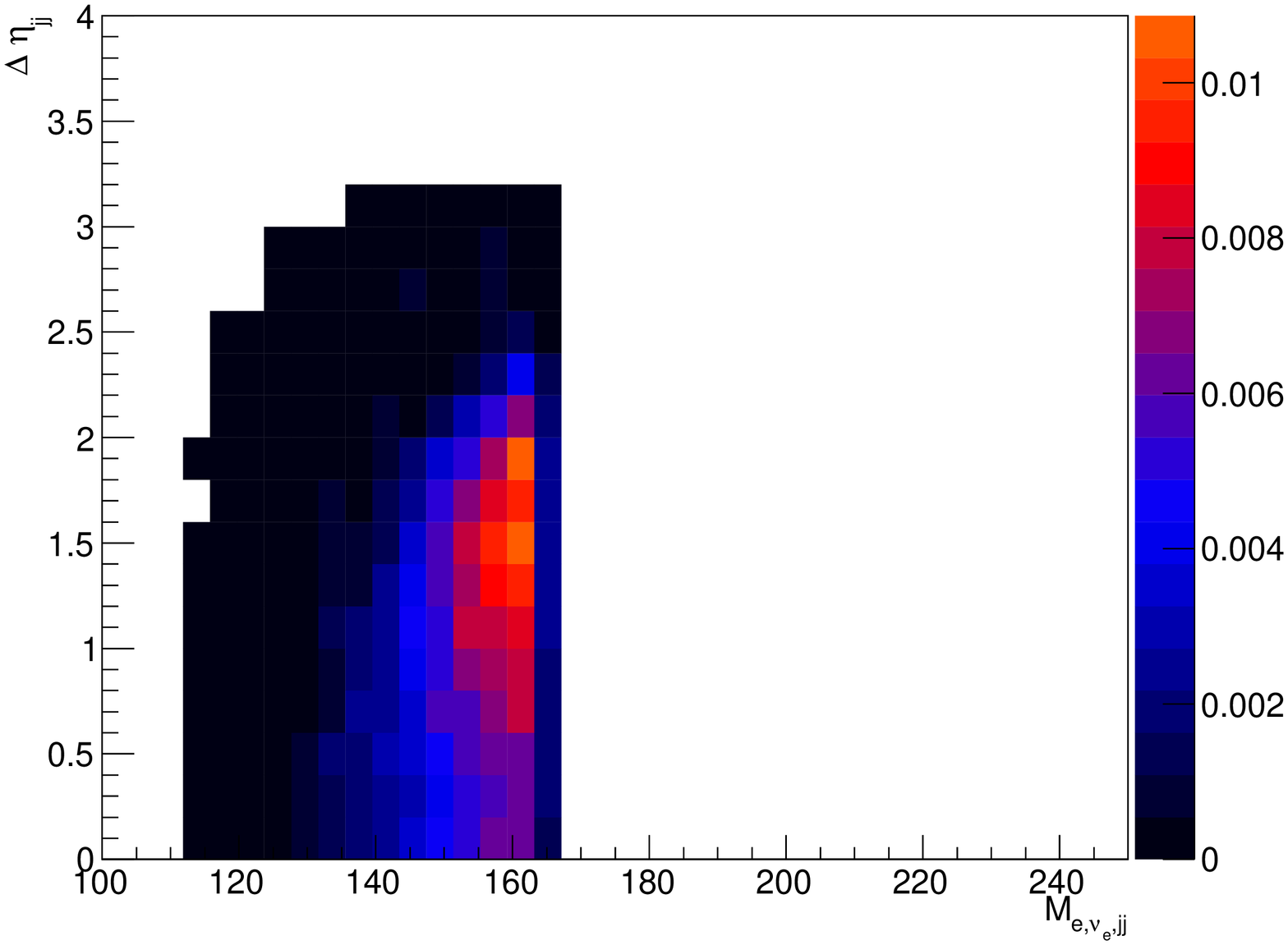}
\includegraphics[clip,width=0.384\columnwidth,angle=-90]{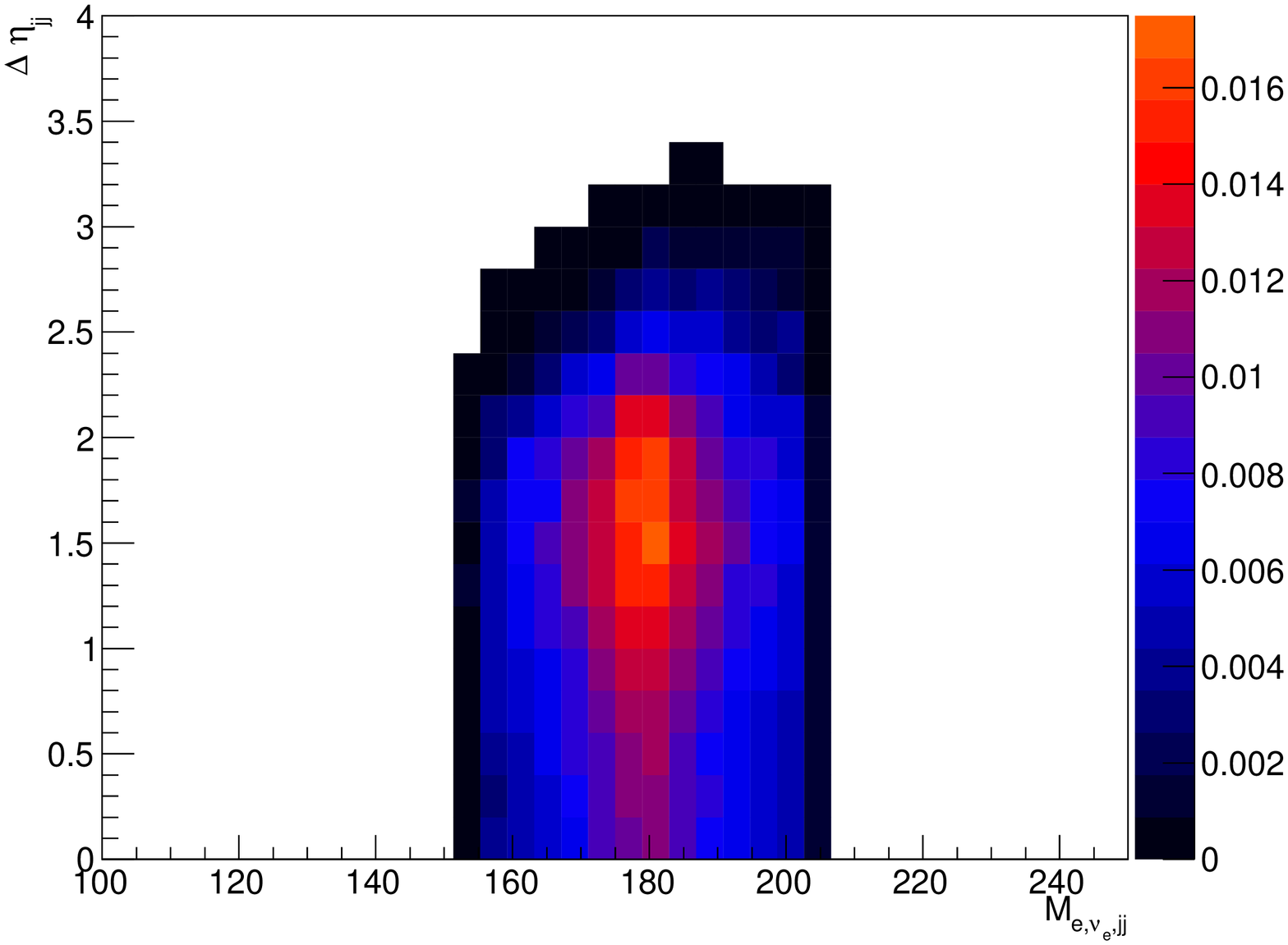}
\caption{\label{fig:2Dpzmw}
  Two-dimensional distributions showing the selected-jet
  pseudo-rapidity difference, $|\Delta\eta_{j,j'}|$, plotted versus
  the reconstructed mass $m_{e\nu_e jj'}$ of the selected
  $\{e,\nu_e,j,j'\}$ combinations. The predictions for $gg\to h\to
  e\nu_e$+jets production obtained after \texttt{pzmw} reconstruction
  of Higgs boson candidates are compared with each other and to the
  corresponding predictions given by the $W(\to e\nu_e)$+jets
  background. The upper (lower) plots represent the signal
  (background) predictions, while the left (right) panes show results
  for $M_h=140\ (180)\ \mrm{GeV}$.}
\end{figure}

\begin{figure}[t!]
\centering\vskip2mm
\includegraphics[clip,width=0.384\columnwidth,angle=-90]{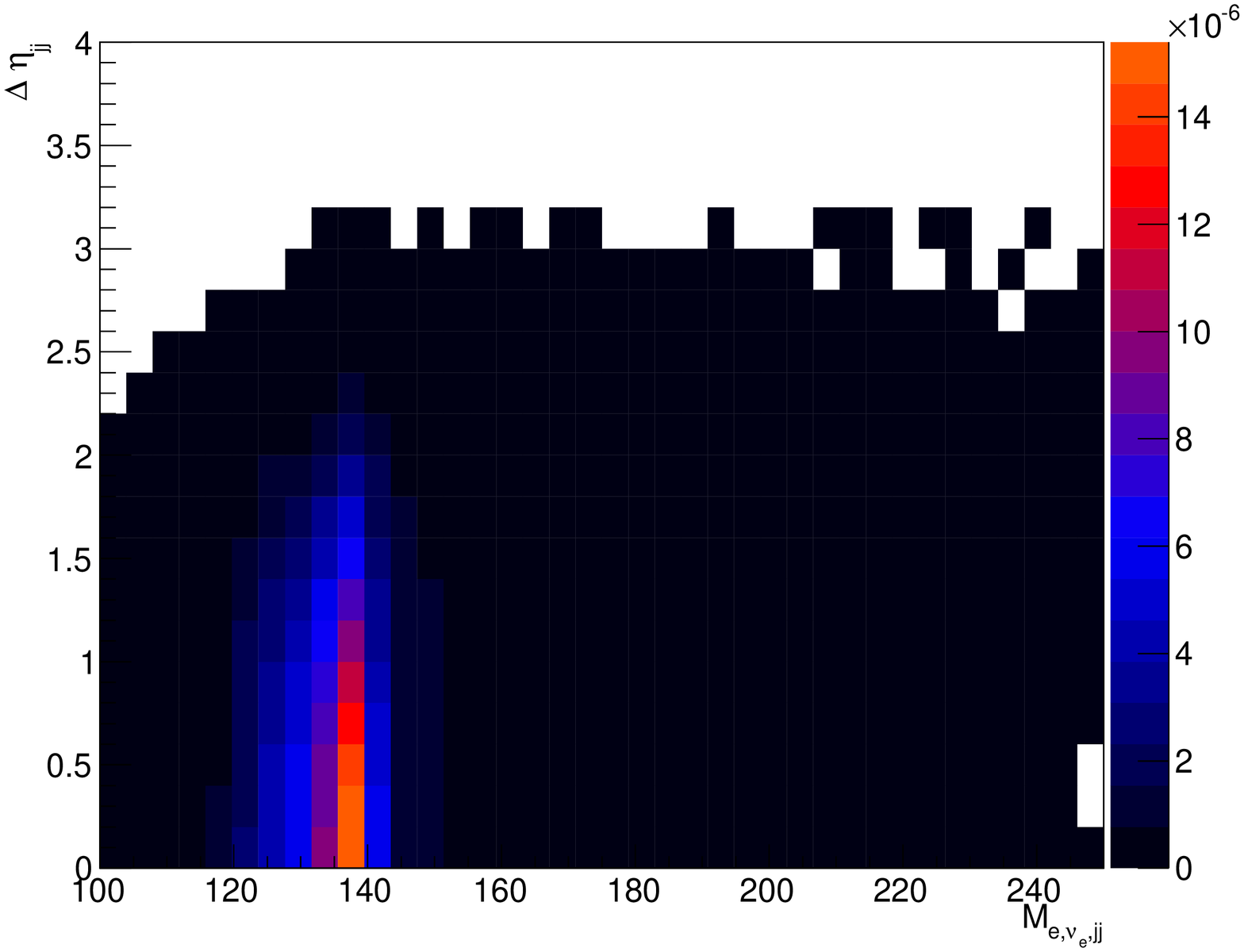}
\includegraphics[clip,width=0.384\columnwidth,angle=-90]{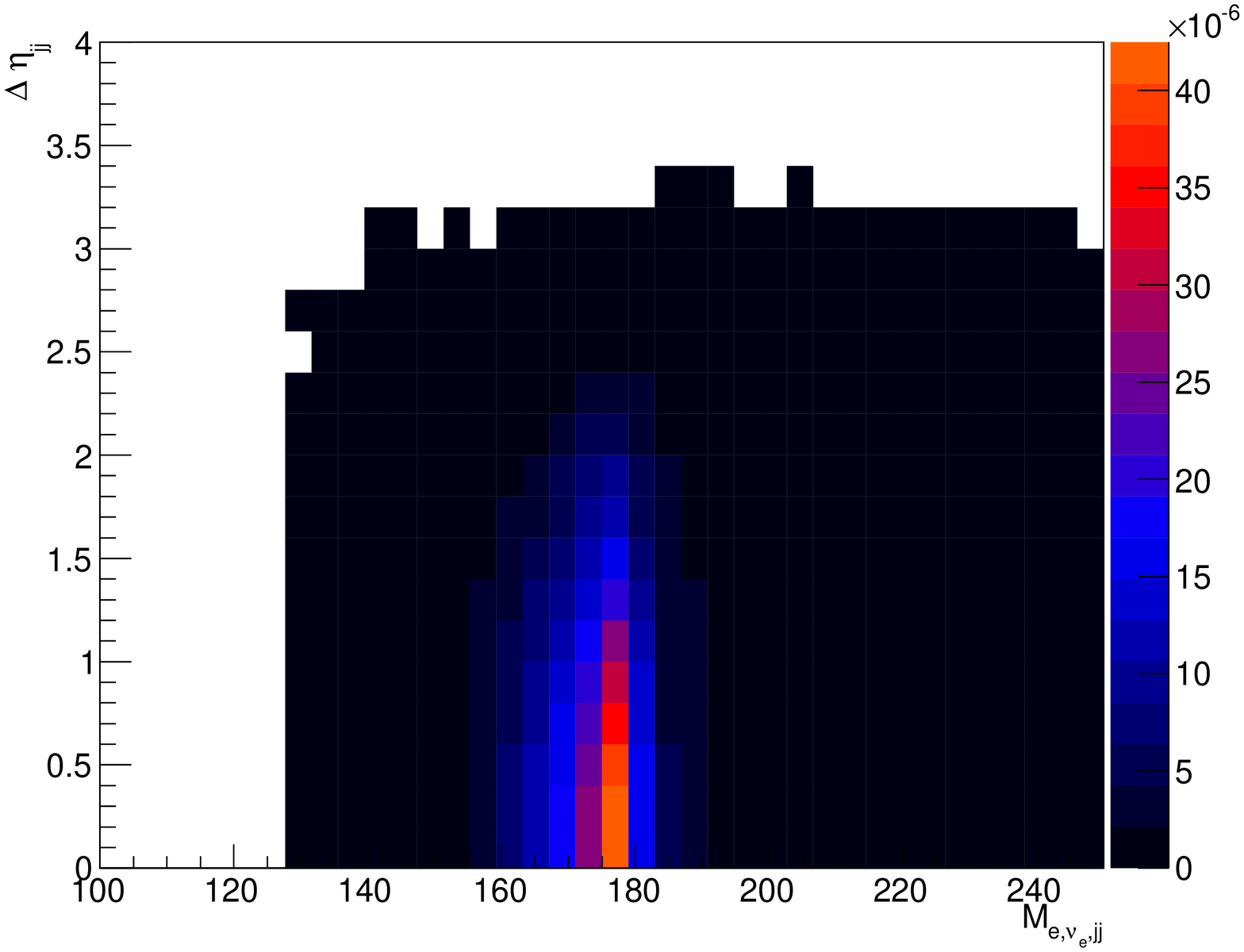}
\includegraphics[clip,width=0.384\columnwidth,angle=-90]{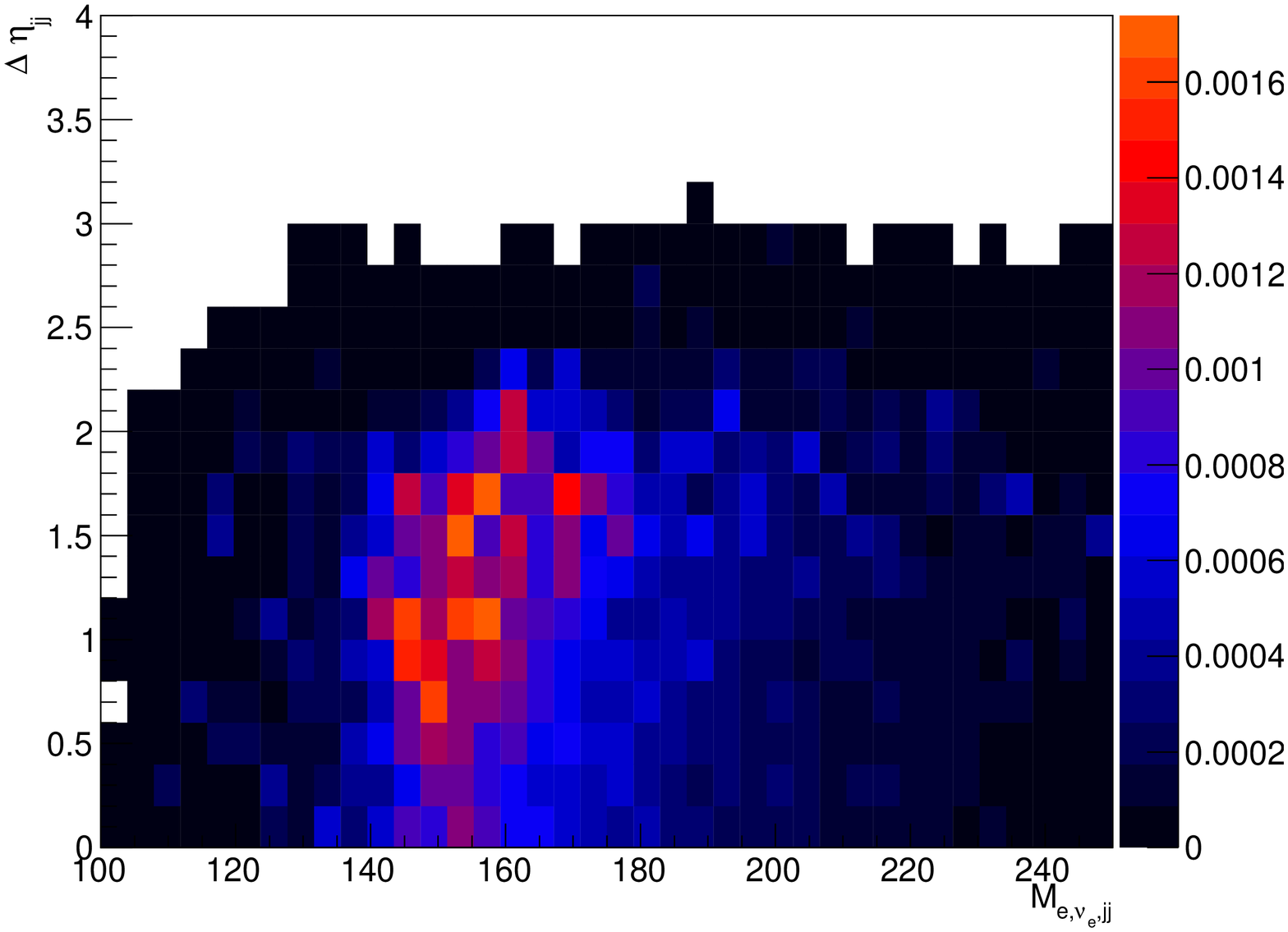}
\includegraphics[clip,width=0.384\columnwidth,angle=-90]{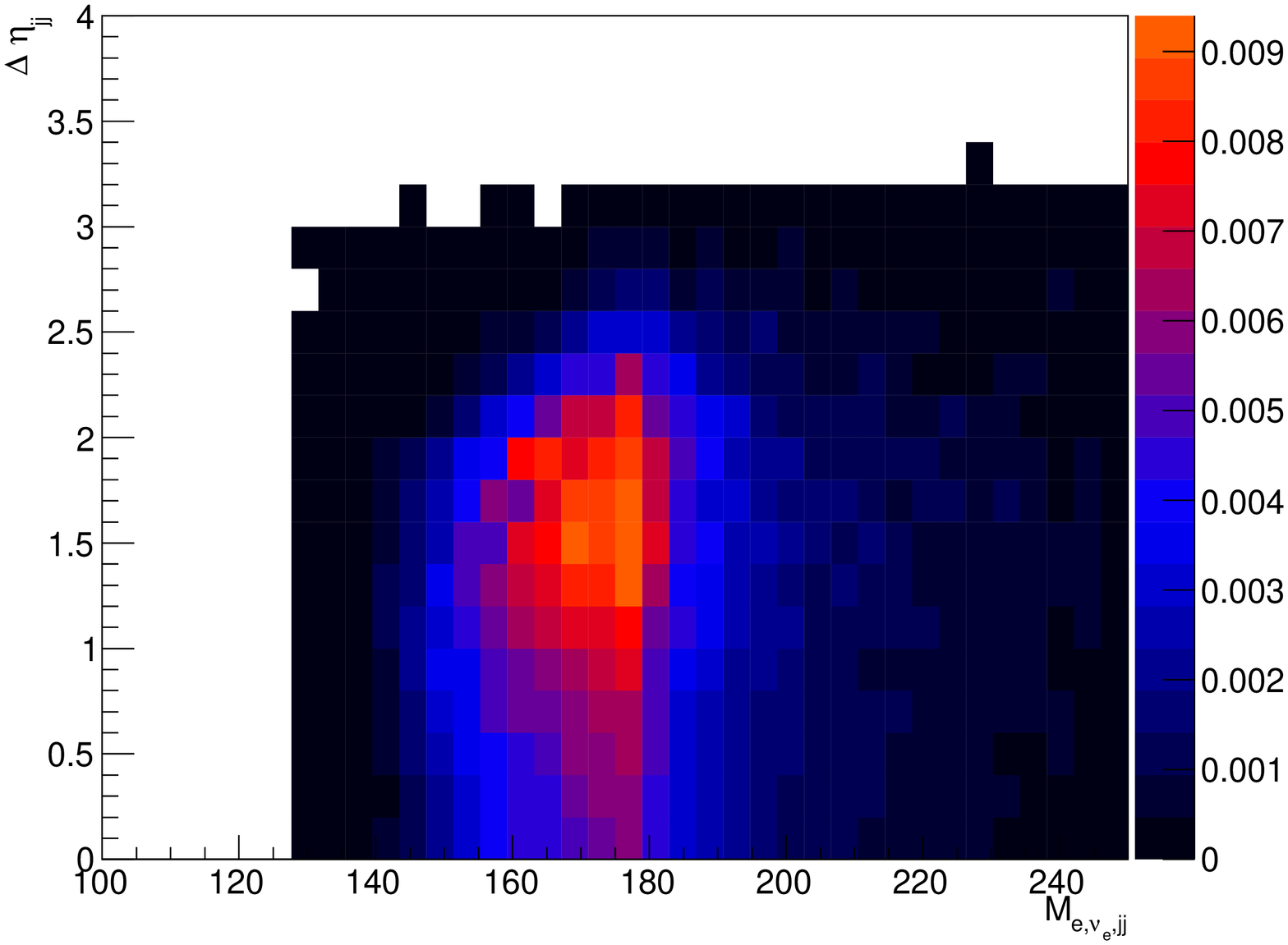}
\caption{\label{fig:2Dmt}
  Two-dimensional distributions showing the selected-jet
  pseudo-rapidity difference, $|\Delta\eta_{j,j'}|$, plotted versus
  the reconstructed mass $m_{e\nu_e jj'}$ of the selected
  $\{e,\nu_e,j,j'\}$ combinations. The predictions for $gg\to h\to
  e\nu_e$+jets production obtained after combinatorial selection
  according to the \texttt{mt} procedure are compared with each other
  and to the corresponding predictions given by the $W(\to
  e\nu_e)$+jets background. The upper (lower) plots represent the
  signal (background) predictions, while the left (right) panes show
  results for $M_h=140\ (180)\ \mrm{GeV}$. Note that for the purpose
  of illustration, the $m_{e\nu_e jj'}$ quantities are reconstructed
  as in the ideal case.}
\end{figure}

\begin{figure}[p!]
\centering\vskip-2mm\hskip15pt
  \includegraphics[clip,width=0.46\columnwidth,angle=0]{%
    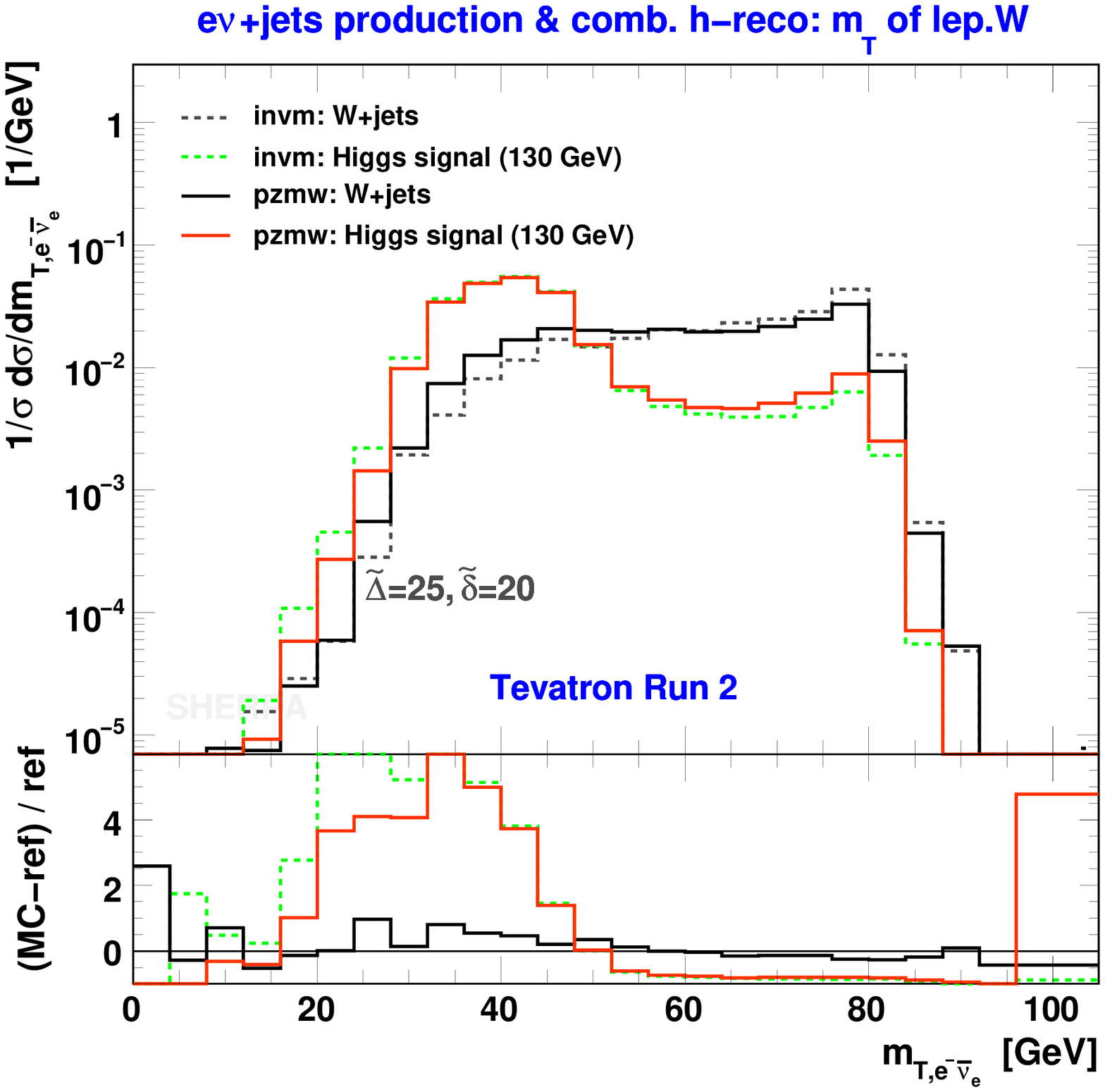}
  \includegraphics[clip,width=0.46\columnwidth,angle=0]{%
    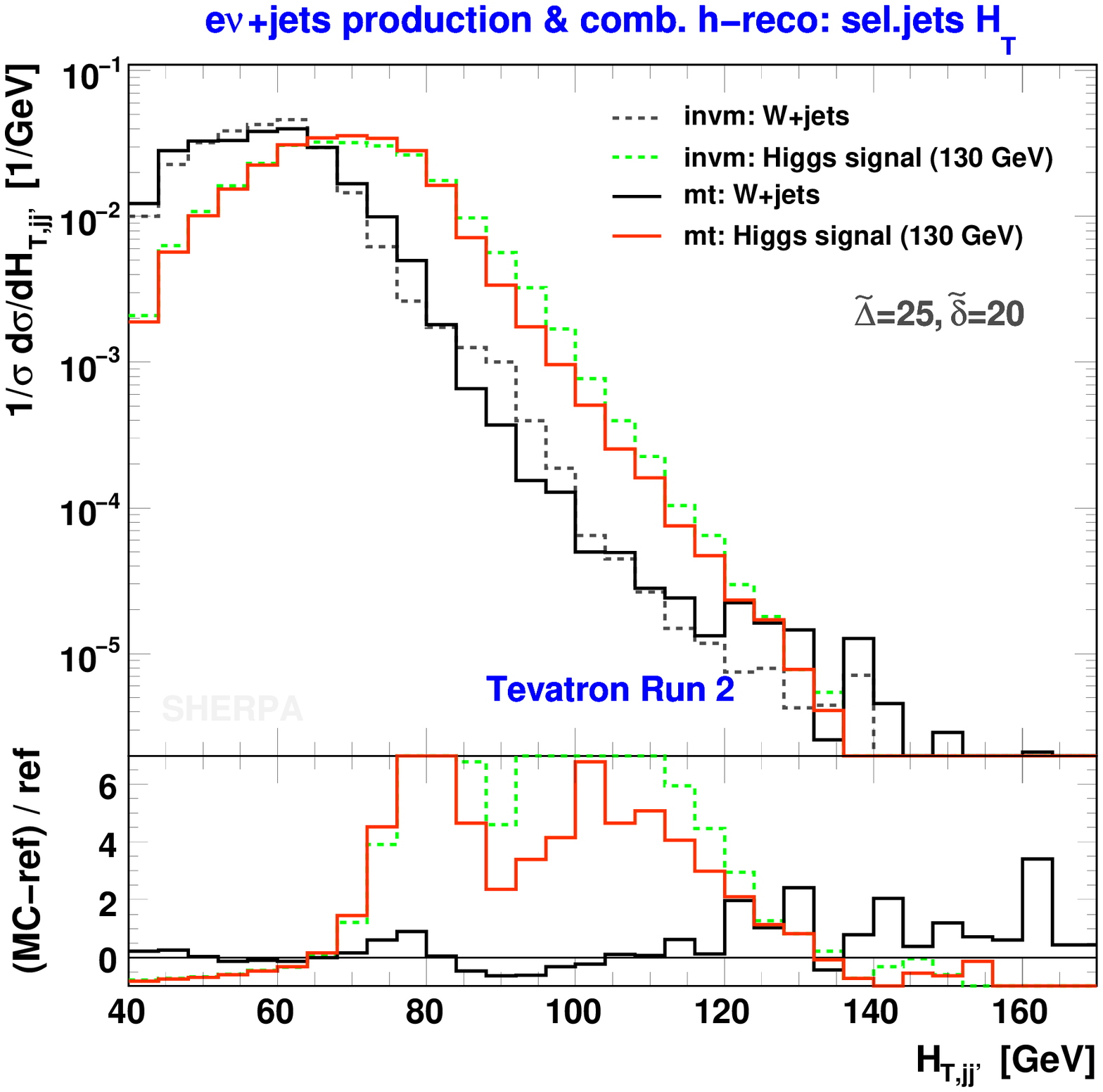}\\[-4mm]\hskip15pt
  \includegraphics[clip,width=0.46\columnwidth,angle=0]{%
    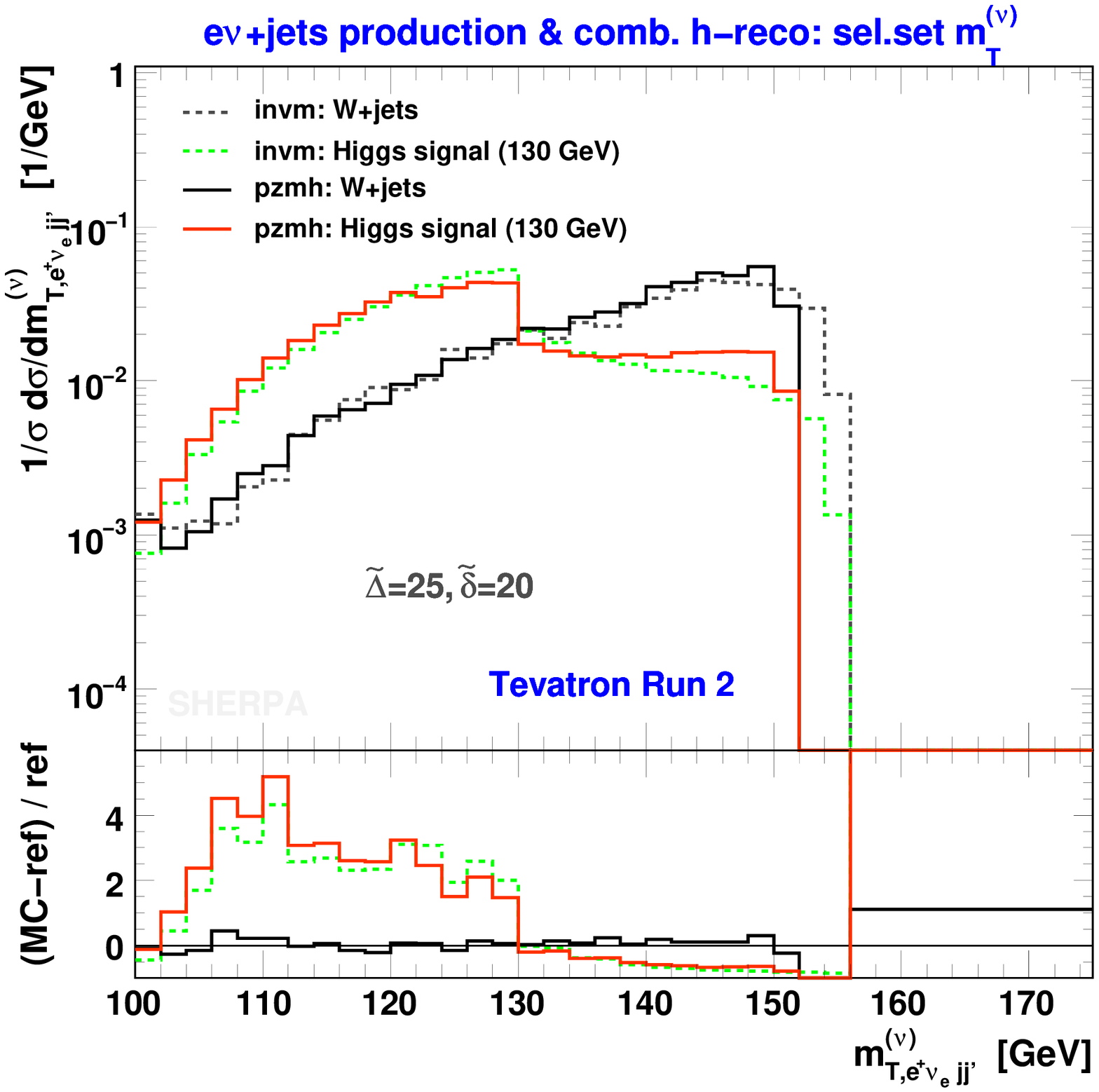}
  \includegraphics[clip,width=0.46\columnwidth,angle=0]{%
    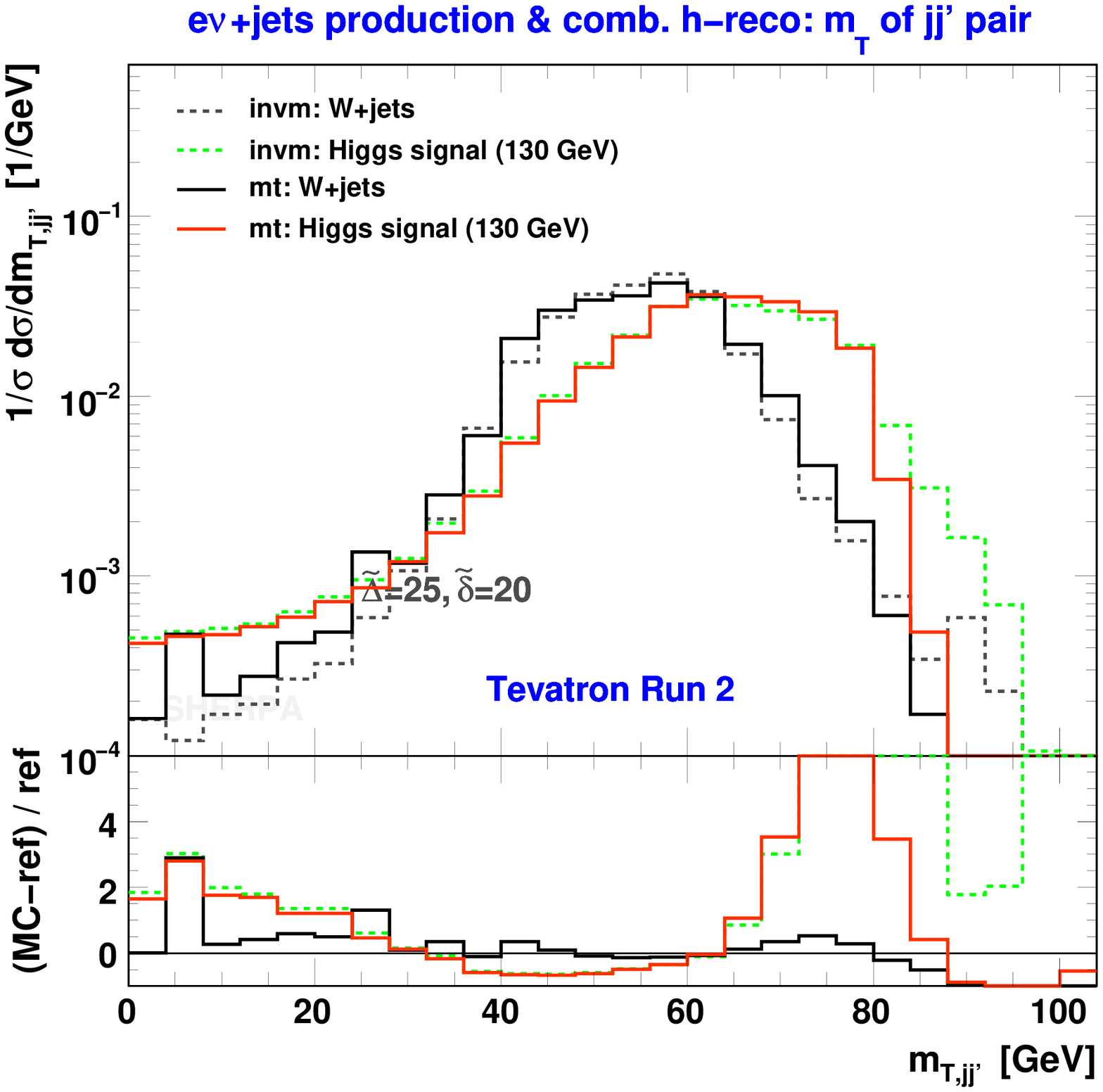}\\[-4mm]\hskip15pt
  \includegraphics[clip,width=0.46\columnwidth,angle=0]{%
    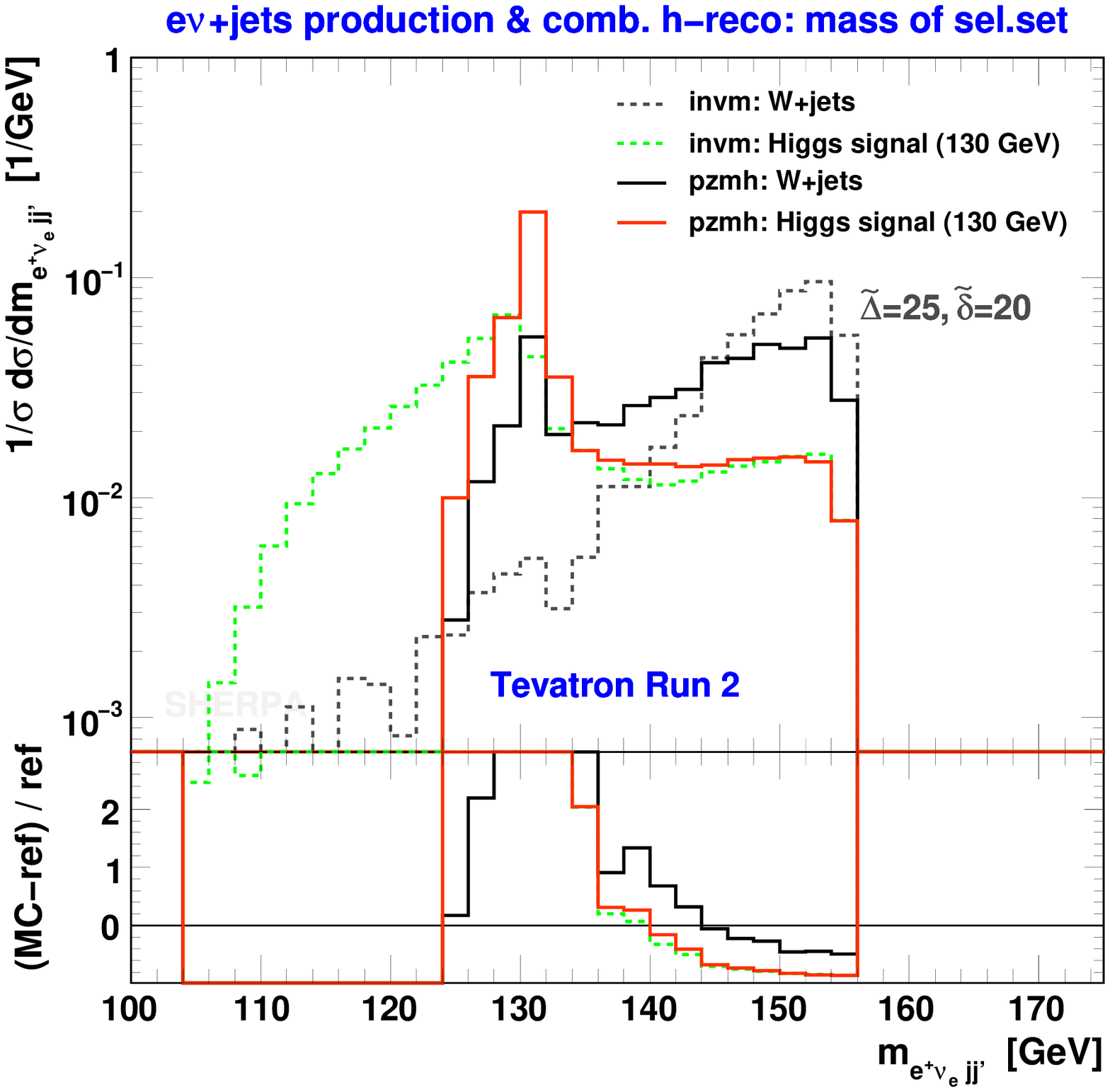}
  \includegraphics[clip,width=0.46\columnwidth,angle=0]{%
    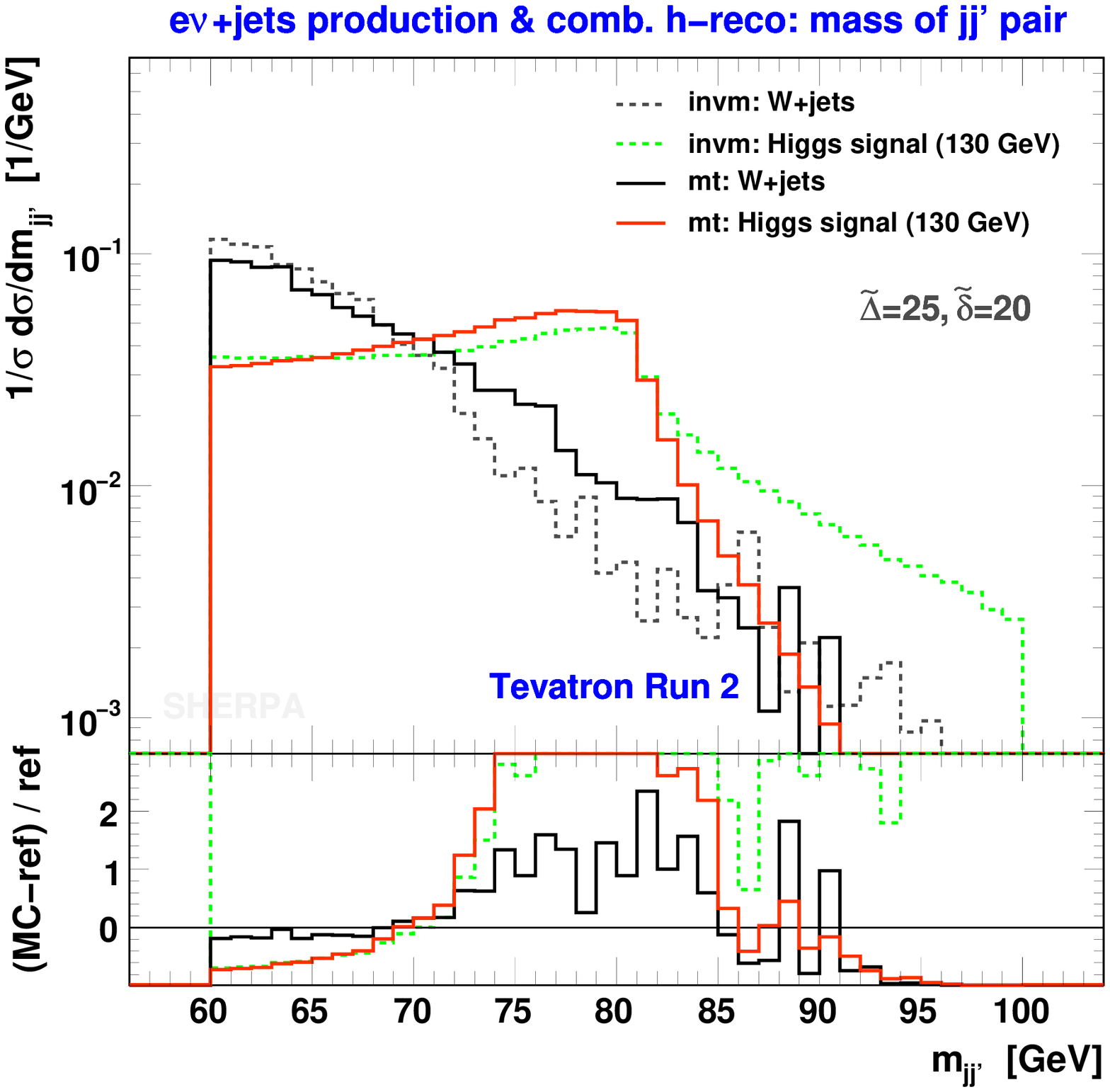}
  \caption{\label{fig:cuts.below}
    Examples of leading and subleading cut observables below the
    $2\,M_W$ threshold, for $M_h=130\ \mrm{GeV}$. Predictions are
    shown for the $gg\to h$\/ and $W$\/ production of $e\nu_e$+jets
    final states using the \texttt{invm} (dashed) and more realistic
    (solid) combinatorial selections. Top to bottom, left panes:
    $m_{T,e^-\bar\nu_e}$ (\texttt{pzmw} leading),
    $m^{(\nu_e)}_{T,e^+\nu_e jj'}$ and $m_{e^+\nu_e jj'}$
    (\texttt{pzmh} leading and subleading); right panes: $H_{T,jj'}$,
    $m_{T,jj'}$ and $m_{jj'}$ (\texttt{mt} leading to subsubleading).}
\end{figure}

\begin{figure}[p!]
\centering\vskip-2mm\hskip15pt
  \includegraphics[clip,width=0.46\columnwidth,angle=0]{%
    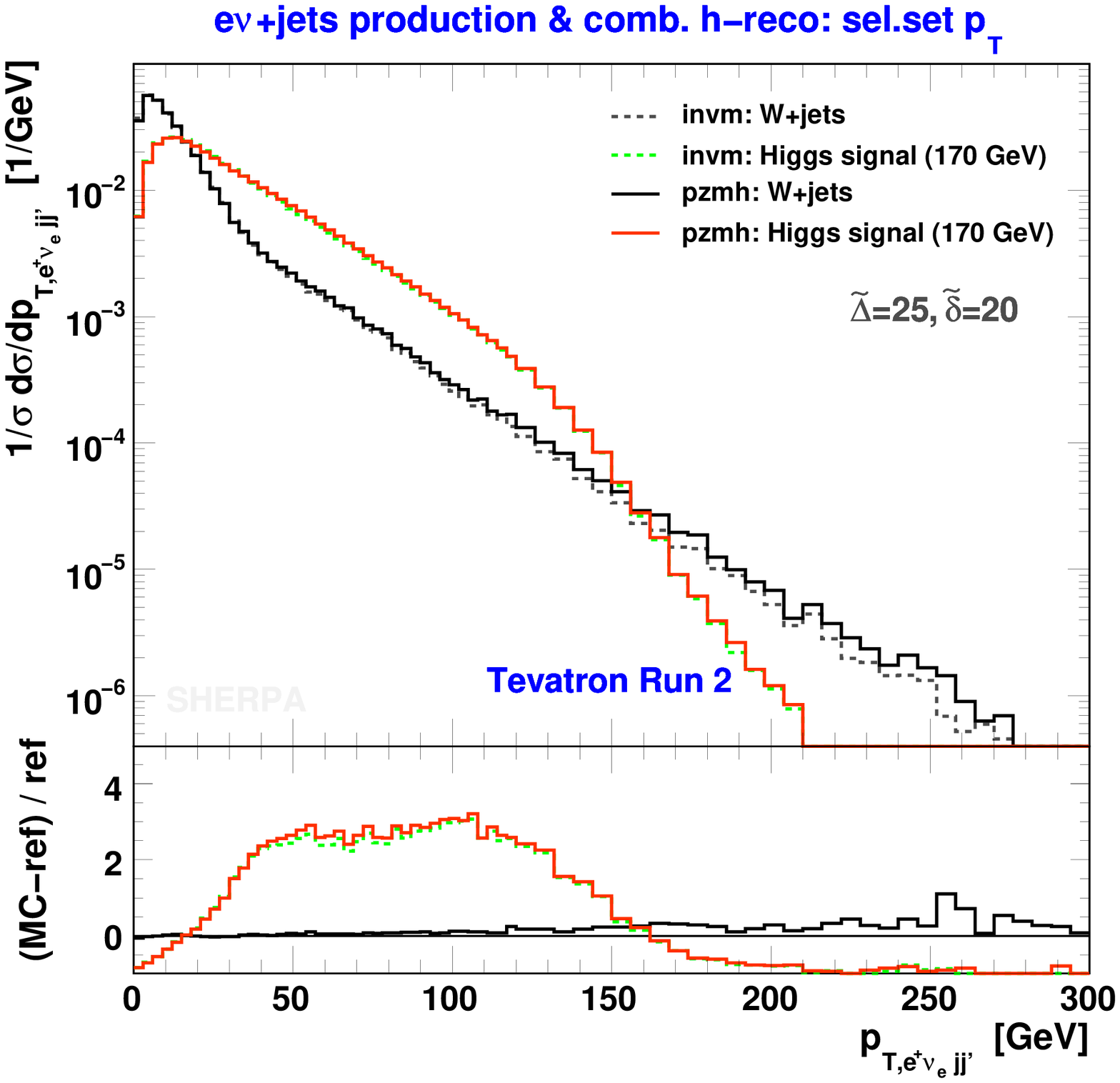}
  \includegraphics[clip,width=0.46\columnwidth,angle=0]{%
    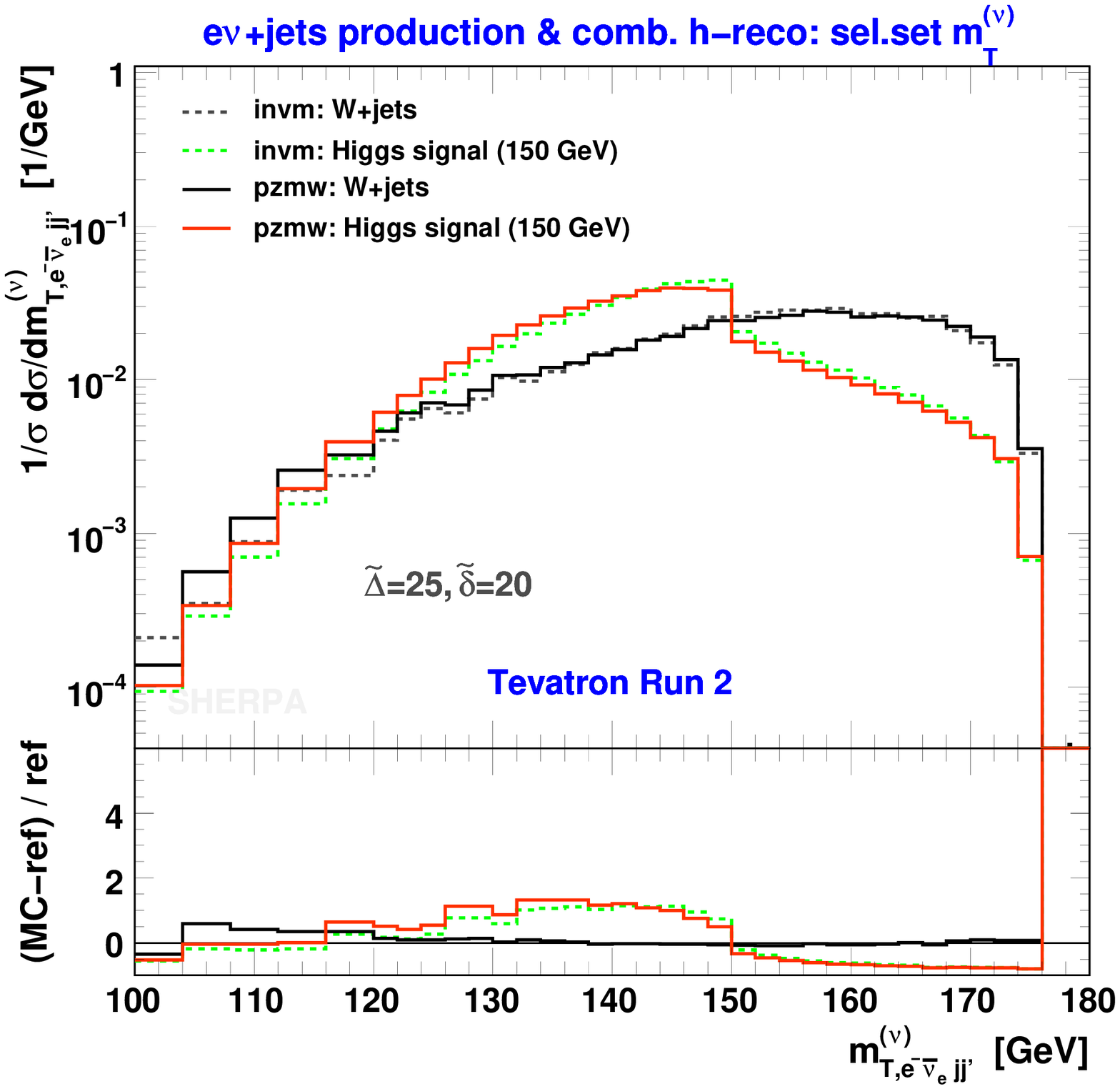}\\[-4mm]\hskip15pt
  \includegraphics[clip,width=0.46\columnwidth,angle=0]{%
    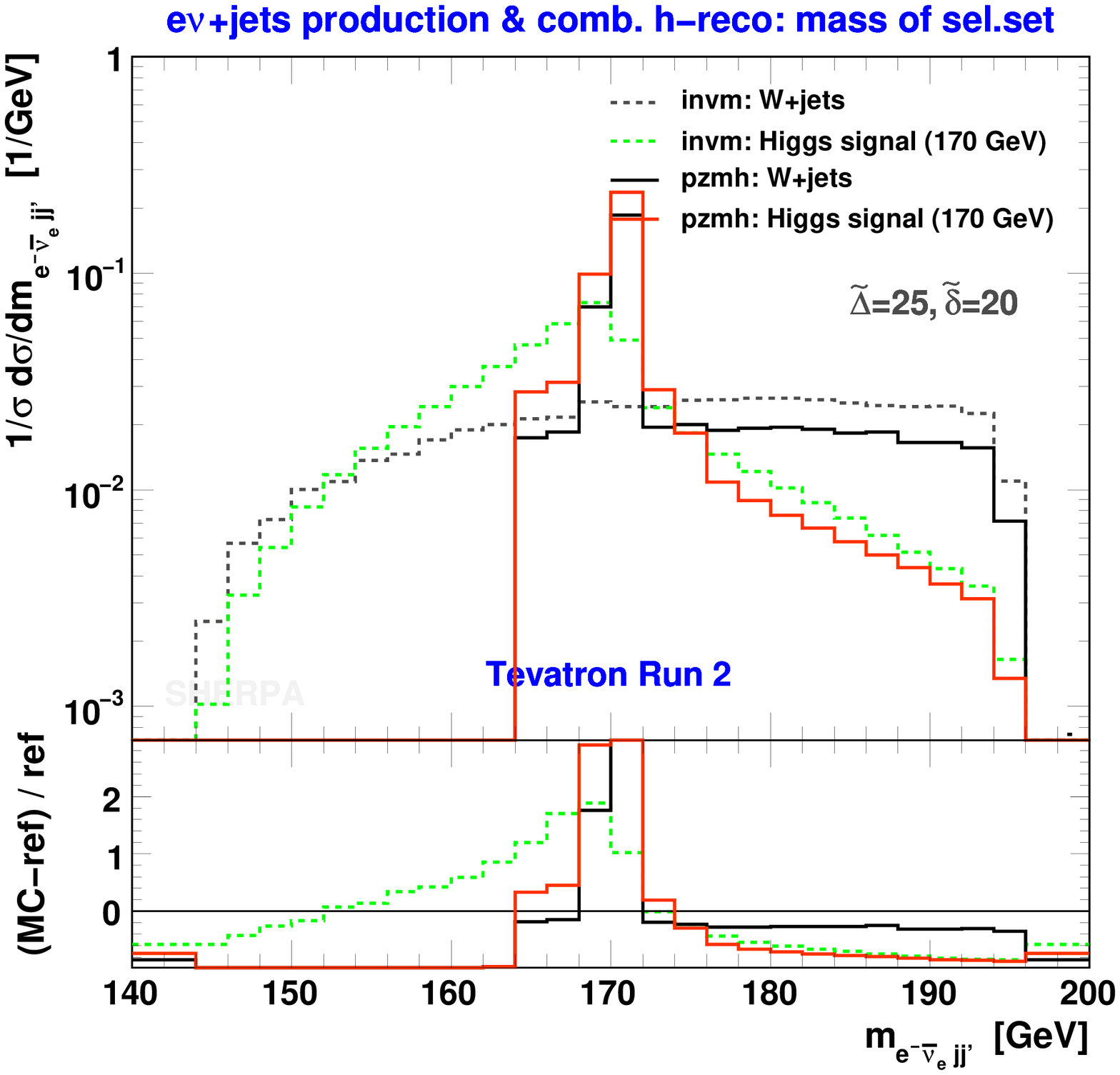}
  \includegraphics[clip,width=0.46\columnwidth,angle=0]{%
    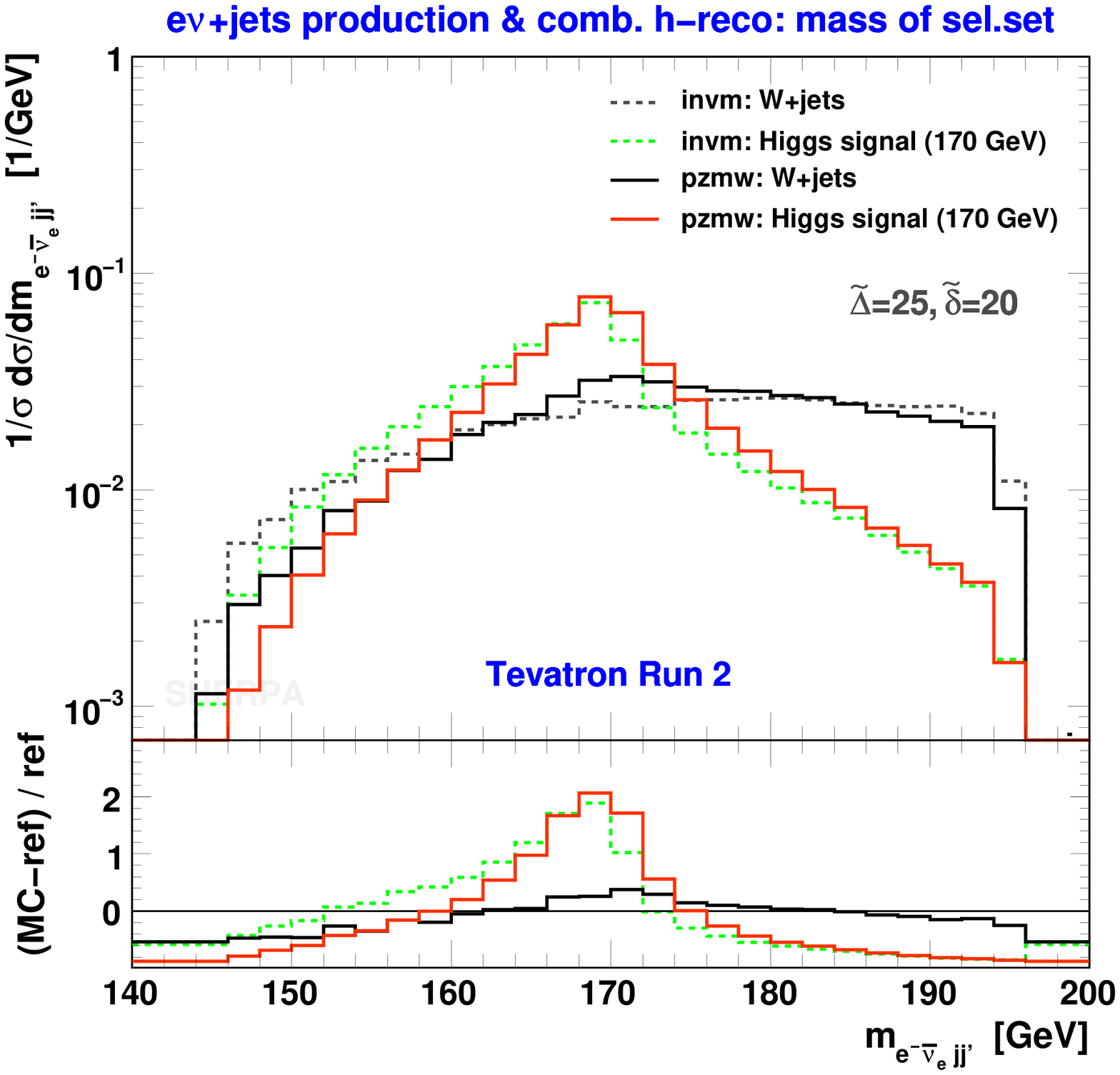}\\[-4mm]\hskip15pt
  \includegraphics[clip,width=0.46\columnwidth,angle=0]{%
    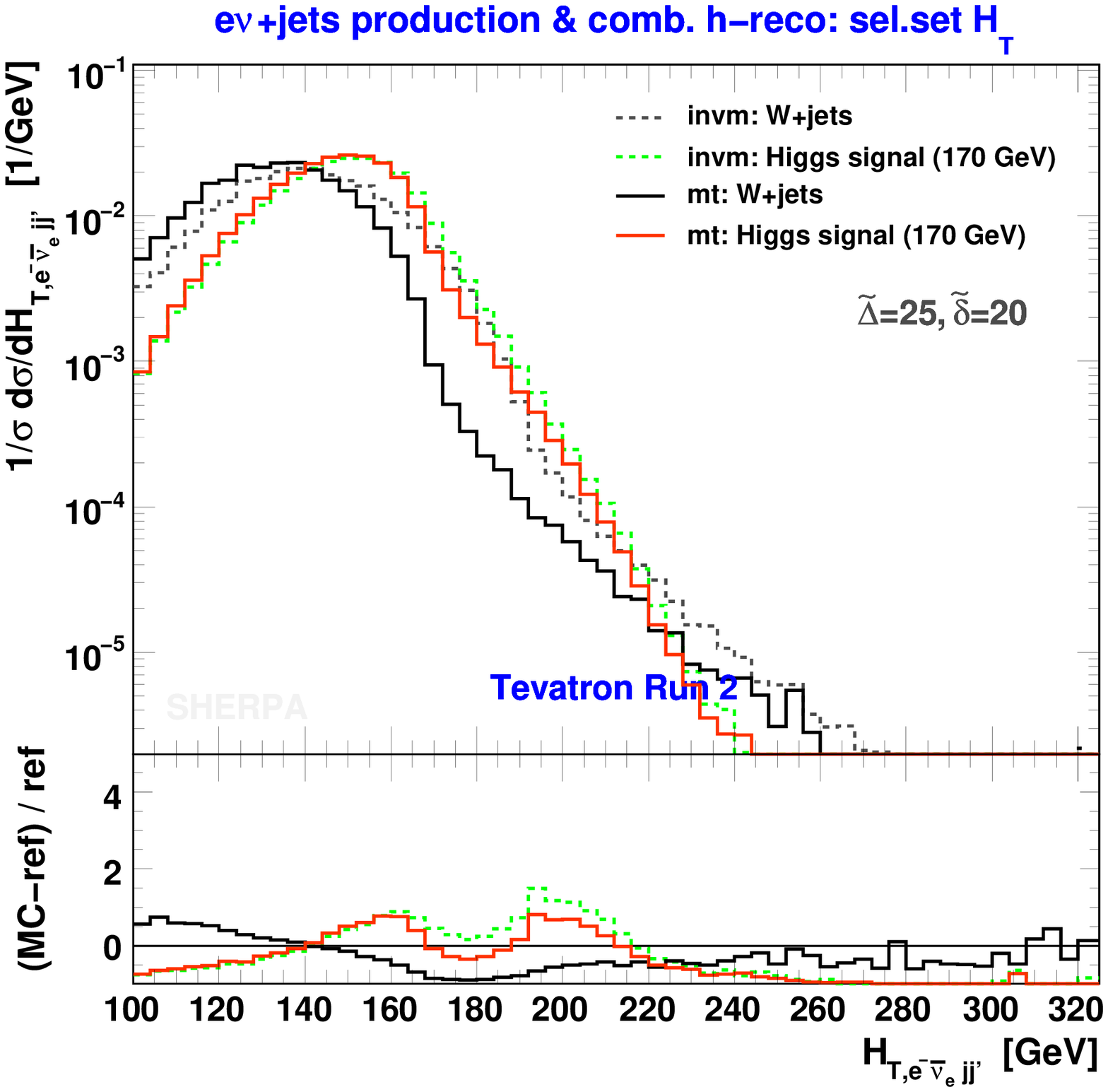}
  \includegraphics[clip,width=0.46\columnwidth,angle=0]{%
    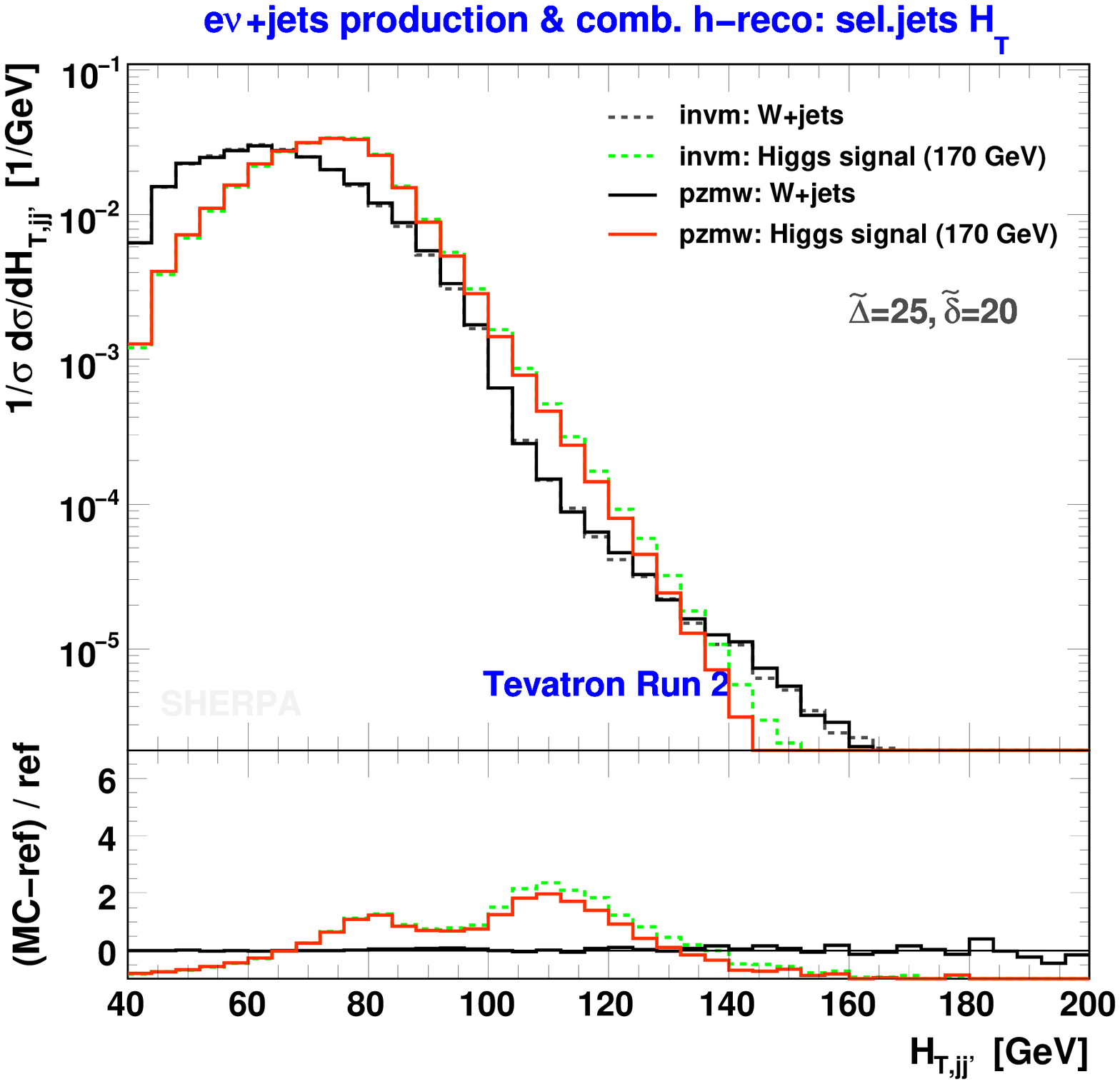}
  \caption{\label{fig:cuts.near}
    Examples of (sub)leading cut observables for Higgs boson masses
    $M_h\sim2\,M_W$. Predictions are shown for the $gg\to h$\/ and
    $W$\/ production of $e\nu_e$+jets final states using the
    \texttt{invm} (dashed) and more realistic (solid) combinatorial
    selections. Top to bottom, left panes:
    $p_{T,e^+\nu_e jj'}$ and $m_{e^-\bar\nu_e jj'}$ (\texttt{pzmh}
    leading and subleading), $H_{T,e^-\bar\nu_e jj'}$ (\texttt{mt}
    subleading); right panes: $m^{(\nu_e)}_{T,e^-\bar\nu_e jj'}$,
    $m_{e^-\bar\nu_e jj'}$ and $H_{T,jj'}$ (\texttt{pzmw} leading
    to subsubleading).}
\end{figure}

\begin{figure}[p!]
\centering\vskip0mm\hskip15pt
  \includegraphics[clip,width=0.46\columnwidth,angle=0]{%
    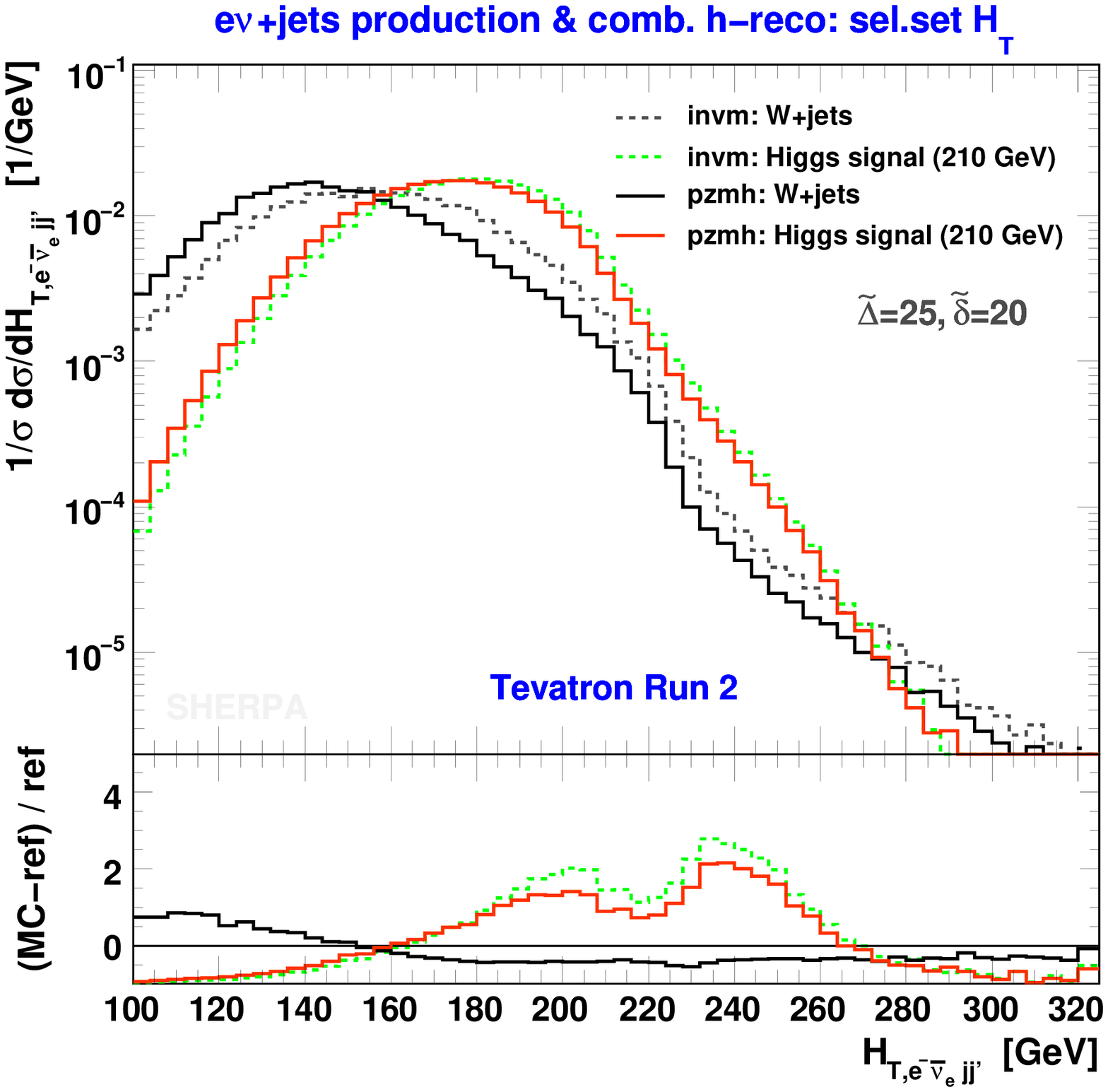}
  \includegraphics[clip,width=0.46\columnwidth,angle=0]{%
    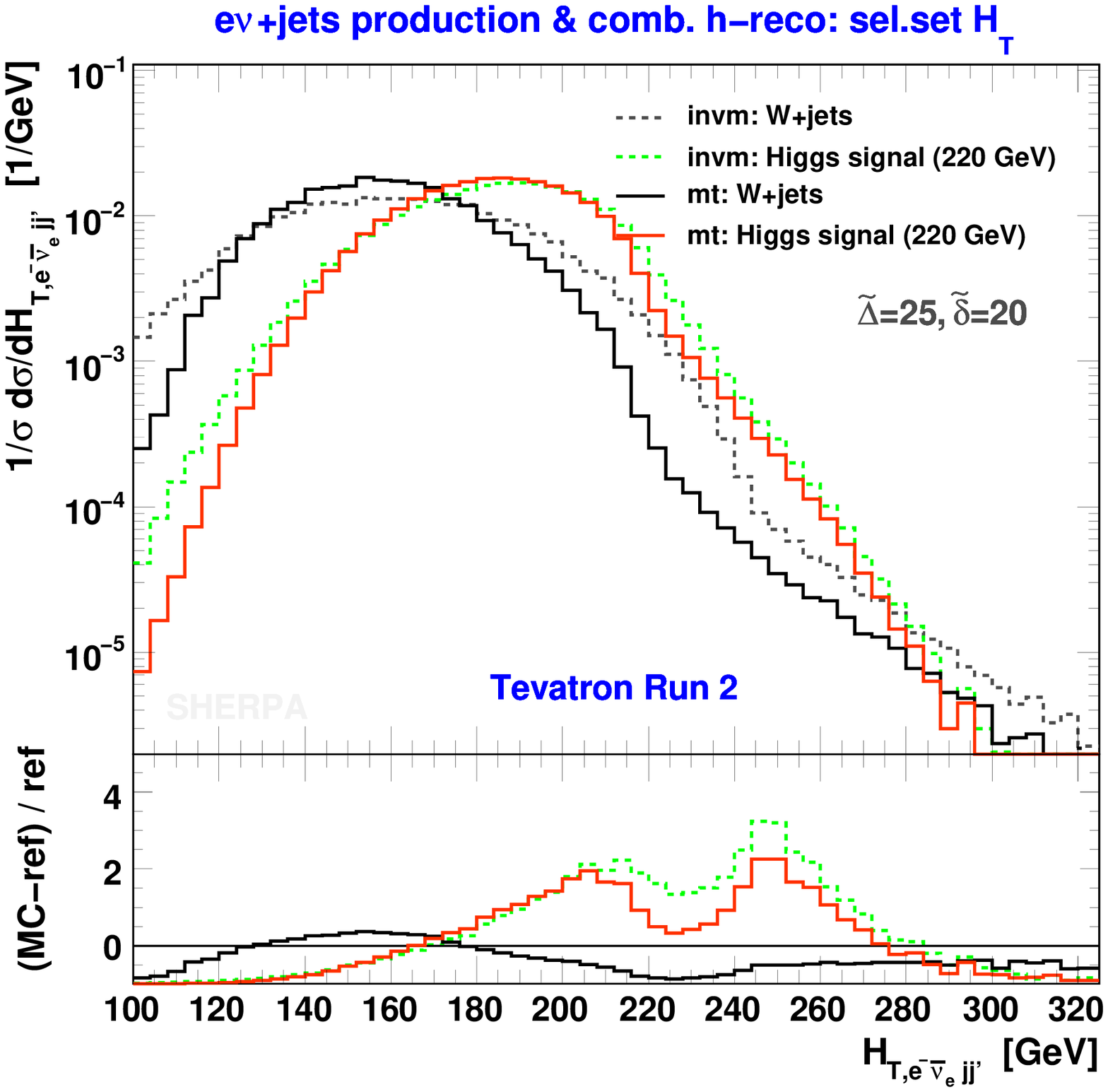}\\[-4mm]\hskip15pt
  \includegraphics[clip,width=0.46\columnwidth,angle=0]{%
    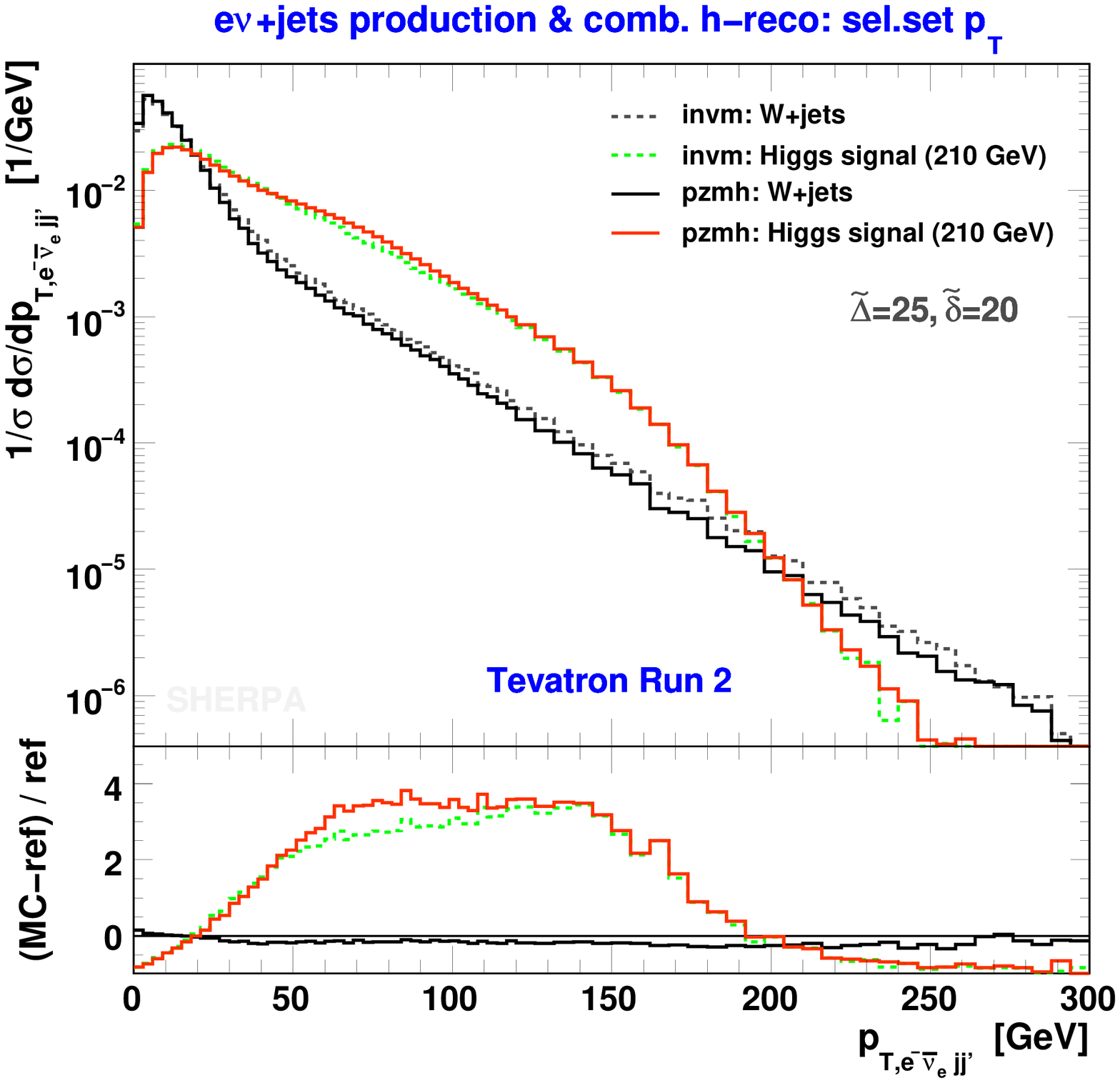}
  \includegraphics[clip,width=0.46\columnwidth,angle=0]{%
    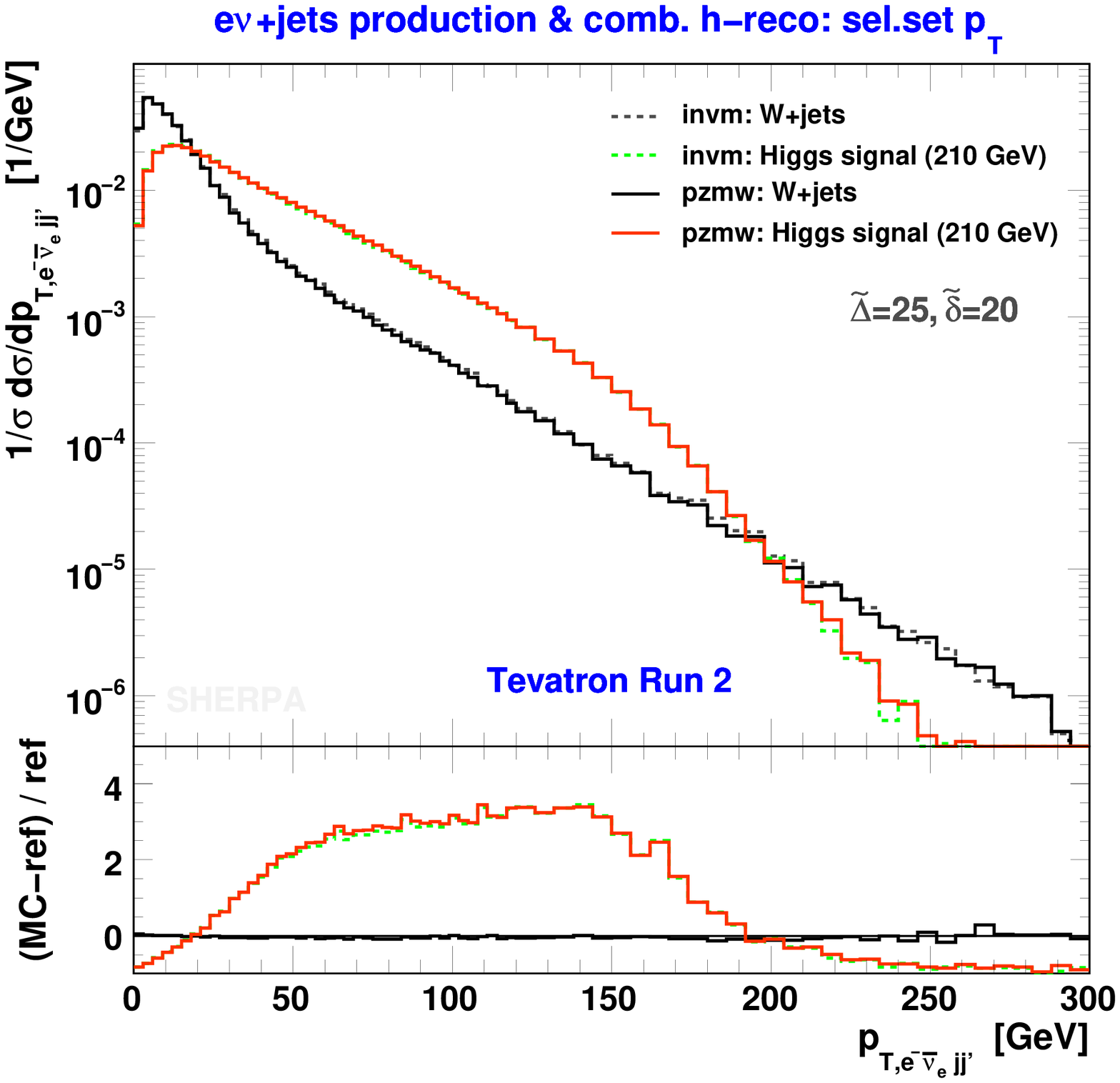}\\[-4mm]\hskip15pt
  \includegraphics[clip,width=0.46\columnwidth,angle=0]{%
    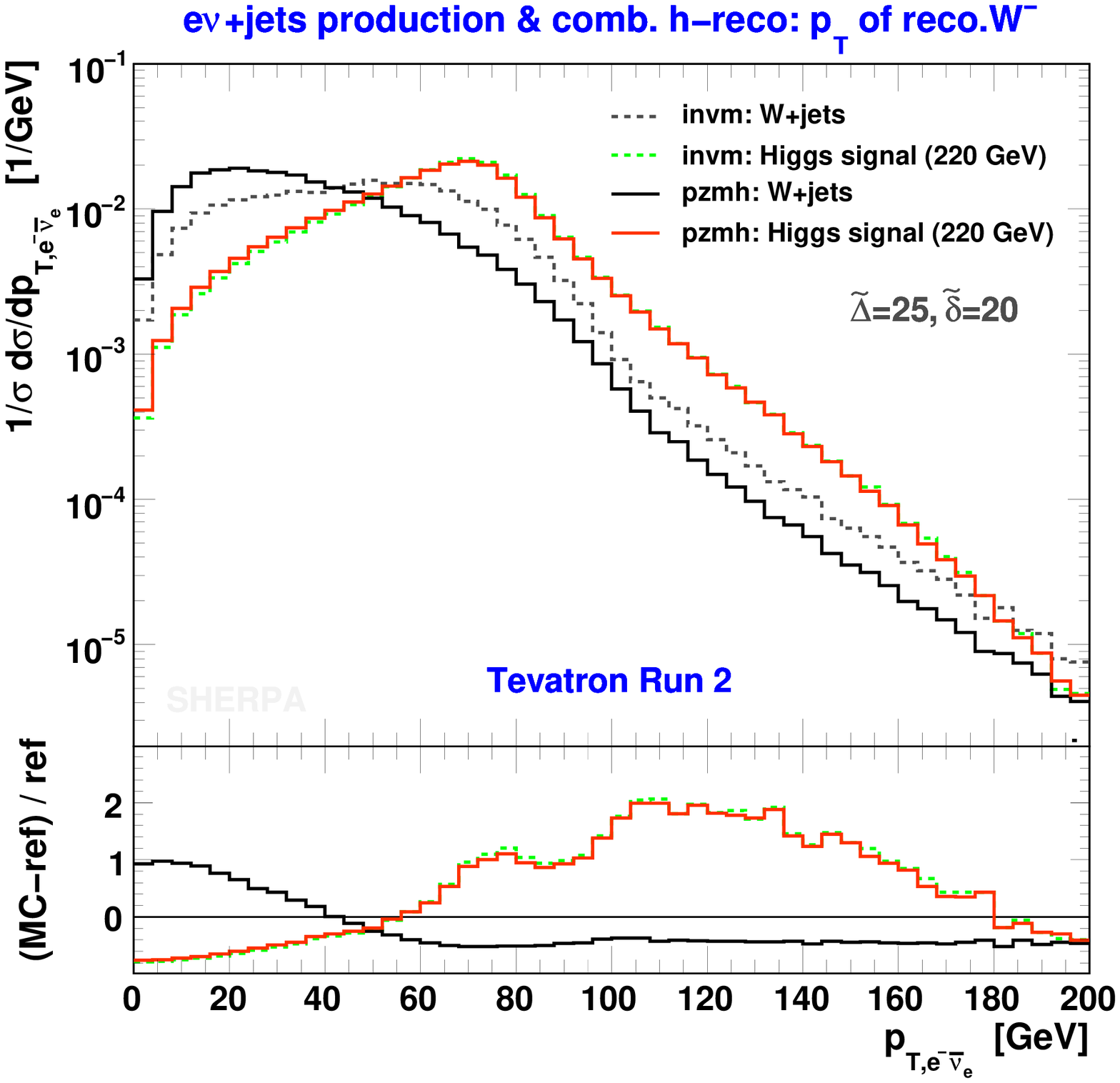}
  \includegraphics[clip,width=0.46\columnwidth,angle=0]{%
    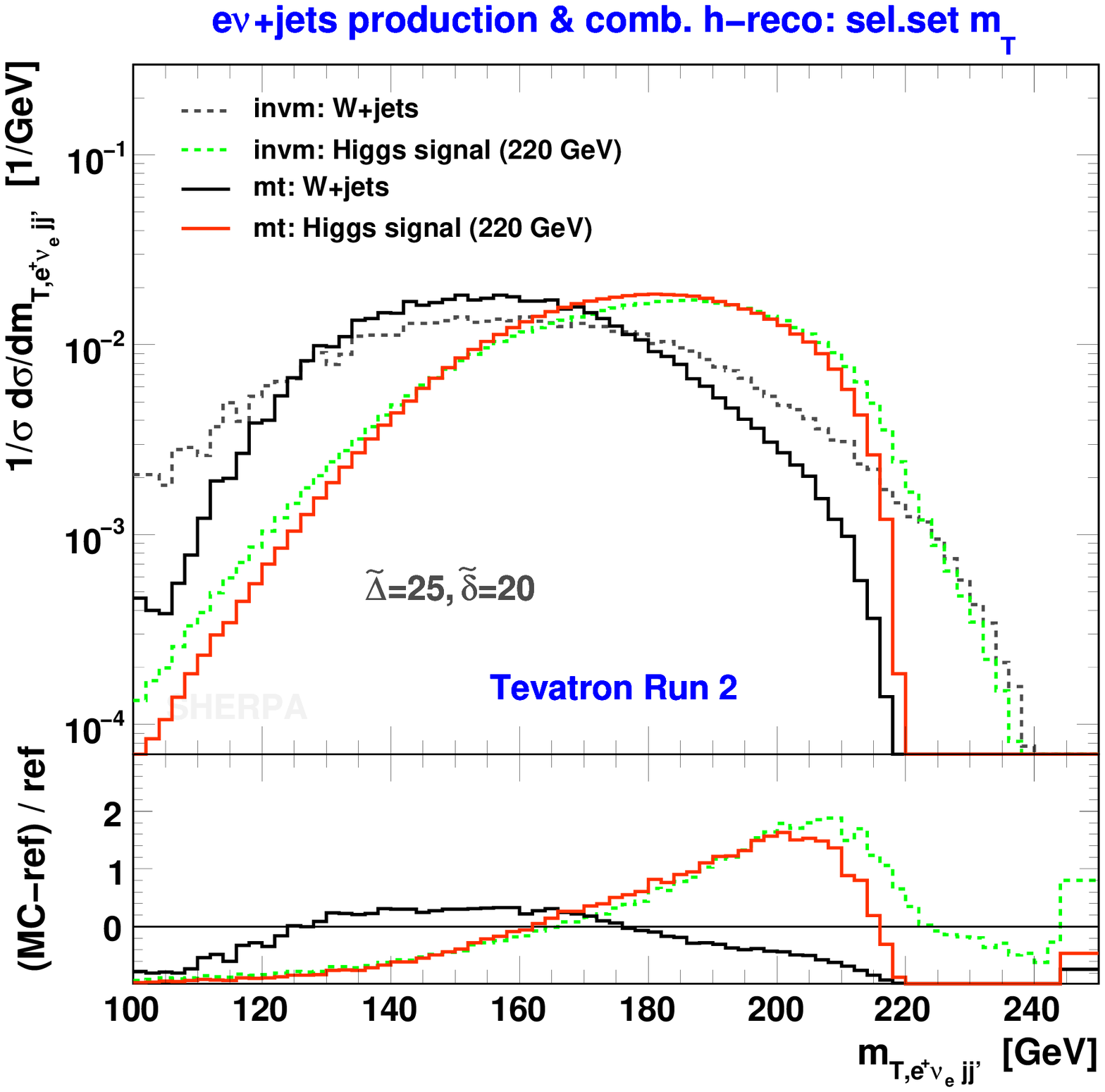}
  \caption{\label{fig:cuts.above}
    Leading and subleading cut observables for $M_h$ choices well
    above the $2\,M_W$ threshold. Predictions are shown for the
    $gg\to h$\/ and $W$\/ production of $e\nu_e$+jets final states
    using the \texttt{invm} (dashed) and more realistic (solid)
    combinatorial selections. Upper panes: $H_{T,e^-\bar\nu_e jj'}$
    (leading, \texttt{pzmh} (left) and \texttt{mt}); center panes:
    $p_{T,e^-\bar\nu_e jj'}$ (subleading, \texttt{pzmh} (left) and
    \texttt{pzmw}); lower panes (subleading), left:
    $p_{T,e^-\bar\nu_e}$ (\texttt{pzmh}), right: $m_{T,e^+\nu_e jj'}$
    (\texttt{mt}).}
\end{figure}

We collect here in this appendix additional material to substantiate
our findings presented in \Sec{sec:optimize}.
\Figs{fig:detajj}--\ref{fig:cuts.above} display a number of
distributions resulting from the baseline plus combinatorial analyses.
In each plot we show four curves, two predictions each for the signal
and the $W$\!+jets background as obtained after the ideal
(\texttt{invm}) and one of the more realistic combinatorial
selections. While we vary the Higgs boson masses and the choice of the
realistic $h$\/ reconstruction method, we keep the window parameters
of the combinatorial selections fixed at $\td\Delta=\Delta/\mrm{GeV}=25$
and $\td\delta=\delta/\mrm{GeV}=20$ over the whole set of spectra
shown in these figures.

\begin{figure}[p!]
\centering\vskip0mm\hskip15pt
  \includegraphics[clip,width=0.46\columnwidth,angle=0]{%
    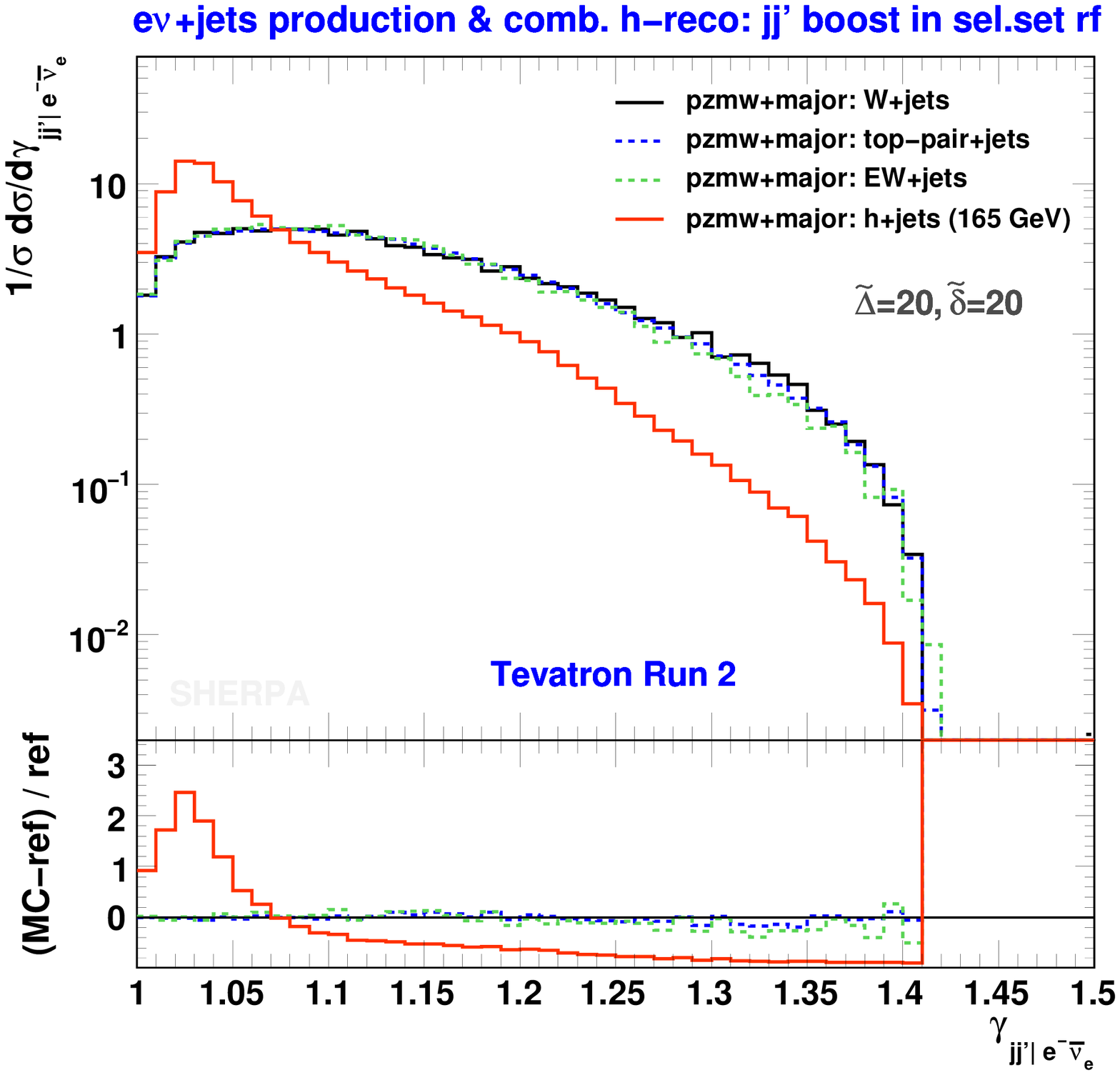}
  \includegraphics[clip,width=0.46\columnwidth,angle=0]{%
    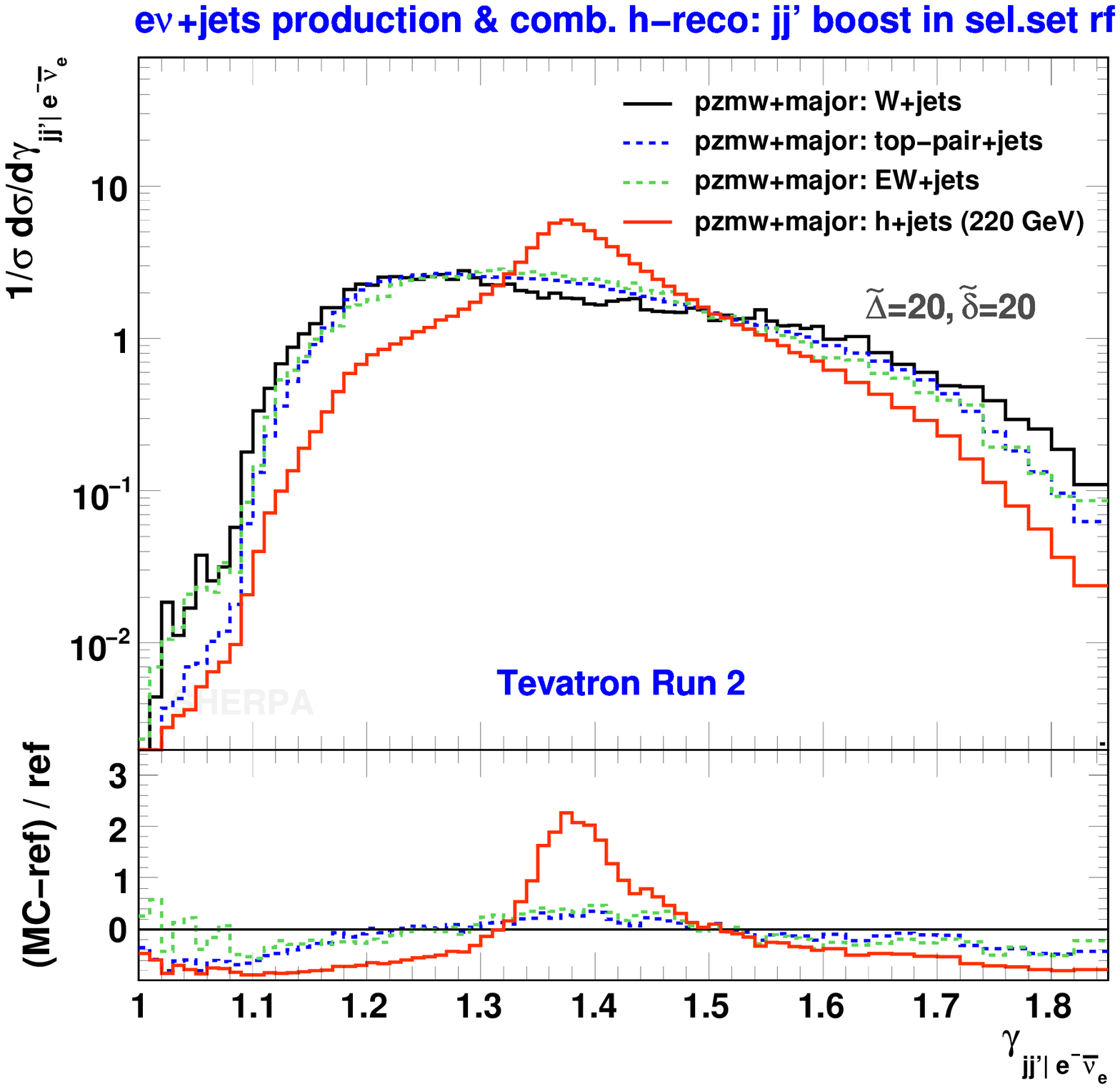}\\[-4mm]\hskip15pt
  \includegraphics[clip,width=0.46\columnwidth,angle=0]{%
    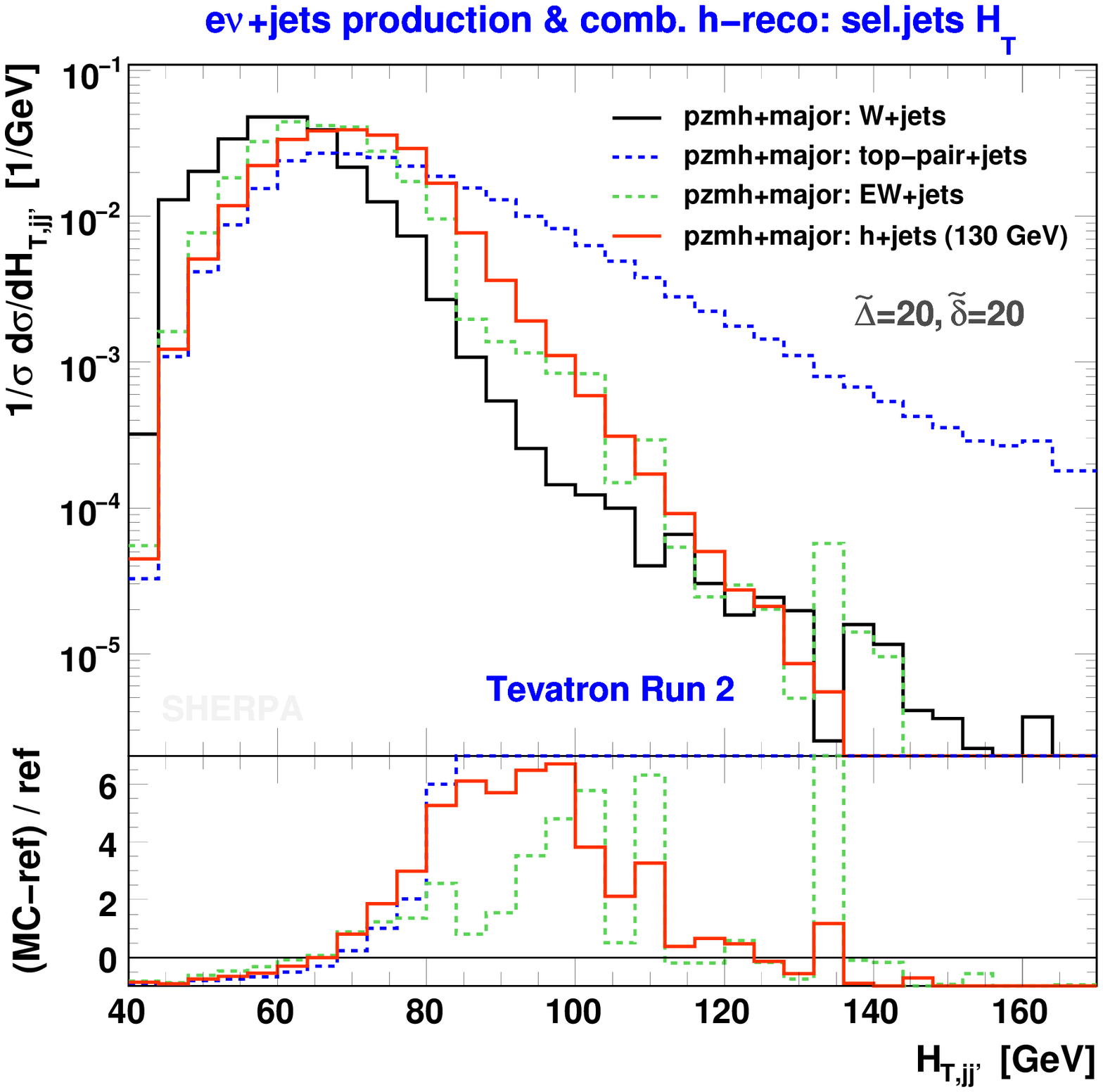}
  \includegraphics[clip,width=0.46\columnwidth,angle=0]{%
    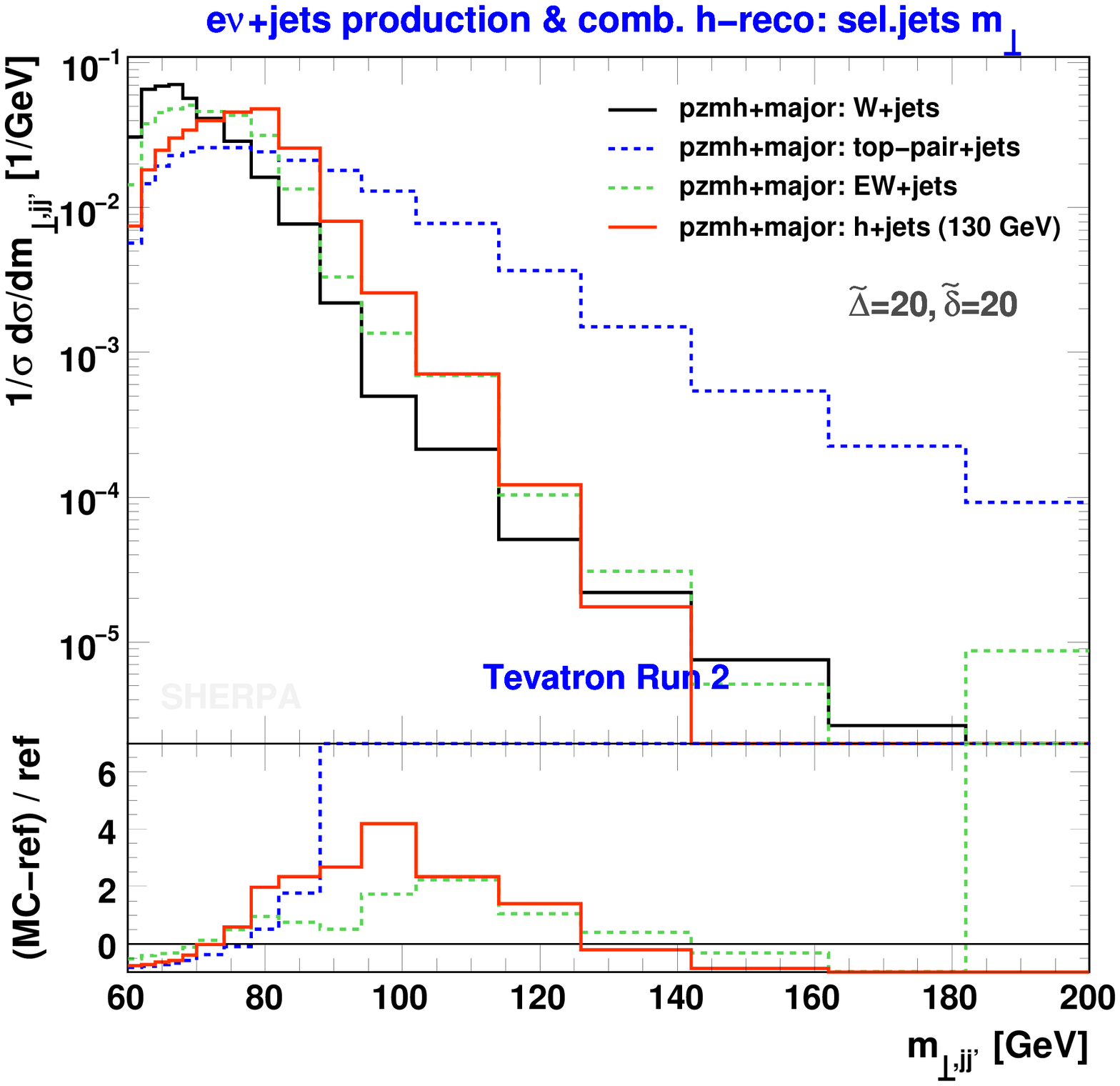}\\[-4mm]\hskip15pt
  \includegraphics[clip,width=0.46\columnwidth,angle=0]{%
    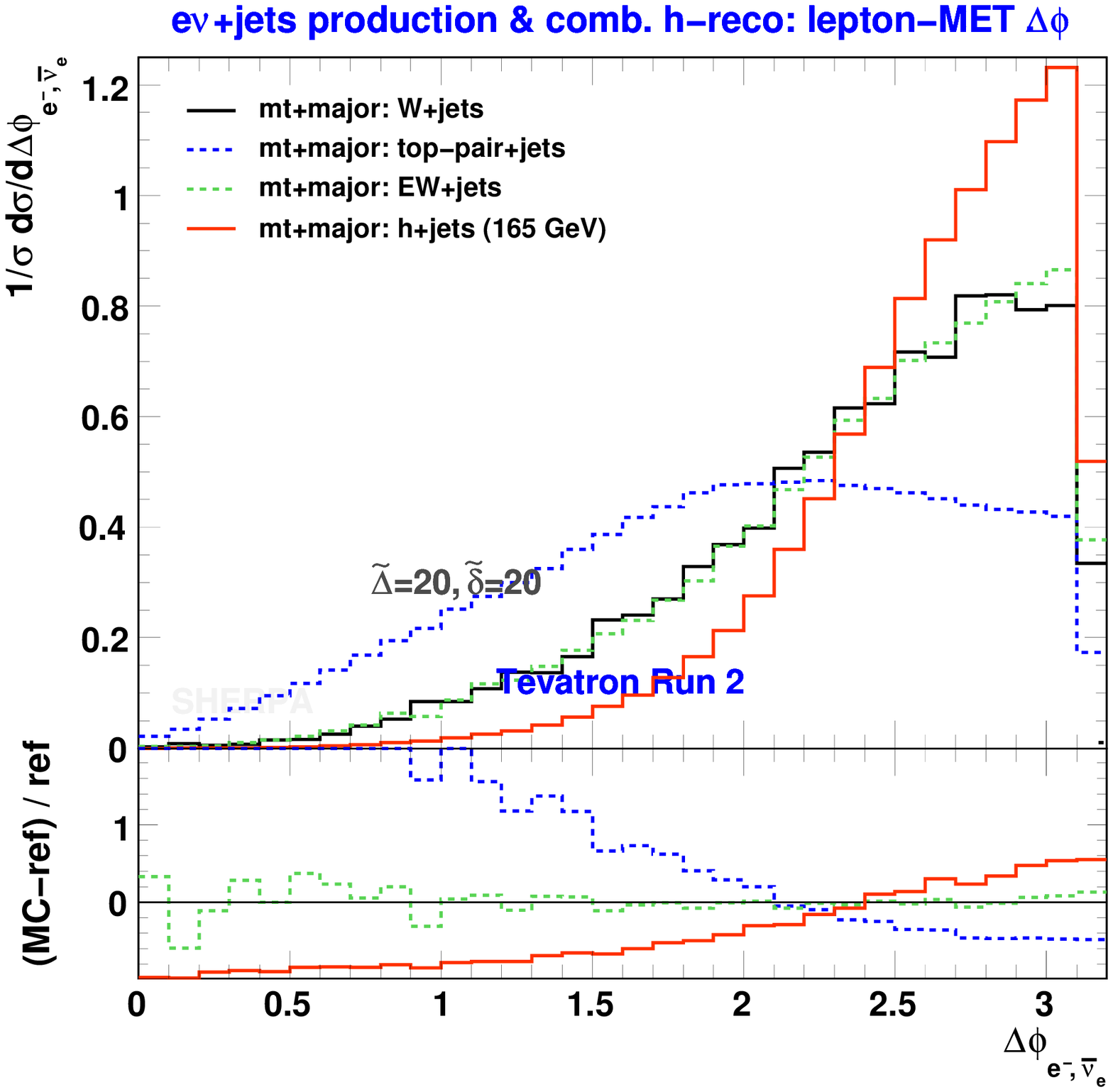}
  \includegraphics[clip,width=0.46\columnwidth,angle=0]{%
    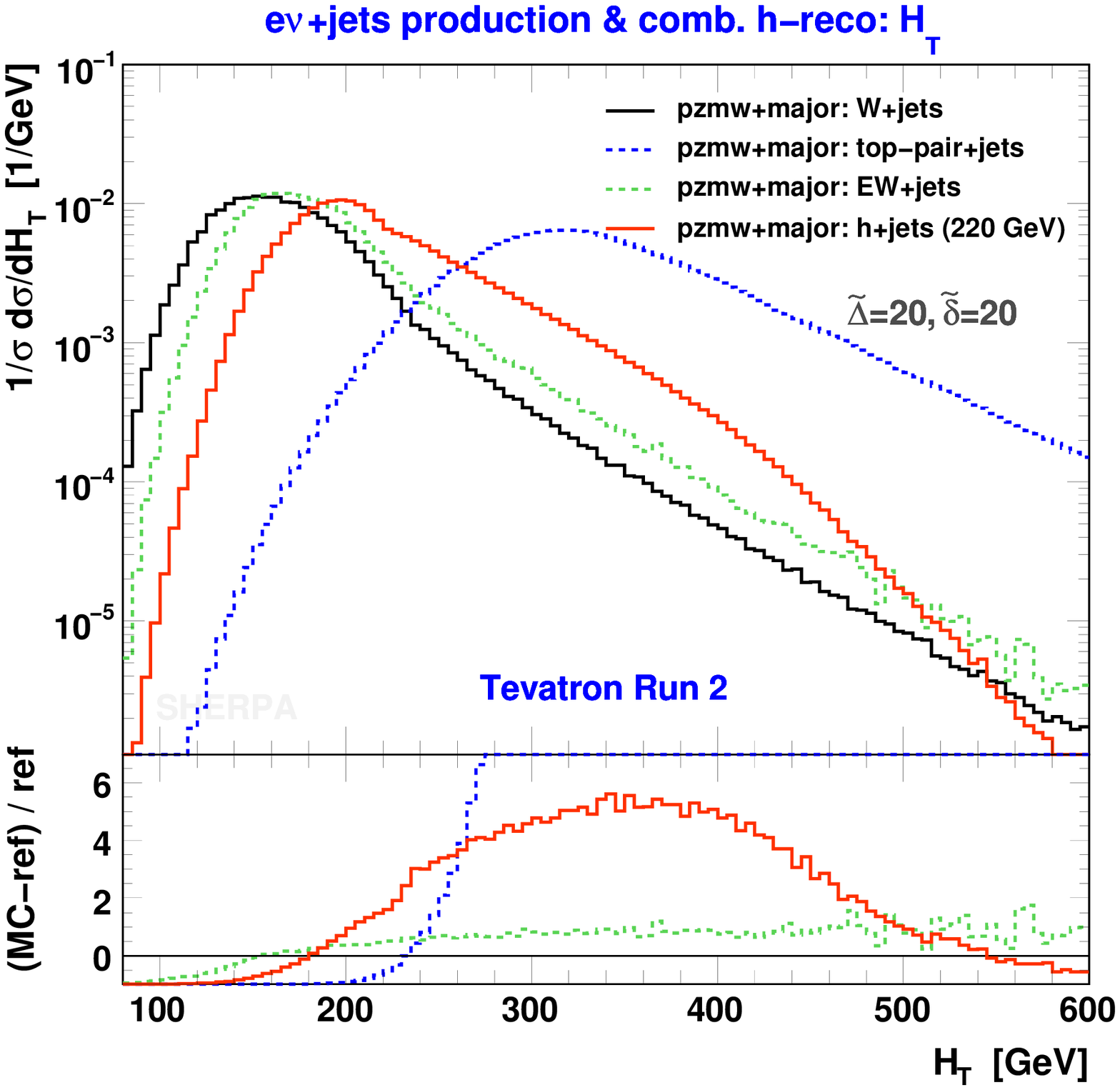}
  \caption{\label{fig:cuts.minor}
    Examples of minor cut observables after application of major cuts
    and rapidity constraints including spectra resulting from
    subdominant backgrounds. Predictions using the more realistic
    combinatorial $h$\/ candidate selections are shown for
    $e\nu_e$+jets final states arising from $gg\to h$\/ (solid red),
    $W$\!+jets (solid black), $t\bar t$\/ (dashed blue) and
    electroweak (dashed green) production at Tevatron Run II. See text
    for more details.}
\end{figure}

In \Fig{fig:cuts.minor} we look at distribution taken at the
intermediate optimization level, including the effects of the major
cuts but before the application of the respective minor cuts as listed
in \Tab{tab:optsig}. In these plots we compare the Monte Carlo
predictions for the Higgs boson signals of different $M_h$ with all,
dominant and subdominant background predictions. All these predictions
follow from using realistic selection procedures where Higgs boson and
dijet mass windows of $\td\Delta=\td\delta=20$ have been employed.

Each one-dimensional distribution is supplemented by an
one-minus-ratio subplot. The first prediction in the respective legend
is always taken as the reference, which we have arranged to be a
$W$\!+jets prediction for all plots shown here. The ratio subplots
nicely visualize why we place the cuts as given in \Tab{tab:optsig}.
Note that in all plots we only compare the shapes, \ie all
distributions are normalized to unit area.

We start by showing the $|\Delta\eta_{j,j'}|$ distributions in
\Fig{fig:detajj}. The differences in the results of the ideal and more
realistic selections are immaterial; there are essentially no
differences in the above-threshold cases. Furthermore, the shapes are
very stable under $M_h$ variations. We see that placing a cut around
$|\Delta\eta_{j,j'}|=1.5$ keeps most of the signal, while it removes a
large fraction of the $W$\!+jets events. We note it is only the
$W$\!+jets background featuring a peak location away from zero, all
other backgrounds (not shown here) behave similarly to the signal.

When working with the reconstruction methods, \texttt{pzmh/w}, we have
access to another longitudinal variable: we can include a cut on the
$h$\/ candidate's pseudo-rapidity to supplement the constraints from
the major and $|\Delta\eta_{j,j'}|$ cuts. \Fig{fig:eta.set} displays
various $\eta_{e\nu_ejj'}$ distributions. Again, all predicted shapes
are rather independent of the choice of the test mass $M_h$. The
$W$\!+jets background (as well as the electroweak background which is
not shown here) tends to preferably populate the forward rapidities
while the signal (and the $t\bar t$ contribution also not shown here)
shows up more central. This leads us to require
$|\eta_{e\nu_ejj'}|\lesssim3.0$ as pointed out in \Sec{sec:optimize}
to achieve additional significance gains. Deviations between the ideal
and more realistic reconstructions become visible; they now are
$\cal O(\mbox{25\%})$, this is clearly because one needs information
about the neutrino to form this observable. We observe that the
$W$\!+jets background receives the larger corrections compared to the
signal.

We add one more comment regarding longitudinal quantities. In this
study, as stated in \Sec{sec:optimize}, the pseudo-rapidity variables,
which we discussed above, occur largely uncorrelated with the
transverse observables as well as invariant masses. Schematically, we
illustrate this on the basis of two-dimensional $|\Delta\eta_{j,j'}|$
versus $m_{e\nu_e jj'}$ distributions for both the Higgs boson signal
and the $W$\!+jets background. In \Fig{fig:2Dpzmw} we show these
distributions as resulting from the combinatorial \texttt{pzmw}
reconstruction for two different Higgs boson masses, below
($M_h=140\ \mrm{GeV}$) and above ($M_h=180\ \mrm{GeV}$) the diboson
mass threshold. Similarly, \Fig{fig:2Dmt} exhibits the results
obtained with the \texttt{mt} selection where the $m_{e\nu_e jj'}$
quantities were reconstructed as in the ideal case. When confronted
with the respective $W$\!+jets backgrounds, we notice that the
predictions originating from the production of Higgs bosons cover
rather different parameter regions in the
$m_{e\nu_e jj'}$ -- $|\Delta\eta_{j,j'}|$ plane. This happens
independent of the value of $M_h$ and the chosen combinatorial
selection. Based on these observations, we expect the total
$S/\sqrt{B}$ increase to almost completely factorize into a product of
single $S/\sqrt{B}$ improvement factors.

In \Fig{fig:cuts.below} we depict examples of potential discriminators
below the diboson mass threshold. From the upper left to the lower
right we present, for the choice of $M_h=130\ \mrm{GeV}$, the
transverse mass $m_{T,e^-\bar\nu_e}$ of the leptonically decaying
$W^-$, the scalar $p_T$ sum $H_{T,jj'}$ of the selected jets, the
transverse masses of the selected 4-particle set,
$m^{(\nu_e)}_{T,e^+\nu_e jj'}$, and of the selected jets, $m_{T,jj'}$,
as well as their corresponding invariant masses, $m_{e^+\nu_e jj'}$
and $m_{jj'}$. The first three plots in that order are the leading cut
variables for \texttt{pzmw}, \texttt{mt} and \texttt{pzmh} followed by
the subleading cut variables for \texttt{mt} and \texttt{pzmh}, cf.\
\Tab{tab:optsig}. In the lower left plot we display the subsubleading
cut variable for the \texttt{mt} method, $m_{jj'}$, which would yield
a significance gain of 19\% if we were to demand
$m_{jj'}\ge72\ \mrm{GeV}$. Unlike the transverse observables depicted
in the upper and center panes of \Fig{fig:cuts.below}, the
invariant-mass distributions are affected by the modifications of the
ideal selection owing to a realistic neutrino treatment. The selected
4-object invariant mass (lower left) exemplifies to what degree shapes
can get distorted by the \texttt{pzmh} approach. The shoulder above
$M_h=130\ \mrm{GeV}$ emerges because complex solutions cannot be
completely avoided in the reconstruction of the neutrino momenta; the
lower tail arises from the $m_{T,e\nu_e jj'}$ constraints on the
target mass $m_{*,e\nu_e jj'}$, see \Sec{sec:realreco}.

The 2-particle transverse masses shown in the upper left and center
right of \Fig{fig:cuts.below} together with the $m_{jj'}$ spectra
document why we exploit the signal's preference for on-shell hadronic
and off-shell leptonic decays of the $W$\/ bosons. All backgrounds
considered in this study disfavor this correlation. The 4-particle
transverse mass (center left) exhibits -- as expected -- a nice
kinematic edge for the signal at $m^{(\nu_e)}_{T,e^+\nu_e jj'}=M_h$,
while all backgrounds are continuous in this variable reaching their
broad maxima above the applied mass window. The features of the
$H_{T,jj'}$ handle (upper right) have been already described in
\Sec{sec:strategy}. For this variable, the electroweak background
turns out signal-like whereas the $t\bar t$\/ background generates (by
far) the hardest tails.

We now discuss some of the near-threshold discriminators where most of
the examples are given for the test point $M_h=170\ \mrm{GeV}$.
\Fig{fig:cuts.near} shows in its top row the two leading cut variables
that we found for this mass region. In the upper right we have the
transverse mass $m^{(\nu_e)}_{T,e^-\bar\nu_e jj'}$ of the selected
$\{e,\nu_e,j,j'\}$ objects as resulting after the \texttt{pzmw}
reconstruction with $M_h=150\ \mrm{GeV}$. We discover properties very
similar to those discussed for the case of low Higgs boson masses,
cf.\ \Fig{fig:cuts.below}. In the upper left we have depicted the
first-rank discriminator for medium Higgs boson masses -- the
selected 4-object transverse momentum distribution. It was introduced
early on in \Sec{sec:strategy}. Here, we present the $p_{T,e^+\nu_e jj'}$
distribution as obtained after the combinatorial \texttt{pzmh}
selection. We remark that the curves of the other two realistic
approaches, \texttt{pzmw} and \texttt{mt}, (both not shown here)
deviate even less from the respective curves of the ideal selection.
Also not shown in \Fig{fig:cuts.near} but worthwhile to mention, the
electroweak background would yield spectra similar to those of the
$W$\!+jets background whereas the top-pair production would turn up
significantly harder than the signal's $p_T$ spectra.

The other example plots of \Fig{fig:cuts.near} are chosen from the set
of second-leading cut variables. The center panes display the
invariant mass distributions $m_{e^-\bar\nu_e jj'}$ resulting from the
\texttt{pzmh/w} selections. As opposed to the -- by construction --
sculpted shapes of the \texttt{pzmh} method, we observe that the
\texttt{pzmw} selection reproduces the invariant mass shapes of the
\texttt{invm} ideal reconstruction to a large extent. The lower panes
demonstrate the potential possessed by scalar transverse momentum
sums that we exemplify by means of the selected-set
$H_{T,e^-\bar\nu_e jj'}$ observable as given by the \texttt{mt}
selection (lower left) and the selected-jet $H_{T,jj'}$ observable
resulting from the \texttt{pzmw} reconstruction (lower right).
Remarkably, based on the 4-object $H_T$, we can achieve an even
clearer separation between the signal and the $W$\!+jets background
once we select $h$\/ candidates according to the \texttt{mt} method.

\Fig{fig:cuts.above} summarizes the types of discriminating
observables as identified in \Tab{tab:optsig} for the region of large
Higgs boson masses; we use $M_h=210$ and $220\ \mrm{GeV}$ in the
example plots. The upper panels represent $H_{T,e^-\bar\nu_e jj'}$
distributions obtained after \texttt{pzmh} (left) and \texttt{mt}
combinatorial selections. Between these two cases we detect only
marginal differences, and similarly between the predictions of the
ideal and \texttt{pzmw} reconstructions (not shown here). At large
Higgs boson masses, the signal develops the peak at considerably
larger $H_T$ values compared to the $W$\!+jets background. This
characterizes the selected-set $H_T$ as the strongest handle we have
above the $WW$\/ threshold. Moreover, the \texttt{pzmh} and
\texttt{mt} realistic selections further enhance the separation
between the two peak regions. For the subdominant backgrounds (not
shown here), we noticed a strong similarity between the $H_T$ spectra
arising from the electroweak production and the $W$\!+jets background.
The $t\bar t$\/ background however yields the hardest spectra both in
terms of the peak position as well as the tail of the $H_T$
distributions.

The center panes of \Fig{fig:cuts.above} show two examples of selected
$h$\/ candidate $p_T$ distributions. These variables do not constitute
the best discriminators anymore, but still quite often rank second
best in separating signal from $W$\!+jets production in the domain of
large $M_h$. As the \texttt{pzmw} reconstruction works extremely well
for heavy Higgs boson decays into on-shell $W$\/ bosons, it gives
$p_{T,e^-\bar\nu_e jj'}$ shapes almost identical with the ideal
selection.

For the $M_h=220\ \mrm{GeV}$ point, we present two different
subleading cut variables in the lower pane of \Fig{fig:cuts.above}. To
the left, one finds the $p_{T,e^-\bar\nu_e}$ distribution of the
reconstructed $W^-$ resulting after selecting Higgs boson candidates
according to \texttt{pzmh}. The purely transverse mass
$m_{T,e^+\nu_e jj'}$ of the \texttt{mt}-selected 4-object set, cf.\
\Eq{eq:mt}, is depicted on the lower right. For these variables, we
recognize similar features as for the $H_{T,e\nu_e jj'}$ spectra
concerning the minor backgrounds and the comparison with the
\texttt{invm} Higgs boson candidate reconstruction.

In \Fig{fig:cuts.minor} we give a brief overview of discriminators
that lead to a further increase in significance after implementing all
major and rapidity constraints discussed in \Sec{sec:optimize}. The
boost factor $\gamma_{jj'|e\nu_e}$, as introduced in \Sec{sec:strategy},
proves very helpful in separating signal from backgrounds over a large
range of above-threshold Higgs boson masses. Since
$\gamma_{jj'|e\nu_e}$ develops a peak for the signal only, it is
advantageous to isolate the boost factor peak region in order to
exploit that a not too broad scalar resonance has been produced over a
multitude of continuous backgrounds. This is exemplified in the upper
plots of \Fig{fig:cuts.minor} where we present two boost factor
$\gamma_{jj'|e^-\bar\nu_e}$ example distributions for
$M_h=165\ \mrm{GeV}$ (left) and $M_h=220\ \mrm{GeV}$ when selecting
via the \texttt{pzmw} method.

For the \texttt{pzmh} method in particular, we identified two
transverse variables that, if constrained from below, are very
yielding in the low Higgs boson mass region, even at this more
involved level of the analysis. We exhibit these variables,
$H_{T,jj'}$ on the left and $m_{\perp,jj'}$ on the right, in the
center panes of \Fig{fig:cuts.minor} choosing $M_h=130\ \mrm{GeV}$.
As in similar cases the significance gain originates from exploiting
the different peak locations associated with the signal, at higher
values, and the dominant background, preferring the low values.

For $M_h$ very close above threshold, we can use the azimuthal angle
between the lepton and MET or the two selected jets (as suggested by
Han and Zhang in Ref.~\cite{Han:1998ma,Han:1998sp}) to further
suppress the backgrounds. A low-$p_T$ Higgs boson produced at $WW$\/
threshold gives rise to two longitudinally moving $W$\/ bosons, which
in turn each decay into two objects oriented almost back-to-back in
the transverse plane. This is demonstrated by the example plot in the
lower left of \Fig{fig:cuts.minor} where signal and background
distributions are shown for the $\Delta\phi_{e^-,\bar\nu_e}$
observable when employing the \texttt{mt} selection for
$M_h=165\ \mrm{GeV}$.

Finally, we display in the lower right of \Fig{fig:cuts.minor} an
example of a global-event observable, namely $H_T$ as calculated from
the entire event, not vetoed by the selection and major cuts. It
illustrates the hierarchy of scales intrinsic to the heavy Higgs boson
signal ($M_h=220\ \mrm{GeV}$) and different background processes; it
also visualizes the leftover potential when considering the global
$H_T$ for the implementation of additional cuts, see
\App{app:addimprovs}.

\subsection{Directions for additional improvements}
\label{app:addimprovs}

The final sets of events surviving our optimized combinatorial
selections of Higgs boson candidates are perfect for use as input
to a full multivariate analysis. This is primarily because the
analysis types presented here were geared towards significance
maximization, so that a sufficiently large number of events can be
preserved.

\bi
\item\underline{$S/B$ improvements:}\quad
  since we are rather safe from a statistics point of view, there is
  in many cases potential to improve $S/B$\/ by simply requiring more
  restrictive constraints accepting (mild) significance losses at the
  same time. The simplest way is to cut harder in the tails of the
  major observables; for instance for \texttt{pzmw} at
  $M_h=220\ \mrm{GeV}$, using $H_{T,e\nu_e jj'}\ge188\ \mrm{GeV}$
  (instead of the bound given in \Tab{tab:optsig}) results in a 40\%
  gain in $S/B$ while the significance only drops by 10\%. Of course,
  a change in variable sometimes is more beneficial for maximizing
  $S/B$\/ and keeping a reasonable significance. Consider for example
  \texttt{mt} at $M_h=140\ \mrm{GeV}$; hardening the $H_{T,jj'}$
  constraint by cutting out the region below $92\ \mrm{GeV}$ maximizes
  $S/B$\/ by doubling it, but reduces the significance by a factor of
  $3.4$. In contrast, using $m_{T,jj'}\ge72\ \mrm{GeV}$ gives a factor
  $1.6$ increase in $S/B$, yet only a factor $1.2$ decrease in
  significance. Certainly one can opt for the (other) extreme and
  totally maximize $S/B$, \eg for the \texttt{pzmw} case just
  mentioned, we find a huge $S/B$\/ gain of 600\% by demanding
  $p_{T,j'}\ge64\ \mrm{GeV}$ but we actually just traded a reasonable
  significance associated with half a percent $S/B$\/ for a $\sim$ 4\%
  $S/B$\/ of very low significance diminished by a factor of $6.5$.
  This behaviour is typical owing to the limited amount of data taken
  at the Tevatron.
\item\underline{Overall $H_T$ cut:}\quad the philosophy of this study
  is to only constrain variables involving the candidate set of
  particles. Allowing cut observables sensitive to the whole event
  structure can lead to additional significance improvements but the
  related uncertainties are larger since such observables are more
  prone to hard radiative corrections that need to be described
  appropriately, often beyond parton-shower modeling. At the level of
  identifying minor cuts, as given in \Tab{tab:optsig}, we found that
  an overall $H_T$ cut on the selected events yields in most cases
  significance gains of the order of 20--40\% near and above
  threshold. This is a conservative estimate considering the larger
  uncertainties on such cuts. An example is shown in the lower right
  of \Fig{fig:cuts.minor} where we see why one benefits by
  constraining $H_T$ from below. In addition one could suppress the
  $t\bar t$\/ background by introducing an upper $H_T$ bound, or
  equally, exploit the fact that the leading jets in $t\bar t$\/
  production yield a substantially harder $H_{T,2}=H_{T,j_1j_2}$
  spectrum.
\item\underline{Asymmetric mass windows for reconstruction methods:}\quad
  as touched during the discussion of \Fig{fig:inj_significances},
  the use of asymmetric test mass windows,
  \ie $(M_h-\Delta_\mrm{low},M_h+\Delta_\mrm{up})$,
  can improve the realistic selections that either approximate or set
  the selected 4-object mass, $m_{e\nu_e jj'}$. For all $M_h$ values,
  it is advantageous to choose $\Delta_\mrm{low}$ larger by a few GeV
  than $\Delta_\mrm{up}$ where $2\,\Delta=\Delta_\mrm{low}+\Delta_\mrm{up}$.
  The results presented in \Fig{fig:sec2} (upper left) and
  \Fig{fig:mass_spectrum} clarify why this is a good idea: first, the
  Higgs boson resonance is deformed by radiative losses and resolution
  effects (there is no perfect jet finding let alone (sub)event
  reconstruction) amounting to a larger portion of cross section below
  the peak; second, the backgrounds are either sufficiently flat or --
  below the $WW$\/ threshold -- steeply falling towards smaller masses
  amplifying the asymmetry effect further. We checked the asymmetric
  window scenario for the \texttt{pzmw} method, where this led to
  significance gains up to 10\%. For the dijet mass windows, the
  effects turned out too weak.
\ei



\bibliographystyle{JHEP}
{\raggedright\bibliography{mcsim}}

\providecommand{\href}[2]{#2}\begingroup\raggedright\begin{thebibliography}{10%
0}

\bibitem{Baak:2011ze}
M.~Baak {\em et.~al.}, {\it {Updated Status of the Global Electroweak Fit and
  Constraints on New Physics}},  \href{http://xxx.lanl.gov/abs/1107.0975}{{\tt
  arXiv:1107.0975}}.

\bibitem{ATLAS-CONF-2011-112}
{\bf ATLAS} Collaboration, {\it {Combination of the Searches for the Higgs
  Boson in $\sim$ 1 fb$^{-1}$ of Data Taken with the ATLAS Detector at 7 TeV
  Center-of-Mass Energy}},  Tech. Rep. ATLAS-CONF-2011-112, CERN, Geneva,
  August, 2011.

\bibitem{CMS-PAS-HIG-11-022}
{\bf CMS} Collaboration, {\it {Combination of Higgs Searches}},  Tech. Rep.
  CMS-PAS-HIG-11-022, CERN, Geneva, 2011.

\bibitem{:2011cb}
{\bf For CDF and D\O, the TEVNPH Working Group} Collaboration, {\it {Combined
  CDF and D\O\ upper limits on Standard Model Higgs boson production with up to
  8.6 fb$^{-1}$ of data}},  \href{http://xxx.lanl.gov/abs/1107.5518}{{\tt
  arXiv:1107.5518}}. FERMILAB-CONF-11-354-E, CDF Note 10606 and D\O\ Note 6226.

\bibitem{EPS:CDFhiggs}
{\bf On behalf of the CDF} Collaboration, A.~Buzatu, {\it {Standard Model Higgs
  boson search combination at CDF}},  in {\em talk presented at the
  International Europhysics Conference on High Energy Physics, Grenoble, 21--27
  July 2011.}

\bibitem{EPS:D0higgs}
{\bf On behalf of the D\O} Collaboration, S.~Greder, {\it {Combined upper
  limits on SM Higgs}},  in {\em talk presented at the International
  Europhysics Conference on High Energy Physics, Grenoble, 21--27 July 2011.}

\bibitem{Han:1998ma}
T.~Han and R.-J. Zhang, {\it {Extending the Higgs boson reach at upgraded
  Tevatron}},  {\em Phys. Rev. Lett.} {\bf 82} (1999) 25--28,
  [\href{http://xxx.lanl.gov/abs/hep-ph/9807424}{{\tt hep-ph/9807424}}].

\bibitem{Han:1998sp}
T.~Han, A.~S. Turcot, and R.-J. Zhang, {\it {Exploiting $h\to W^*W^*$ decays at
  the upgraded Fermilab Tevatron}},  {\em Phys. Rev.} {\bf D59} (1999) 093001,
  [\href{http://xxx.lanl.gov/abs/hep-ph/9812275}{{\tt hep-ph/9812275}}].

\bibitem{Carena:2000yx}
{\bf Higgs Working Group} Collaboration, M.~S. Carena {\em et.~al.}, {\it
  {Report of the Tevatron Higgs working group}},
  \href{http://xxx.lanl.gov/abs/hep-ph/0010338}{{\tt hep-ph/0010338}}.

\bibitem{Stirling:1985bi}
W.~Stirling, R.~Kleiss, and S.~Ellis, {\it {$W^+W^-$ Pair Production in
  High-Energy Hadronic Collisions: Signal Versus Background}},  {\em Phys.
  Lett.} {\bf B163} (1985) 261.

\bibitem{Gunion:1985vt}
J.~Gunion, Z.~Kunszt, and M.~Soldate, {\it {A Background to Higgs Detection}},
  {\em Phys. Lett.} {\bf B163} (1985) 389.

\bibitem{Gunion:1985dj}
J.~Gunion, P.~Kalyniak, M.~Soldate, and P.~Galison, {\it {S}earching for the
  intermediate mass {H}iggs boson},  {\em Phys. Rev.} {\bf D34} (1986) 101.

\bibitem{Gunion:1986cc}
J.~Gunion and M.~Soldate, {\it {Overcoming a Critical Background to Higgs
  Detection}},  {\em Phys. Rev.} {\bf D34} (1986) 826.

\bibitem{Benjamin:2011sv}
{\bf For CDF and D\O, the TEVNPH Working Group} Collaboration, D.~Benjamin,
  {\it {Combined CDF and D\O\ upper limits on $gg\to H\to W^+W^-$ and
  constraints on the Higgs boson mass in fourth-generation fermion models with
  up to 8.2 fb$^{-1}$ of data}},  \href{http://xxx.lanl.gov/abs/1108.3331}{{\tt
  arXiv:1108.3331}}.

\bibitem{CDFnote10599:2011}
{\bf CDF} Collaboration, T.~Aaltonen {\em et.~al.}, {\it {Search for $H\to
  WW^*$ Production at CDF using 8.2 fb$^{-1}$ of data}},  {\em {\rm CDF Note}}
  {\bf {\rm 10599}} (2011).

\bibitem{D0Note6219:2011}
{\bf D\O} Collaboration, V.~M. Abazov {\em et.~al.}, {\it {Search for Higgs
  boson production in dilepton plus missing transverse energy final states with
  8.1 fb$^{-1}$ of $p\bar p$\/ collisions at $\sqrt{s}$ = 1.96 TeV}},  {\em
  {\rm D\O\ Note}} {\bf {\rm 6219-CONF}} (2011).

\bibitem{Aaltonen:2010yv}
{\bf CDF and D\O} Collaboration, T.~Aaltonen {\em et.~al.}, {\it {Combination
  of Tevatron searches for the Standard Model Higgs boson in the $W^+W^-$ decay
  mode}},  {\em Phys. Rev. Lett.} {\bf 104} (2010) 061802,
  [\href{http://xxx.lanl.gov/abs/1001.4162}{{\tt arXiv:1001.4162}}].

\bibitem{Aaltonen:2010cm}
{\bf CDF} Collaboration, T.~Aaltonen {\em et.~al.}, {\it {Inclusive search for
  Standard Model Higgs boson production in the $WW$ decay channel using the CDF
  II detector}},  {\em Phys. Rev. Lett.} {\bf 104} (2010) 061803,
  [\href{http://xxx.lanl.gov/abs/1001.4468}{{\tt arXiv:1001.4468}}].

\bibitem{Abazov:2010ct}
{\bf D\O} Collaboration, V.~M. Abazov {\em et.~al.}, {\it {Search for Higgs
  boson production in dilepton and missing energy final states with 5.4
  fb$^{-1}$ of $p\bar p$ collisions at $\sqrt{s}$ = 1.96 TeV}},  {\em Phys.
  Rev. Lett.} {\bf 104} (2010) 061804,
  [\href{http://xxx.lanl.gov/abs/1001.4481}{{\tt arXiv:1001.4481}}].

\bibitem{Iordanidis:1997vs}
K.~Iordanidis and D.~Zeppenfeld, {\it {Searching for a heavy Higgs boson via
  the $H\to\ $lepton neutrino jet jet mode at the CERN LHC}},  {\em Phys. Rev.}
  {\bf D57} (1998) 3072--3083,
  [\href{http://xxx.lanl.gov/abs/hep-ph/9709506}{{\tt hep-ph/9709506}}].

\bibitem{Djouadi:2005gi}
A.~Djouadi, {\it {The Anatomy of electro-weak symmetry breaking. I: The Higgs
  boson in the standard model}},  {\em Phys. Rept.} {\bf 457} (2008) 1--216,
  [\href{http://xxx.lanl.gov/abs/hep-ph/0503172}{{\tt hep-ph/0503172}}].

\bibitem{Mellado:2007fb}
B.~Mellado, W.~Quayle, and S.~L. Wu, {\it {Feasibility of Searches for a Higgs
  Boson using $H\to WW\to\ell\ell$ + MET and High PT Jets at the Tevatron}},
  {\em Phys. Rev.} {\bf D76} (2007) 093007,
  [\href{http://xxx.lanl.gov/abs/0708.2507}{{\tt arXiv:0708.2507}}].

\bibitem{Dobrescu:2009zf}
B.~A. Dobrescu and J.~D. Lykken, {\it {Semileptonic decays of the standard
  Higgs boson}},  {\em JHEP} {\bf 1004} (2010) 083,
  [\href{http://xxx.lanl.gov/abs/0912.3543}{{\tt arXiv:0912.3543}}].

\bibitem{Gleisberg:2003xi}
T.~Gleisberg {\em et.~al.}, {\it {SHERPA 1.alpha, a proof-of-concept version}},
   {\em JHEP} {\bf 02} (2004) 056,
  [\href{http://xxx.lanl.gov/abs/hep-ph/0311263}{{\tt hep-ph/0311263}}].

\bibitem{Gleisberg:2008ta}
T.~Gleisberg {\em et.~al.}, {\it {Event generation with SHERPA 1.1}},  {\em
  JHEP} {\bf 02} (2009) 007, [\href{http://xxx.lanl.gov/abs/0811.4622}{{\tt
  arXiv:0811.4622}}].

\bibitem{D0note6095:2010}
{\bf D\O} Collaboration, V.~M. Abazov {\em et.~al.}, {\it {A search for the
  Standard Model Higgs boson in $H\to WW\to\ $leptons + jets in 5.4 fb$^{-1}$
  of $p\bar p$\/ collisions at $\sqrt{s}$ = 1.96 TeV}},  {\em {\rm D\O\ Note}}
  {\bf {\rm 6095-CONF}} (July, 2010).

\bibitem{Abazov:2011bc}
{\bf D\O} Collaboration, V.~M. Abazov {\em et.~al.}, {\it {Search for the
  Standard Model Higgs Boson in the $H\to WW\to\ $lepton + neutrino + $q'\bar
  q$\/ Decay Channel}},  {\em Phys. Rev. Lett.} {\bf 106} (2011) 171802,
  [\href{http://xxx.lanl.gov/abs/1101.6079}{{\tt arXiv:1101.6079}}].

\bibitem{cms:2011}
{\bf CMS} Collaboration \:\!\!\!, according to CMS Statistics Committee,
  September, 2011.

\bibitem{Krauss:2001iv}
F.~Krauss, R.~Kuhn, and G.~Soff, {\it {AMEGIC++ 1.0: A Matrix element generator
  in C++}},  {\em JHEP} {\bf 02} (2002) 044,
  [\href{http://xxx.lanl.gov/abs/hep-ph/0109036}{{\tt hep-ph/0109036}}].

\bibitem{Gleisberg:2008fv}
T.~Gleisberg and S.~H{\"o}che, {\it {Comix, a new matrix element generator}},
  {\em JHEP} {\bf 12} (2008) 039,
  [\href{http://xxx.lanl.gov/abs/0808.3674}{{\tt arXiv:0808.3674}}].

\bibitem{:2011gs}
{\bf For CDF and D\O, the TEVNPH Working Group} Collaboration, {\it {Combined
  CDF and D\O\ upper limits on Standard Model Higgs boson production with up to
  8.2 fb$^{-1}$ of data}},  \href{http://xxx.lanl.gov/abs/1103.3233}{{\tt
  arXiv:1103.3233}}. FERMILAB-CONF-11-044-E, CDF Note 10441 and D\O\ Note 6184.

\bibitem{Dittmaier:2011ti}
{\bf LHC Higgs Cross Section Working Group} Collaboration, S.~Dittmaier {\em
  et.~al.}, {\it {Handbook of LHC Higgs Cross Sections: 1. Inclusive
  Observables}},  \href{http://xxx.lanl.gov/abs/1101.0593}{{\tt
  arXiv:1101.0593}}.

\bibitem{Spira:1997dg}
M.~Spira, {\it {QCD effects in Higgs physics}},  {\em Fortsch. Phys.} {\bf 46}
  (1998) 203--284, [\href{http://xxx.lanl.gov/abs/hep-ph/9705337}{{\tt
  hep-ph/9705337}}].

\bibitem{Djouadi:1997yw}
A.~Djouadi, J.~Kalinowski, and M.~Spira, {\it {HDECAY: A program for Higgs
  boson decays in the Standard Model and its supersymmetric extension}},  {\em
  Comput. Phys. Commun.} {\bf 108} (1998) 56--74,
  [\href{http://xxx.lanl.gov/abs/hep-ph/9704448}{{\tt hep-ph/9704448}}].

\bibitem{Butterworth:2010ym}
{\bf Les Houches 2009 Tools and Monte Carlo Working Group} Collaboration, J.~M.
  Butterworth {\em et.~al.}, {\it {The Tools and Monte Carlo working group:
  Summary Report}},  \href{http://xxx.lanl.gov/abs/1003.1643}{{\tt
  arXiv:1003.1643}}.

\bibitem{Denner:2011mq}
A.~Denner, S.~Heinemeyer, I.~Puljak, D.~Rebuzzi, and M.~Spira, {\it {Standard
  Model Higgs-Boson Branching Ratios with Uncertainties}},  {\em Eur. Phys. J.}
  {\bf C71} (2011) 1753, [\href{http://xxx.lanl.gov/abs/1107.5909}{{\tt
  arXiv:1107.5909}}].

\bibitem{Amsler:2008zzb}
{\bf Particle Data Group} Collaboration, C.~Amsler {\em et.~al.}, {\it {Review
  of particle physics}},  {\em Phys. Lett.} {\bf B667} (2008) 1.

\bibitem{Anastasiou:2008tj}
C.~Anastasiou, R.~Boughezal, and F.~Petriello, {\it {Mixed QCD-electroweak
  corrections to Higgs boson production in gluon fusion}},  {\em JHEP} {\bf 04}
  (2009) 003, [\href{http://xxx.lanl.gov/abs/0811.3458}{{\tt
  arXiv:0811.3458}}].

\bibitem{Keung:2009bs}
W.-Y. Keung and F.~J. Petriello, {\it {Electroweak and finite quark-mass
  effects on the Higgs boson transverse momentum distribution}},  {\em Phys.
  Rev.} {\bf D80} (2009) 013007, [\href{http://xxx.lanl.gov/abs/0905.2775}{{\tt
  arXiv:0905.2775}}].

\bibitem{Actis:2008ug}
S.~Actis, G.~Passarino, C.~Sturm, and S.~Uccirati, {\it {NLO Electroweak
  Corrections to Higgs Boson Production at Hadron Colliders}},  {\em Phys.
  Lett.} {\bf B670} (2008) 12--17,
  [\href{http://xxx.lanl.gov/abs/0809.1301}{{\tt arXiv:0809.1301}}].

\bibitem{Actis:2008ts}
S.~Actis, G.~Passarino, C.~Sturm, and S.~Uccirati, {\it {NNLO Computational
  Techniques: The Cases $H\to\gamma\gamma$ and $H\to gg$}},  {\em Nucl. Phys.}
  {\bf B811} (2009) 182--273, [\href{http://xxx.lanl.gov/abs/0809.3667}{{\tt
  arXiv:0809.3667}}].

\bibitem{deFlorian:2009hc}
D.~de~Florian and M.~Grazzini, {\it {Higgs production through gluon fusion:
  updated cross sections at the Tevatron and the LHC}},  {\em Phys. Lett.} {\bf
  B674} (2009) 291--294, [\href{http://xxx.lanl.gov/abs/0901.2427}{{\tt
  arXiv:0901.2427}}].

\bibitem{Catani:2003zt}
S.~Catani, D.~de~Florian, M.~Grazzini, and P.~Nason, {\it {Soft gluon
  resummation for Higgs boson production at hadron colliders}},  {\em JHEP}
  {\bf 0307} (2003) 028, [\href{http://xxx.lanl.gov/abs/hep-ph/0306211}{{\tt
  hep-ph/0306211}}].

\bibitem{Moch:2005ky}
S.~Moch and A.~Vogt, {\it {Higher-order soft corrections to lepton pair and
  Higgs boson production}},  {\em Phys. Lett.} {\bf B631} (2005) 48--57,
  [\href{http://xxx.lanl.gov/abs/hep-ph/0508265}{{\tt hep-ph/0508265}}].

\bibitem{Martin:2009iq}
A.~D. Martin, W.~J. Stirling, R.~S. Thorne, and G.~Watt, {\it {Parton
  distributions for the LHC}},  {\em Eur. Phys. J.} {\bf C63} (2009) 189--285,
  [\href{http://xxx.lanl.gov/abs/0901.0002}{{\tt arXiv:0901.0002}}].

\bibitem{:2010ar}
{\bf For CDF and D\O, the TEVNPH Working Group} Collaboration, {\it {Combined
  CDF and D\O\ upper limits on Standard Model Higgs boson production with up to
  6.7 fb$^{-1}$ of data}},  \href{http://xxx.lanl.gov/abs/1007.4587}{{\tt
  arXiv:1007.4587}}. FERMILAB-CONF-10-257-E, CDF Note 10241 and D\O\ Note 6096.

\bibitem{Baglio:2010um}
J.~Baglio and A.~Djouadi, {\it {Predictions for Higgs production at the
  Tevatron and the associated uncertainties}},  {\em JHEP} {\bf 1010} (2010)
  064, [\href{http://xxx.lanl.gov/abs/1003.4266}{{\tt arXiv:1003.4266}}]. [{\tt
  Addendum~arXiv:1009.1363}].

\bibitem{Baglio:2011wn}
J.~Baglio, A.~Djouadi, S.~Ferrag, and R.~Godbole, {\it {The Tevatron Higgs
  exclusion limits and theoretical uncertainties: A Critical appraisal}},  {\em
  Phys. Lett.} {\bf B699} (2011) 368--371,
  [\href{http://xxx.lanl.gov/abs/1101.1832}{{\tt arXiv:1101.1832}}].

\bibitem{Baglio:2011hc}
J.~Baglio, A.~Djouadi, and R.~Godbole, {\it {Clarifications on the impact of
  theoretical uncertainties on the Tevatron Higgs exclusion limits}},
  \href{http://xxx.lanl.gov/abs/1107.0281}{{\tt arXiv:1107.0281}}.

\bibitem{Ahrens:2008qu}
V.~Ahrens, T.~Becher, M.~Neubert, and L.~L. Yang, {\it {Origin of the Large
  Perturbative Corrections to Higgs Production at Hadron Colliders}},  {\em
  Phys. Rev.} {\bf D79} (2009) 033013,
  [\href{http://xxx.lanl.gov/abs/0808.3008}{{\tt arXiv:0808.3008}}].

\bibitem{Ahrens:2008nc}
V.~Ahrens, T.~Becher, M.~Neubert, and L.~L. Yang, {\it {Renormalization-Group
  Improved Prediction for Higgs Production at Hadron Colliders}},  {\em Eur.
  Phys. J.} {\bf C62} (2009) 333--353,
  [\href{http://xxx.lanl.gov/abs/0809.4283}{{\tt arXiv:0809.4283}}].

\bibitem{Ahrens:2010rs}
V.~Ahrens, T.~Becher, M.~Neubert, and L.~L. Yang, {\it {Updated Predictions for
  Higgs Production at the Tevatron and the LHC}},  {\em Phys. Lett.} {\bf B698}
  (2011) 271--274, [\href{http://xxx.lanl.gov/abs/1008.3162}{{\tt
  arXiv:1008.3162}}].

\bibitem{Anastasiou:2004xq}
C.~Anastasiou, K.~Melnikov, and F.~Petriello, {\it {Higgs boson production at
  hadron colliders: Differential cross sections through next-to-next-to-leading
  order}},  {\em Phys. Rev. Lett.} {\bf 93} (2004) 262002,
  [\href{http://xxx.lanl.gov/abs/hep-ph/0409088}{{\tt hep-ph/0409088}}].

\bibitem{Anastasiou:2005qj}
C.~Anastasiou, K.~Melnikov, and F.~Petriello, {\it {Fully differential Higgs
  boson production and the di-photon signal through next-to-next-to-leading
  order}},  {\em Nucl. Phys.} {\bf B724} (2005) 197--246,
  [\href{http://xxx.lanl.gov/abs/hep-ph/0501130}{{\tt hep-ph/0501130}}].

\bibitem{Petriello.priv:2010}
F.~J. Petriello \:\!\!\!, private communication, July, 2010.

\bibitem{Nadolsky:2008zw}
P.~M. Nadolsky {\em et.~al.}, {\it {Implications of CTEQ global analysis for
  collider observables}},  {\em Phys. Rev.} {\bf D78} (2008) 013004,
  [\href{http://xxx.lanl.gov/abs/0802.0007}{{\tt arXiv:0802.0007}}].

\bibitem{Dixon:2003yb}
L.~J. Dixon and M.~Siu, {\it {Resonance continuum interference in the diphoton
  Higgs signal at the LHC}},  {\em Phys. Rev. Lett.} {\bf 90} (2003) 252001,
  [\href{http://xxx.lanl.gov/abs/hep-ph/0302233}{{\tt hep-ph/0302233}}].

\bibitem{Binoth:2005ua}
T.~Binoth, M.~Ciccolini, N.~Kauer, and M.~Kramer, {\it {Gluon-induced WW
  background to Higgs boson searches at the LHC}},  {\em JHEP} {\bf 0503}
  (2005) 065, [\href{http://xxx.lanl.gov/abs/hep-ph/0503094}{{\tt
  hep-ph/0503094}}].

\bibitem{Binoth:2006mf}
T.~Binoth, M.~Ciccolini, N.~Kauer, and M.~Kramer, {\it {Gluon-induced W-boson
  pair production at the LHC}},  {\em JHEP} {\bf 0612} (2006) 046,
  [\href{http://xxx.lanl.gov/abs/hep-ph/0611170}{{\tt hep-ph/0611170}}].

\bibitem{Campbell:2011cu}
J.~M. Campbell, R.~Ellis, and C.~Williams, {\it {Gluon-gluon contributions to
  W+ W- production and Higgs interference effects}},  {\em JHEP} {\bf 1110}
  (2011) 005, [\href{http://xxx.lanl.gov/abs/1107.5569}{{\tt
  arXiv:1107.5569}}].

\bibitem{Campbell:Ellis}
J.~M. Campbell and R.~K. Ellis, {\em {MCFM -- Monte Carlo for FeMtobarn
  processes}}.
\newblock http://mcfm.fnal.gov.

\bibitem{Campbell:2002tg}
J.~M. Campbell and R.~K. Ellis, {\it {Next-to-leading order corrections to $W$
  + 2-jet and $Z$ + 2-jet production at hadron colliders}},  {\em Phys. Rev.}
  {\bf D65} (2002) 113007, [\href{http://xxx.lanl.gov/abs/hep-ph/0202176}{{\tt
  hep-ph/0202176}}].

\bibitem{Campbell:2003hd}
J.~M. Campbell, R.~K. Ellis, and D.~L. Rainwater, {\it {Next-to-leading order
  QCD predictions for $W$ + 2-jet and $Z$ + 2jet production at the CERN LHC}},
  {\em Phys. Rev.} {\bf D68} (2003) 094021,
  [\href{http://xxx.lanl.gov/abs/hep-ph/0308195}{{\tt hep-ph/0308195}}].

\bibitem{Catani:2007vq}
S.~Catani and M.~Grazzini, {\it {An NNLO subtraction formalism in hadron
  collisions and its application to Higgs boson production at the LHC}},  {\em
  Phys. Rev. Lett.} {\bf 98} (2007) 222002,
  [\href{http://xxx.lanl.gov/abs/hep-ph/0703012}{{\tt hep-ph/0703012}}].

\bibitem{Grazzini:2008tf}
M.~Grazzini, {\it {NNLO predictions for the Higgs boson signal in the $H\to WW
  \to\ell\nu\ell\nu$ and $H\to ZZ\to\ $4$\ell$ decay channels}},  {\em JHEP}
  {\bf 02} (2008) 043, [\href{http://xxx.lanl.gov/abs/0801.3232}{{\tt
  arXiv:0801.3232}}].

\bibitem{Catani:2001cc}
S.~Catani, F.~Krauss, R.~Kuhn, and B.~R. Webber, {\it {QCD Matrix Elements +
  Parton Showers}},  {\em JHEP} {\bf 11} (2001) 063,
  [\href{http://xxx.lanl.gov/abs/hep-ph/0109231}{{\tt hep-ph/0109231}}].

\bibitem{Krauss:2002up}
F.~Krauss, {\it {Matrix elements and parton showers in hadronic interactions}},
   {\em JHEP} {\bf 08} (2002) 015,
  [\href{http://xxx.lanl.gov/abs/hep-ph/0205283}{{\tt hep-ph/0205283}}].

\bibitem{Krauss:2004bs}
F.~Krauss, A.~Sch{\"a}licke, S.~Schumann, and G.~Soff, {\it {Simulating $W/Z$ +
  jets production at the Tevatron}},  {\em Phys. Rev.} {\bf D70} (2004) 114009,
  [\href{http://xxx.lanl.gov/abs/hep-ph/0409106}{{\tt hep-ph/0409106}}].

\bibitem{Krauss:2005nu}
F.~Krauss, A.~Sch{\"a}licke, S.~Schumann, and G.~Soff, {\it {Simulating $W/Z$ +
  jets production at the CERN LHC}},  {\em Phys. Rev.} {\bf D72} (2005) 054017,
  [\href{http://xxx.lanl.gov/abs/hep-ph/0503280}{{\tt hep-ph/0503280}}].

\bibitem{Alwall:2007fs}
J.~Alwall {\em et.~al.}, {\it {Comparative study of various algorithms for the
  merging of parton showers and matrix elements in hadronic collisions}},  {\em
  Eur. Phys. J.} {\bf C53} (2008) 473--500,
  [\href{http://xxx.lanl.gov/abs/0706.2569}{{\tt arXiv:0706.2569}}].

\bibitem{Winter:2007zza}
J.~Winter, {\em {QCD jet evolution at high and low scales}}.
\newblock PhD thesis, Technische Universit{\"a}t Dresden, 2008.

\bibitem{Lenzi:2009fi}
P.~Lenzi and J.~M. Butterworth, {\it {A study on Matrix Element corrections in
  inclusive $Z/\gamma^\ast$ production at LHC as implemented in PYTHIA, HERWIG,
  ALPGEN and SHERPA}},  \href{http://xxx.lanl.gov/abs/0903.3918}{{\tt
  arXiv:0903.3918}}.

\bibitem{Binoth:2010ra}
{\bf Les Houches 2009 SM and NLO Multileg Working Group} Collaboration, J.~R.
  Andersen {\em et.~al.}, {\it {The SM and NLO multileg working group: Summary
  report}},  \href{http://xxx.lanl.gov/abs/1003.1241}{{\tt arXiv:1003.1241}}.

\bibitem{D0note5066:2006}
{\bf D\O} Collaboration, V.~M. Abazov {\em et.~al.}, {\it {$Z$+jet production
  in the D\O\ experiment: A comparison between data and the PYTHIA and SHERPA
  Monte Carlos}},  {\em {\rm D\O\ Note}} {\bf {\rm 5066-CONF}} (March, 2006).

\bibitem{Abazov:2008ez}
{\bf D\O} Collaboration, V.~M. Abazov {\em et.~al.}, {\it {Measurement of
  differential $Z/\gamma^{*}$ + jet + X cross sections in $p\bar{p}$ collisions
  at $\sqrt{s}$ = 1.96 TeV}},  {\em Phys. Lett.} {\bf B669} (2008) 278--286,
  [\href{http://xxx.lanl.gov/abs/0808.1296}{{\tt arXiv:0808.1296}}].

\bibitem{Hoeche:2009rj}
S.~H{\"o}che, F.~Krauss, S.~Schumann, and F.~Siegert, {\it {QCD matrix elements
  and truncated showers}},  {\em JHEP} {\bf 05} (2009) 053,
  [\href{http://xxx.lanl.gov/abs/0903.1219}{{\tt arXiv:0903.1219}}].

\bibitem{Abazov:2009av}
{\bf D\O} Collaboration, V.~M. Abazov {\em et.~al.}, {\it {Measurements of
  differential cross sections of $Z/\gamma^\ast$ + jets + X events in proton
  anti-proton collisions at $\sqrt{s}$ = 1.96 TeV}},  {\em Phys. Lett.} {\bf
  B678} (2009) 45--54, [\href{http://xxx.lanl.gov/abs/0903.1748}{{\tt
  arXiv:0903.1748}}].

\bibitem{Abazov:2009pp}
{\bf D\O} Collaboration, V.~M. Abazov {\em et.~al.}, {\it {Measurement of
  $Z/\gamma^\ast$ + jet + X angular distributions in $p\bar{p}$ collisions at
  $\sqrt{s}$ = 1.96 TeV}},  {\em Phys. Lett.} {\bf B682} (2010) 370--380,
  [\href{http://xxx.lanl.gov/abs/0907.4286}{{\tt arXiv:0907.4286}}].

\bibitem{Abazov:2010kn}
{\bf D\O} Collaboration, V.~M. Abazov {\em et.~al.}, {\it {Measurement of the
  normalized $Z/\gamma^\ast\to\mu^+\mu^-$ transverse momentum distribution in
  $p\bar{p}$ collisions at $\sqrt{s}$ = 1.96 TeV}},
  \href{http://xxx.lanl.gov/abs/1006.0618}{{\tt arXiv:1006.0618}}.

\bibitem{Berger:2009ep}
C.~F. Berger {\em et.~al.}, {\it {Next-to-Leading Order QCD Predictions for $W$
  + 3-Jet Distributions at Hadron Colliders}},  {\em Phys. Rev.} {\bf D80}
  (2009) 074036, [\href{http://xxx.lanl.gov/abs/0907.1984}{{\tt
  arXiv:0907.1984}}].

\bibitem{Berger:2010vm}
C.~F. Berger {\em et.~al.}, {\it {Next-to-Leading Order QCD Predictions for
  $Z,\gamma^*$ + 3-Jet Distributions at the Tevatron}},
  \href{http://xxx.lanl.gov/abs/1004.1659}{{\tt arXiv:1004.1659}}.

\bibitem{Gleisberg:2005qq}
T.~Gleisberg, F.~Krauss, A.~Sch{\"a}licke, S.~Schumann, and J.~Winter, {\it
  {Studying $W^+ W^-$ production at the Fermilab Tevatron with SHERPA}},  {\em
  Phys. Rev.} {\bf D72} (2005) 034028,
  [\href{http://xxx.lanl.gov/abs/hep-ph/0504032}{{\tt hep-ph/0504032}}].

\bibitem{Hoche:2010pf}
S.~H{\"o}che, F.~Krauss, M.~Sch{\"o}nherr, and F.~Siegert, {\it {Automating the
  POWHEG method in Sherpa}},  {\em JHEP} {\bf 1104} (2011) 024,
  [\href{http://xxx.lanl.gov/abs/1008.5399}{{\tt arXiv:1008.5399}}].

\bibitem{Hoche:2010kg}
S.~H{\"o}che, F.~Krauss, M.~Sch{\"o}nherr, and F.~Siegert, {\it {NLO matrix
  elements and truncated showers}},  {\em JHEP} {\bf 1108} (2011) 123,
  [\href{http://xxx.lanl.gov/abs/1009.1127}{{\tt arXiv:1009.1127}}].

\bibitem{Aaltonen:2009vh}
{\bf CDF} Collaboration, T.~Aaltonen {\em et.~al.}, {\it {Measurement of the
  $WW$ + $WZ$ Production Cross Section Using the Lepton + Jets Final State at
  CDF II}},  {\em Phys. Rev. Lett.} {\bf 104} (2010) 101801,
  [\href{http://xxx.lanl.gov/abs/0911.4449}{{\tt arXiv:0911.4449}}].

\bibitem{Abazov:2009tr}
{\bf D\O} Collaboration, V.~Abazov {\em et.~al.}, {\it {Measurement of
  trilinear gauge boson couplings from $WW+WZ\to\ell\nu jj$ events in p anti-p
  collisions at $\sqrt{s}$ = 1.96 TeV}},  {\em Phys. Rev.} {\bf D80} (2009)
  053012, [\href{http://xxx.lanl.gov/abs/0907.4398}{{\tt arXiv:0907.4398}}].

\bibitem{Campbell:1999ah}
J.~M. Campbell and R.~K. Ellis, {\it {An update on vector boson pair production
  at hadron colliders}},  {\em Phys. Rev.} {\bf D60} (1999) 113006,
  [\href{http://xxx.lanl.gov/abs/hep-ph/9905386}{{\tt hep-ph/9905386}}].

\bibitem{Campbell:2011gp}
J.~M. Campbell, A.~Martin, and C.~Williams, {\it {NLO predictions for a lepton,
  missing transverse momentum and dijets at the Tevatron}},  {\em Phys. Rev.}
  {\bf D84} (2011) 036005, [\href{http://xxx.lanl.gov/abs/1105.4594}{{\tt
  arXiv:1105.4594}}].

\bibitem{Cacciari:2008zb}
M.~Cacciari, S.~Frixione, M.~L. Mangano, P.~Nason, and G.~Ridolfi, {\it
  {Updated predictions for the total production cross sections of top and of
  heavier quark pairs at the Tevatron and at the LHC}},  {\em JHEP} {\bf 0809}
  (2008) 127, [\href{http://xxx.lanl.gov/abs/0804.2800}{{\tt
  arXiv:0804.2800}}].

\bibitem{Blazey:2000qt}
G.~C. Blazey {\em et.~al.}, {\it {Run II jet physics}},
  \href{http://xxx.lanl.gov/abs/hep-ex/0005012}{{\tt hep-ex/0005012}}.

\bibitem{Neubauer:2011pp}
{\bf For the ATLAS} Collaboration, M.~S. Neubauer, {\it {Search for the
  Standard Model Higgs Boson in the Lepton + Missing Transverse Energy + Jets
  Final State in ATLAS}},  \href{http://xxx.lanl.gov/abs/1110.2265}{{\tt
  arXiv:1110.2265}}.

\bibitem{Ball:2007zza}
{\bf CMS} Collaboration, G.~Bayatian {\em et.~al.}, {\it {CMS technical design
  report, volume II: Physics performance}},  {\em J. Phys. G} {\bf G34} (2007)
  995--1579.

\bibitem{Yang.priv:2010}
T.~Becher and L.~L. Yang \:\!\!\!, private communication, March, 2010.

\bibitem{Hoeche:2009xc}
S.~H{\"o}che, S.~Schumann, and F.~Siegert, {\it {Hard photon production and
  matrix-element parton-shower merging}},  {\em Phys. Rev.} {\bf D81} (2010)
  034026, [\href{http://xxx.lanl.gov/abs/0912.3501}{{\tt arXiv:0912.3501}}].

\bibitem{Carli:2010cg}
T.~Carli, T.~Gehrmann, and S.~H{\"o}che, {\it {Hadronic final states in
  deep-inelastic scattering with Sherpa}},  {\em Eur. Phys. J.} {\bf C67}
  (2010) 73--97, [\href{http://xxx.lanl.gov/abs/0912.3715}{{\tt
  arXiv:0912.3715}}].

\bibitem{Hoeche:2011fd}
S.~H{\"o}che, F.~Krauss, M.~Sch{\"o}nherr, and F.~Siegert, {\it {A critical
  appraisal of NLO+PS matching methods}},
  \href{http://xxx.lanl.gov/abs/1111.1220}{{\tt arXiv:1111.1220}}.

\bibitem{Campbell:2010ff}
J.~M. Campbell and R.~Ellis, {\it {MCFM for the Tevatron and the LHC}},  {\em
  Nucl. Phys. Proc. Suppl.} {\bf 205-206} (2010) 10--15,
  [\href{http://xxx.lanl.gov/abs/1007.3492}{{\tt arXiv:1007.3492}}].

\bibitem{Berger:2008sj}
C.~Berger, Z.~Bern, L.~Dixon, F.~Febres~Cordero, D.~Forde, {\em et.~al.}, {\it
  {An Automated Implementation of On-Shell Methods for One-Loop Amplitudes}},
  {\em Phys. Rev.} {\bf D78} (2008) 036003,
  [\href{http://xxx.lanl.gov/abs/0803.4180}{{\tt arXiv:0803.4180}}].

\bibitem{Berger:2010gf}
C.~Berger, Z.~Bern, L.~J. Dixon, F.~Febres~Cordero, D.~Forde, {\em et.~al.},
  {\it {Vector Boson + Jets with BlackHat and Sherpa}},  {\em Nucl. Phys. Proc.
  Suppl.} {\bf 205-206} (2010) 92--97,
  [\href{http://xxx.lanl.gov/abs/1005.3728}{{\tt arXiv:1005.3728}}].

\bibitem{Binoth:2010xt}
T.~Binoth, F.~Boudjema, G.~Dissertori, A.~Lazopoulos, A.~Denner, {\em et.~al.},
  {\it {A Proposal for a standard interface between Monte Carlo tools and
  one-loop programs}},  {\em Comput. Phys. Commun.} {\bf 181} (2010)
  1612--1622, [\href{http://xxx.lanl.gov/abs/1001.1307}{{\tt
  arXiv:1001.1307}}]. Dedicated to the memory of, and in tribute to, Thomas
  Binoth, who led the effort to develop this proposal for Les Houches 2009.

\bibitem{Aaltonen:2007ip}
{\bf CDF} Collaboration, T.~Aaltonen {\em et.~al.}, {\it {Measurement of the
  cross section for $W$ boson production in association with jets in $p\bar p$
  collisions at $\sqrt{s}$ = 1.96 TeV}},  {\em Phys. Rev.} {\bf D77} (2008)
  011108, [\href{http://xxx.lanl.gov/abs/0711.4044}{{\tt arXiv:0711.4044}}].

\bibitem{Schalicke:2005nv}
A.~Sch{\"a}licke and F.~Krauss, {\it {Implementing the ME+PS merging
  algorithm}},  {\em JHEP} {\bf 07} (2005) 018,
  [\href{http://xxx.lanl.gov/abs/hep-ph/0503281}{{\tt hep-ph/0503281}}].

\end{thebibliography}\endgroup


\end{document}